\documentclass[a4paper,11pt]{book}
\usepackage[a4paper, total={6in, 8in}]{geometry}
\usepackage{amsmath,amssymb}
\usepackage{verbatim}
\usepackage{graphicx}
\usepackage{hyperref}
\usepackage{color}
\usepackage{footnote}
\usepackage{enumitem}

\DeclareFontFamily{OT1}{rsfs}{}
\DeclareFontShape{OT1}{rsfs}{m}{n}{ <-7> rsfs5 <7-10> rsfs7 <10->rsfs10}{} 
\DeclareMathAlphabet{\mycal}{OT1}{rsfs}{m}{n}

\newcommand{\be}[1]{ \begin{equation}\label{#1} }
\newcommand{\ee}{\end{equation}}
\newcommand{\bea}[1]{\begin{eqnarray}\label{#1} }
\newcommand{\eea}{\end{eqnarray}}

\newcommand{\<}{\langle}
\renewcommand{\>}{\rangle}

\newcommand{\gam}{\gamma_{ij}^{(1)}}

\newcommand{\eq}[2]{\begin{equation} #1 \label{#2} \end{equation}}

\newcommand{\al}{\alpha}
\newcommand{\ga}{\gamma}
\newcommand{\de}{\delta}

\newcommand{\si}{\sigma}

\DeclareMathOperator{\extdm}{d}
\newcommand{\extd}{\extdm \!}

\newcommand{\ms}[1]{\textrm{\tiny $#1$}}
\newcommand{\LO}{\ms{(0)}}
\newcommand{\FO}{\ms{(1)}}
\newcommand{\SO}{\ms{(2)}}
\newcommand{\TO}{\ms{(3)}}

\newcommand{\m}{\mu}
\newcommand{\n}{\nu}

\newcommand{\kvze}{\xi^{(0)}}

\newcommand{\kvth}{\xi^{(3)}}
\newcommand{\kvfo}{\xi^{(4)}}
\newcommand{\kvfi}{\xi^{(5)}}

\def\deltabar{{\mathchar '26\mkern -10mu\delta}}

\newcounter{rowcount}
\usepackage{etoolbox}
\makeatletter
\patchcmd{\chapter}{\if@openright\cleardoublepage\else\clearpage\fi}{}{}{}
\makeatother

\usepackage{titlesec}
\titleformat{\chapter}[display]
{\normalfont\huge\bfseries}{\chaptertitlename\ \thechapter}{20pt}{\Huge}
\titlespacing*{\chapter}{0pt}{0pt}{40pt}

\begin{document}
\clearpage\thispagestyle{empty}

\vspace{1cm}
\begin{center}
\textbf{DISSERTATION}
\end{center}
\vspace{1cm}


\begin{center}\textbf{
{\huge Classical and holographic aspects of conformal gravity in four dimensions}}
\end{center}
\vspace{0.8cm}

\begin{center}
\noindent Ausgef\"uhrt zum Zwecke der Erlangung des akademischen Grades einer Doktorin der Naturwissenschaften unter der Leitung von
\end{center}
\vspace{0.6cm}
\begin{center}
\large{\textbf{Daniel Grumiller}} \\
\textbf{E136}\\
\textbf{Institute for Theoretical Physics} 
\end{center}
\vspace{0.4cm}

\begin{center}
eingereicht an der Technischen Universit\"at Wien
\end{center}
\vspace{0.2cm}
\begin{center}
\textbf{Faculty of Physics}
\end{center}
\vspace{0.6cm}
\begin{center}
\large{\textbf{Iva Lovrekovi\'c}} \\
\textbf{e1049640}
\end{center}
\noindent\begin{center}
\begin{align}
\hspace{-3.6cm} \text{Wien, am} &  &
\nonumber
\end{align}
\end{center}
\newpage

\tableofcontents
\mainmatter
\chapter*{Acknowledgements}

I would like to thank a supervisor Daniel Grumiller that made possible my arrival and stay in Vienna and start of working on the research for the doctoral thesis, availability for discussions and guidance in the scientific research as well as financial support. I would like to thank parents for the financial support to be able to work on the PhD thesis.  R. McNees for explanations, help during the work and fruitful collaboration, S. Stricker that was always available for discussions and answering the small questions newly started PhD student encounters. F. Preis for useful discussions and successful collaboration, D. Vassilevich for guidance in the research, useful discussions and reading and commenting the drafts of my articles. I would also like to thank F. Bruenner for discussions about various physics topics, T. Zojer, for useful Skype conversations about the holographic renormalisation, J. Rosseel for discussions and being a bridge between the knowledge of professor and new PhD student, H. Afshar for discussions about canonical analysis, F. Schoeller for discussions about intrinsic physical properties, S. Prohazka and J. Salzer for discussions related to partition function, M. Gary for discussions. I would like to thank A. Tesytlin for very useful discussions about the partition function and invited talk at the Imperial College London and hospitality, M. Beccaria for useful communications about the partition function, and E. Bergshoeff for visiting Vienna to be an external referee at my thesis defence. 
At the end I would like to thank friends. 
The financial support was provided by  START project Y~435-N16 of the Austrian Science Fund (FWF) and the FWF project I~952-N16, and Forschungsstipendien 2015 of Technische Universit\"at Wien.


\thispagestyle{empty}
\clearpage

\titlespacing*{\chapter}{0pt}{30pt}{90pt}

\chapter*{Abstract} 
We formulate new boundary conditions that prove well defined variational principle and finite response functions for conformal gravity (CG). 
In the Anti--de Sitter/conformal field theory framework, gravity theory that is considered in the bulk gives information about the corresponding boundary theory. The metric is split in the holographic coordinate, used to approach the boundary, and the metric at the boundary. One can consider the quantities in the bulk  perturbing the (one dimension lower) boundary metric  in  holographic coordinate. 
The response functions to fluctuations of the boundary metric are Brown--York stress energy tensor sourced by the leading term in the expansion of the boundary metric and a  Partially Massless Response, specific for CG and  sourced by the subleading term in the expansion of the boundary metric. 
 They formulate boundary charges that define the asymptotic symmetry algebra or Lie algebra of the diffeomorphisms that preserve the boundary conditions of the theory.  We further analyse CG via canonical analysis constructing the gauge generators of the canonical charges that agree with Noether charges, while the charge associated to Weyl transformations, vanishes. 
Asymptotic symmetry algebra is determined by the leading term in the expansion of the boundary metric and for the asymptotically Minkowski,  $R\times S^2$ and the boundaries related by conformal rescaling, defines conformal algebra. 
The key role is played by the subleading term in the expansion of the metric forbidden by Einstein gravity equations of motion, however allowed in CG.  We classify the subalgebras of conformal algebra restricted by this term and use them to deduce the global solutions of CG. The largest subalgebra is five dimensional and extrapolates to plane wave (or geon) global solution of CG. 
Further, we compute the one loop partition function of CG in four and six dimensions  and supplement the theoretical computations with the computation of thermodynamical quantities and observables, black holes and Mannheim--Kazanas--Riegert solution which is the most general spherically symmetric solution of conformal gravity analogous to Schwarzschild solution of Einstein gravity. 

\thispagestyle{empty}
\newpage

\chapter{Introduction}
Conformal gravity (CG) is an effective, low energy, theory of gravity. To justify it, we first have to approach the known issues of CG and recognise the issues of Einstein gravity (EG) that are solved within the CG. The overall goal is to provide a bridge for CG 
that one can use as a map from the string theory, more correctly from limiting case of string theory to quantum field theories. In general, that framework is known and it is called Anti--de Sitter/conformal field theory (AdS/CFT) correspondence. 

 It is important to notice, before proceeding further, that based on the current knowledge of EG vs. CG, much more is known about EG since it is used for the description of our Universe and experimentally verified, which encourages  its further investigations. Although the experimental results and the fact it has been investigated the most turn us to investigate it further, its unresolved issues that exist for long time have turned some researchers to study other theories of gravity.  Another reason to search for the alternative theory of gravity is cosmological constant. 
Cosmological constant appears naturally in EG as the lowest order term in a derivative expansion that is compatible with all symmetries, therefore the issue is not how to add it to  EG, however to argue its small value which is $10^{-123}$ in natural units. This may motivate to consider theories where the cosmological constant is not a fixed parameter in the action.
 From observing the anisotropies of cosmic microwave background (CMB), structure formation, data from the clusters, galaxy rotation curves and others, it is evident that there should be additional matter in the Universe which is not visible, which is therefore called dark matter \cite{Bertone:2004pz}. Dark matter however has not been found yet which additionally motivates the consideration of the modified gravity theories. Big caveat is that the most modifications that would replace the dark matter encounter a conflict with the solar system precision tests or other tests of EG.

Higher derivative theories are prime candidates for the alternative theory of gravity. It was shown \cite{Utiyama:1962sn,Stelle:1976gc,Fradkin:1981iu} that with addition of the higher derivative terms to EG one can obtain renormalizable theory.  If one expands the gravitational potential in power series in the gravitation constant $\kappa$, each term will correspond to a Feymann diagram, that consists (in case of energy momentum tenosr) of a loop with matter lines within, that are responsible for the divergencies of three different types $\infty^4$, $\infty^2$ and $\log\infty$. The divergencies $\infty^2$ and $\log\infty$ are new in comparison with electrodynamics. First one can be treated by renormalisation of the gravitation constant while the second one introducing a counterterm that can be derived from the Lagrangian with quadratic Riemann tensor\footnote{The $\infty^4$ divergency is removed by introducing a term analogous to cosmological constant, refer to "cosmological term" \cite{Utiyama:1962sn}}.
That, together with the nonrenomalizability of GR \cite{Deser:1974xq} led to a conclusion that gravitational actions with quadratic curvature tensor are renormalizable \cite{Stelle:1976gc}.

The main aims of the thesis are to prove the following points.
\begin{itemize}
\item CG has well defined variational principle and finite response functions without addition of the generalised Gibbons--Hawking--York term or holographic counterterms.
 \item The Noether charges obtained from the response functions agree with the charges obtained by canonical analysis of CG, where charge associated to Weyl symmetry vanishes. The algebra defined by the canonical charges is isomporhic to the Lie algebra of the boundary conditions preserving diffeomorphisms along the boundary.
 \item The asymptotic symmetry algebra defined by the charges has a rich boundary structure. It can be classified into subalgebras defined by the asymptotic solutions.  Such largest subalgebra is five dimensional and defines global pp wave (or geon) solution. 
 \item The one loop partition function of CG around the $AdS_4$ background is not negligible, neither as the classical contribution. It consists of the partition function of EG, conformal ghost and partially masses mode, analogously to structure of partition function in three dimensions.
 \end{itemize}

While higher derivative theories of gravity are superior to EG when it comes to renormalizability properties, their main disadvantage is that generically these theories suffer from ghosts, i.e. states of negative energy. Which is the case for pure Weyl theory \cite{Hooft:1974bx}.  
Therefore, before studying CG one needs to be aware that CG solves the two loop non-renormalizability issue of EG, however it introduces one of its own, existence of ghosts. 
They have been treated from several different approaches. Pais-Uhlenbeck oscillator approach \cite{Bender:2007wu}  defines the parameter space on which negative energy states which identify ghost states, do not appear. 
 Mannheim's approach treats non-unitarity proposing non-hermiticity of the certain operators that do not affect computation of remaining observables \cite{Mannheim:2011ds}. 
Aspects that favour studyng CG as an effective theory of gravity come from the fact that within the AdS/CFT framework, CG arrises as a conuterterm from five dimensional EG \cite{Liu:1998bu,Balasubramanian:2000pq}, and from the twistor string theory \cite{Berkovits:2004jj}. It has been studied in a series of articles by t'Hooft who suggested that conformal symmetry might be the key for understanding the physics at the Planck scale \cite{Hooft:2009ms, Hooft:2010ac, Hooft:2010nc, Hooft:2014daa}. 
Phenomenologically, CG has been studied as a theory that could describe the galactic rotation curves without the addition of dark matter in series of articles by Mannheim \cite{Mannheim:1988dj, Mannheim:2006rd, Mannheim:2010xw, Mannheim:2011ds, Mannheim:2012qw}. It was also used in the cosmological model in which particular terms responsible for the inflation were defined by conformal invariance \cite{Jizba:2014taa}. \newline The thesis is structured as follows. We give brief introduction in GR in the second chapter, after which we focus on studying CG. In the third chapter we describe the holographic renormalisation procedure of the CG action and outline its main result that is a proof  that CG action in its initial form without addition of boundary counterterms of the "Gibbons-Hawking-York type" or holographic counterterms has well defined variational principle and finite response functions. The result is obtained by imposition of the suitable boundary conditions which conserve the gauge transformations and define the asymptotic symmetry algebra at the boundary. 
We continue the study of the CG via canonical analysis in the following chapter, where we find the canonical charges that agree with the Noether charges from the third chapter.
  In addition, we learn that the Weyl charge vanishes, which is analogous to three dimensional Chern-Simons action that exhibits conformal invariance. The analogy exists when the Weyl factor is kept fixed. In four dimensions, Weyl charge is for the freely varying charge vanishing, while the discrepancy with the 3D analogy vanishes. 
The fifth chapter analyses the richness of the structure defined by the asymptotic symmetry algebra obtained from the charges (in the fourth chapter) and from the boundary conditions that conserve gauge transformations (in the third chapter).
The subalgebra defined by the Schwarzschild analogue of CG solution is four dimensional $\mathbb{R}\times o(3)$ algebra, while the highest subalgerba we find is the five dimensional subalgebra that extends to global geon or pp-wave solution. 
The sixth and the final chapter consists of the computation of the one loop partition function of CG in four and six dimensions, using the heat kernel mechanism and the group theoretic approach for the evaluation of the traced heat kernel. 
\newpage

\chapter{General Relativity and AdS/CFT}

\section{Preliminaries} 

 To consider CG as an effective theory of gravity one has to introduce the mathematical framework. In this chapter we introduce the mathematical framework and the crucial concepts on the example of EG.

Assume we have a manifold $\mathcal{M}$ on which we define the metric $g_{\mu\nu}$, fields build from the metric and the derivatives of the metric. 
 The manifold that is curved is described by the curvature tensors, defined by the parallel transport. If a vector $V^{\rho}$, parallel transported (appendix: General Relativity and AdS/CFT: Paralel Transport) along the loop defined by two vectors $A^{\mu}$ and $B^{\nu}$ for the distances $\delta a$ and $\delta b$, is not the same when it comes back to its initial position, the manifold $\mathcal{M}$ is curved. The vector changes by the value $\delta V^{\rho}$ 
\begin{equation}
\delta V^{\rho}=(\delta a)(\delta b)A^{\nu}B^{\mu}R^{\rho}{}_{\sigma\nu\mu}V^{\sigma}, 
\end{equation}
 where we define the {\it curvature} or  {\it Riemann tensor } $R^{\rho}{}_{\sigma\nu\mu}$  antisymmetric in the $\mu\nu$ indices. Interchanging the vectors would mean traveling the loop in the other direction, that would give the inverse of the original answer. The Riemann tensor can be conveniently expressed with by adding "Christoffel symbols" and "covariant derivative" which we introduce below.

 Covariant derivative is a curved space generalisation of the partial derivative on the flat space. 
In flat space, partial derivative maps $(k,l)$ tensor fields into $(k,l+1)$ tensor fields, acts linearly on its arguments and obeys the Leibnitz rule. On the curved background, covariant derivate preforms this but in the coordinate independent way. 
We define covariant derivative $\nabla$ as a map from  $(k,l)$ into $(k,l+1)$ tensor fields such that it satisfies 
\begin{enumerate}
\item linearity: $\nabla(T+S)=\nabla T+ \nabla S$, and
\item Leibnitz product rule: $\nabla(T\times S)=(\nabla T)\times S+T\times (\nabla S)$
\end{enumerate}
This translates into: if the covariant derivative in the direction $\mu$ acts on the vector pointed in the direction $\nu$, it can be expressed in terms of the partial derivative and the correction 
defined by the $n\times n\times n$ quantity $\Gamma_{\mu}{}^{\rho}{}_{\sigma} $, for $n$  dimensional manifold.  That is called a Christoffel symbol, or connection coefficient \footnote{For more detailed definition of the connection coefficients see \cite{Wald:1984rg}, \cite{Carroll:1997ar}}. This reads
\begin{equation}
\nabla_{\mu}V^{\nu}=\partial_{\mu}V^{\nu}+\Gamma^{\nu}_{\mu\lambda}V^{\lambda}
\end{equation}
 and under change of coordinates transforms as
 \begin{equation}
 \Gamma^{\nu'}{}_{\mu'\lambda'}=\frac{\partial x^{\mu}}{\partial x^{\mu'}}\frac{\partial x^{\lambda}}{\partial x^{\lambda'}}\frac{\partial x^{\nu'}}{\partial x^{\nu}}\Gamma^{\nu}{}_{\mu\lambda}-\frac{\partial x^{\mu}}{\partial x^{\mu'}}\frac{\partial x^{\lambda}}{\partial x^{\lambda'}}\frac{\partial^2 x^{\nu'}}{\partial x^{\mu}\partial x^{\lambda}},
 \end{equation}
from which we see that connection is not a tensor. 
 However assuming that we have another connection $\tilde{\Gamma}^{\lambda}_{\mu\nu}$, their difference $S_{\mu\nu}{}^{\lambda}=\Gamma_{\mu\nu}^{\lambda}-\tilde{\Gamma}_{\mu\nu}^{\lambda}$ will transform as a tensor.
To any curvature connection, we can as well associate the torsion tensor 
\begin{equation}
T_{\mu\nu}^{\lambda}=\Gamma_{\mu\nu}^{\lambda}-\Gamma_{\nu\mu}^{\lambda}=2\Gamma_{[\mu\nu]}^{\lambda}
\end{equation}
which is in standard effective gravity theory taken to be zero.
Therefore, on can define the unique connection with $g_{\mu\nu}$ with two more properties 
\begin{itemize}
\item $\Gamma^{\lambda}_{\mu\nu}=\Gamma^{\lambda}_{(\mu\nu)}$ (that is torsion free)
\item $\nabla_{\rho}g_{\mu\nu}=0$ (and that is metric compatible)
\end{itemize}

Expanding the equation for the metric compatibility for the three different permutations, one can derive the expression for the connection in terms of the metric tensor and show the uniqueness and the existence of exactly one torsion-free connection on a given manifold, that is compatible with some given metric on that manifold
\begin{equation}
\Gamma^{\lambda}_{\mu\nu}=\frac{1}{2}g^{\rho\sigma}(\partial_{\mu}g_{\nu\rho}+\partial_{\nu}g_{\rho\mu}-\partial_{\rho}g_{\mu\nu}).
\end{equation}
 Using the commutation of the two covariant derivatives we can write the Riemann tensor with 
\begin{equation}
[\nabla_{\mu},\nabla_{\nu}]V^{\rho}=R^{\rho}{}_{\sigma\mu\nu}V^{\sigma}-T_{\mu\nu}{}^{\lambda}\nabla_{\lambda}V^{\rho}
\end{equation}
 where $T_{\mu\nu}{}^{\lambda}$ is a torsion, in our conventions zero, and the Riemann tensor is defined with 
 \begin{equation}
 R^{\rho}{}_{\sigma\m\n}=\partial_{\m}\Gamma^{\rho}_{\n\sigma}-\partial_{\n}\Gamma^{\rho}_{\m\sigma}+\Gamma^{\rho}_{\m\lambda}\Gamma^{\lambda}_{\n\sigma}-\Gamma^{\rho}_{\n\lambda}\Gamma^{\lambda}_{\m\sigma}.
\end{equation}
 Contracting the indices of the Riemann tensor $R^{\lambda}{}_{\mu\lambda\nu}=R_{\mu\nu}$, one obtains Ricci tensor, and contracting the indices $R^{\lambda}_{\lambda}=R$, Ricci scalar, both of which enter the definition of the Einstein tensor.   

Einstein tensor defines the Einstein field equations that govern the metric response to the energy and momentum. One could introduce it in two ways, by using the variational principle, or following Einstein's way of introducing them. To generalise the physical laws for the curved space-times, in principal, one has to use tensor fields and covariant derivative instead of the partial derivative that is used on flat spacetimes. 
There is no unique way since writting such physical laws causes ambiguities in literature dealt with via various prescriptions, as remembering to preserve gauge invariance for electromagnetism. There can be more than one way to adapt  a physical law to curved spacetimes, and right alternative can ultimately be decided by an experiment.
 We want to find an equation analogous to the Poisson equation for Newtonian potential, 
\begin{equation}
\nabla^2 \phi=4\pi G \rho,
\end{equation}
with $\rho$ mass density, and $\nabla^2$ Laplacian in flat space. The equation connects the Laplacian acting on the gravitational potential with the mass distribution, and according to prescription to obtain relativistic equation (in curved spacetime) we need relation between tensors. On the right hand side (RHS) we need energy-momentum tensor and on the left hand side (LHS), metric tensor. 

To deduce whether the covairantized laws are correct, we want to know their Newtonian limit. This is defined with the requirements that the particles are moving slowly (with the respect to the speed of light) and the gravitational field is weak and static (does not change in time). 
The equation we expect to get in the Newtonian limit would reproduce the $\nabla^2h_{00}=-8\pi GT_{00}$ for $T_{00}=\rho$, G Newton's constant and $h_{00}$ the $00$ component of a small perturbation around flat metric.
Which leads to expected Newton's potential. (see appendix: General Relativity and AdS/CFT: Newton Potential for Small Perturbation Around the Metric and \cite{Carroll:1997ar}). 
 To obtain the covariant expression, for the $\nabla^2$ we can assume to be D'Alambert  operator that acts on the metric - and in order to obtain non-vanishing result for second derivative of a tensor, instead of the metric we take Riemann tensor. That quantity should be proportional to the stress energy tensor $T_{\m\n}$. From the \textbf{Principle of Equivalence}, energy conservation \begin{align}\nabla^{\m} T_{\m\n}=0\label{encon}\end{align} in combination with the Bianchi identity \begin{equation}\nabla_{\m}R_{\m\n}=\frac{1}{2}\nabla_{\n}R\end{equation} implies that covariant derivative acting on the Ricci tensor cannot be zero, however there is a tensor constructed from the second derivatives of the metric, Ricci tensor and Ricci scalar, which obeys $\nabla^{\m}G_{\m\n}=0$. That is Einstein tensor
\begin{equation}
G_{\m\n}=R_{\m\n}-\frac{1}{2}R g_{\m\n}.
\end{equation} Its generalisation gives Einstein equations of motion (EOM) with matter
\begin{equation}
G_{\m\n}=\kappa T_{\m\n}
\end{equation} reproduces correct result in Newtonian limit and in comparison with it,  defines $\kappa=8\pi G$.

The approach of obtaining EOM, common in gravity theories, is from the variational principle.

\section{Variational Principle}
We start with an  action consisted of the integral over spacetime over Lagrange density. According to Hilbert the simplest possible choice for the Lagrangian and the action is  only independent scalar that can be constructed form the Riemann tensor, Ricci scalar  
\begin{equation}
S=\int d^nx \sqrt{-g}R,\label{egndim}
\end{equation}
where we label the spacetime dimension with {\it n}.
Varying the action we obtain boundary terms and EOM \begin{equation}R_{\m\n}-\frac{1}{2}Rg_{\m\n}=0,\label{eeq}\end{equation}
  i.e., Einstein's equations in vacuum.
Adding the properly normalised matter to the action $S=\frac{1}{8\pi G}S_{H}+S_{M}$ and we can recover the Einstein's non-vacuum equations 
 \begin{equation}
 \frac{1}{\sqrt{-g}}\frac{\delta S}{\delta g^{\m\n}}=\frac{1}{8\pi G}\left( R_{\m\n} -\frac{1}{2}Rg_{\m\n}\right)+\frac{1}{\sqrt{-g}}\frac{\delta S_{M}}{\delta g^{\m\n}}=0
 \end{equation}
 in which we set \begin{equation}T_{\m\n}=-\frac{1}{\sqrt{-g}}\frac{\delta S_{M}}{\delta g^{\m\n}}.\end{equation}
 
 To think of the Einstein equations without specification of the theory from which $T_{\m\n}$ is derived our real concern is existence of solutions for Einstein's equations when there are present realistic sources of energy and momentum. The most common property is that $T_{\m\n}$ represents positive energy densities, negative masses are not allowed. If we allow the action constructed from scalars up to two derivatives in the metric, the first term to add is constant. By itself, it does not lead to interesting dynamics, however it gives an important role to EOM 
\begin{equation}R_{\m\n}-\frac{1}{2}Rg_{\m\n}+\lambda g_{\m\n}=0\end{equation} 
where $\lambda$ is "energy density of the vacuum"  $T_{\m\n}=-\lambda g_{\m\n}$,  energy and momentum present in the universe even in the absence of matter. 

In quantum mechanics the minimum of classical energy $E_0=0$ of an harmonic oscillator with frequency $\omega$ has, upon quantisation a ground state $E_0=\frac{1}{2}\hbar \omega$. Each of the modes contribute to the ground state. The result is infinity and must be regularised using a cutoff at high frequencies. For the cosmological constant, the final vacuum energy, which is the regularised sum of the energies for the ground state oscillations of all the fields in the theory, is expected to have a natural scale 
\begin{equation}\lambda \sim m_P^4.\end{equation} The prediction of the theory considers the Planck mass $m_P\sim 10^{19}GeV$ which differs from the observations of the Universe on the large scale by at least a factor of $10^{123}$. This convinces people that the  "cosmological constant problem" is one of the most important unsolved issues in the physics today. 
\section{Conformal Gravity}

Allowing higher derivatives, one generalisation of the Einstein-Hilbert action  is the action of CG
\begin{equation}
S_{CG}=\alpha_{CG}\int d^4 x \sqrt{|g|}g_{\al\m}g^{\beta\n}g^{\gamma\lambda}g^{\de\tau}C^{\alpha}{}_{\beta\gamma\delta}C^{\m}{}_{\n\lambda\tau}. \label{scg}
\end{equation}
It is consisted of the Weyl squared term, in n dimensions given by
\begin{equation}C_{\rho\sigma\m\n}=R_{\rho\sigma\m\n}-\frac{2}{n-2}\left(g_{\rho[\mu}R_{\n]\sigma}-g_{\sigma[\m}R_{\n]\rho}\right)+\frac{2}{(n-1)(n-2)}Rg_{\rho[\m}g_{\n]\sigma}\end{equation}
that inherits the properties of Riemann tensor
\begin{align}
C_{\rho\sigma\m\n}&=C_{[\rho\sigma][\m\n]} \\
C_{\rho\sigma\m\n}&=C_{\m\n\rho\sigma} \\
C_{\rho[\sigma\m\n]}&=0.
\end{align}
In addition, Weyl tensor is invariant under the Weyl rescalings  of the metric \begin{equation}g_{\m\n}\rightarrow \Omega(x)^2g_{\m\n}\label{wresc}.\end{equation}  
The bulk action (\ref{scg}) is therefore unique, it is the only action polynomial in curvature invariants that enjoys not just the diffeomorphism invariance but also Weyl invariance. The factor that comes from the square root determinant of the metric is exactly cancelled by the factor from the contributions of the metric in (\ref{scg}).

Up to now, we have introduced basic concepts used in the general relativity (GR). To describe the further research and the first step in verification whether the theory of gravity can be considered as correct effective theory, i.e. whether its variational principle is well defined and the response functions finite, 
 we continue with introduction of the partition function, variational principle, correlators and the AdS/CFT correspondence. 
 To obtain the EOM of CG, we use variational principle. After variation of action, in general, one would expect to obtain the EOM and the boundary terms, however, as we will see explicitly, CG does not require such additional terms that are called boundary terms.

\section{Partition Function, Variational Principle and Correlators}

Let us introduce the one of the key functions in the AdS/CFT correspondence, partition function.

Spectrum of energy levels is convenient to compute in the form of a trace $Tr \exp(\beta H)$, where $H$ is Hamiltonian of the considered action.
Including a conserved angular momentum $J$ that generates a rotation at infinity of the asymptotical space and commutes with H,  we can write the partition function as 
\begin{equation}
Z(\beta,\theta)=\text{ Tr } exp(-\beta H-i\theta J),\label{pt0}
\end{equation}
where $\theta$  is the angular chemical potential (rotation chemical potential). The partition function (\ref{pt0}) is standardly computed using the Euclidean path integral according to the formal recipe. In general, Euclidean quantum gravity path integral is not convergent because the action is not bounded from below \cite{Maloney:2007ud}. One approaches that issue, by expanding from below around the classical solution and obtain a perturbatively meaningful result. However, it is important to mention that it is not clear whether the topologies that do not admit classical solutions contribute to the Eucliedan path integral  or not. There is as well no known method to evaluate the contributions in case they do exist. We are focused on the four dimensions in which the classical solutions are not completely described. (Which differs currently from the lower dimensional cases, in particular three, where one can completely describe the partition function due to knowing the classical solutions and the fact that the perturbation theory around them terminates with one-loop term. One can in that case write the complete sum of the known contributions to the path integral.) In addition to the contribution from the classical solutions, there is a possibility of the contributions from the excitations described by cosmic strings or the contribution from complex, and not just real, saddle points\footnote{see below for the description of saddle points} which were considered for the three cases. In our, four dimensional case, such additional contribution could occur from the solutions that describe as well cosmic string solution, or solution such as geon. 

The path integral 
\begin{equation}
\mathcal{Z}=\int \mathcal{D} g \exp \left(-\frac{1}{\hbar} I[g]\right) \label{pathint}
\end{equation}
is evaluated by imposing the boundary conditions on the fields and summing over the relevant spacetimes ($\mathcal{M},g$) using the weighted sum. 
The semi-classical limit is dominated by the stationary points of the action \cite{Bergamin:2007sm,Grumiller:2007ju}, so one considers the saddle point approximation. 
 The meaningful expansion around the classical solution 
\begin{equation}
I[g_{cl}+\delta g] = I[g_{cl}]+\delta I[g_{cl},\delta g]+\frac{1}{2}\delta^2 I[g_{cl},\delta g]+...
\end{equation}
verifies that. Here, $\delta I$ and $\delta^2 I$ are linear and quadratic terms in the Taylor expansion and the saddle point approximation 

\begin{equation}
\mathcal{Z}\sim \exp\left(-\frac{1}{\hbar} I[g_{cl}]\right)\int \mathcal{D}\delta g \exp\left(-\frac{1}{2\hbar}\delta^{2}I[g_{cl},\delta g]\right) \label{saddleap}
\end{equation}
is defined with the requirements that 
\begin{enumerate}[label=(\alph*),ref=(\alph*)]
\item the on-shell actions is bounded from below \label{a}
\item the first variation of the action vanishes on shell for all the variations of the metric that preserve the boundary conditions. \label{firstvar} 
\item the second variation has the correct sign of convergence of the Gaussian in (\ref{saddleap}). \label{c}
\end{enumerate}
In the gravity cases the transition from the (\ref{pathint}) to  (\ref{saddleap}) is complicated when the action does not posses the requirements \ref{a}, \ref{firstvar} and \ref{c}. The fact that the  on-shell gravity action can diverge is commonly solved by the "background subtraction" technique \cite{Gibbons:1976ue,Liebl:1996ti}. The non-vanishing of the linear term appears when the boundary terms are not considered in detail, or when one is interested into the EOM. If one is interested in the response functions, one-, two- or three-point functions, one needs to treat properly the boundary terms. The proper treatment of the boundary terms is called holographic renormalisation procedure that we explain below. 
The third issue that can arise is that the Gaussian integral is divergent. Then, the canonical partition function is not well-defined and it does not describe the thermodynamics of the stable system but the information about the decay rates between the field configurations with specified boundary conditions \cite{Gross:1982cv}.
The third issue dictates the thermodynamic stability of the system and  is usually treated in the same manner as in \cite{York:1986it}. The density of states grows so fast that the canonical ensemble is not defined. The black hole is put inside the cavity and the system is coupled to a thermal reservoir, with the boundary conditions fixed at the wall of the cavity. The canonical ensemble obtained after the procedure is well defined if and only if the specific heat of the system is positive.

From the above consideration one can notice the reason that the action is required to have well defined variational principle. In four dimensions, using the normalisation from \cite{poisson}, variation of the action 
\begin{equation}
\delta S=\frac{1}{16\pi}\int d^4 x \delta(\sqrt{-g}R)\label{eh4}
\end{equation}
beside to EOMs, leads to boundary term that has to be canceled by adding an appropriate boundary term to action (\ref{egndim}) (for n=4), which is called Gibbons-Hawking-York boundary term. 
Variation of (\ref{eh4}) is 
\begin{equation}
\delta S=\int d^{n} x \left[\sqrt{-g}(g^{\m\n}\delta R_{\m\n}+R_{\m\n}\delta g^{\m\n})+R\delta\sqrt{-g}\right].\label{eh}
\end{equation}
To vary the second term in (\ref{eh}) we have to vary the metric with upper indices
\begin{equation}
\delta(g^{\m\n})=-g^{\m\alpha}g^{\n\beta}\delta g_{\alpha\beta}\label{vargup}
\end{equation}
while for the third term we use the matrix property that 
\begin{equation}
Tr (\ln M)=\ln (\det M) 
\end{equation}
 where $ exp (ln M)=M$ and the variation is \begin{equation} Tr(M^{-1}\delta M)=\frac{1}{det M}\delta (det M).\end{equation}
The variation of the third term brings to
\begin{equation}
\delta \sqrt{g^{-1}}=-\frac{1}{2}\sqrt{-g}g_{\m\n}\delta g^{\m\n}.
\end{equation}
  Plugging the results of variation of the acton (\ref{eh}) we obtain 
\begin{equation}
\delta S=\int \left(R_{\m\n}-\frac{1}{2}Rg_{\m\n}\right)\delta g^{\m\n}\sqrt{-g}d^4x+\int g^{\m\n}\delta R_{\m\n}\sqrt{-g}d^4x.\label{vareh}
\end{equation}
With variation of Ricci tensor $\delta R_{\m\n}$, which can be in more detail found in the appendix: General Relativity and AdS/CFT: Summary of the Conventions (\ref{varricten}), one can write
\begin{align}
g^{\m\n}\delta R_{\m\n}=\deltabar v^{\m}{}_{;\m} & \deltabar v^{\m}=g^{\alpha\beta}\delta\Gamma^{\m}_{\alpha\beta}-g^{\alpha\m}\delta\Gamma^{\beta}_{\alpha\beta}
\end{align}
in which we denoted with  $\deltabar v^{\m}$ that $\deltabar$ is not the variation of the quantity $v^{\m}$.
The second integral in (\ref{vareh}) is 
\begin{align}
\int_{\mathcal{M}} g^{\m\n}\delta R_{\m\n}\sqrt{-g}d^4x&=\int \deltabar v^{\m}_{;\m}\sqrt{-g}d^4x \nonumber \\
& =\oint_{\partial\mathcal{M}}\deltabar v^{\m}d\Sigma_{\m}  \nonumber \\
&=\oint_{\partial\mathcal{M}}\epsilon\deltabar v^{\m}n_{\m}\sqrt{|\gamma|}d^3x
\end{align}
in which $n_{\m}$ is the unite normal to $\partial\mathcal{M}$, $\epsilon\equiv n^{\m}n_{\m}\pm1$, $\gamma$ is metric on the three dimensional manifold $\partial\mathcal{M}$ and $d\Sigma_{\m}$ infinitesimal element of the hyper surface $\Sigma$. Following the conventions of \cite{poisson}, the hypersurface $\Sigma$ partitions spacetime in two regions $\mathcal{M}^{\pm}$ defined with  the metric $g_{\m\n}^{\pm}$ and the coordinates $x_{\pm}^{\m}$. Now, one has to evaluate $\deltabar v^{\m}n_{\m}$. While on $\partial\mathcal{M}$, $\delta g_{\m\n}=\delta g^{\m\n}=0$. Under that conditions 
\begin{equation}
\delta \Gamma^{\m}_{\alpha\beta}|_{\partial\mathcal{M}}=\frac{1}{2}g^{\m\n}\left(\delta g_{\n\alpha,\beta}+\delta g_{\n\beta,\alpha}-\delta g_{\alpha\beta,\n}\right)
\end{equation}
where "$,$" denotes partial derivative "$\partial$". It follows $\deltabar v_{\m}=g^{\alpha\beta}\left(\delta g_{\m\beta,\alpha}-\delta g_{\alpha\beta,\m}\right)$ and
\begin{align}
n^{\m}\deltabar v_{\m}|_{\partial\mathcal{M}}&=n^{\m}\left(\epsilon n^{\alpha} n^{\beta}+\gamma^{\alpha\beta}\right)\left(\delta g_{\m\beta,\alpha}-\delta g_{\alpha\beta,\m}\right) \nonumber \\
& n^{\m}\gamma^{\alpha\beta}\left(\delta g_{\m\beta,\alpha}-\delta g_{\alpha\beta,\m}\right).
\end{align}
To obtain the second line, we have used the completeness relation $g^{\m\n}=\epsilon n^{\m}n^{\n}+\gamma^{\m\n}$ and multiplied $n^{\alpha}n^{\mu}$ with the antisymmetric quantity in the brackets. Next, we observe that the tangential derivative of $\delta g_{\m\n}$ must vanish since $\delta g_{\m\n}$ vanishes everywhere on $\partial \mathcal{M}$, which means $\gamma^{\alpha\beta}\delta g_{\m\beta,\alpha}=0$ and one obtains
\begin{equation}
n^{\m}\deltabar v_{\m}|_{\partial\mathcal{M}}=-\gamma^{\alpha\beta}\delta g_{\alpha\beta,\m}n^{\m},
\end{equation}
which is nonzero since $\delta g_{\alpha\beta}$ can contain non-vanishing $\textbf{normal}$ derivative on hyper surface.
One can write the variation of the action with 
\begin{equation}
\delta S=\int_{\mathcal{M}}G_{\alpha\beta}\delta g^{\alpha\beta}\sqrt{-g}d^4x-\oint_{\partial\mathcal{M}}\epsilon \gamma^{\alpha\beta}\delta g_{\alpha\beta,\m}n^{\m}\sqrt{|\gamma|}d^3x\label{vareh1}
\end{equation}
where the second term is canceled by adding 
\begin{equation}
\delta S_{B}=\frac{1}{8 \pi}\int_{\partial\mathcal{M}}\epsilon K\sqrt{|\gamma|}d^3x
\end{equation}
and $K$ is a trace of the extrinsic curvature which can be written
\begin{align}
K&=n^{\alpha}_{;\alpha}=(\epsilon n^{\alpha}n^{\beta}+\gamma^{\alpha\beta})n_{\alpha;\beta} =\gamma^{\alpha\beta}n_{\alpha;\beta}=\gamma^{\alpha\beta}\left(n_{\alpha,\beta}-\Gamma^{\gamma}_{\alpha\beta}n_{\gamma}\right)
\end{align}
whose variation is
\begin{align}
\delta K&=-\gamma^{\alpha\beta}\delta^{\gamma}_{\alpha\beta}n_{\gamma}=\frac{1}{2}\gamma^{\alpha\beta}\delta g_{\alpha\beta,\m}n^{\m}.
\end{align}
Here, we used that the tangential derivatives from $\delta g_{\m\n}$ vanish on $\partial\mathcal{M}$. That leads precisely to the second integral in (\ref{vareh1}). 
Adding that term to the entire action (\ref{eh4}) leads to the first variation that vanishes when EOM are evaluated, which is in agreement with the requirement $(b)$.

\section{Anti de Sitter/Conformal Field Theory Correspondence}
Anti de Sitter/Conformal field theory ($AdS/CFT$) correspondence is the framework 
that relates the gravity theory in one dimension higher with the quantum field theory in one dimension lower, to which it is often referred to as gauge/gravity correspondence. Since its proposal in 1997 \cite{Maldacena:1997zz} it has been generalised to wider framework that fits the name "gauge/gravity" correspondence.

The duality has been discovered in the context of string theory and it has been extended over the different domains, for example, analysis of the strong coupling dynamics of QCD and the electroweak theories, quantum gravity and physics of black holes, relativistic hydrodynamics, applications in condensed matter physics (for example holographic superconductors, quantum phase transitions and cold atoms...).

The fields in AdS correspond to sources of operators on the field theory side and by analysing the dynamic of the sources in the curved space we can learn about the dual operators. 

Let us introduce the notion of sources of the operators.
Taking an particular example in analysing the systems on lattice (also called Kadandoff-Willson renormalisation group approach \cite{Ramallo:2013bua}), one may consider a gravitational system in a lattice with a Hamiltonian
\begin{equation}
H=\sum_{x,i}J_i(x,a)\mathcal{O}^i(x)
\end{equation}
for $a$ lattice spacing, $x$ different lattice sites and $i$ labels of operators $\mathcal{O}^i$. $J_{i}(x,a)$ are the coupling constants that are called sources,  for the operators defined at the point $x$. Using particular computational method \cite{Ramallo:2013bua} the operators are appropriately weighed while Hamiltonian retains its form. Therefore, the couplings change at each step and acquire a dpenedence on the scale
\begin{align}
J_{i}(x,a)\rightarrow J_i(x,2a)\rightarrow J_i(x,4a)\rightarrow..
\end{align}
which can be written as $J_i(x,u)$ for $u=(a,2a,4a,..)$ a length scale at which we probe a system. The evolution of the couplings with the scale is defined with equations
\begin{equation}
u\frac{\partial}{\partial u}J_i(x,u)=\beta_i(J_j(x,u),u)
\end{equation}
for $\beta_i$ a $\beta$ function of the $i^{th}$ coupling constant. 
$\beta_{i}$'s can at weak coupling be determined in perturbation theory, while at strong coupling, $AdS/CFT$ suggests  to consider $u$ as an extra dimension.  This way, one may consider successive lattices at different $u$-s, as layers of a new higher-dimensional space, while $J_i(x,u)$ are considered as fields in a space with one extra dimension. One may write \begin{equation}
J_i(x,u)=\phi_i(x,u),
\end{equation}
where the dynamics of sources is governed by defined action which is in  $AdS/CFT$ particular gravity theory.
This way, one may think of holographic duality as a geometrization of the quantum dynamics defined by the renormalisation group. 

Couplings of the theory at  UV  are identified with values of the bulk fields at the boundary of the higher dimensional space. 
 The source $\phi_i$ on the gravity side need to have equal tensor structure of the corresponding dual operator $\mathcal{O}^i$ of the field theory so that  $\phi_i\mathcal{O}^i$ is scalar. $A_{\m}$ is dual to current $J^{\m}$, spin two field $g_{\m\n}$ to symmetric second order tensor $T_{\m\n}$, identified with the energy momentum tensor $T_{\m\n}$ of the field theory.

The often usage of the correspondence is in compuation of the correlation functions. One may compute the correlation functions 
\begin{equation}
\langle\mathcal{O}(x_1)...\mathcal{O}(x_n)\rangle
\end{equation}
in Euclidean space from the gravity theory.
In the field theory the correlators can be computed from
\begin{equation}
\mathcal{L}\rightarrow\mathcal{L}+J(x)\mathcal(O(x))\equiv\mathcal{L}+\mathcal{L}_J,
\end{equation}
where Lagrangian $\mathcal{L}$ is perturbed by the source term $J(x)$, and perturbation of the Lagrangian denoted with $\mathcal{L}_J$.
The generating functional 
\begin{equation}
Z_{QFT}[J]=\langle exp\left[\int \mathcal{L}_J\right]\rangle
\end{equation}
 defines the connected correlators 
\begin{equation}
\langle \prod_{i}\mathcal{O}(x_i)\rangle=\prod_i\frac{\delta}{\delta J(x_i)}\log Z_{QFT}[J]|_{J=0} .
\end{equation}
For a bulk field $\phi(z,x)$ that fluctuates in AdS  we define $\phi_0$ as a value of $\phi$ at the boundary
\begin{equation}
\phi_0(x)=\phi(z=0,x)=\phi|_{\partial AdS}(x),
\end{equation}
where field $\phi_0$ is related to a source for the dual operator $\mathcal{O}$ in QFT.
The  value of $\phi$ at $z=0$ is actually the limit
\begin{equation}
lim_{z\rightarrow0}z^{\Delta-d}\phi(z,x)=\psi(x)
\end{equation}
in which $\Delta$ is defined as a dimension of the dual operator and determined from the largest root of the equation
\begin{equation} 
(\Delta-p)(\Delta+p-d)=m^2L^2
\end{equation}
for $p$ denoting indices of antisymmetric tensor $A_{\mu_1....\mu_p}$ (in our case scalar field $\phi$), its mass $m$ and $L$ radius of the AdS space. 
It reads
\begin{equation}
\Delta=\frac{d}{2}+\sqrt{\left(\frac{d-2p}{2}\right)^2+m^2L^2}.
\end{equation}
AdS/CFT then claims \cite{Gubser:1998bc,Witten:1998qj}
\begin{equation}
Z_{QFT}[\phi_0]=\langle exp[\int\phi_0\mathcal{O}]\rangle_{QFT}=Z_{gravity}[\phi\rightarrow\phi_0]
\end{equation}
for $Z_{gravity}[\phi\rightarrow\phi_0]$ partition function (path integral) of the gravity theory evaluated for the functions with value $\phi_0$ at the boundary 
\begin{equation}
Z_{gravity}[\phi\rightarrow\phi_0]=\sum_{\{\phi\rightarrow\phi_0\}}e^{S_{gravity}}.
\end{equation}
In the limiting case of the classical gravity, the sum can be approximated with the classical solution term. That term, which contains on-shell gravity action, is usually divergent and must be holographically renormalised \cite{Henningson:1998gx}, see chapter below, 
and the classical action is replaced with the renormalised one. One may write for the generating functional
\begin{equation}
\log Z_{QFT}=S^{ren}_{grav}[\phi\rightarrow\phi_0]
\end{equation}
and the $n-$point function is obtained from 
\begin{equation}
\langle\mathcal{O}(x_1)...\mathcal{O}(x_n)\rangle=\frac{\delta^{(n)}S^{ren}_{grav}[\phi]}{\delta\psi(x_1)...\delta\psi(x_n)}\vert_{\phi=0}.
\end{equation}
We will compute this explicitly on the example for CG and one-point function, which is for an operator $\mathcal{O}$  in the presence of the source $\phi$ written as
\begin{equation} \langle\mathcal{O}(x)\rangle_{\phi}=\frac{\delta S^{ren}_{grav}[\phi]}{\delta \psi (x)}.\end{equation}

\chapter{Holographic Renormalisation} 

\section{Variation of Conformal Gravity Action and Boundary Conditions}

 To obtain the equation of motion of the CG we follow the above described procedure. We vary the action (\ref{scg}) and obtain 
\begin{equation}
\delta S_{CG}=\alpha_{CG}\int d^4x \sqrt{|g|}\left(EOM^{\m\n}\delta g_{\m\n}+\nabla_{\sigma}J^{\sigma}\right)\label{var1}
\end{equation}
in which $EOM^{\m\n}$ denotes EOM of CG, and $\nabla_{\sigma}J^{\sigma}$ boundary terms. $\alpha_{CG}$ is dimensionless coupling constant and only coupling constant of the theory. 
The EOM of the CG require vanishing of the Bach tensor \cite{bachr}
\begin{equation}
\left(\nabla^{\de}\nabla_{\ga}+\frac{1}{2}R^{\de}_{\ga}\right)C^{\gamma}{}_{\alpha\delta\beta}=0\label{bach},
\end{equation}
where the computation of finding EOM is preformed as described on the Einstein case, while one can verify it using the computer program xAct \cite{xAct}, convenient for the application on higher derivative actions. We use it in particular for obtaining EOM, while upon introducing certain auxiliary tensor in variations one can obtain the boundary terms as well. 

EOM of CG are equations of the fourth order in derivatives and it is not straightforward to obtain their most general solution.
The equations consist of the coupled partial differential equations. The same issue arises even in the EOM of EG. 
Then, one searches for the perturbative or numerical solutions.
 In general, in perturbative approach in the AdS/CFT framework, one splits the metric in the holographic coordinate $\rho$ using which is approached to the boundary, and the boundary metric. 
 That is generalised Fefferman-Graham expansion of the metric, which describes the boundary conditions. 
 We introduce the length scale $\ell$ that is related to cosmological constant with $\Lambda=3\sigma/\ell^2$ where $\sigma=-1$ for AdS and $\sigma=+1$ for dS in which the asymptotic $(0<\rho<<\ell)$ line-element is
\begin{equation}
ds^2=\frac{\ell^2}{\rho^2}\left(-\sigma d\rho^2+\gamma_{ij}dx^idx^j\right).\label{le}
\end{equation}
Here, we have partially fixed the gauge and used Gaussian coordinates. 
Near the conformal boundary, at $\rho=0$ on the three dimensional manifold, $\gamma_{ij}$ is 
\begin{equation}
\gamma_{ij}=\gamma_{ij}^{(0)}+\frac{\rho}{\ell}\gam+\frac{\rho^2}{\ell^2}\gamma_{ij}^{(2)}+\frac{\rho^{3}}{\ell^3}\gamma_{ij}^{(3)}+...\label{expansiongamma}
\end{equation}
The coefficients in the expansion, $\gamma_{ij}^{(n)}$ matrices, can depend on the coordinates on the boundary of the manifold, and the boundary metric $\gamma^{(0)}_{ij}$ needs to be invertible. The EOM in each order of the expansion in the holographic coordinate give condition on the terms in the expansion of the boundary metric. 

We will be interested in the asymptotic and full solutions of the CG EOM that we use as examples. Full solutions of CG, one can classify in \begin{itemize}
\item most general solutions. To this class belongs the most general spherically symmetric CG solution 
\begin{equation}
ds^2=-k(r)dt^2+\frac{dr^2}{k(r)}+r^2d\Omega^2_{S^2} \label{MKR1}
\end{equation}
for $d\Omega_{S^2}^2$ line-element of the 2-sphere and 
\begin{equation}
k(r)=\sqrt{1-12aM}-\frac{2M}{r}-\Lambda r^2+2ar^2
\end{equation}
in which for $a=0$ one obtains the Schwarzschild-(A)dS solution. In a lower dimensional effective model for gravity at large distances, the obtained solution corresponds to ours when $aM<<1$. From phenomenological aspect for $\Lambda\approx10^{-123}$, $a\approx10^{-61}$, $M\approx10^{38}M_{\odot}$ with $M_{\odot}=1$ for the Sun, one obtains $aM\approx10^{-23}M_{\odot}<<1$ for black holes or galaxies in the Universe \cite{Grumiller:2010bz}.

\item conformally flat solutions, which automatically makes them satisfy of the Bach equation.
\item Einstein metrics, in which $R_{\alpha\beta}\propto g_{\alpha\beta}$.  That makes solutions of EG a subset of the broader class of solutions of CG. 
\end{itemize} In the perturbative expansion approach described above, the restrictions from the EOM do not appear until the fourth order in the expansion of EOM. 
These restrictions, are important when they affect the results evaluated "on-shell" ("on shell"=when restrictions from EOM are taken into an account).

In order for the first variation of the action to vanish, in general gravitational theories, one requires boundary conditions as part of the definition, as we have seen when considering the variation of the partition function \ref{firstvar}.
Often, "natural" boundary conditions consist of the rapid fall-off of the fields as approaching the boundary in an asymptotic region.  That is not the case for gravitational theories, because the metric should not be zero. An example for that is in AdS/CFT correspondence where boundary conditions define the dual field theory on the boundary. De Sitter space similarly, requires boundary conditions which have been defined for EG in four dimensions by Starobinsky \cite{Starobinsky:1982mr}, and further worked out in \cite{Anninos:2010zf}, \cite{Anninos:2011jp}. Precisely imposing the right boundary conditions, Maldacena reduced CG solutions to solutions of EG \cite{Maldacena:2011mk}.

In our case boundary conditions are imposed by fixing  the leading and the first-order terms in (\ref{expansiongamma}) on $\partial\mathcal{M}$. They are fixed up to a local Weyl rescalings
\begin{align} 
\delta\gamma_{ij}^{(0)}|_{\partial\mathcal{M}}=2\lambda\gamma_{ij}^{(0)},& & \delta\gamma_{ij}^{(1)}|_{\partial\mathcal{M}}=\lambda\gamma_{ij}^{(1)}\label{bcs}
\end{align} for $\lambda$ regular function on $\partial\mathcal{M}$ and second and higher order terms that are allowed to vary.

For the set of boundary conditions to be consistent, on general grounds, one may expect to require adding an analog of Gibbons-Hawking-York term, that was, as we have seen, in EG played by extrinsic curvature \cite{York:1972sj,Gibbons:1977mu},
which would prove that the variational principle is well defined and produces the desired boundary value problem. 
The additional terms that may be required, are the holographic counterterms \cite{Henningson:1998ey,Balasubramanian:1999re,Emparan:1999pm,Kraus:1999di,deHaro:2000vlm,Papadimitriou:2005ii}. Their assignment is to make the response functions (in the AdS/CFT language) finite. 
Below, we will show that
for CG however, these counterterms are not required. This one might have anticipated based on the computation of the on-shell action.  On-shell action 
\begin{equation}
\Gamma_{CG}=S_{CG}=\int_{\partial\mathcal{M}}d^4x\sqrt{|g|}C^{\lambda}{}_{\m\sigma\n}C_{\lambda}{}^{\m\sigma\n}\label{action},
\end{equation}for any metric of the form (\ref{expansiongamma}) and (\ref{le}) when evaluated on compact region for which $\rho_c\leq \rho$ is finite when $\rho_c\rightarrow0$. In addition, the free energy obtained from the on-shell action (\ref{action}) agrees wit the Arnowitt-Deser-Misner mass and the definition of the entropy according to Wald \cite{Wald:1993nt}. Free energy that agrees with the on-shell action can imply that boundary terms that are added to the action (\ref{action}) should vanish on-shell. The simplest answer is that the terms themselves are zero. 

To verify this claim rigorously one computes consistency of variational principle and finiteness of response functions.
For that,  first we rewrite the action in the form 
\begin{equation}
S_{CG}=\int_{M}d^4x\sqrt{-g}\left(32\pi^2\epsilon_4+2R_{\m\n}R^{\m\n}-\frac{2}{3}R^2\right)
\end{equation}
for $\epsilon_4$ Euler density in four dimensions, see appendix: General Relativity and AdS/CFT: Summary of the Conventions  
and normalization $\chi(S^4)=2$.  By adding that surface term to the bulk integral of $\epsilon_4$, one obtains a topological invariant on a space with boundary. Adding and subtracting that surface term leads to action separated into a topological part consisted from the Euler characteristic $\chi(\mathcal{M})$, and part consisted from Ricci squared term and the boundary terms
\begin{align}
\Gamma_{CG}&=\int_{\mathcal{M}}d^4x\sqrt{|g|}\left(2R^{\m\n}R_{\m\n}-\frac{2}{3}R^2\right)+32\pi^2\chi(\mathcal{M}) \nonumber \\ &+\int_{\partial\mathcal{M}}d^3x\sqrt{|\gamma|}\left(-8\sigma\mathcal{G}^{ij}K_{ij}+\frac{4}{3}K^3-4KK^{ij}K_{ij}+\frac{8}{3}K^{ij}K_{j}^kK_{ki}\right).\label{ac2}
\end{align}
Where boundary terms cancel similar terms from the Euler characteristic for spacetimes with conformal boundary \cite{Myers:1987yn}, and $\mathcal{G}^{ij}$ is 3D Einstein tensor on the 3D surface $\partial\mathcal{M}$ for the metric $\gamma_{ij}$. 
%
The extrinsic curvature is \begin{equation}K_{ij}=-\frac{\sigma}{2}\pounds_{n}\gamma_{ij}\end{equation}
for $\pounds$ Lie derivative and $n^{\m}$ outward (future) pointing unit vector $n^{\mu}$ normal to $\partial\mathcal{M}$.
Using the auxiliary Lagrangian, one can rewrite the action  %
\begin{align}
S_{CG}+32\pi^2\chi(\mathcal{M})&=-\int_{\mathcal{M}}d^4x\sqrt{-g}\left(f^{\m\n}G_{\m\n}+\frac{1}{8}f^{\m\n}f_{\m\n}-\frac{1}{8}f^{\m}_{\m}f^{\n}_{\n}\right) \nonumber \\ 
& +\int_{\partial\mathcal{M}}d^3x\sqrt{|\gamma|}\big(-8\sigma\mathcal{G}^{ij}K_{ij}+\frac{4}{3}K^3-4KK^{ij}K_{ij}\nonumber\\&+\frac{8}{3}K^{ij}K_{j}^kK_{ki}\big),\label{ac3}
\end{align}
in which the variation of the bulk action was somewhat simplified using the auxiliary field $f_{\m\n}$. 
The fields in the first integral, after the variation further have to be first decomposed into $3+1$ metric, in Gaussian normal coordinates 
while under the second integral all the quantities are already defined on the three dimensional manifold. 
Variation of the first integral in (\ref{ac3}) requires variation of three terms $I_1=f^{\m\n}G_{\m\n}$, $I_2=\frac{1}{8}f^{\m\n}f_{\m\n}$, $I_3=-\frac{1}{8}f^{\m}_{\m}f^{\n}_{\n}$
\begin{align}
\delta (f^{\m\n}G_{\m\n})&= \delta f_{\m\n}G^{\m\n}-2 f^{\m\n}\delta G_{\m\n}
 \nonumber \\ 
\frac{1}{8}\delta f^{\m\n}f_{\m\n}&=\frac{1}{4}\left(f^{\m\n}\delta f_{\m\n}-f_{\kappa}^{\m}f^{\n\kappa}\delta g_{\m\n}\right) \nonumber \\ 
-\frac{1}{8}\delta{\left(f^2\right)}&=-\frac{1}{4}fg^{\m\n}\delta f_{\m\n}+\frac{1}{4}ff^{\m\n}\delta g_{\m\n}\label{term3}
\end{align}
where we used (\ref{vargup}).  Variation of $G_{\m\n}$ is brought to variation of the Ricci tensor and Ricci scalar given in the appendix: General Relativity and AdS/CFT: Summary of the Conventions, 
(\ref{varricsc}) and (\ref{varricten}), that is 
\begin{align}
-\int d^4x \sqrt{|q|} 2f^{\m\n}\delta G_{\m\n} =&-\int d^4x \sqrt{|g|} 2f^{\m\n}\delta\left(R_{\m\n}-\frac{1}{2}R g_{\m\n}\right) \nonumber \\
 =&-\int d^4x\sqrt{|g|} f^{\m\n}\bigg(\big( \nabla^{\lambda}\nabla_{\m}\delta g_{\n\lambda}\nonumber\\ +&\nabla^{\lambda}\nabla_{\n}\delta g_{\m\lambda}-g^{\lambda\sigma}\nabla_{\m}\nabla_{\n}\delta g_{\lambda\sigma}-\nabla^2\delta g_{\m\n} \big)
\nonumber \\ \nonumber +&\big(-R^{\lambda\sigma}\delta g_{\lambda\sigma}+\nabla^{\lambda}(\nabla^{\sigma}\delta g_{\lambda\sigma}\nonumber \\  - &g^{\kappa\delta}\nabla_{\lambda} \delta g_{\kappa\delta}) \big)g_{\m\n} +R\delta g_{\m\n} \bigg). \label{varfg}
\end{align}
Obviously we have to partially integrate analogously to case with EG which will lead to terms that define EOM, and boundary terms that define the response functions, which we demonstrate on the first term under the integral (\ref{varfg})
\begin{align}
\int d^4x\sqrt{g} f^{\m\n}\nabla^{\lambda}\nabla_{\m}\delta g_{\n\lambda} &= \int d^4x \sqrt{g}\nabla^{\lambda}\left(f^{\m\n}\nabla_{\m}\delta g_{\n\lambda} \right)-\int d^4x\sqrt{g} \nabla^{\lambda}f^{\m\n}\nabla_{\m}\delta g_{\n\lambda}\nonumber \\& =\int d^4x \sqrt{g}\nabla^{\lambda}\left(f^{\m\n}\nabla_{\m}\delta g_{\n\lambda} \right)-\int d^4x\sqrt{g}\nabla_{\m}(\nabla^{\lambda}f^{\m\n}\delta g_{\n\lambda})\nonumber \\&+\int d^4x \sqrt{g}\nabla_{\m}\nabla^{\lambda}f^{\m\n}\delta g_{\n\lambda}\label{pokaz}
\end{align}
in which we perform partial integration in the first line, and partial integration of the second term on the RHS when going form the first to the second line. 
In second partial integration  the non-trivial part is commutation of the covariant derivatives. Both terms in second line on the RHS, that participate in the partial integration have contribution from Christoffel symbols that appear in  commutation, however rewriting explicitly the covariant derivative before preforming the partial integration, shows that the Christoffel symbols remained, combine with the ones required for writing the covariant derivatives. The new required Christoffels that have to be added to first partial derivative to form it into covariant derivative are exactly equal to those that have to be subtracted from the second partial derivative in order to make it covariant.

From the equation (\ref{pokaz}) we may observe which of the terms upon the transformation to the GNC contribute to EOM, and which to the boundary terms. To boundary terms contribute obviously two terms of (\ref{pokaz}) in the second line, while to EOM, the term in the third line.
The EOM do not contribute with conditions on $\gam$ matrix or conditions up to fourth order in the $\rho$ expansion, we provide them in the appendix: Holographic Renormalisation: Equations of Motion in Conformal Gravity.

\section{Boundary Terms}

We can write the first variation of the action as
\eq{
\delta\Gamma_{\textrm{\tiny CG}} = \textrm{EOM} + \int_{\partial{\cal M}}\!\!\!\!\extd^3x\sqrt{|\ga|}\,\big(\pi^{ij}\,\de\gamma_{ij} + \Pi^{ij}\,\de K_{ij}\big)~,
}{eq:CG13}
where the boundary terms are momenta $\pi^{ij}$ and $\Pi^{ij}$ that read 
\begin{align}
\pi^{ij}& = \tfrac{\sigma}{4}(\ga^{ij} K^{kl}-\ga^{kl} K^{ij}) f_{kl} + \tfrac{\sigma}{4} f^{\rho}{}_{\rho} (\ga^{ij} K - K^{ij}) 
 - \tfrac12 \ga^{ij} {\cal D}_{k}(n_{\rho} f^{k \rho})  + \tfrac{1}{2} {\cal D}^{i}(n_{\rho} f^{\rho j}) \nonumber \\
&- \tfrac{1}{4} (\ga^{ik} \ga^{jl} - \ga^{ij} \ga^{kl}) \pounds_{n} f_{kl} + \sigma\,\big(2 K {\cal R}^{ij} - 4 K^{ik} {\cal R}_{k}{}^{j}  + 2 \ga^{ij}  K_{kl} {\cal R}^{kl} - \ga^{ij} K {\cal R}  \nonumber \\
& + 2 {\cal D}^{2} K^{ij}- 4 {\cal D}^{i} {\cal D}_{k} K^{kj} 
+ 2 {\cal D}^{i} {\cal D}^{j} K  + 2 \ga^{ij} ({\cal D}_{k} {\cal D}_{l} K^{kl} - {\cal D}^{k} {\cal D}_{k} K) \big)\nonumber \\ & + \tfrac{2}{3} \ga^{ij} K^{k}{}_{m} K^{lm} K_{kl} - 4 K^{i k} K^{jl} K_{kl} + 2 K^{ij} K^{kl} K_{kl} + \tfrac{1}{3} \ga^{ij} K^3 - 2 K^{ij} K^2\nonumber \\
& -\ga^{ij} K K^{kl} K_{kl} + 4 K K^{i}{}_{k} K^{jk} 
 + i \leftrightarrow j\
\label{eq:CG14}
\end{align} 
and
\begin{align}
\Pi^{ij} &= -8 \,\sigma\, {\cal G}^{ij} - \sigma \,\big(f^{ij} - \ga^{ij}f^k{}_k \big) 
+ 4\ga^{ij} \big(K^2 - K^{kl}K_{kl}\big)\nonumber \\
& - 8 K K^{ij} + 8K^i{}_k K^{kj} \, , 
\label{eq:CG15}
\end{align} 
respectively. Where we can vary independently the boundary metric and the extrinsic curvature.

To obtain the response functions that correspond to the sources $\delta \gamma^{(0)}_{ij}$ and $\delta\gamma_{ij}^{(1)}$ we insert the expansion of the curvatures and extrinsic curvature in (\ref{eq:CG14}) and (\ref{eq:CG15}). We obtain for $\Pi_{Kij}$
\begin{align}
\Pi_{K}&=\rho^2 (\frac{4 R[D]^{ij}}{\ell^2} - \frac{4 \gamma^{ij} R[D]}{3 \ell^2} - \frac{\gamma^{(1)ij} \gamma^{(1)k}{}_{k}}{\ell^4} + \frac{\gamma^{ij} \gamma^{(1)k}{}_{k} \gamma^{(1)l}{}_{l}}{3 \ell^4} \nonumber \\ & + \frac{2 \gamma^{(2)ij}}{\ell^4} - \frac{2 \gamma^{ij} \gamma^{(2)k}{}_{k}}{3 \ell^4}) + \rho^3 (- \frac{2 R[D] \gamma^{(1)ij}}{3 \ell^3} + \frac{2 R[D]^{jk} \gamma^{(1)i}{}_{k}}{\ell^3} + \frac{2 R[D]^{ik} \gamma^{(1)j}{}_{k}}{\ell^3} \nonumber \\ & - \frac{8 \gamma^{ij} R[D]^{kl} \gamma^{(1)}{}_{kl}}{3 \ell^3} - \frac{4 R[D]^{ij} \gamma^{(1)k}{}_{k}}{\ell^3} + \frac{2 \gamma^{ij} R[D] \gamma^{(1)k}{}_{k}}{\ell^3} + \frac{\gamma^{(1)ij} \gamma^{(1)}{}_{kl} \gamma^{(1)kl}}{\ell^5} \nonumber \\ &+ \frac{2 \gamma^{(1)ik} \gamma^{(1)j}{}_{k} \gamma^{(1)l}{}_{l}}{\ell^5} - \frac{\gamma^{(1)ij} \gamma^{(1)k}{}_{k} \gamma^{(1)l}{}_{l}}{3 \ell^5} - \frac{2 \gamma^{ij} \gamma^{(1)k}{}_{k} \gamma^{(1)}{}_{lm} \gamma^{(1)lm}}{3 \ell^5}\nonumber \\ & - \frac{\gamma^{(1)k}{}_{k} \gamma^{(2)ij}}{\ell^5} - \frac{2 \gamma^{(1)j}{}_{k} \gamma^{(2)ik}}{\ell^5} - \frac{2 \gamma^{(1)ik} \gamma^{(2)j}{}_{k}}{\ell^5} + \frac{2 \gamma^{ij} \gamma^{(1)kl} \gamma^{(2)}{}_{kl}}{3 \ell^5} - \frac{\gamma^{(1)ij} \gamma^{(2)k}{}_{k}}{3 \ell^5} \nonumber \\ & + \frac{2 \gamma^{ij} \gamma^{(1)k}{}_{k} \gamma^{(2)l}{}_{l}}{3 \ell^5} + \frac{2 \gamma^{(3)ij}}{\ell^5} - \frac{2 \gamma^{ij} \gamma^{(3)k}{}_{k}}{3 \ell^5} + \frac{2 D^{i}D_{k}\gamma^{(1)jk}}{\ell^3}  - \frac{2 D^{j}D^{i}\gamma^{(1)k}{}_{k}}{\ell^3} \nonumber \\ & + \frac{2 D^{j}D_{k}\gamma^{(1)ik}}{\ell^3} - \frac{2 D_{k}D^{k}\gamma^{(1)ij}}{\ell^3} - \frac{4 \gamma^{ij} D_{l}D_{k}\gamma^{(1)kl}}{3 \ell^3}  + \frac{4 \gamma^{ij} D_{l}D^{l}\gamma^{(1)k}{}_{k}}{3 \ell^3}).
\end{align}
Analogously, inserting the expansions of tensors for the auxiliary fields and unphysical fields (\ref{fijexp1}), (\ref{fijexp2}), (\ref{vexp}) and (\ref{wexp}) from the appendix: Holographic Renormalisation: EOM for CG, while keeping in mind the order of $\frac{\rho}{\ell}$ in which the fields appear, one obtains 
\begin{align}
\pi_{\tilde{g}}&=\rho^2 (- \frac{4 R[D]^{ij}}{\ell^3} + \frac{4 \gamma^{ij} R[D]}{3 \ell^3} + \frac{\gamma^{(1)ij} \gamma^{(1)k}{}_{k}}{\ell^5} - \frac{\gamma^{ij} \gamma^{(1)k}{}_{k} \gamma^{(1)l}{}_{l}}{3 \ell^5} - \frac{2 \gamma^{(2)ij}}{\ell^5}\nonumber \\ & + \frac{2 \gamma^{ij} \gamma^{(2)k}{}_{k}}{3 \ell^5}) + \rho^3 (\frac{7 R[D] \gamma^{(1)ij}}{3 \ell^4} - \frac{7 R[D]^{jk} \gamma^{(1)i}{}_{k}}{\ell^4} - \frac{7 R[D]^{ik} \gamma^{(1)j}{}_{k}}{\ell^4} + \frac{19 \gamma^{ij} R[D]^{kl} \gamma^{(1)}{}_{kl}}{3 \ell^4} \nonumber \\ & + \frac{2 \gamma^{(1)ik} \gamma^{(1)jl} \gamma^{(1)}{}_{kl}}{\ell^6} + \frac{8 R[D]^{ij} \gamma^{(1)k}{}_{k}}{\ell^4} - \frac{4 \gamma^{ij} R[D] \gamma^{(1)k}{}_{k}}{\ell^4} - \frac{3 \gamma^{(1)ij} \gamma^{(1)}{}_{kl} \gamma^{(1)kl}}{2 \ell^6} \nonumber \\ & - \frac{2 \gamma^{ij} \gamma^{(1)}{}_{k}{}^{m} \gamma^{(1)kl} \gamma^{(1)}{}_{lm}}{3 \ell^6} - \frac{3 \gamma^{(1)ik} \gamma^{(1)j}{}_{k} \gamma^{(1)l}{}_{l}}{\ell^6} + \frac{2 \gamma^{(1)ij} \gamma^{(1)k}{}_{k} \gamma^{(1)l}{}_{l}}{3 \ell^6}\nonumber \\ & + \frac{13 \gamma^{ij} \gamma^{(1)k}{}_{k} \gamma^{(1)}{}_{lm} \gamma^{(1)lm}}{12 \ell^6}  - \frac{\gamma^{ij} \gamma^{(1)k}{}_{k} \gamma^{(1)l}{}_{l} \gamma^{(1)m}{}_{m}}{12 \ell^6} + \frac{3 \gamma^{(1)k}{}_{k} \gamma^{(2)ij}}{2 \ell^6} \nonumber \\ & + \frac{\gamma^{(1)j}{}_{k} \gamma^{(2)ik}}{\ell^6} + \frac{\gamma^{(1)ik} \gamma^{(2)j}{}_{k}}{\ell^6}  + \frac{\gamma^{ij} \gamma^{(1)kl} \gamma^{(2)}{}_{kl}}{6 \ell^6} + \frac{\gamma^{(1)ij} \gamma^{(2)k}{}_{k}}{6 \ell^6} - \frac{5 \gamma^{ij} \gamma^{(1)k}{}_{k} \gamma^{(2)l}{}_{l}}{6 \ell^6} - \frac{\gamma^{(3)ij}}{\ell^6} \nonumber \\ & + \frac{\gamma^{ij} \gamma^{(3)k}{}_{k}}{3 \ell^6} - \frac{4 D^{i}D_{k}\gamma^{(1)jk}}{\ell^4}  + \frac{3 D^{j}D^{i}\gamma^{(1)k}{}_{k}}{\ell^4} \nonumber \\ & - \frac{4 D^{j}D_{k}\gamma^{(1)ik}}{\ell^4}  + \frac{5 D_{k}D^{k}\gamma^{(1)ij}}{\ell^4} + \frac{8 \gamma^{ij} D_{l}D_{k}\gamma^{(1)kl}}{3 \ell^4} - \frac{8 \gamma^{ij} D_{l}D^{l}\gamma^{(1)k}{}_{k}}{3 \ell^4}).
\end{align}

The response functions, we are interested in, arise  as tensor fields multiplying $\delta\gamma_{ij}^{(0)}$ and $\delta\gamma_{ij}^{(1)}$. 
Therefore we have to express the variation of 
extrinsic curvature as
\begin{equation}
\delta K_{ij}=\left(\frac{\ell}{\rho}\right)^2\left(\delta \theta_{ij}-\frac{1}{\ell}\delta\gamma_{ij}\right)
\end{equation}
since expansion of $\theta_{ij}$ is given explicitly in terms of $\gamma_{ij}$ (\ref{thetadef}).
The variation of action
\begin{align}
\delta\Gamma_{CG}&=\int_{\partial_{\mathcal{M}}}\sqrt{\tilde{g}}\big[\pi_g^{ij}\left(\frac{\ell^2}{\rho^2}\right)^2\delta\gamma_{ij}+\Pi_K^{ij}\left(\frac{\ell}{\rho}\right)^2\left(\delta\theta_{ij}-\frac{1}{\ell}\delta\gamma_{ij}\right) \big]\nonumber \\
&=\int_{\partial\mathcal{M}}d^3x\sqrt{\gamma}\left(\frac{\ell}{\rho}\right)^5\left(\left( \pi_g^{ij}-\frac{1}{\ell}\Pi_{K}^{ij}\right)\delta\gamma_{ij}+\Pi_{K}\delta\theta_{ij}\right).\label{varac1}
\end{align}
in which we have written $\tilde{g}_{ij}=\frac{\ell^2}{\rho^2}\gamma_{ij}$ for $\tilde{g}_{ij}$ three dimensional part of the metric $g_{\m\n}$ (defined on the $\partial\mathcal{M}$ manifold), combines both $\pi_g^{ij}$ and $\Pi^{ij}_K$ into one response function. We express the tensors from (\ref{varac1}) in unphysical variables
\begin{align}
\delta\Gamma_{CG}&=\int_{\partial\mathcal{M}}d^3x\sqrt{\gamma}\left(\frac{\ell}{\rho}\right)^3\left(\pi_{\gamma}^{ij}\delta\gamma_{ij}+\Pi_{\theta}^{ij}\delta\theta_{ij}\right)
\end{align}
for $\pi_{\gamma}^{ij}=\left(\frac{\ell^2}{\rho^2}\right)^2\left(\pi_{\tilde{g}}^{ij}-\frac{1}{\ell}\Pi_{K}^{ij}\right)$ and $\Pi_K^{ij}=\left(\frac{\rho}{\ell}\right)^2\pi_{\theta}^{ij}$,
and expand the variations 
\begin{align}
\delta \gamma_{ij}&=\delta\gamma_{ij}^{(0)}+\left(\frac{\rho}{\ell}\right)\delta\gamma_{ij}^{(1)}+...\\
\delta\theta_{ij}&=\frac{\rho}{\ell}\frac{1}{2\ell}\delta
\gamma_{ij}^{(1)}+...
\end{align}
We obtain that the most important equation in this section that is variation of action 
\begin{align}
\delta \Gamma_{CG}=\int_{\partial{M}}d^3x \sqrt{\gamma^{(0)}}\left(\tau^{ij}\delta \gamma_{ij}^{(0)}+P^{ij}\delta\gamma_{ij}^{(1)}\right)\label{finres1}
\end{align}vanishes up to $\mathcal{O}(\rho^{0})$, which means that response functions $\tau^{ij}$ and $P^{ij}$  are finite as $\rho_c\rightarrow 0$. The result for the $\tau_{ij}$ and $P_{ij}$ response functions can be found below in (\ref{eq:CG17}) and (\ref{eq:CG18}), respectively.
Here, the result did not require Weyl invariance. The response function $\tau^{ij}$ plays a role of stress energy tensor, which in the case of EG corresponds to a response function of the source $\delta \gamma_{ij}^{(0)}$. While $P^{ij}$ is a response function specific for CG.
Response functions $\tau^{ij}$ and $P^{ij}$ satisfy the conditions 
\begin{align}
\gamma_{ij}^{(0)}\tau^{ij}+\frac{1}{2}\psi_{ij}^{(1)}P^{ij}=0, && \gamma_{ij}^{(0)}P^{ij}=0 \label{tracecond}
\end{align}
for $\psi_{ij}^{(1)}$ traceless $\gamma^{(1)}_{ij}$ matrix as defined in (\ref{tracelessmet}). The first variation therefore vanishes on shell for the satisfied boundary conditions (\ref{bcs}), which proves a well-defined variational principle. To write the response functions, it is convenient to define the electric $E_{ij}$ and magnetic $B_{ijk}$ part of the Weyl tensor
\begin{align}
E_{ij}&=n_{\m}n^{\n}C^{\m}{}_{i\n j}\label{elc} \\ 
B_{ijk}&=n_{\m}C^{\m}{}_{ijk} \label{magc}
\end{align}
which are as well expanded
\begin{align}
& B_{ijk}^{\FO} = \tfrac{1}{2\ell}\,\big({\cal D}_j\psi^{\FO}_{ik}-\tfrac12\,\ga_{ij}^{\LO}\,{\cal D}^l\psi^{\FO}_{kl}\big) - j \leftrightarrow k \label{eq:CG24} \\
& E_{ij}^{\SO} =  - \tfrac{1}{2\ell^2} \psi_{ij}^{\SO}  + \tfrac{\sigma}{2}\, \big({\cal R}_{ij}^{\LO} - \tfrac13 \ga_{ij}^{\LO}{\cal R}^{\LO}\big) + \tfrac{1}{8\ell^2} \gamma^{\FO} \psi_{ij}^{\FO} \label{eq:CG23} \\
& E^{\TO}_{ij} = -\tfrac{3}{4\ell^2}\,\psi^{\TO}_{ij} -\tfrac{1}{12\ell^{2}}\,\ga_{ij}^{\LO}\,\psi^{kl}_{\FO} \, \psi_{kl}^{\SO}
-\tfrac{1}{16\ell^{2}}\,\psi^{\FO}_{ij}\,\psi^{\FO}_{kl}\,\psi_{\FO}^{kl} \nonumber - \tfrac{\si}{12}\,\big(\mathcal{R}^{\LO}\,\psi_{ij}^{\FO}\\ &-\ga _{ij}^{\LO}\,\mathcal{R}_{kl}^{\LO}\,\psi^{kl}_{\FO}
 +\ga _{ij}^{\LO}\,\mathcal{D}_{l}\,\mathcal{D}_{k}\,\psi^{kl}_{\FO} \nonumber  +\tfrac{3}{2}\,\mathcal{D}_{k}\,\mathcal{D}^{k}\,\psi_{ij}^{\FO}
-3\,\mathcal{D}_{k}\,\mathcal{D}_{i}\,\psi^{\FO k}_{j} 
\big)\\
& + \tfrac{1}{24\ell^2} \bigg( \ga_{\FO}\,(3\,\psi^{\SO}_{ij}+\tfrac12\,\ga_{ij}^{\LO}\,
\psi_{kl}^{\FO}\,\psi^{kl}_{\FO}       - \gamma_{\FO}\,\psi_{ij}^{\FO})   + 5\,\ga _{\SO}\,\psi_{ij}^{\FO} \nonumber\\
&- \si\ell^2\,(\mathcal{D}_{j}\,\mathcal{D}_{i}\,\ga_{\FO} 
-\tfrac{1}{3}\,\ga _{ij}^{\LO}\,\mathcal{D}^{k}\,\mathcal{D}_{k}\,\ga _{\FO})\,\bigg) + i \leftrightarrow j.
\label{eq:CG26a}. 
\end{align}
The response function $\tau_{ij}$ expressed in terms of the electric and magnetic part of the Weyl tensor when $\rho_c\rightarrow0$ reads
\begin{align}
\tau_{ij} &= \sigma \big[\tfrac{2}{\ell}\,(E_{ij}^{\TO}+ \tfrac{1}{3} E_{ij}^{\SO}\ga^{\FO}) -\tfrac4\ell\,E_{ik}^{\SO}\psi^{\FO k}_j
+ \tfrac{1}{\ell}\,\ga_{ij}^{\LO} E_{kl}^{\SO}\psi_{\FO}^{kl} 
+ \tfrac{1}{2\ell^3}\,\psi^{\FO}_{ij}\psi_{kl}^{\FO}\psi_{\FO}^{kl}
\nonumber\\&- \tfrac{1}{\ell^3}\,\psi_{kl}^{\FO}\,\big(\psi^{\FO k}_i\psi^{\FO l}_j-\tfrac13\,\ga^{\LO}_{ij}\psi^{\FO k}_m\psi_{\FO}^{lm}\big)\big] 
- 4\,{\cal D}^k B_{ijk}^{\FO} + i\leftrightarrow j\,,
\label{eq:CG17}
\end{align}
while $P_{ij}$ is partially massless response obtained in the following way. The response function sourced by $\delta\gamma^{(1)}_{ij}$ 
\begin{equation}
P_{ij}=-\tfrac{4\,\sigma}{\ell}\,E_{ij}^{\SO}\,
\label{eq:CG18}\end{equation}
is finite like $\tau_{ij}$ and  does not require adding counterterms. Its definition as  partially massleless response (PMR) is in a sense of 
Deser, Nepomechie and Waldron \cite{Deser:1983mm,Deser:2001pe}. That means that the tensor does not contain full rank,  for example, if we decompose the tensor into transverse and traceless part the "partial massless" means that not all modes are present. 

 When we plug in $P^{ij}$ into linearised CG EOM, around (A)dS background we obtain partial masslessness. That behaviour is expected comparing with the behaviour in 3D \cite{Afshar:2011yh} and on general grounds when one thinks of Weyl invariance \eqref{wresc} as a non-linear completion of the gauge enhacement at the linearised level caused by partial masslesness \cite{Deser:2013bs, Deser:2013uy}. That kind of non-perturbative completion does not appear in general for partial masslessness in higher derivative theories \cite{Deser:2012ci}.

\section{Ward identity} 

This is a good point to explain the concepts of the Ward identity, Noether charge and entropy, which we use in analysis of the response functions obtained from CG and observables that can be analysed.

Most common example of Ward identity is using the photon polarisations, where we follow the the description of Peskin and Schr\v{o}der \cite{Peskin:1995ev}. The sum over electron polarisations can be done using the identity $\sum u(p)\overline{u}(p)=p+m$, where u(p) are electron wave functions, p its momenta and m its mass,   similarly, for photon polarisations one performs
$\sum_{polarizations}\epsilon_{\m}^*\epsilon_{\n}\rightarrow -g_{\m\n}$ where  $\epsilon_{\m}$ denote photon polarisation, and $g_{\m\n}$ spacetime metric. If for simplicity we orient the impulse $k$ of the vectors in the $z$ direction $k^{\m}=(k,0,0,k)$,  the transverse polarisation vectors can be be chosen as $\epsilon_1^{\m}=(0,1,0,0)$ and $\epsilon_2^{\m}=(0,0,1,0)$ that one can use to write
cross section with a QED amplitude of arbitrary QUD process that involves external photon with momentum $k$ as $\mathcal{M}(k)\equiv\mathcal{M}^{\m}(k)\epsilon_{\m}^*(k)$
\begin{equation} 
\sum_{\epsilon}|\epsilon_{\m}^{*}(k)\mathcal{M}^{\m}(k)|^2=\sum_{\epsilon}\epsilon_{\m}^{*}\epsilon_{\n}\mathcal{M}^{\m}(k)\mathcal{M}^{\n*}(k)=|\mathcal{M}^1(k)|^2+|\mathcal{M}^2(x)|^2.
\end{equation}
Since one expects $\mathcal{M}^{\m}(k)$ to be given by the matrix element of the Heisenberg field $j^{\m}$
\begin{equation}
\mathcal{M}^{\m}(k)=\int d^4x e^{ikx}\langle f| j^{\m}(x)|i \rangle \label{wi0}
\end{equation}
for $f$ and $i$ final and initial states tagat include all particles except photon in question, respectively, and $j^{\m}$ Dirac vector current $j^{\m}=\overline{\psi}\gamma^{\m}\psi$ where $\psi$ is a wave function and $\gamma^{\m}$ in this (and only this) context Dirac gamma matrix.
Since EOM  tell us that the current $j^{\m}$ is conserved  $\partial_{\m}j^{\m}(x)=0$, assuming the property holds for quantum theory, it follows from (\ref{wi0})
\begin{equation} k_{\m}\mathcal{M}^{\m}(k)=0. \end{equation} I.e., description for the vanishing of the amplitude $\mathcal{M}$ when polarisation vector $\epsilon_{\m}(k)$ is replaced by $k_{\m}$ and is called as "Ward identity". It states the current conservation, a consequence of gauge symmetry \begin{align}\psi(x)\rightarrow e^{i\alpha(x)}\psi(x),&& A_{\m}\rightarrow A_{\m}-\frac{1}{e}\partial_{\m}\alpha(x) \end{align} for $\alpha(x)$ local phase, $A_{\m}$ electromagnetic vector potential and $e$ charge of an electron. 

The general form of this identity is Ward-Takanashi identity. It states that for an external photon of momentum $k$, $n$ electrons in an initial state with momenta $p_1,...,p_n$ and $n$ electrons with momenta $q_1,...,q_n$ in the final state, one may write
\begin{align}
k_{\m}\mathcal{M}^{\m}(k;p_1,...,p_n;q_1,...,q_n)&=e\sum_i\big[\mathcal{M}_0(p_1,...,p_n;q_1,...(q_i-k),...q_n) \nonumber \\&-\mathcal{M}_0(p_1,...(p_i+k),...,p_n;q_1,...,q_n) \big].
\end{align}
If the external electrons are on-shell, the particles on the right have one external particle off-shell and do not contribute, such that for all external electrons on-shell one obtains Ward identity.\footnote{For proof of the the identity one may take a look at \cite{Peskin:1995ev}}

Similarly in EG Ward identity will yield  $\nabla^j\langle \tau_{ij}\rangle=0$ which we regarded as energy conservation (\ref{encon}). Depending on the considered action, Ward identities yield corresponding results. If we couple EG to matter we notice the change in the Ward identity (\ref{encon}) \cite{deHaro:2000xn}.
For the action 
\begin{align}
S&=S_{gr}+S_{M} \nonumber \\ 
& =\frac{1}{16\pi G_{M}}\big[ \int_{\mathcal{M}}d^{d+1}x\sqrt{g}(R-R \Lambda)-\int_{\partial \mathcal{M}}d^dx\sqrt{\gamma}2K \big]
\nonumber \\&+\frac{1}{2}\int_{\mathcal{M}} d^{d+1}x \sqrt{g}(g^{\m\n}\partial_{\m}\Phi\partial_{\n}\Phi+m^2\Phi^2) \label{smat}
\end{align}
where $\Phi$ is a scalar of mass $m$ defined with
\begin{align}
\Phi(x,\rho)=\rho^{(d-\Delta)/2}\phi(x,\rho), && \phi(x,\rho)=\phi_{(0)}+\phi_{(2)}\rho+...
\end{align}
and $\Delta$ conformal dimension of the dual operator,
 the regulated on-shell value of (\ref{smat}) reads \cite{deHaro:2000xn}
\begin{align}
S_{reg}(bulk)&=\int_{\rho\leq\epsilon}d\rho d^dx\sqrt{g}\frac{1}{\rho}\sqrt{\gamma(x,\rho)}\left[\frac{d}{16\pi G_{N}}\rho^{-d/2}-\frac{m^2}{2(d-1)}\phi^2(x,\rho)\rho^{-k}\right]
\end{align}
for $k=\Delta-d/2$, and $g$ and $\gamma$, defined with 
\begin{align}
ds^2&=g_{\m\n}dx^{\m}dx^{\n}=\ell^2\left(\frac{d\rho^2}{4\rho}^2+\frac{1}{\rho}\gamma_{ij}(x,\rho)dx^idx^j\right) \\
\gamma(x,\rho)&=\gamma_{(0)}+.....+\rho^{d/2}\gamma_{(d)}+h_{(d)}\rho^{d/2}\log \rho+...
\end{align}
Here, the part with logarithm appears for even $d$\footnote{When $d$ is even, one obtains conformal anomalies, while when $k$ is positive integer, one obtains matter conformal anomalies.}.
The expectation value of boundary stress energy tensor,  is not conserved with existing, sources but it satisfies Ward identity relating covariant divergence and expectation value of the operators coupling the sources. 
For generating functional
\begin{equation}
Z_{CFT}[\gamma_{(0)},\phi_{(0)}]=\langle \exp \int d^dx \sqrt{\gamma_{(0)}}\left[\frac{1}{2} \gamma^{ij}_{(0)}\tau_{ij}-\phi_{(0)}O\right] \rangle
\end{equation}
for $\langle O(x)\rangle=-\frac{1}{\sqrt{det \gamma_{(0)}}}\frac{\delta S_{M,ren}}{\delta\phi_{(0)}}$, the obtained Ward identity is
\begin{equation}
\nabla^j\langle \tau_{ij}\rangle=\langle O\rangle \partial_i\phi_{(0)}\label{wieg}.
\end{equation}
It yields from the invariance under infinitesimal diffeomorphisms
\begin{equation}
\delta \gamma_{(0)ij}=\nabla_i\xi_j+\nabla_j\xi_i.
\end{equation}
Analogously to the (\ref{wieg}) we obtain the relation that is similar to the energy condition (\ref{encon}) however since it considers CG it is modified. Interesting findings about Ward identities in axial gauge are shown in \cite{Capper:1981rc,Capper:1981rd}.

\subsection{Application to Conformal Gravity in Four Dimensions}

From the equations (\ref{eq:CG17}) and (\ref{eq:CG18}) the (\ref{tracecond}), trace condition on the response functions are result of the identities
\begin{align}
\gamma_{(0)}^{ij}E_{ij}^{(3)}&=\psi_{(1)}^{ij}E_{ij}^{(2)}, \\
 \gamma_{(0)}^{ij}E_{ij}^{(2)}&=\gamma_{(0)}^{ij}B_{ijk}^{(1)}=0
\end{align}
obtained by the tracelessness of the electric and magnetic
 parts of the Weyl tensor. For Starobinsky boundary conditions  only the Brown-York stress energy tensor is traceless, while in general that is true only for PMR. From the obtained current 
\begin{align}
J^i=\left(2\tau^j{}_j+2P^{il}\gamma_{lj}^{(1)}\right)\xi^{j}, \label{currentcg}
\end{align}
with $\xi^j$ a boundary diffeomorphism that contributes in the definition of the asymptotic symmetry of CG,  we may obtain the conserved charges. To ensure that conformal boundary $ \partial \mathcal{M}$ is timelike, we set $\sigma=-1$ and consider $AdS$ case. Which implies considering a constant time surface $\mathcal{C}$ in $\partial\mathcal{M}$ 
\begin{align}
Q[\xi]=\int_{\mathcal{C}}d^2x \sqrt{h}u_iJ^i\label{charge1} \end{align} defines the charge, where $u^i$ is the future pointing unit normal vector to $\mathcal{C}$ and $h$ a metric on $\mathcal{C}$. These charges we continue to analyse in the "Canonical Analysis" chapter and prove that they generate the asymptotic symmetres. 
The combination of the response functions and $\gamma_{ij}^{(1)}$ that paperers in $J^i$ corresponds to the modified stress energy tensor in a sense of Hollands, Ishibashi and Marlof \cite{Hollands:2005ya}.
Modified stress energy tensor satisfies the covariant divergence
\begin{align}
\mathcal{D}(2\tau^{ij}+2P^i{}_{l}\gamma^{(1)lj})=P^{il}\mathcal{D}^j\gamma_{il}^{(1)},\label{ident}
\end{align}
which is responsible for the difference in charges on surfaces $\mathcal{C}_1$ and $\mathcal{C}_2 $ bounded by a region $\mathcal{V}\subset\partial\mathcal{M}$ 
\begin{align}
\Delta Q[\xi]=\int_{\mathcal{V}}d^3x\sqrt{|\gamma^{(0)}|}
\left(\tau^{ij}\pounds_{\xi}\gamma_{ij}^{(0)}+P^{ij}\pounds_{\xi}\gamma^{(1)}_{ij}\right).\label{dq}\end{align}
The difference of the charges (\ref{dq}) vanishes for the asymptotic symmetries.

\subsection{Alternate Boundary Conditions}

Conformal gravity action (\ref{action}) can be modified by adding a Weyl invariant boundary term
\begin{align}
\tilde{\Gamma}_{CG}=\Gamma_{CG}+8\int_{\partial \mathcal{M}}d^3x\sqrt{|\gamma|}K^{ij}E_{ij}
\end{align}
which performs a Lagandre transformation of the action. Written in this form, the action is also finite on-shell, however its first variation 
\begin{equation} 
\delta \tilde{\Gamma}_{CG}=\int_{\partial\mathcal{M}}d^3x\sqrt{|\gamma|}\left(\tilde{\tau}_{ij}\delta\gamma_{ij}^{(0)}+\tilde{P}^{ij}\delta E_{ij}^{(2)}\right) \label{ltr1var}
\end{equation}
is an expression that contains exchanged roles of the source and the response function. The role of the source is played by the $E_{ij}^{(2)}$, while in (\ref{finres1}) that role belongs to $\gamma_{ij}^{(1)}$. The response function to $\gam$ in (\ref{finres1})  is $P_{ij}$ which is proportional to $E_{ij}^{(2)}$ (\ref{eq:CG18}) and in (\ref{ltr1var}) the response function $\tilde{P}_{ij}$ is proportional to $\gam$, precisely
\begin{align}
\tilde{P}_{ij}=\frac{4\sigma}{\ell}\gamma_{ij}^{(1)}.
\end{align}
The response function to a $\delta\gamma_{ij}^{(0)}$ in (\ref{ltr1var}) is stress tensor
\begin{align}
\tilde{\tau}_{ij} &= \tau_{ij}+\tfrac{2\sigma}{\ell} \,E_{\SO}^{kl}\psi^{\FO}_{kl}\,\ga^{\LO}_{ij} + \tfrac{8\sigma}{3\ell} \, E^{\SO}_{ij}\ga^{\FO} \nonumber\\
&\quad - \tfrac{4\sigma}{\ell}\,\big(E^{\SO}_{ik}\psi^{\FO k}_j + E^{\SO}_{jk}\psi^{\FO k}_i\big),
\end{align}
which now interestingly has zero trace, $\tilde{\tau}_i^i=0$.
In further consideration, when referring to the response functions, we will primarily think of the response functions from the action (\ref{action}).
In the following chapter we apply these results on the three prominent examples.

\section{Black hole solutions}

The charges and the response functions, one can compute explicitly on the asymtptocally (A)dS black hole solution of EG with cosmological constant, i.e. Schwarzschild black hole, MKR solution and the rotating black hole. 
The solution that obeys the Starobinsky boundary conditions, $\gam=0$ includes solutions of EG with cosmological constant, which are asymptotically (A)dS. It follows from the EOM that $E_{ij}^{(2)}=0$, which implies vanishing of the PMR. The stress energy tensor becomes 
\begin{align}
\tau_{ij}=\frac{4\sigma}{\ell}E^{(3)}_{ij}.
\end{align}
That agrees with the traceless and conserved stress energy tensor of EG \cite{deHaro:2000xn}, Maldacena's analysis and the work by Deser and Tekin \cite{Deser:2002jk}.

An interesting example is MKR solution that does not have vanishing  $\gam$ matrix in the FG expansion. In the equation (\ref{le}) we set $\sigma=-1$ and from the MKR solution (\ref{MKR1}) transform to the FG form. 
The transition from the original MKR solution to the FG form of the metric is performed by transformation of the coordinate $r(\rho)$ into $\rho$ with 
\begin{equation}
r(\rho)=\frac{a_{-1}}{\rho}+a_0+a_1\rho+a_2\rho^2+a_3\rho^3+a_4\rho^4. \label{expan}
\end{equation}
We insert the new coordinate in 
\begin{equation}
\frac{d\left(r(\rho)\right)^2}{V\left[r(\rho)\right]}=\frac{1}{\rho^2} \label{cond},
\end{equation}
 demand for the equality to hold, and read out the coefficients $a_i$, $i=(-1,0,1,2,3,4)$. We insert the coordinate $r(\rho)$ in the remaining components of the metric, expand them in the $\rho$ coordinate, and read out the matrices
 in the FG expansion
\begin{align}
\gamma_{ij}^{(0)}=diag\left(-1,1,1\right)
\end{align}
\begin{align}
\gam=diag\left(0,-2a,-2a\right)\label{gama1mkr}
\end{align}
\small
\begin{align}
\gamma_{ij}^{(2)}=  diag\left(\frac{1}{2}\left(a^2-\sqrt{1-12aM}\right),\frac{3a^2}{2}-\frac{1}{2}\sqrt{1-12aM} ,\frac{3a^2}{2}-\frac{1}{2}\sqrt{1-12aM}\right) \nonumber
\end{align}
\normalsize
\begin{align}
\gamma^{(3)}_{ij}=diag\left(\frac{4M}{3},\frac{1}{6}\left(-3a^2+4M+3a\sqrt{1-12aM}\right),\frac{1}{6}\left(-3a^2+4M+3a\sqrt{1-12aM}\right)\right)\nonumber
\end{align}
and for the response functions 
\begin{align}
\tau_{11}&= \frac{4 \left(a \left(\sqrt{1-12 a M}-1\right) \ell^2+6 M\right)}{3 \ell^4}\nonumber \\
\tau_{22}&=\frac{4}{3} \left(\frac{3 M}{\ell^2}+a \left(\sqrt{1-12 a M}-1\right)\right)\nonumber \\
\tau_{33}&=\frac{4 \left(a \left(\sqrt{1-12 a M}-1\right) \ell^2+3 M\right) \sin ^2(\theta )}{3 \ell^2} 
\end{align}
\begin{align}
P_{ij}=\left(
\begin{array}{ccc}
 \frac{4 \left(\sqrt{1-12 a M}-1\right)}{3 \ell^3} & 0 & 0 \\
 0 & \frac{2 \left(\sqrt{1-12 a M}-1\right)}{3 \ell} & 0 \\
 0 & 0 & \frac{2 \left(\sqrt{1-12 a M}-1\right) \sin ^2(\theta )}{3 \ell} \\
\end{array}
\right),
\end{align}
where the off-diagonal elements of the $\tau_{ij}$ are vanishing.
One can notice that the Rindler acceleration $a$ appears linearly in the partially massless response and makes it non-vanishing. Non-vanishing Rindler acceleration, appears as well quadratically in the trace of the stress tensor 
\begin{equation}
\tau_i^i=\frac{4a(-1+\sqrt{1-12aM})}{3\ell^2}
\end{equation}
and leads using the equation (\ref{currentcg}) 
to the charge 
\begin{equation}
Q_{ij}=\left(
\begin{array}{ccc}
 \frac{8 \left(A \left(\sqrt{1-12 A M}-1\right) L^2+6 M\right)}{3 L^4} & 0 & 0 \\
 0 & \frac{8 M}{L^2} & 0 \\
 0 & 0 & \frac{8 M \sin ^2(\theta )}{L^2} \\
\end{array}
\right) \label{chargemkrsph}.
\end{equation}
Conserved charge associated with the Killing vector $\partial_t$, using (\ref{charge1}) 
with normalisation of action $\alpha_{CG}=\frac{1}{64\pi}$ gives
\begin{equation}
Q[\partial_t]=\frac{M}{\ell^2}-a(1-\sqrt{1-12aM}).\label{holrenmkrcharge}
\end{equation}
Using the Wald's approach the on-shell action gives for the entropy 
\begin{equation}
S=\frac{A_h}{4\ell^2}
\end{equation}
for $A_h=4\pi r_h^2$ where $r_h$ is area of the horizon $k(r_h)=0$. 
Where we notice that the area law is obeyed  despite the fact that we are considering higher-derivative gravity theory. 

On the rotating black hole example we consider solution in AdS with Rindler hair. The solution is parametrised with Rindler acceleration $\mu$ and rotation parameter $\tilde{a}$, however the mass parameter vanishes.
The fact that the mass parameter vanishes leads to vanishing of the PMR, $P_{ij}=0$, which means that in order for existence of PMR $\gam\neq0$ is necessary but not sufficient.
The conserved energy is \begin{align}
E=-\frac{\tilde{a}^2\mu}{\ell^2\left(1-\frac{\tilde{a}^2}{\ell^2}\right)^2}.
\end{align}

\newpage	
\chapter{Canonical Analysis of Conformal Gravity}

The main goal of this chapter is to present canonical analysis of CG. Analysis of CG using the holographic renormalisation as described in the first chapter is complemented and supplemented via canonical analysis. That way one obtains detail insight into the boundary charges. 
Canonical analyses of higher derivative gravities in four dimensions have been done earlier \cite{Kluson:2013hza} pointing out that theory with most symmetries is CG. It has also been applied to lower dimensional gravitational theories. In conformally invariant three dimensional Chern-Simons gravity \cite{Afshar:2011qw}, the charge that describes conformal invariance vanishes as well as in four dimensions. In three dimensions, the result depends on the Weyl factor, if the Weyl factor varies freely the corresponding charge does not vanish. That leads to an enhancement of the algebra at the boundary which is consisted of the two copies of Virasoro algebra, with current $U(1)$ algebra.  In four dimensions, however, as we shall demonstrate below, Weyl charge vanishes even in the case of freely varying Weyl factor. 

The physical system that is complicated and non-linear, however it contains global symmetries and conserved quantities, can be considered using canonical analysis of conserved quantities as one of the most useful analytic tools to understand the system. The conserved quantities in the ADM split have been studied for asymptotically flat dynamical spacetimes exploring the subtleties in diffeomorphism invariance. 
The notion of global symmetry is given by asymptotic symmetries, equivalence classes of diffeomorphisms that exhibit analogous asymptotic behaviour at infinity. 
In other words, asymptotic symmetrys are defined as gauge transformations that leave the field configurations that are considered, asymptotically invariant. Furthermore, they are essential to define the total ("global") charges. \cite{Brown:1986nw, Abbott:1981ff, Abbott:1982jh}.

The notion of asymptotic symmetry naturally depend vastly on the boundary conditions. The imposition of boundary conditions causes true gauge symmetries to be merely a subset of the entire diffeomeophism group that allows for the non-trivial asymptotic symmetries. Three most prominent reasons to study asymptotic symmetries and corresponding conserved charges of AdS spacetimes are
\begin{enumerate}
\item Simply to gain further insight into the asymptotic symmetry in gravity. Empty AdS is maximally symmetric solution and studying  asymtpotically AdS  spaces is simple and natural choice.
\item The found structure in the AdS is richer then one obtained for the asymptotically flat space, which is connected too the fact that multiple moments of a field in AdS decay at the equal rate at infinity \cite{Fischetti:2012rd,Horowitz:2000fm}. The asymptotically flat spacetime is dominated by monopoles, while AdS equally admits higher multipoles. Therefore, one is interested not just into global charges, (e.g. total energy), rather into local densities of the charges at the boundary. Actually, it is natural to study entire boundary stress energy tensor.
\item Conserved charges have fundamental reason to the AdS/CFT correspondence, most oftenly used in these times.
\end{enumerate}
\footnote{Computation of the variations using the constraints, as a general method, has been introduced by Ter Haar in 1971.}
Therefore, we devote this chapter to analysis of the CG charges, while the analysis of the asymptotic symmetry algebra and richness of its structure is a theme of the following chapter.

From CG Lagrangian 
\begin{equation}
\mathcal{L}=-\frac{1}{4}\omega_{g}C^a{}_{bcd}C_a{}^{bcd} \label{langcg}
\end{equation}
where $\omega_g$ is $\sqrt{q}d^4x$ a volume form, we split the Lagrangian (\ref{langcg}) in the Arnowitt-Deser-Misner (ADM) decomposition. We introduce a more general formalism which is not defined in a given basis, while the traditional ADM formalism, in coordinate system, is presented in the appendix: Canonical Analysis of Conformal Gravity: ADM Decomposition. 

Conisder a function t on a manifold $\mathcal{M}$ which we call time. We assume it to foliate the manifold with spatial hypersurfaces $\sigma_t$ on which $t=const$. Kernel of the one-form $\nabla_a t$ for $\nabla$ Levi-Civita connection defined on the manifold, defines a tangent bundle $\tau\Sigma$. The spacial hyper surfaces are defined when 
\begin{equation}
g^{ab}\nabla_at\nabla_bt<0,
\end{equation}
 the future pointing normal vector is $n_{\alpha}=\alpha\nabla_at$ with $\alpha$ a normalisation constant completely defined in the terms of so called lapse function N. 
A congruence of curves and $t^a$, their tangent vector fields, are related as \begin{equation}
t^a\nabla_at=1,
\end{equation}
$t^a$ can be decomposed in
\begin{equation}
t^a=Nn^a+N^a.
\end{equation}
N measures a tick rate for a physical observer that follows normal $n^a$, while $N^a$ is defined as shift vector. If we imagine a manifold in terms of the coordinate grid, drag describes its shift orthogonally to $n^{a}=0$. That leads to 
\begin{align}
t^an_a=\alpha=-N\\
n_a=-N\nabla_at.
\end{align}
One can write the decomposition of the metric $g_{ab}$ with the metric $h_{ab}$ on $\Sigma$ and  the normal vectors
\begin{equation}
g_{ab}=-n_an_b+h_{ab},
\end{equation}
which is called 3+1 or ADM decomposition of the metric, while the Levi-Civita connection on the boundary is called $D_a$.  With the boundary metric $h_{ab}$ in the form $h_a^b$ and the normal vector $n_a$, 
one can split the tensor fields defined on the manifold $\mathcal{M}$. When we have split the four dimensional tensor field expressing it solely in the terms of the boundary indices, we say that the tensor field has been projected to the boundary. 
We denote the projection of the tensor with the $h_{a}{^b}$ metric with 
\begin{align}
P=\perp\mathcal{P}.
\end{align}
Where tensors $P$ on $\Sigma$ can be obtained from the tensor fields $\mathcal{P}$ on the manifold $\mathcal{M}$.
The relation of the Levi-Civita connections reads
\begin{equation}
DP=\perp[\nabla (\perp\mathcal{P})].
\end{equation}
while the decomposition of the determinant is
\begin{align}
\sqrt{g}=N\sqrt{h}.
\end{align}
The bending of the surface and curves with respect to the space in which they are embedded, defined by the change of the normal vector projected on the hypersurface, defines extrinsic curvature $K_{ab}$.
Extrinsic curvature  of the spatial hypsersurfaces is 
\begin{equation}
K_{ab}=h_a^{\ c}\nabla_cn_b= \frac{1}{2}\,^{\ms{(4)}}\pounds_n h_{ab},
\end{equation}
for ${}^{(4)}\pounds_n$ Lie derivative in the $n^a$ direction, while we reserve the symbol $\pounds$ without prefix for the Lie derivative on $\Sigma$. Normal of the 4D Lie derivative of the covariant spatial tenor field  on $\Sigma$
\begin{equation}
n^{a_i}\,^{\ms{(4)}}\pounds_n P_{a_1\cdots a_i \cdots a_n}=- P_{a_1\cdots a_i \cdots a_n}(n\nabla)n^{a_i}+ P_{a_1\cdots a_i \cdots a_n}(n\nabla)n^{a_i}=0,
\end{equation}
is spatial, while the the 4D Lie derivative along the spatial vector $V^a$ becomes spatial (3D) Lie derivative on $\Sigma$ when projected to the tensor bundle on $\Sigma$
\begin{equation}
\perp \,^{\ms{(4)}}\pounds_V P_{a_1\cdots a_i \cdots a_n}=\pounds_V P_{a_1\cdots a_i \cdots a_n}.\label{proj1}
\end{equation}
The relation (\ref{proj1}) plays a key role in definition of velocities 
\begin{eqnarray}
\dot{h}_{ab}&=&\perp \,^{\ms{(4)}}\pounds_t h_{ab}=N\,^{\ms{(4)}}\pounds_n h_{ab}+\pounds_N h_{ab}=\nonumber\\
&=&2\left(NK_{ab}+D_{(a}N_{b)}\right),
\end{eqnarray}
it measure the change of the spatial quantity  when $t$ changes on the spatial slice. With the definition of the ADM decomposition of the curvatures in the appendix: Canonical Analysis of Conformal Gravity: ADM Decomposition of Curvatures 
 we obtain the decomposed Lagrangian of CG
\begin{equation}
\mathcal{L}=N\omega_h\left(\perp n^eC_{ebcd}\perp n_fC^{fbcd}-2\perp n^en^fC_{aecf}\perp n_gn_hC^{agch}\right).
\end{equation}

As we already know, CG is gravity theory of the fourth order in derivatives, while the terms quadratic in curvature are of the second order in time derivatives.
In other words, our Lagrangian contains acceleration of $h_{ab}$, i.e. velocity of $K_{ab}$, which is in contrast to GR in ADM form that contains first  order in time derivatives. 
The Hamiltonian formulation, defines only first order time derivatives 
$\frac{df}{dt}=\{f,H\}$. In order to be able to use Hamiltonian formulation, we define an additional constraint. 
We consider $K_{ab}$ as a canonical coordinate independent on $h_{ab}$ and relate it with $\dot{h}_{ab}$ via constraint with corresponding Lagrange multiplier, $\lambda^{ab}$. The Lagrangian of CG in the ADM decomposition then reads
\begin{align}
\mathcal{L}&=N\omega_h\bigg\{ -\frac{1}{2}\mathcal{T}^{abcd}\left[R_{ab}+K_{ab}K-\frac{1}{N}\left(\dot{K}_{ab}-\pounds_NK_{ab}-D_aD_bN\right)\right]\nonumber\\ &\times\left[R_{cd}+K_{cd}K-\frac{1}{N}\left(\dot{K}_{cd}-\pounds_NK_{cd}-D_cD_dN\right)\right]\nonumber\\ 
&+B_{abc}B^{abc}+\lambda^{ab}\left[\frac{1}{N}\left(\dot{h}_{ab}-\pounds_Nh_{ab}\right)-2K_{ab}\right]\bigg\}.\label{Lagrangean1}
\end{align}
where 
\begin{align}
\mathcal{T}^{abcd}=\frac{1}{2}(h^{ac}h^{bd}+h^{ad}h^{bc})-\frac{1}{3}h^{ab}h^{cd}
 \end{align}
 denotes DeWitt metric. 
Lagrangian (\ref{Lagrangean1}), function of the variables and velocities
\begin{align}
\mathcal{L}(N,N^a,h_{ab},\partial_th_{ab},K_{ab},\partial_t K_{ab},\lambda^{ab})
\end{align}
allows us to immediately notice primary constraints
\begin{align}
\Pi_N&=\frac{\partial\mathcal{L}}{\partial (\partial_t N)}\approx0 &&\Pi_a=\frac{\partial \mathcal{L}}{\partial (\partial_t N^a)}\approx0 \label{prim1} \\ 
\Pi^{\lambda}_{ab}&=\frac{\mathcal{L}}{\partial(\partial_t\lambda^{ab} )}\approx0&& \label{prim2}
\end{align}
To write the Lagrangian in the Hamiltonian formulation (\ref{hc0}) one needs to identify the momenta and corresponding canonical variables, that requires analysis of the constraints and for the consistency conditions (\ref{cc1}, \ref{cc2}, \ref{cc3}). Since that procedure requires introducing the Dirac brackets, it is convenient to first inspect whether one can read out the momenta and corresponding variables directly from the Lagrangian following the method of \cite{Faddeev:1988qp}. Since the Lagrangian (\ref{Lagrangean1}) allows for identification of the momenta conjugate to $h_{ab}$ and $K_{ab}$, we denote them with $\Pi_{h}^{ab}$ and $\Pi_{K}^{ab}$
\begin{align}
\Pi_h^{ab}&=\frac{\partial \mathcal{L}}{\partial(\partial_th_{ab})}=\sqrt{h}
\lambda^{ab}\label{primh}\\
\Pi_K^{ab}&=\frac{\partial\mathcal{L}}{\partial(\partial_tK_{ab})}=\sqrt{h}2\alpha C^a{}_{\textbf{n}}{}^b{}_{\textbf{n}}\label{primk}
\end{align}
 respectively. For the projection of the Weyl tensor
 \begin{align}
 \Pi_{K}^{ab}=-\alpha\sqrt{h}\mathcal{T}^{abcd}\bigg(\mathcal{L}K_{cd}-R_{cd}-K_{cd}K-\frac{1}{N}D_cD_dN\bigg),
 \end{align}
that can be recognised from (\ref{Lagrangean1}).
 Since the DeWitt metric and the projection of the Weyl tensor are traceless we will have to ensure that $\Pi_K^{ab}$ is traceless and define one more primary constraint.  
  We can rewrite the Lagrangian (\ref{Lagrangean1}) via (\ref{prim1}), (\ref{prim2}), (\ref{primk}) and (\ref{primh}) and the canonical variables
\begin{eqnarray}
\mathcal{L}&=&\Pi_K^{ab}\dot{K}_{ab}+\Pi_h^{ab}\dot{h}_{ab}+N\bigg[\omega_h^{-1}\frac{\Pi_K^{ab}\Pi^K_{ab}}{2}-\Pi_K^{ab}\left(R_{ab}+K_{ab}K\right)+\omega_hB_{abc}B^{abc}\nonumber\\ &&-2\Pi_h^{ab}K_{ab}\bigg]
-\Pi_K^{ab}D_aD_bN-\Pi_K^{ab}\pounds_NK_{ab}-\Pi_h^{ab}\pounds_Nh_{ab}-\lambda_P\Pi_K^{ab}h_{ab}.\label{Lagrangean2}
\end{eqnarray}
To write the Lagrangian in the form that manifestly contains the constraints of the Hamiltonian, using partial integration we rewrite the Lagrangian in the form that there is no lapse or shift under covariant derivatives 
\begin{align}
L&=\int_\Sigma \bigg\{\Pi_K^{ab}\dot{K}_{ab}+\Pi_h^{ab}\dot{h}_{ab}-N\bigg[-\omega_h^{-1}\frac{\Pi_K^{ab}\Pi^K_{ab}}{2}+\Pi_K^{ab}\left(R_{ab}+K_{ab}K\right)-\omega_hB_{abc}B^{abc} \nonumber\\
&+2\Pi_h^{ab}K_{ab}+D_aD_b\Pi_K^{ab}\bigg]-N^c\left[\Pi_K^{ab}D_cK_{ab}-2D_a\left(\Pi_K^{ab}K_{bc}\right)-D_a\Pi_h^{ab}h_{bc}\right]-\lambda_P\Pi_K^{ab}h_{ab}\bigg\} \nonumber\\
&-\int_{\partial\Sigma}\ast\left[\Pi_K^{ab}D_bN-D_b\Pi_K^{ab}N+2N^c\left(\Pi_h^{ab}h_{bc}+\Pi_K^{ab}K_{bc}\right)\right].\label{Lagrangean3}
\end{align}
Where the $\ast$ denotes contraction with the free index that belongs to one of the indices of the differential form hidden in tensor densities that build momentum variables.
The term that is multiplied with the Lagrange multiplier $\lambda_P$ is the term that ensures the tracelessness of $\Pi_K^{ab}$ and new primary constraint. Demanding that primary constraints $\Pi_N$ and $\Pi_{a}^{\vec{N}}$ are conserved in time we can from the (\ref{Lagrangean3}) identify the constraints  
\begin{align}
\mathcal{H}_\perp&=-\omega_h^{-1}\frac{\Pi_K^{ab}\Pi^K_{ab}}{2}+D_aD_b\Pi_K^{ab}+\Pi_K^{ab}\left(R_{ab}+K_{ab}K\right)\nonumber \\ &-\omega_hB_{abc}B^{abc}+2\Pi_h^{ab}K_{ab},
\end{align}
the Hamiltonian constraint that is multiplied by $N$, and 
\begin{align}
\mathcal{V}_c&=\Pi_K^{ab}D_cK_{ab}-2D_a\left(\Pi_K^{ab}K_{bc}\right)-D_a\Pi_h^{ab}h_{bc},
\end{align}
vector constraint that is multiplied with $N^a$. The constraint that ensures tracelessness and is multiplied with $\lambda_P$ we define with $\mathcal{P}\equiv \Pi_K^{ab}h_{ab}$. 
The  constraints $N$ and $N^a$ can be considered as Lagrange multipliers, however we consider them to be canonical coordinates, since they multiply  secondary constraints $\mathcal{H}_{\perp}$ and $\mathcal{V}_c$. 
This step may seem superficial, however, it ensured that the gauge generators found via  Castellani algorithm have accurate space-time interpretation. This is important in considering the asymptotic symmetry algebra of CG. 
\section{Total Hamiltonian of Conformal Gravity}
We can write the total Hamiltonian (\ref{ht}), from the canonical Hamiltonian (\ref{Lagrangean3}) expressing the terms using the constraints, 
\eqref{Lagrangean3} now reads
\begin{eqnarray}
H_T=\int_{\Sigma}\left(\lambda_N\Pi_N+\lambda_{\vec{N}}^a\Pi^{\vec{N}}_{a}+\lambda_P\mathcal{P}+N\mathcal{H}_\perp+N^a\mathcal{V}_a\right)+\int_{\partial\Sigma}\left(\mathcal{Q}_\perp+\mathcal{Q}_D\right).\label{scw}
\end{eqnarray}
where we have denoted the surface terms 
\begin{align}
\mathcal{Q}_{\perp}&=\ast\left[\Pi_K^{ab}D_bN-D_b\Pi_K^{ab}N\right]
\\
\mathcal{Q}_D&=\ast\left[2N^c\left(\Pi_h^{ab}h_{bc}+\Pi_K^{ab}K_{bc}\right)\right]. 
\end{align}
with $\mathcal{Q}_{\perp}$ and $\mathcal{Q}_D$.
Surface terms appear because of the integration by parts of 
\begin{align}
\int_{\Sigma}d^3x \Pi_K^{ab}D_aD_bN&=\int_{\Sigma}\left(D_a\big(Pi_K^{ab}D_bN\big)-D_b\big(D_a\Pi_K^{ab}N\big)+ND_aD_b\Pi_K^{ab}\right)\\
&=\int_{\Sigma}d^3xND_aD_b\Pi_K^{ab}+\oint_{\partial\Sigma}\ast(D_bN\Pi_K^{ab}-ND_a\Pi_K^{ab}),
\end{align}
and integration by parts of the vector constraint.
One can define the canonical pairs $(h_{ab},\Pi_h^{cd})$, $(K_{ab},\Pi_K^{cd})$,
 $(N^a,\Pi_c^{\vec{N}})$ and $(N,\Pi_{N})$ and define the canonical Poisson bracket with 
\begin{align}
\{g_A(x),p^B(x')\}=\delta(x-x')\delta_A^B\label{krondel}
\end{align}
where $\delta_A^B$ denotes symmetrized product of delta Kronecker symbols.
From (\ref{krondel}) and following the consistency conditions (\ref{cc1}, \ref{cc2},\ref{cc3}), we can define one further secondary constraint, that we denote with $\mathcal{W}$
\begin{align}
\left\{H_T,\mathcal{P}\right\}&=N\left(\Pi_K^{ab}K_{ab}+2\Pi_h^{ab}h_{ab}\right)+NK\mathcal{P}-D_c\left(N^c\mathcal{P}\right)\\&\approx N\left(\Pi_K^{ab}K_{ab}+2\Pi_h^{ab}h_{ab}\right)\equiv N\mathcal{W}.
\end{align} 
To find the gauge generators, improved generators and their algebra, we have to compute the Poisson bracket algebra among the constraints.
For that, we define smeared function on an example of a momentum constraint. 
The smeared momentum constraint can be written as a functional 
\begin{align}
V[\vec{X}]=\int_{\Sigma}d^3x X^aV_a\label{smear1}
\end{align}
for $\vec{X}$ an arbitrary test vector on $\Sigma$.
In this sense, one can alternatively write the momentum constraint with 
\begin{align}
V[\vec{X}]=\int_{\Sigma}d^3x(\Pi_h^{ab}\mathcal{L}_{\vec{X}}h_{ab}+\Pi_K^{ab}\mathcal{L}_{\vec{X}}K_{ab})-\oint_{\partial\Sigma}\ast(X_b\Pi_h^{ab}+X^{a}\Pi_K^{bc}K_{bc})
\end{align}
for $\mathcal{L}_{\vec{X}}h_{ab}=2D_{(a}X_{b)}$.
Where the terms under the first integral, for $\vec{X}=\vec{N}$, read
\begin{align}
\int_{\Sigma}\Pi_h^{ab}\pounds_Nh_{ab}&=-2\int_{\Sigma}D_a\Pi_h^{ab}h_{bc}N^c+2\int_{\partial\Sigma}\ast\Pi_h^{ab}h_{bc}N^c\\
\int_{\Sigma}\Pi_K^{ab}\pounds_NK_{ab}&=\int_{\Sigma}\big[\Pi_K^{ab}D_cK_{ab}-2D_a\left(\Pi_K^{ab}K_{bc} \right)\big]N^c+2\int_{\partial\Sigma}\ast\Pi_{K}^{ab}K_{bc}N^c.
\end{align}
Vector constraint satisfies the Lie algebra 
\begin{equation}
\big\{V[\vec{X}],V[\vec{Y}]\big\}=V[[\vec{X},\vec{Y}]],
\end{equation}
that is obeyed since the Lie derivative has the property 
\begin{align}
\mathcal{L}_{\vec{X}}\mathcal{L}_{\vec{Y}}-\mathcal{L}_{\vec{Y}}\mathcal{L}_{\vec{X}}=\mathcal{L}_{[\vec{X},\vec{Y}]}
\end{align}
where $\vec{X}$ and $\vec{Y}$ satisfy
\begin{align}
[\vec{X},\vec{Y}]^a=X^{b}\partial_bY^a-Y^b\partial_bX^a.
\end{align}
Under spatial diffeomorphisms, the variables $N, N^a,h_{ab}$ and $K_{ab}$ are scalar or tensor fields while the corresponding canonical momenta are scalar or tensor densities with unit weight. \footnote{Under spatial diffeomorphisms, all the constraints are scalar or tensor densities.}
For the Hamiltonian constraint, the smeared function acts as one of the scalars
\begin{align}
\mathcal{H}_{\perp}[\epsilon]=\int_{\Sigma}d^3x\epsilon\mathcal{H}_{\perp}\label{smear2}
\end{align}
for $\epsilon$ an arbitrary function on $\Sigma$.

If we write the total Hamiltonian using that conventions we can write
\begin{equation}
H_T=H_0[N]+V[\vec{N}]+P[\psi]+\sum_{i=1}^4 C_i[\psi^{(i)}]+Q_{\perp}[N]+Q_{D}[\vec{N}] \label{htw2}
\end{equation}
where we define the functionals in the form of (\ref{smear1}) and (\ref{smear2}). The terms in the Hamiltonian are the following
\begin{align}
\sum_{i=1}^4C_i[\phi^{(i)}]\equiv\int_{\Sigma}\phi^{(1)}_{ab}\big(\Pi_h^{ab}-\omega_h\lambda^{ab}\big)+\phi_{(2)}^{ab}\Pi_{ab}^{\lambda}+\phi_a^{(3)}\Pi_{ab}^{\lambda}+\phi_a^{(3)} \Pi_{a}^{\vec{N}}+\phi^{(4)}\Pi_N
\end{align}
for $\phi^{(2)},\phi^{(3)},\phi^{(4)}$ Lagrange multipliers of $\lambda^{ab},N^a,N$ respectively. 
The remaining functionals read
\begin{align}
H_0[N]&\equiv\int_{\Sigma}N\bigg[-\omega_h^{-1}\frac{\Pi_K^{ab}\Pi_{ab}^{K}}{2}+D_aD_b\Pi_K^{ab}+\Pi_K^{ab}(R_{ab}K_{ab}K)+2\Pi_h^{ab}K_{ab}\nonumber \\&-N\omega_hB_{abc}B^{abc}\bigg]\\
V[\vec{N}]&\equiv \int_{\Sigma}N^c\big[\Pi_K^{ab}D_cK_{ab}-2D_a\left(\Pi_K^{ab}K_{bc}\right)-D_a\Pi_h^{ab}h_{ab}\big]\\
P[\psi]&\equiv\int_{\Sigma}\psi\Pi_{K}^{ab}h_{ab}.
\end{align}
Note that in the Hamiltonian (\ref{htw2}), in comparison to the Hamiltonian (\ref{scw}) we have two additional constraints. Namely, the constraint that ensures that the momentum from the $\lambda^{ab}$ vanishes, and the constraint that ensures that the momentum of $h^{ab}$ variable is proportional to $\lambda^{ab}$. These, are exactly the constraints that can be immediately identified, as we did when considering (\ref{htw2}) or treated with Dirac brackets, as we show below. 

\subsection{Poisson Bracket Algebra}
To consider the Poisson bracket algebra, constraints need to satisfy consistency conditions. For the $\Pi_{ab}^{\lambda}$ and $\Pi_{h}^{ab}-\omega_h\lambda^{ab}$ the Poisson brackets 
\begin{align}
\big\{H_T,\Pi_{ab}^{\lambda} \big\}&=-\omega_{h}\phi^{(1)}\approx0\\
\big\{H_T,\Pi_h^{ab}-\omega_h\lambda^{ab}\big\}&=\frac{\delta H_T}{\delta h_{ab}}+\frac{1}{2}\omega_hh^{cd}\frac{\delta H_T}{\delta_h^{cd}}+\omega_h\phi^{ab}_{(2)}\approx0.
\end{align}
define  $\phi_{ab}^{(1)}$ and $\phi_{ab}^{(2)}$. That implies that 
$\Pi_{ab}^{\lambda}$ and $\Pi_{h}^{ab}-\omega_h\lambda^{ab}$ are second class constraints which need to be considered using the Dirac brackets.  Here, we set them strongly to zero. The Poisson brackets with the remaining primary constraints give 
\begin{align}
\{H_T,\Pi_N\}=\mathcal{H}_0\approx0 \\
\{H_{T},\Pi_{\vec{N}}\}=\mathcal{V}_a\approx0
\end{align}
where consistency for the third constraint $\mathcal{P}$ that results with the new constraint $\mathcal{W}$ was verified in (\ref{scw}) \footnote{see appendix: Canonical Analysis of Conformal Gravity: Variations}. 
The diffeomorphism constraint $\{\cdot,V[\vec{X}]\}$ of arbitrary tensor density on the phase space defined with $(h,K,\Pi_h,\Pi_K)$ is defined with 
\begin{align}
\{\Phi,V[\vec{X}]\}=\pounds_{\vec{X}}\Phi\label{eqpb}
\end{align}
where the change under diffeomorphisms, of the canonical coordinate $h_{ab}$ and its momenta reads
\begin{align}
h_{ab}\rightarrow h_{ab}+\pounds_{\vec{X}}h_{ab}, && \Pi_h^{ab}\rightarrow\Pi_h^{ab}-\pounds_{\vec{X}}\Pi_h^{ab}.
\end{align}
To compute this bracket one needs to consider the scalar density $\psi$ as a form that has a maximal degree on a manifold
\begin{align}
\pounds_{\vec{\lambda}} \Psi=\mathrm{d}(\iota_{\vec{\lambda}}\Psi)+\iota_{\vec{\lambda}}\mathrm{d}\Psi=\mathrm{d}(\iota_{\vec{\lambda}}\Psi).
\end{align}
which means that identity
\begin{equation}
\int_\Sigma Y_{a_1\cdots a_n}\pounds_{\vec{\lambda}} \Psi^{a_1\cdots a_n}=-\int_\Sigma \pounds_{\vec{\lambda}} Y_{a_1\cdots a_n}\Psi^{a_1\cdots a_n},
\end{equation}
holds up to boundary terms. That allows us to treat the Lie derivative as partial integration. The Poisson brackets for the diffeomorphism constraint then read 
\begin{align}
\left\{V\left[\vec{X}\right],V\left[\vec{Y}\right]\right\}&=V\left[\pounds_{\vec{X}}\vec{Y}\right],\nonumber\\
\left\{V\left[\vec{X}\right],H_\perp[\epsilon]\right\}&=H_\perp\left[\pounds_{\vec{X}}\epsilon\right],\nonumber\\
\left\{V\left[\vec{X}\right],P[\epsilon]\right\}&=P\left[\pounds_{\vec{X}}\epsilon\right],\nonumber\\
\left\{V\left[\vec{X},\right],W[\epsilon]\right\}&=W\left[\pounds_{\vec{X}}\epsilon\right].
\end{align}
The brackets for the $P$ constraint are
\begin{align}
\left\{P[\epsilon],W[\eta]\right\}&=P[\epsilon\eta],\nonumber\\
\left\{P[\epsilon],H_\perp[\eta]\right\}&=-W[\epsilon\eta]-P[\epsilon\eta K],
\end{align}
while the one for $\mathcal{W}$ and $H_{0}$ are
\begin{align}
\left\{W[\epsilon],H_\perp[\eta]\right\}&=H_\perp[\epsilon\eta]+P[D^2\epsilon\eta+\epsilon D^2\eta-D\epsilon\cdot D\eta)],\nonumber\\
\left\{H_\perp[\epsilon],H_\perp[\eta]\right\}&=V\left[\epsilon D^a\eta-\eta D^a\epsilon\right]+P\left[\left(\epsilon D^a\eta-\eta D^a\epsilon\right)\left(D_cK^c_{\ a}-D_cK\right)\right].\label{eq:DiracAlg}
\end{align}
Now, we can count the degrees of freedom. 
 Among the 32 phase space coordinates we found 10 constraints that are first class, that eliminates $2\times10$ coordinates from phase space. The remaining number of the physical degrees of freedom is $12/2=6$. CG degrees of freedom are divided in 2 degrees of freedom that describe massless graviton, and 4 degrees of freedom that belong to partially massless graviton.

\subsection{Gauge Generators of Conformal Gravity}

To obtain the generators of CG we follow the procedure described in section "Castellani algorithm".
Since the algorithm uses PFCs for the start of the Castellani procedure, the start is determined with PFCs
\begin{align}
\Pi_N\approx0,&&\Pi_{\vec{N}_i}\approx0,&& \mathcal{P}\approx0.
\end{align}
Consider $\mathcal{G}_1=\Pi_N$. Castellani algorithm then suggests 
\begin{align} 
\mathcal{G}_1&=\Pi_N,\\
\mathcal{G}_0+\{\Pi_N,H_T\}&=PFC, \\
\{\mathcal{G}_0,H_T\}&=PFC. 
\end{align}
The ansatz for the linear combination (\ref{castel}) is
\begin{align}
PFC(x)=\int_{\Sigma}\left(\alpha_1(x,y)\Pi_{\vec{N}_i}(y)+\alpha_2(x,y)\Pi_N(y)+\alpha_3(x,y)\mathcal{P}(y)\right).\label{ansatz1}
\end{align}
Determination of the coefficients leads to
\begin{align}
\label{eq:c1}
\alpha_1^{a}(x, y)&=\delta^{3}(x-y)D^{a}N(y)+N(y)\gamma^{ab}D_{b}\delta^{3}(x-y)\\
\alpha_2(x, y)&=N^{a}(y)\partial_{a}\delta^{3}(x-y)\,\,\,,\,\, \\\alpha_3(x, y)&=\frac{\lambda_{\mathcal{P}}}{N}(y)\delta^{3}(x-y)
\end{align}
which writting in the form (\ref{geng}) allows us to 
write the canonical gauge generator for the diffeomorphisms that are orthogonal to the hypersurface 
\begin{equation}
\label{eq:g1}
G_{\perp}[\epsilon, \dot{\epsilon}]= \int \! \left[\dot{\epsilon}^{\FO} \Pi_{N}+ \epsilon \left(\mathcal{H}+\pounds_{\vec{N}}\Pi_{N}+{\Pi_{\vec{N}}}_{a} D^{a}N+D_{a}(\Pi_{\vec{N}}^{a}N)+\frac{\lambda_{\mathcal{P}}}{N}\mathcal{P}\right)\right].
\end{equation}
Choosing that $\mathcal{G}_{1a}=\Pi_{\vec{N}a}$ we obtain the recursion relations
\begin{align}
\mathcal{G}_{1a}&=\Pi_{\vec{N}_a}\\
\mathcal{G}_{0a}+\{\Pi_{\vec{N}a},H_T\}&=PFC_a\\
\{\mathcal{G}_{0a},H_T\}&=PFC_a
\end{align}
that with an ansatz 
\begin{align}
PFC_a(x)&=\int_{y}\left(\alpha_{1a}(x,y)\Pi_{\vec{N}_i}(y)+\alpha_{2a}(x,y)\Pi_N(y)+\alpha_{3a}(x,y)\mathcal{P}(y)\right).
\end{align}
 lead  to coefficients
\begin{align}
\alpha_{1a}^b(x,y)&=\delta^3(x-y)D_aN^b(y)+N^c(y)\delta_a^bD_c\delta^3(x-y)\\
\alpha_2(x,y)&=\delta^3(x-y)D_aN(y)\\
\alpha_3(x,y)&=0,
\end{align}
and generator for spatial diffeomorphisms
\begin{align}
G_D[\epsilon^{a}, \dot{\epsilon}^{a}]= \int \left[ \dot{\epsilon}^{\FO a} {\Pi_{\vec{N}}}_{a}+ \epsilon^{a} \left(\mathcal{V}_{a}+\Pi_{N} D_{a}N+\pounds_{\vec{N}}{\Pi_{\vec{N}}}_{a}\right)\right]\label{eq:g2}.
\end{align}
For $\mathcal{G_1}=\mathcal{P}$ Castellani algorithm reads
\begin{align}
\mathcal{G}_1&=\mathcal{P}\\
\mathcal{G}_0+\{\mathcal{P},H_T\}&=PFC \\
\{\mathcal{G}_0,H_T\}&=PFC
\end{align}
for the ansatz generator of the form equal to (\ref{ansatz1}).
The coefficients for this case read
\begin{align}
\alpha_1^a(x,y)&=0\\
\alpha_2(x,y)&=N^2(y)\delta^3(x-y) \\
\alpha_3(x,y)&=N^a(y)D_a\delta^3(x-y)+\frac{\lambda_N}{N}(y)\delta^3(x-y),\end{align}
that inserting in (\ref{geng}) lead to 
\begin{align}
G_{W}[w, \dot{w}]= \int \! \left[ \frac{\dot{w}^{\FO }}{N}\mathcal{P} (x)+ w \left( \mathcal{W}+N\Pi_{N}+\pounds_{\vec{N}}\frac{\mathcal{P}}{N}\right)\right].\label{eq:g3}
\end{align}
One can compare the generators of the diffeomeorphisms orthogonally and in the direction of the spatial hypersurface to the ones from GR \cite{Castellani:1981us} and notice the same structure apart from the terms that involve $\mathcal{P}$. Naturally, the generator involving the Weyl symmetry does not appear among generators in EG.

The relation of the generators of the diffeomorphism orthogonal and transversal to the hypersurface and the diffeomorphisms generated with a vector field $\xi^a$ on the manifold $\mathcal{M}$, is 
\begin{align}
\xi^a&=\epsilon_{\perp}n^a+\epsilon^a \label{decompx}\\
\text{ for } \epsilon^a&=h^a_b\xi^b \text{ and }\epsilon_{\perp}=-n_a\xi^a. \label{decompdif}
\end{align}
The generators (\ref{eq:g1}), (\ref{eq:g2}) and (\ref{eq:g3}) generate Weyl rescalings and diffeomorphisms. For the ADM decomposition of the metric 
$g_{tt}=-N^2+N^aN^bh_{ab}$, $g_{tb}=N^ah_{ab}$ and $g_{ab}=h_{ab}$ and the identification of the diffeomorphisms transversal  to the hypersurface $\epsilon_{\perp}=N\xi^t$ and along the hyper surface $\epsilon^a=\xi^a+N^a\xi^a$ it follows
\begin{align}
\{g_{\m\n},G_W[\omega]\}&=2\omega g_{\m\n}\\
\{g_{\m\n},G_{\perp}[\epsilon_{\perp}]+G_{D}[\vec{\epsilon}]\}&=\pounds_{\xi}g_{\m\n}
\end{align}
These generators differ from the generators in \cite{Kluson:2013hza} evaluated on the full phase space.  The generator of Weyl transformations does not change the shift vector field $N^i$ and takes into account the constraint $\mathcal{P}$ responsible for the correct transformation of $K_{ab}$ and $\Pi_{h}^{ab}$.
Gauge generators modify the surface deformation algebra which one can see from the Poisson bracket algebra of the constraints $H_{\perp}$ and $V$, (\ref{eq:DiracAlg})
\begin{align}
\left[\xi,\chi\right]_{\mathrm{SD}}^\perp&=\pounds_\epsilon\eta_\perp,\nonumber\\
\left[\xi,\chi\right]_{\mathrm{SD}}^a&=h^{ab}\left(\epsilon_\perp D_b\eta_\perp- \eta_\perp D_b\epsilon_\perp\right)+\pounds_\epsilon\eta^a,
\end{align}
 for the decomposition of $\xi^a$ as in (\ref{decompdif}, \ref{decompx}) and analogously for $\chi^a$. This modification appears because we consider the action of the PFCs. Poisson brackets of the generators
\begin{align}
G[\xi]\equiv G_\perp[\epsilon_\perp]+G_D[\vec{\epsilon}] && G[\chi]\equiv G_\perp[\eta_\perp]+G_D[\vec{\eta}]
\end{align} 
close the algebra
\begin{equation}
\left\{G[\xi],G[\chi]\right\}=G[[\xi,\chi]]+PFC,\label{eq:CastellaniAlg}
\end{equation}
where
\begin{align}
\left[\xi,\chi\right]^\perp&=n_a\ ^{\ms{(4)}}\pounds_\xi\chi^a,\nonumber\\
\left[\xi,\chi\right]^a&=h^a_b\ ^{\ms{(4)}}\pounds_\xi\chi^b,
\end{align}
and we have set $\dot{N}=\lambda_N$ and $\dot{N}^a=\lambda_{\vec{N}}^a$ to accurately treat $\dot{\epsilon}_{\perp}$ and $\dot{\epsilon}^a$.

\section{Boundary Conditions}

In order to be able to find the boundary charges we have to define the boundary conditions and the asymptotic expansion at the boundary.  We consider the Gaussian coordinates and asymptotically AdS space using the metric
\begin{equation}
ds^2=\frac{\ell^2e^{2\omega}}{\rho^2}\left(d\rho^2+\gamma_{ij}dx^{i}dx^{j} \right)
\end{equation}
for $i,j..=0,1,2$.  Which is equal to the expansion (\ref{le}) for $\sigma=-1$ and up to a term $e^{2\omega}$ with which we can multiply the metric since it is conformally invariant. The Fefferman-Graham expansion of the boundary metric is equal to the one in (\ref{expansiongamma}) with $\ell$ set to one, it reads
\begin{equation}
\gamma_{ij}=\gamma_{ij}^{(0)}+\rho\gamma_{ij}^{(1)}+\rho^2\gamma_{ij}^{(2)}+\rho^3\gamma_{ij}^{(3)}+...
\end{equation}
where the metric in ADM variables near the boundary $\partial \Sigma$ is of the form
\begin{align}
h_{ab}=\Omega^2\overline{h}_{ab},&& N=\Omega\overline{N}, && N^a=\overline{N}^a \,\text{    for    }\,\Omega\equiv\frac{\ell e^{\omega}}{\rho}.
\end{align} therefore\begin{equation}
\omega_h=\Omega^3\omega_{\overline{h}}.
\end{equation}
 $h_{ab}$ metric can be further split 
\begin{align}
\overline{h}_{ab}=\partial_a\rho\partial_b\rho+\gamma_{IJ}\partial_ax^{I}\partial_bx^J,&& \overline{N}^I=\gamma^{IJ}\gamma_{J0},&&\overline{N}^3=0, && \overline{N}=\sqrt{-\frac{1}{\gamma^{00}}}
\end{align}
for $a,b,...=1,2,3$ and $I,J,..=1,2$.
Evaluating on shell EOM for $h_{ab}, $ $K_{ab} $ and $\Pi_{K}^{ab}$ (and $\mathcal{P}$) lead to new expressions for $K_{ab}$, $\Pi_{K}^{ab}$ and $\Pi_{h}^{ab}$. 
EOM for $h_{ab}$ lead to  
\begin{align}
K_{ab}&=\frac{1}{2N}\left(\partial_t-\pounds_{\vec{N}}\right)h_{ab}=\nonumber\\
&=\Omega\left[\frac{\overline{h}_{ab}}{\overline{N}}\left(\partial_t-\pounds_{\vec{N}}\right)\ln\Omega+\overline{K}_{ab}\right]
\end{align}
Using the transformation properties for the connection 
\begin{align}
C^a_{bc}&=2\delta^a_{(b}\overline{D}_{c)}\ln\Omega-\overline{h}_{bc}\overline{D}^a\ln\Omega+\overline{C}^a_{bc}
\end{align}
Ricci tensor
\begin{align}
R_{ab}&=-\overline{D}_a\overline{D}_b\ln\Omega-\overline{h}_{ab}\overline{h}^{cd}\overline{D}_c\overline{D}_d\ln\Omega+\nonumber\\
&+\overline{D}_a\ln\Omega\overline{D}_b\ln\Omega-\overline{h}_{ab}\overline{h}^{cd}\overline{D}_c\ln\Omega\overline{D}_d\ln\Omega+\overline{R}_{ab}
\end{align}
and the double covariant derivative of the momenta 
\begin{align}
\frac{1}{N}D_aD_bN&=\overline{D}_a\overline{D}_b\ln\Omega-\overline{D}_a\ln\Omega\overline{D}_b\ln\Omega+\nonumber\\
&+\overline{h}_{ab}\overline{h}^{cd}\overline{D}_c\ln\Omega\overline{D}_d\ln\Omega+\frac{1}{\overline{N}}\overline{h}_{ab}D^c\ln\Omega\overline{D}_c\overline{N}+\frac{1}{\overline{N}}\overline{D}_a\overline{D}_b\overline{N},
\end{align}
one computes $\Pi_{K}^{ab}$ from the equation of motion for $K_{ab}$ and requirement that $\mathcal{P}$=0
\begin{align}
\Pi_K^{ab}&=\omega_h\mathcal{T}^{abcd}\left[R_{cd}+K_{cd}K+\frac{1}{N}D_aD_bN-\frac{1}{N}\left(\partial_t-\pounds_{\vec{N}}\right)K_{ab}\right]\nonumber\\
&=\Omega^{-1}\overline{\Pi}_K^{ab}.
\end{align}
Which agrees with the relation obtained from the rescaling of projection for four dimensional Weyl 
\begin{equation}
\Pi_K^{ab}=\omega_h n^cn^dC^a{}_b{}^c{}_d 
\end{equation}
 with $n_a=\Omega\overline{n}_a$ and $C^a{}_{bcd}=\overline{C}^a{}_{bcd}$.
 That is similar to the projection of the magnetic part of the Weyl
 \begin{equation}
 B_{abcd}=\Omega \overline{B}_{abcd}.
 \end{equation}
From $\Pi_{K}^{ab}$, one can compute the Weyl rescaling for the $\Pi_h^{ab}$ momenta, 
\begin{align}
\Pi_h^{ab}&=-\frac{1}{2N}\left(\partial_t-\pounds_{\vec{N}}\right)\Pi_K^{ab}-\frac{2}{N}D_c\left(N\omega_hB^{c(ab)}\right)-\frac{1}{2}\left(\Pi_K^{ab}K+\Pi_K^{cd}K_{cd}h^{ab}\right)\nonumber\\
&=\Omega^{-2}\left[-\frac{1}{\overline{N}}\left(\partial_t-\pounds_{\vec{N}}\right)\ln\Omega\overline{\Pi}_K^{ab}+\overline{\Pi}_h^{ab}\right].
\end{align}
The allowed variations near the boundary are accordingly
\begin{align}
\delta h_{ab}&=\Omega^2\left(2\delta\ln\Omega\overline{h}_{ab}+\delta\overline{h}_{ab}\right),\\
\delta K_{ab}&=\Omega\delta\ln\Omega\left[\frac{\overline{h}_{ab}}{\overline{N}}\left(\partial_t-\pounds_{\vec{N}}\right)\ln\Omega+\overline{K}_{ab}\right]
\Omega\frac{\overline{h}_{ab}}{\overline{N}}\left(\partial_t-\pounds_{\vec{N}}\right)\delta\ln\Omega\nonumber\\
&\quad \Omega\left\{\frac{\delta\overline{h}_{ab}}{\overline{N}}\left(\partial_t-\pounds_{\vec{N}}\right)\ln\Omega+\delta\overline{K}_{ab}+\right.\nonumber\\
&\quad \left.+\frac{\overline{h}_{ab}}{\overline{N}}\left[-\left(\partial_t-\pounds_{\vec{N}}\right)\ln\Omega\frac{\delta\overline{N}}{\overline{N}^2}-\delta N^a\overline{D}_a\ln\Omega\right]\right\},\\
\delta \Pi_h^{ab}&=-\frac{2\delta\ln\Omega}{\Omega^2}\left[-\frac{1}{\overline{N}}\left(\partial_t-\pounds_{\vec{N}}\right)\ln\Omega\overline{\Pi}_K^{ab}+\overline{\Pi}_h^{ab}\right] -\frac{1}{\Omega^2\overline{N}}\left(\partial_t-\pounds_{\vec{N}}\right)\delta\ln\Omega\overline{\Pi}_K^{ab}\nonumber\\
&\quad \Omega^{-2}\left\{-\frac{1}{\overline{N}}\left(\partial_t-\pounds_{\vec{N}}\right)\ln\Omega\delta\overline{\Pi}_K^{ab}+\delta\overline{\Pi}_h^{ab}\right.\nonumber\\
&\quad \left.+\frac{\overline{\Pi}_h^{ab}}{\overline{N}}\left[\left(\partial_t-\pounds_{\vec{N}}\right)\ln\Omega\frac{\delta\overline{N}}{\overline{N}^2}+\delta N^a\overline{D}_a\ln\Omega\right]\right\},\\
\delta \Pi_K^{ab}&=\Omega^{-1}\left(-\delta\ln\Omega+\delta\overline{\Pi}_K^{ab}\right).\label{eq:WeylVari}
\end{align}
Where the variations of the quantities that have been rescaled (those that contain an overline) are set to 
\begin{align}
\delta \overline{h}_{ab}\vert_{\partial\Sigma}=D_c\delta \overline{h}_{ab}\vert_{\partial\Sigma}=0,\nonumber\\
\delta \overline{N}\vert_{\partial\Sigma}=D_c\delta \overline{N}\vert_{\partial\Sigma}=0,\nonumber\\
\delta N^a\vert_{\partial\Sigma}=D_c\delta N^a\vert_{\partial\Sigma}=0
\end{align}
that leads to the requirement that 
\begin{equation}
\delta \overline{K}_{ab}\vert_{\partial\Sigma}=0,
\end{equation}
and that the variations of the momenta $\delta\overline{Pi}_K^{ab}|_{\partial\Sigma}$ and $\delta\overline{\Pi}_h^{ab}|_{\partial\Sigma}$ at the boundary are arbitrary but finite. 
These boundary conditions are preserved by the gauge transformation defined by bulk diffeomorphisms $\xi^a$
\begin{equation} 
\pounds_{\xi}\overline{g}_{ab}=2\lambda\overline{g}_{ab},
\end{equation}
and arbitrary rescalings of the metric (Weyl rescalings)
\begin{equation}
\delta_{\omega}g_{ab}=2\omega g_{ab}.
\end{equation}
The scalings we use for variations are
\begin{align}
\delta \overline{h}_{ab}&=2\delta\omega\overline{h}_{ab}\\
\delta\overline{N}&=\delta\omega\overline{N}\\\
\delta\omega&\sim\mathcal{O}(\rho)\\
\delta N^a&\sim\mathcal{O}(\rho^2)
\end{align}
that are consistent with the scalings
\begin{align}
\delta\overline{K}&\sim\mathcal{O}(\rho)\\
\delta\overline{\Pi}_{K}^{ab}&\sim\mathcal{O}(1)\\
\delta \overline{\Pi}_h^{ab}&\sim \mathcal{O}(1).
\end{align}
\section{Canonical Charges}

To compute the charges, we follow the prescription outlined in chapter "Gauge Generators". We have to define the generators that are functionally differentiable, by solving an imposed boundary value problem. One may refer to that as searching for the well defined action for the canonical generators. 
To gauge generators, G, we have to add boundary terms to make them integrable. 

In general, the mechanism to obtain the global charges of certain gauge theory using the Hamiltonian analysis is well known \cite{Brown:1986nw}. First, one needs to define the boundary conditions in the spatial infinity that should be obeyed by the fields which we have done in the above chapter "Boundary Conditions", and then identify asymptotic symmetries conserving that asymptotic behaviour. 
To be able to use the Hamiltonian formulation, one needs to to convert the boundary conditions on the space-time metric into boundary conditions on the canonical variables. The asymptotic symmetries define the allowed surface deformation vectors $\xi^{\mu}$ ($\mu=\perp,i$) for considered space like hypersurfaces.

\subsection{Boundary Terms from Weyl Constraint}
The Weyl charge is given by its boundary integral that is associated to generator (\ref{eq:g3}). To render the generator finite, we have to add to it a term whose total variation will correspond to boundary term of the generator, that corresponds to a charge. 
 Boundary terms for $G_W$ are
\begin{align}
-\int_{\Sigma}\pounds_{\vec{N}} \omega+\int_{\partial \Sigma}\ast\omega N^aP.
\end{align}
The generator is modified due to the charge $Q_W$ 
\begin{equation}
\Gamma[\omega,\dot{\omega}^{(1)}]=G_W[\omega,\dot{\omega}^{(1)}]-Q_{W}[\omega,\dot{\omega}^{(1)}] 
\end{equation}
for $Q_W$ 
\begin{equation}
Q_W=\int_{\partial\Sigma}\ast \omega N^aP.
\end{equation}
Which vanishes on shell because of the $P$ constraint. An improved generator $\Gamma_{W}$, therefore keeps the form of (\ref{eq:g3}) earlier obtained generator. 
\subsection{Boundary Terms from Diffeomorphism Constraint}
Evaluation of the boundary terms on shell leads to vanishing of the terms that involve $\mathcal{P}=0$ and $\mathcal{W}=0$. Among these contributions are all the terms that include $\Omega$.  That allows us to replace the variables $h_{ab}$, $K_{ab}$, $\Pi_{h}^{ab}$ and $\Pi_K^{ab}$ in (\ref{dhdiffbound}) to (\ref{dPiKdiffbound}) with their finite values 
$\overline{h}_{ab}$, $\overline{K}_{ab}$, $\overline{\Pi}_h^{ab}$ and $\overline{\Pi}_K^{ab}$. For $\epsilon^{\rho}\sim\mathcal{O}(\rho)$ and $\delta\overline{h}_{ab}\sim\mathcal{O}(\rho)$ and $\delta\overline{K}_{ab}\sim\mathcal{O}(\rho)$ the term
\begin{align}\int_{\partial\Sigma}\ast \epsilon^{c}\left(\overline{\Pi_K^{ab}}\delta\overline{K}_{ab}+\overline{\Pi}_h^{ab}\delta\overline{h}_{ab}\right)
\end{align} vanishes. 
And remaining part
\begin{equation}
-2\int_{\partial\Sigma}\ast \left(\overline{\Pi}_K^{ca}\epsilon^b\delta\overline{K}_{ab}+\delta\overline{\Pi}_K^{ca}\epsilon^bK_{ab}+\overline{\Pi}_h^{ca}\epsilon^b\delta\overline{h}_{ab}+\delta\overline{\Pi}_h^{ca}\epsilon^b\overline{h}_{ab}\right) \label{chwoov}
\end{equation}
is integrated into an on-shell charge of the spatial diffeomorphism
\begin{equation}
\overline{Q}_D[\epsilon]=2\int_{\partial\Sigma}\ast\epsilon^c\left(\overline{\Pi}_h^{ab}\overline{h}_{bc}+\overline{\Pi}_K^{ab}\overline{K}_{bc}\right),\label{eq:diffchargeol}
\end{equation}
where we denoted the charge expressed with "overlined" values with $\overline{Q}_D$.
The charge is finite because the tensors that constitute it are $\mathcal{O}(1)$ and $\epsilon^I\sim\mathcal{O}(1)$. The terms $\overline{h}_{ab}$ and $\overline{K}_{ab}$ are obtained by insertion of the background metric $\gamma_{ij}$ and its expansion including the terms $\gamma_{ij}^{(0)}$ and $\gamma_{ij}^{(1)}$. The electric part of the Weyl tensor, $\overline{\Pi}_{K}^{ab}$, is determined  from the terms from the expansion up to $\gamma_{ij}^{(2)}$, while the terms in $\overline{\Pi}_{h}^{ab}$ include even $\gamma_{ij}^{(3)}$.
The charge \begin{equation}
Q_D[\epsilon]=2\int_{\partial\Sigma}\ast\epsilon^c\left(\Pi_h^{ab}h_{bc}+\Pi_K^{ab}K_{bc}\right).\label{eq:diffcharge}
\end{equation}
 is as well finite since including the boundary conditions 
 one obtains on-shell equivalence
\begin{equation}
\Pi_h^{ab}h_{bc}+\Pi_K^{ab}K_{bc}=\overline{\Pi}_h^{ab}\overline{h}_{bc}+\overline{\Pi}_K^{ab}\overline{K}_{bc}.
\end{equation}

That leads to a generator 
\begin{equation}
\Gamma_D[\epsilon]=\int{\Sigma}\left(\dot{\epsilon}^{a} {\Pi_{\vec{N}}}_{a}+\Pi_K^{ab}\pounds_\epsilon K_{ab}+\Pi_h^{ab}\pounds_\epsilon h_{ab}+ \Pi_{N}\pounds_\epsilon N+{\Pi_{\vec{N}}}_{a}\pounds_\epsilon N^a\right).\label{eq:imprD}
\end{equation}
whose variation with the included charge vanishes on the constraint surface (or incorporating the boundary conditions)
\begin{align}
\delta \Gamma_D[\epsilon]&\approx\int_{\partial\Sigma}\ast \epsilon^a \left(\Pi_K^{bc}\delta K_{bc}+\Pi_h^{bc}\delta h_{bc}\right)+\ast\xi^t\delta N^c\left(\Pi_h^{ab}h_{bc}+\Pi_K^{ab}K_{bc}\right)\approx\nonumber\\
&\approx\int_{\partial\Sigma}\ast \epsilon^a \left(\overline{\Pi}_K^{bc}\delta \overline{K}_{bc}+\overline{\Pi}_h^{bc}\delta \overline{h}_{bc}\right)+\ast\xi^t\delta N^c\left(\overline{\Pi}_h^{ab}\overline{h}_{bc}+\overline{\Pi}_K^{ab}\overline{K}_{bc}\right)=0.
\end{align}
We proceed with the charge coming form the Hamiltonian constraint.

\subsection{Boundary Terms from Hamiltonian Constraint}

In order to render the variation of the improved generator of $G_{\perp}[\epsilon,\dot{\epsilon}]$ vanish, and obtain the corresponding charge, we compute the boundary term that involves $\Pi_{\vec{N}}$ and $\Pi_{\vec{N}_a}$ (PFCs) and variation of the Hamiltonian constraint (\ref{dhH0bound}), (\ref{dKH0bound}) and (\ref{dPiKH0bound}). The terms that are PFCs vanish on shell, while the term coming from variation of Hamiltonian constraint, with asymptotic on-shell relations (boundary conditions)  $h_{ab}=\Omega^2\overline{h}_{ab}$, $\epsilon=\Omega\overline{\epsilon}$ and $\mathcal{P}=0$ leads to
\begin{equation}
-\int_{\partial\Sigma}\ast \overline{\epsilon}^cD^c\ln\Omega\left(\overline{\Pi}_K^{ab}\delta\overline{h}_{ab}+\delta\overline{\Pi}_K^{ab}\overline{h}_{ab} \right).
\end{equation}
Remaining contributions are equal to those from (\ref{dhH0bound}), (\ref{dKH0bound}) and (\ref{dPiKH0bound}) only with variables replaced with overlined ones. Therefore, we can drop the terms proportional to $\delta\overline{h}_{ab}\sim\mathcal{O}(\rho)$, $\delta\overline{K}_{ab}\sim\mathcal{O}(\rho)$ however not the terms with $\delta\overline{C}^c_{ed}\sim\mathcal{O}(1)$  because of the derivatives that act on $\delta\overline{h}_{ab}$.
These terms
\begin{align}
\int_{\partial\Sigma}\ast \bigg[ -\overline{\epsilon}\overline{D}^cln\Omega\left(\overline{\Pi}_K^{ab}\delta\overline{h}_{ab}+\delta\overline{\Pi}_{K}^{ab}\overline{h}_{ab}\right) \bigg]+\overline{\epsilon}\overline{D}_b\delta\overline{\Pi}_K^{cd}+\overline{\epsilon}\delta\overline{C}^c_{ab}\overline{\Pi}_K-\overline{D}_b\overline{\epsilon}\delta\overline{\Pi}_K^{cd},
\end{align}
are cancelled by varying the counterterm that we can obtain from 
\begin{align}
\int_{\partial\Sigma}\ast\left[\epsilon\delta C^c_{ab}\Pi_K^{ab}+\epsilon D_b\delta\Pi_K^{cb}-D_b\epsilon\delta\Pi_K^{cb} \right].
\end{align}
This, functionally integrated gives an on-shell finite term
\begin{align}
Q_{\perp}[\epsilon]&=\int_{\partial\Sigma}\ast\left[\epsilon D_b\Pi_K^{cb}-D_b\epsilon\Pi_K^{cb}\right] \label{qperp1} \\ 
&=\int_{\partial\Sigma}\ast \left[ \overline{\epsilon}\overline{D}_b\overline{\Pi}_K^{cb}-\overline{D}_b\overline{\epsilon}\overline{\Pi}_K^{cb}-\overline{\epsilon}\overline{D}^c ln\Omega\overline{\mathcal{P}}\right].\label{qperp}
\end{align}
However, in the variation of the  $G_{\perp}$ also appears  
\begin{align}
\int_{\partial\Sigma}\ast \left[ \overline{\epsilon}\left( \delta\overline{C}^c_{ab}\overline{\Pi}_{K}^{ab}-\delta \overline{C}^a_{ba}\overline{\Pi}_K^{cb} \right)\right]
\end{align}
which is finite when one allows $\delta\overline{h}_{ab}=2\delta\omega\overline{h}_{ab}$ 
\begin{align}
-\int_{\partial\Sigma}\ast\left(\overline{\epsilon}\overline{\Pi}_K^{cd}D_d\delta\omega\right).
\end{align}
To cancel it, we take into account that $\epsilon=\xi^tN$ contains variation $\delta \overline{N}=\delta\omega\overline{N}$, such that
$\delta_{\overline{N}}Q_{\perp}$ leads to 
$\delta_{\overline{N}}O_{\perp}=-\int_{\partial\Sigma}\ast \left(\overline{\epsilon}\overline{\Pi}_K^{cd}D_d\delta\omega \right)$.
This has proven that 
\begin{align}
\Gamma_{\perp}[\epsilon_{\perp}]=G_{\perp}[\epsilon_{\perp}] +Q_{\perp}[\epsilon_{\perp}]
\end{align}
with $Q_{\perp}[\epsilon]$ from (\ref{qperp}), is the required modified gauge generator (to which we refer to as well as an "improved generator").
The finiteness of the charges (\ref{eq:diffcharge}) and (\ref{qperp}), beside by using the property of conformal invariance, one may show by direct insertion of the expanded boundary metric. 

\subsection{Asymptotic Symmetry Algebra of the Improved Generators}

In case $\xi^a$ and $\chi^a$ are gauge generators  we can use relation for the Poisson brackets among the generators (\ref{eq:CastellaniAlg})  and write analogous relation for the improved generators
\begin{equation}
\left\{\Gamma[\xi],\Gamma[\chi]\right\}=\Gamma[[\xi,\chi]]+PFC.\label{eq:imprCastellaniAlg}
\end{equation}
This is true  \cite{Brown:1986ed}  for  $\xi^a$ and $\chi^a$ small diffeomorphisms that generate boundary condition preserving gauge transformations.
An improved generator is a functionally differentiable generator that has an action compatible with corresponding boundary conditions.
According to \cite{Henneaux:1985tv,Brown:1986nw}, fixing the gauge would turn first class into second class constraints, that need to strongly vanish while the Poisson brackets are required to be turned into Dirac brackets.  Wince the evaluation of improved generators on shell gives charges, the relation (\ref{eq:imprCastellaniAlg}) in terms of Dirac brackets coverts to
\begin{equation}
\left\{Q[\xi],Q[\chi]\right\}^\ast=Q[[\xi,\chi]].
\end{equation}
That leads to the isomorphism among the Dirac algebra of the charges and the Lie algebra of the boundary condition preserving gauge transformations.

\section{Time Conservation of Charges}
To prove the time conservation of charges, one may use one of the three approaches. 
\begin{itemize}
\item By clever inspection set $\xi^a=t^a$.
\item Prove that upon straightwforwardly acting on charges with $\partial_t$ they remain finite and keep they value.
\item Prove the equivalence of the canonical charges obtained using Hamiltonan procedure and the Noether charges obtained in \cite{Grumiller:2013mxa}. 
\end{itemize}
We will present first and the third method.

\subsection{Time Conservation of Charges Using the Method $\xi^a=t^a$}
 First method requires to set in (\ref{eq:imprCastellaniAlg}) that $\xi^{a}=t^a$. Even though $t^a$ is not a gauge generator that preserves a boundary condition, 
the functional derivative of $\Gamma[t]$ is well defined, and denoting $\dot{N}=\lambda_N$ and $\dot{N}^a=\lambda_{\vec{N}}^a$, leads to $\Gamma[t]\equiv H_T$. That proves functional differentiability of $H_T$ since Hamiltonian EOM do not require additional boundary terms. This agrees with the first chapter in which we have seen that CG does not require boundary terms for the action to have well defined variational principle  \cite{Grumiller:2013mxa}. For $\Gamma[\chi]$ an improved generator, and $\chi^a$ generator of the diffeomorphisms 
\begin{equation}
\left\{H_T,\Gamma[\chi]\right\}=\Gamma[-\pounds_\chi t^a]+PFC=\Gamma[\dot{\chi}]+PFC.\label{eq:timecharge}
\end{equation}
The Poisson bracket can be turned to Dirac bracket by fixing the gauge, and the equation (\ref{eq:timecharge}) can be understood as a time evolution equation for $-Q[\chi]$, where the action $H_T$ is not influenced by $\chi$. To obtain the total time derivative we add $-Q[\dot{\chi}]$ to (\ref{eq:timecharge}) 
\begin{equation}
\frac{dQ[\chi]}{dt}=Q[\dot{\chi}]-\left\{H_T,Q[\chi]\right\}^\ast=0,
\end{equation}
that proves time conservation of  charges.

\section{Asymtpotic Symmetry Algebra and MKR Solution}

We want to consider the canonical charges on the particular example of the CG solution, Mannheim-Kazanas-Riegert solution. 
To be able to do that, first we briefly introduce the asymptotic symmetry algebra that is in more detail analysed in the following chapter. 
Consider the space-time foliation of the manifold with a function $\rho$ that defines timelike hypersurfaces $\rho=const.$, with the boundary at $\rho=0$. The metric is
\begin{equation}
ds^2=\frac{e^{2\omega}\ell^2}{\rho^2}\left(d\rho^2+\gamma_{ij}dx^idx^j\right),\label{eq:BC4}
\end{equation}
for the expansion of $\gamma_{ij}$
\begin{equation}
\gamma_{ij}=\sum_{n=0}\gamma_{ij}^{\ms{(n)}}\left(\frac{\rho}{\ell}\right)^n,\label{eq:BC5}
\end{equation}
where $\omega$ is arbitrary and $\gamma^{(0)}$ and $\gamma^{(1)}$
are fixed.
 (\ref{eq:BC4}) is conserved by $\pounds_{\xi}$ and $\delta_{\omega}$ to leading and the subleading order in $\rho$ up to rescaling with $e^{2\omega}$ (that means $\pounds_{\xi}$ and $\delta_{\omega}$ do not change the metric (\ref{eq:BC4}) when they act on the prefactor). The demand that remains is
\begin{equation}
\pounds_\xi \frac{\ell^2}{\rho^2}\overline{g}_{\mu\nu}=2\lambda\frac{\ell^2}{\rho^2}\overline{g}_{\mu\nu}.\label{rembc}
\end{equation}
We insert the expansion
\begin{eqnarray}
\xi^\rho=\xi^\rho_{\LO}+\rho\xi^\rho_{\FO}+\mathcal{O}(\rho^2),\nonumber\\
\xi^i=\xi^i_{\LO}+\rho \xi^i_{\FO}+\mathcal{O}(\rho^2),\nonumber\\
\lambda=\lambda_{\LO}+\rho\lambda_{\FO}+\mathcal{O}(\rho^2),
\end{eqnarray}
 of the small diffeomorphism generators $\xi^i$, $\xi^{\rho}$ and $\lambda$ coefficient, in the equation (\ref{rembc}),
and obtain 
\begin{equation}
\ ^{\ms{(2+1)}}\pounds_{\xi^k_{\LO}}\gamma^{\LO}_{ij}=2\lambda_{\LO}\gamma^{\LO}_{ij},\label{eq:LOCKV}
\end{equation}
with  requirements $\xi^\rho_{\LO}=0$ and $ \xi^i_{\FO}=0$.
At the leading order  $\xi^\rho_{\FO}=\lambda_{\LO}=\frac{1}{3} \mathcal{D}_i\xi^i_{\LO}$ for $\mathcal{D}$ covariant derivative corresponding to the metric $\gamma_{ij}^{(0)}$. The subleading order \footnote{see following chapter for more details} imposes condition 
\begin{equation}
\pounds_{\xi^k_{\LO}}\gamma_{ij}^{\FO}-\frac{1}{3}\,\gamma_{ij}^{\FO}\mathcal{D}_{k}\epsilon^{k}_{\LO}+4\lambda^{\FO}\gamma_{ij}^{\LO}=0,\label{eq:FOCKV}
\end{equation}
here $\lambda^{(1)}$ is obtained from the trace of (\ref{eq:FOCKV}). Rewriting the MKR metric (\ref{MKR1}) (that for $a=0$ becomes Schwarzschild -(A)dS metric),
in the  FG from, we obtain $\gamma_{ij}^{(0)}$ and $\gam$ matrices $\gamma_{ij}^{(0)}=diag(-1,1,1)$ and (\ref{gama1mkr}) respectively. Leading order Killing equation (\ref{eq:LOCKV}) admits 10 Killing vectors of conformal algebra (\ref{kvsph0})-(\ref{sphkv}), see appendix: Canonical Analysis of Conformal Gravity: Killing Vectors for Conformal Algebra on Spherical Background, 
while the subleading order, conserves a subset of 4 KVs
   \begin{align}
 \xi^{\LO a}_{1}&=(0,0,1),\\
\xi^{\LO a}_{2}&=(0, \sin (\phi), \cot (\theta)\cos (\phi)),\\
\xi^{\LO a}_{3}&=(0, -\cos (\phi), \cot (\theta)\sin (\phi)), \\
\xi^{\LO a}_{4}&=(1,0,0).
 \end{align}that close the asymptotic symmetry algebra $\mathbb{R}\times o(3)$, one of the subalgebras of conformal algebra. 
The conserved charge that does not vanish is $Q[\xi_4^{(0)i}]=Q_{\perp}[N]$, in agreement with the \cite{Grumiller:2013mxa} 
\begin{equation}
Q[\partial_t]=\frac{M}{\ell^2}-a\frac{(1-\sqrt{1-12aM})}{6},
\end{equation}
MKR charge.

\section{Equivalence of Canonical and Noether Charges}

We want to demonstrate that canonical charges (\ref{qperp1}), (\ref{eq:diffcharge}) (that we write here for convenience)
\begin{align}
Q_{\perp}[\epsilon]&=\int_{\partial\Sigma}\ast\left[\epsilon D_b\Pi_K^{cb}-D_b\epsilon\Pi_K^{cb}\right] \label{qperp1} \\
Q_D[\epsilon]&=2\int_{\partial\Sigma}\ast\epsilon^c\left(\Pi_h^{ab}h_{bc}+\Pi_K^{ab}K_{bc}\right).\label{eq:diffcharge}
\end{align}
are equivalent to Noether charges 
\begin{align}
Q=\int d^2x \sqrt{\sigma}n_iJ^i,
\end{align} where we take into account that the second line of the (\ref{eq:CG17})
\begin{align}
\tau_{ij} &= \sigma \big[\tfrac{2}{\ell}\,(E_{ij}^{\TO}+ \tfrac{1}{3} E_{ij}^{\SO}\ga^{\FO}) -\tfrac4\ell\,E_{ik}^{\SO}\psi^{\FO k}_j
+ \tfrac{1}{\ell}\,\ga_{ij}^{\LO} E_{kl}^{\SO}\psi_{\FO}^{kl} 
+\nonumber\\& \tfrac{1}{2\ell^3}\,\psi^{\FO}_{ij}\psi_{kl}^{\FO}\psi_{\FO}^{kl}
- \tfrac{1}{\ell^3}\,\psi_{kl}^{\FO}\,\big(\psi^{\FO k}_i\psi^{\FO l}_j-\tfrac13\,\ga^{\LO}_{ij}\psi^{\FO k}_m\psi_{\FO}^{lm}\big)\big] 
\nonumber\\&- 4\,{\cal D}^k B_{ijk}^{\FO} + i\leftrightarrow j\,,
\label{eq:CG17n}
\end{align}
vanishes due to Cayley-Hamilton theorem that for traceless $\gamma^{(1)}_{ij}$ matrix read
\begin{align}
\frac{1}{2}\psi_{ij}^{(1)}\psi_{lk}^{(1)}\psi^{(1)lk}-\psi_{lk}^{(1)}\psi^{(1)k}_i\psi^{(1)l}_j+\frac{1}{3}\psi^{(1)}_{lk}\psi^{(1)k}_{m}\psi_{(1)}^{lm}\gamma_{ij}^{(0)}=0
\end{align}
One decomposes the finite part of the metric near the boundary with respect to the timelike unit normal $n^a$ and the unit normal $u_a=\nabla_a\rho$
\begin{align}
\overline{g}_{ab}=-n_an_b+u_au_b+\sigma_{ab}\end{align}
where $\sigma_{ab}$ is the induced metric on $\partial\Sigma$.
Electric and magnetic parts of the Weyl tensor can be decomposed with respect to $u_a$ as
\begin{align}
\mathcal{\varepsilon}=\perp_au_cu^dC^c{}_{adb}, && \mathcal{B}_{abc}=\perp_uu_cC^c{}_{abc}.
\end{align}
The fully projected Weyl tensor consists of a polynomial of the extrinsic curvature of the timelike hypersurface $\rho=const.$ which is irrelevant due to Cayley-Hamilton theorem, and a second one that we can write in terms of the electric part of the Weyl tensor.

 Close to the boundary we can decompose the momentum $\Pi_K^{ab}$ as
 \begin{align}
\Pi_K^{ab}&=2u\wedge\omega_\sigma\left(u^au^b\mathcal{E}_{nn}+2u^{(a}\sigma^{b)c}\mathcal{B}_{ncn}+\sigma^{ac}\sigma^{bd}\mathcal{E}_{cd}-\sigma^{ab}\mathcal{E}_{nn}\right).
\end{align}
Let us consider the first term in $Q_{\perp}$ and note that $\epsilon_{\perp}=N\xi^t$ and $\xi^t=\xi_{(0)}^t+\mathcal{O}(\rho^2)$ at the boundary
\begin{align}
\int_{\partial\Sigma}\ast D_b\epsilon_\perp\Pi_K^{ba}=\int_{\partial\Sigma}2\omega_\sigma\epsilon_\perp\left(\mathcal{E}_{nn}u\cdot\partial\ln N-\ ^{(2)}D_b\mathcal{B}^{nbn}\right),
\end{align}
for ${}^{(2)}D_b$ covariant derivative at the boundary and we obtained the second term by partial integration. The second term in the $Q_{\perp}$ reads
\begin{align}
-\int_{\partial\Sigma}\ast\epsilon_\perp D_b\Pi_K^{ba}=-\int_{\partial\Sigma}&2\omega_\sigma\epsilon_\perp\left(n^cn^du\cdot\partial\mathcal{E}_{cd}+2\mathcal{E}_{nn}\mathcal{K}-4\mathcal{E}^{nb}\mathcal{K}_{bn}-\mathcal{E}^{cd}\mathcal{K}_{cd}\right)\nonumber\\
&-2\omega_\sigma\ ^{(2)}D_b\mathcal{B}^{nbn}-2\omega_\sigma\epsilon_\perp\mathcal{E}_{nn}u\cdot\partial\ln N
\end{align}
 where we used
\begin{align}
\ ^{(2)}D_b\mathcal{B}^{nbn}&=-n_an_c\mathcal{D}_b\mathcal{B}^{(ac)b}+\mathcal{B}^{acn}k_{ac},\nonumber\\
\mathcal{E}_{ab}&\equiv\mathcal{E}^{\SO}_{ab}+\mathcal{O}(\rho)\nonumber\\
u\cdot\partial\mathcal{E}_{ab}&\equiv\mathcal{E}^{\TO}_{ab}+\mathcal{O}(\rho)\nonumber\\
\mathcal{B}_{abc}&\equiv-\mathcal{B}^{\FO}_{abc}+\mathcal{O}(\rho)
\end{align}
with $k_{ab}$ extrinsic curvature on $\partial\Sigma$ in $\partial\mathcal{M}$.
This leads to 
\begin{align}
Q_\perp=\int_{\partial\Sigma}-2\omega_\sigma n_c\xi^g\left(-n_gn^b\right)&\left(\mathcal{E}^c_{\TO b}+\mathcal{E}^c_{\SO b}\gamma^{\FO}-2\mathcal{E}^c_{\SO d}\gamma^{\FO d}_b-\mathcal{E}^{\SO}_{bd}\gamma^{ad}_{\FO}+\frac{1}{2}\gamma_{\LO b}^{c}\mathcal{E}^{cd}_{\SO}\gamma^{\FO}_{cd}\right.\nonumber\\
&\left.+2\omega_\sigma\gamma^{\LO}_{be}\mathcal{D}_d\mathcal{B}^{(ce)d}_{\FO}\right)-4\omega_\sigma \epsilon_\perp\mathcal{B}^{acn}k_{ab}.
\end{align}
To obtain the $Q_D$ charge, we first need via EOM obtain 
\begin{align}
\Pi_h^{ab}&=K\Pi_K^{ab}-2\Pi_K^{e(a}K^{b)}_{\ e}+\frac{1}{2}\Pi_K^{cd}K_{cd}h^{ab}\\&+\omega_h\perp\left(n_en_dn\nabla C^{aebd}-2n_d\nabla_c C^{c(ab)d}\right).
\end{align}
We obtain for the decomposition of $Q_D$
\begin{align}
Q_D=\int_{\partial\Sigma}-2\omega_\sigma n_c\xi^g\sigma_g^{\ b}&\left(\mathcal{E}^c_{\TO b}+\mathcal{E}^c_{\SO b}\gamma^{\FO}-2\mathcal{E}^c_{\SO d}\gamma^{\FO d}_b-\mathcal{E}^{\SO}_{bd}\gamma^{ad}_{\FO}\right.\nonumber\\
&\left.+2\omega_\sigma\gamma^{\LO}_{be}\mathcal{D}_d\mathcal{B}^{(ce)d}_{\FO}\right)+4\omega_\sigma \epsilon_b\ ^{(2)}D_d\perp_n\mathcal{B}^{(bd)n}.
\end{align}
Integrating the last term by part reads
\begin{align}
\int_{\partial\Sigma}4\omega_\sigma \epsilon_b\ ^{(2)}D_d\perp_n\mathcal{B}^{(bd)n}=-\int_{\partial\Sigma}4\omega_\sigma \ ^{(2)}D_{(d}\epsilon_{b)}\mathcal{B}^{bdn}.
\end{align}
Decomposition of the metric $\gamma_{ab}$ with the generator with respect to $n_a$ and $\sigma_{ab}$ gives for spatially projected part
\begin{align}
\sigma_a^{\ c}\sigma_b^{\ d}\pounds_\xi\gamma_{cd}=2\epsilon_\perp k_{ab}+2\ ^{(2)}D_{(d}\epsilon_{b)}=2\lambda^{\LO}\sigma^{\LO}_{ab}+\mathcal{O}(\rho).
\end{align}
With $\sigma^{\LO}_{ab}\mathcal{B}^{abn}_{\FO}=0$ one can notice that the last term in the $Q_D$ is equal to the last term in $Q_{\perp}$ and they cancel. The sum of charges therefore reads
\begin{align}
Q[\xi]=\int_{\partial\Sigma}-2\omega_\sigma n_c\xi^{b}&\left(\mathcal{E}^c_{\TO b}+\mathcal{E}^c_{\SO b}\gamma^{\FO}-2\mathcal{E}^c_{\SO d}\gamma^{\FO d}_b-\mathcal{E}^{\SO}_{bd}\gamma^{ad}_{\FO}+\frac{1}{2}\gamma_{\LO b}^{c}\mathcal{E}^{cd}_{\SO}\gamma^{\FO}_{cd}\right.\nonumber\\
&\left.+2\omega_\sigma\gamma^{\LO}_{be}\mathcal{D}_d\mathcal{B}^{(ce)d}_{\FO}\right).
\end{align}
This charge agrees with the one from the first chapter up to an over all factor 4 that is the difference in the initial action that we started with. 

\newpage
\chapter{Classification}
\section{Introduction to Classification}
Higher derivative theories of gravity, as CG, lead to very complicated sets of partial differential equations. In order for them to be solved in full generality one reaches for the simplifications imposing physically interesting conditions such as spherical, axial or particular kind of symmetry \cite{Breitenlohner:2004fp}, which in combination with restriction on the coordinate dependency other than radial coordinate,  can make equations analytically solvable. The other two approaches are numerical one that focuses on certain set of solutions  and the bottom-up approach that can be imagined analogously to the reverse procedure of Klauza-Klein reduction  \cite{Grumiller:2006ww}.
The latter approach arises from the analysis of the asymptotic symmetry algebra. 

The exemplary case is the algebra of Einstein Hilbert action. When $\Lambda<0$ matter free Einstein equations have a solution with a maximally symmetric AdS space and group $O(3,2)$ which is for $\Lambda=0$ analogous to Minkowski. The fields that form the action need to asymptotically approach AdS configuration, that requires
\begin{itemize}
\item proving the invariance of asymptotic conditions under the AdS group action
\item well defined canonical generators of the symmetry (which we have proven in the previous chapter)
\item included physically interesting asymptotically AdS solutions
\item boundary conditions written in terms of the spacetime metric components \cite{Henneaux:1985tv}.
\end{itemize}
We have seen that computing the canonical boundary charges one finds the algebra that agrees with the one obtained from the boundary conditions preserving diffeomorphisms.

The boundary conditions imposed on the metric are the AdS boundary conditions. This asymptotic symmetry algebra correspond to the algreba formed by the canonical charges that one obtains with the boundary conditions that are background independent.
 
 Boundary conditions are generated by infinitesimal diffeomorphism \begin{equation}x^{\mu}\rightarrow x^{\m}+\xi^{\mu}\label{diff}\end{equation} with vector field $\xi_{\mu}$ and Weyl rescalings of the metric $g_{\m\n}\rightarrow  e^{2\omega}g_{\m\n}$, where $\omega$ is Weyl factor. The transformation of the metric is 
\begin{equation}
\delta g_{\mu\n}=\left(e^{2\omega}-1\right)g_{\m\n}\pounds_{\xi}g_{\m\n}\label{trafo}.
\end{equation}
The boundary conditions needed to be conserved, introduced in chapter 1, are the form of the metric $ds^2=\frac{\ell^2}{\rho^2}\left(-\sigma d\rho^2+\gamma_{ij}dx^idx^j\right)$ (\ref{le}) with the expansion  $\gamma_{ij}=\gamma_{ij}^{(0)}+\frac{\rho}{\ell}\gam+\frac{\rho^2}{\ell^2}\gamma_{ij}^{(2)}+\frac{\rho^{(3)}}{\ell^3}\gamma_{ij}^{(3)}+...$ (\ref{expansiongamma}) at the boundary, the variations 
\begin{enumerate}
\item $\delta g_{\rho\rho}=0$, 
\item $\delta g_{\rho i}=0$,  \label{varcond2}
\item $\delta \gamma_{ij}^{(0)}=2\overline{\lambda} \gamma_{ij}^{(0)}$ 
\item and $\delta \gamma_{ij}^{(1)}=\overline{\lambda(x)} \gamma_{ij}^{(1)}$ (\ref{bcs}),  for $\overline{\lambda(x)}$ arbitrary function of the boundary coordinates. 
\end{enumerate}
We want to find the transformations that preserve that. Assuming the expansion of the Weyl factor 
\begin{equation}
\omega=\omega^{(0)}+\frac{\rho}{\ell}\omega^{(1)}+\frac{1}{2}\left(\frac{\rho}{\ell}\right)^2\omega^{(2)}+...\label{omexp}
\end{equation}
for the $\rho\rho$ component of equation (\ref{trafo}) the obtained condition 
\begin{align}
\delta g_{\rho\rho}=\left(e^{2\omega}-1\right)g_{\rho\rho}+\xi^{\mu}\partial_{\mu}g_{\rho\rho}+2g_{\rho\rho}\partial_{\rho}\xi^{\rho}
\end{align}
\begin{align}
(e^{2\omega}-1)-2\frac{\xi^{\rho}}{\rho}+2\partial_{\rho}\xi^{\rho}=0.\label{eqr}
\end{align}
dictates the allowed form of $\xi^{\rho}$. To consider the contribution of each term in the expansion (\ref{omexp}) we compute the differential equation (\ref{eqr}) in $\rho$ for $\xi$. To weaken the restrictions from equation (\ref{eqr}) we set the RHS to be equal to constant $c$ and consider the restrictions of the expansion (\ref{omexp}) term by term.

\begin{itemize}
\item When only the first term in the expansion is taken, i.e. $\omega=\omega_0$, one obtains
\begin{equation}
\xi^{\rho}=c_1 \rho+\frac{1}{2}\rho\left(1+c-e^{2\omega_0}\right)\rho \ln(\rho)
\end{equation}
\item for $\omega$ equal to the second term in the expansion, $\omega=\omega_1 \rho$, the result of the equation is
\begin{equation}
\xi^{\rho}=c_1 \rho+\frac{1}{2}\rho\left(-Ei(2\rho\omega_1)+(1+c)\ln(\rho)\right),
\end{equation}
where $Ei(z)=-\int_{-z}^{\infty}\frac{e^{-t}}{t}dt$,
\item while $\omega$ that is equal to one of the higher terms, $\omega=\omega_n \rho^n$ leads to
\begin{equation}
\xi^{\rho}=c_1 \rho+\frac{1}{2}\rho\left(-\frac{1}{n}Ei(2\rho^n\omega_n)+(1+c)\ln(\rho)\right).
\end{equation}
\end{itemize}
Let us now consider general forms of $\xi^{\rho}$ and $\omega$. The solution for $\xi^{\rho}$ to the (\ref{eqr})
 reads 
\begin{equation}
\xi^{\rho}=c_1r+r\int_1^r-\frac{-k_1+e^{2\omega(k_1)}k_1}{2k_1^2}dk_1\label{integr}
\end{equation}
with $k_1$ integration parameter. We want general expansion of $\xi^{\rho}$ for which  we have to take into account particular conditions that may appear. One of them being not allowing logarithmic terms. The simplest way to restrict this is to require from under integral function in (\ref{integr}) to be the form $\frac{constant}{k_1}$ and impose the condition that 
\begin{equation}
\frac{-k_1+e^{2\omega(k_1)}k_1}{2k_1^2}\neq c.
\end{equation}
That leads to $\omega(k_1)\neq c'$ where $c'=\frac{1}{2}\ln(c+1)$ and implies that the first condition we need to impose in the expansion (\ref{omexp}) is that the first term is vanishing, as we will also see below. If we consider expansion for $\xi^{\rho}$ 
\begin{equation}
\xi^{\rho}=\xi^{(0)\rho}+\rho\xi^{(1)\rho}+\rho^2\xi^{(2)\rho}+\rho^3\xi^{(3)\rho}+...
\end{equation} and for $\omega$ (\ref{omexp}), and insert in the equation for $\xi^{\rho}$ (\ref{eqr}) we obtain requirements $\omega_0=\xi^{(0)\rho}=0$, $\xi^{(2)\rho}=-\omega^{(1)}$, $\omega_{2}=-2\xi^{(3)\rho}-\omega_1^2$, $\xi^{(3)\rho}=\frac{\sqrt{-12\xi^{(5)\rho}+2\omega_1^4-6\omega_1\omega_3-3\omega_4}}{2\sqrt{3}}$, and $\xi^{(4)\rho}=\frac{1}{9}(12 \xi^{(3)\rho}\omega_1+4\omega_1^3-3\omega_3)$...
They satisfy the equation (\ref{eqr}) to $\mathcal{O}(\rho^5)$ and define $\omega$ and $\xi^{\rho}$ which read
\begin{align}
\omega&=\frac{\rho}{\ell}\omega^{(1)}+\frac{1}{2}\frac{\rho^2}{\ell^2}\omega^{(2)}+..\\
\xi^{\rho}&=\rho\xi^{(0)\rho}-\frac{\rho^2}{\ell}\omega^{(1)}+...
\end{align}
Here $"..."$ denote higher order terms and $\xi^{(0)\rho}\equiv\lambda(x)$ for $\lambda(x)$ allowed to depend on the boundary coordinates. 

The second condition on variation of the metric $\delta g_{\rho i}=0$ and equation for transformation of the metric (\ref{trafo}) lead to the equation 
\begin{align}
0&=g_{ij}\partial_{\rho}\xi^j+g_{\rho\rho}\partial_i\xi^{\rho}\nonumber \\
&=\gamma_{ij}\partial_{\rho}\xi^{i}-\sigma\xi^{\rho}
\end{align}
which rewritten in $\partial_{\rho}\xi^k=\sigma\gamma^{ki}\partial_j\xi^{\rho}$ has a solution
\begin{equation}
\xi^{i}=\xi^{i(0)}+\int d\rho\left(\gamma^{kj}\partial_j\xi^{\rho} \right)
\end{equation}
that can be integrated collecting orders of $\rho$
\begin{align}
\xi^i&=\xi^{i(0)}+\int d\rho\big[\gamma^{(0)ij}-\gamma^{(1)ij}\rho+\big(-\gamma^{(0)jk}\gamma_j^{m(2)}+\gamma_j^{(1)m}\gamma^{(1)jk}\big) \big]\partial_j(\rho\xi^{(1)\rho}\nonumber \\ &+\rho^2\xi^{(2)\rho}+\rho^3\xi^{(3)\rho})\nonumber \\
&=\xi^{(0)i}+\int d\rho\left[ \rho \gamma^{(0)ij}\partial_j\xi^{(1)\rho}+ \rho^2\left(\gamma^{(0)ij}\partial_j\xi^{(2)\rho}-\gamma^{(1)ij}\partial_j\xi^{(1)\rho}\right) \right].\label{eq:1}
\end{align}
\noindent We are interested into the expansion up to first order. Integrating the first term in (\ref{eq:1}) leads to $\int d\rho \rho \gamma^{(0)ij}\partial_j\xi^{(1)\rho}=\frac{1}{2}\gamma^{(0)ij}\partial_j\lambda\rho^2$ of $\mathcal{O}(\rho^2)$ that defines $\xi^i$:
\begin{equation}
\xi^{i}=\xi^{(0)i}+\frac{1}{2}\sigma\rho^2\mathcal{D}^i\lambda
\end{equation}
where with $\mathcal{D}_i$ is covariant derivative along the boundary, compatible with $\gamma_{ij}^{(0)}$.
The linear term in the expansion of the Killing vector (KV) $\xi^{i}$ does not exist what will allow classification of the subalgebras of the conformal algebra coming from $ij$ compnent of equation for the transformation of the metric (\ref{trafo}). From $ij$ component of (\ref{trafo}) (with $\ell=1$) one obtains 
\begin{align}
\delta g_{ij}&=(e^{2\omega}-1)g_{ij}+\pounds_{\xi}g_{ij}\\
\frac{1}{\rho^2}\delta\gamma_{ij}&=(e^{2\omega}-1)\frac{1}{\rho^2}\gamma_{ij}+\left(\xi^{\alpha}\partial_{\alpha}\left(\frac{1}{\rho^2}\gamma_{ij}\right)+\frac{1}{\rho^2}\gamma_{\alpha i}(\partial_j\xi^{\alpha})+\frac{1}{\rho^2}\gamma_{\alpha j}(\partial_i\xi^{\alpha}) \right) \nonumber \\
&=\frac{1}{\rho^2}\left[(e^{2\omega}-1)\gamma_{ij}+\pounds_{\xi^k}\gamma_{ij}+\xi^{\rho}\left(-\frac{2}{\rho}\gamma_{ij}+\partial_{\rho}\gamma_{ij}\right)\right] \label{eq:2}
\end{align}
which expanded (\ref{eq:2}) in $\rho$ 
\begin{align}
\delta \left(\gamma_{ij}^{(0)}+\rho \gamma_{ij}^{(1)}+...\right)&=\left(2\omega^{(1)} \rho+2\left(\omega^{(1)}+\omega^{(2)}\right)\rho^2 \right)\left(\gamma_{ij}^{(0)}+\rho \gamma_{ij}^{(1)}+...\right)\nonumber \\ &+[ \left(\xi^{(0)l}+\frac{1}{2}\sigma\rho^2\mathcal{D}^l\lambda \right)\mathcal{D}_{l}\left(\gamma_{ij}^{(0)}+\rho\gam\right) \nonumber \\ &+\left(\gamma_{il}^{(0)}+\rho\gamma_{il}^{(1)}\right)\mathcal{D}_j\left(\xi^{(0)l}+\frac{1}{2}\sigma\rho^2\mathcal{D}^l\lambda\right)\nonumber \\ &+\left(\gamma_{lj}^{(0)}+\rho\gamma_{lj}^{(1)}\right)\mathcal{D}_i\left(\xi^{(0)l}+\frac{1}{2}\sigma\rho^2\mathcal{D}^l\lambda\right)  ] \nonumber \\ &+ \left(\rho\xi^{(0)\rho}-\frac{\rho^2}{\ell}\omega^{(1)} \right)\partial_{\rho}\left(\gamma_{ij}^{(0)}+\rho \gamma_{ij}^{(1)}+...\right) \nonumber \\ &-\frac{2}{\rho}\left(\rho\xi^{(0)\rho}-\frac{\rho^2}{\ell}\omega^{(1)} \right)\left(\gamma_{ij}^{(0)}+\rho \gamma_{ij}^{(1)}+...\right)
\end{align}\noindent in the leading and the next to leading order read
\begin{align}
\delta \gamma_{ij}^{(0)}&=\mathcal{D}_i\xi_j^{(0)}+\mathcal{D}_j\xi^{(0)}_i-2\lambda\gamma_{ij}^{(0)}\label{eqnn1}\\ 
\delta\gamma_{ij}^{(1)}&=\pounds_{\xi^{k}_{(0)}}\gamma_{ij}^{(1)}+4\omega^{(1)}\gamma_{ij}^{(0)}-\lambda\gamma_{ij}^{(1)}.\label{eqnn2}
\end{align} Since the boundary conditions of the theory allow variations  $\delta\gamma_{ij}^{(0)}=2\overline{\lambda}(x)\gamma_{ij}^{(0)}$ and $\delta \gamma_{ij}^{(1)}=\overline{\lambda}(x)\gamma_{ij}^{(1)}$ from above, for $\overline{\lambda}\neq\lambda$, 
 the trace of the condition (\ref{eqnn1}) gives
$\gamma_{il}^{(0)}\mathcal{D}^{i}\xi^{(0)i}=3\lambda+3\overline{\lambda}$
and defines $\overline{\lambda}$ \begin{equation}\overline{\lambda}=\frac{1}{3}\gamma_{il}^{(0)}\mathcal{D}^i\xi^{l}-\lambda.\end{equation}
Inserting it back in (\ref{eqnn1})
gives
\begin{equation}
\mathcal{D}_{i}\xi^{(0)}_j+\mathcal{D}_j\xi^{(0)}_i=\frac{2}{3}\gamma_{ij}^{(0)}\mathcal{D}_{k}\xi^{(0)k}\label{lo}
\end{equation}
that defines the asymptotic symmetry algebra (ASA) of the theory at the boundary. Further restrictions that define subalgebras one determines from (\ref{eqnn2}).

The trace of the (\ref{eqnn2}), the boundary condition $\delta\gamma_{ij}^{(1)}=\overline{\lambda}\gamma_{ij}^{(1)}$ and relation between $\overline{\lambda}$ and $\lambda$ from (\ref{lo}) determine 
\begin{equation}
\omega^{(1)}=\frac{1}{12}\left[-\pounds_{\xi^{(0)l}}\gamma^{(1)}+\frac{1}{3}\gamma^{(1)}\mathcal{D}_l\xi^{(0)k} \right]\label{om1}
\end{equation}
that introduced in equation (\ref{eqnn2})  
read
\begin{equation}
\pounds_{\xi^{(0)l}}\gamma_{ij}^{(1)}=\frac{1}{3}\gamma_{ij}^{(1)}\mathcal{D}_l\xi^{(0)l}-4\omega^{(1)}\gamma_{ij}^{(0)}.\label{nloke}
\end{equation}
 
 \section{Flat, Spherical and Linearly Combined Killing Vectors}
 
The equation that defines ASA (\ref{lo}), defines the leading order Killing equation dependent only on the first term in the expansion of the Killing vector $\xi^{(0)i}$, and on the background metric $\gamma^{(0)}_{ij}$.  We will consider two background metrics 
\begin{enumerate}
\item
flat background metric
\begin{equation}\ga_{ij}^{\LO}=\eta_{ij}=diag(-1,1,1)_{ij}\label{eqga0}\end{equation}
with coordinates $(t,x,y)$ defined on $\partial\mathcal{M}$,  
\item
and the spherical $\mathbb{R}\times S^2$ background metric
\begin{equation}\ga_{ij}^{\LO}=\eta_{ij}=diag(-1,1,\sin(\theta)^2)_{ij},\label{eqga0}\end{equation}
with coordinates $(t,\theta,\phi)$.
\end{enumerate}
 To compute the Killing vectors $\xi^{(0)i}$, we follow the procedure from \cite{DiFrancesco:1997nk}, and from now on write quantities with indices $\m,\n,\kappa...$ since the computation relates to d dimensions. 

\noindent The conformal transformation $g'_{\m\n}(x')=\Omega(x) g_{\m\n}$ is locally equivalent to a (pseudo) rotation and a dilation. The group that is formed by a set of conformal transformations contains Poincare group as a subgroup when $\Omega\equiv1$. The name "conformal" comes from the fact that the angle between two arbitrary curves that cross each other at some point, is unaffected, and the angles are preserved. 
The consequences of $g'_{\m\n}(x')=\Omega(x)g_{\m\n}(x)$ on an infinitesimal transformation (\ref{diff}) of the metric are that in the first order of $\xi^{\m}$ one obtains
\begin{equation}
g_{\m\n}\rightarrow g_{\m\n}-(\partial_{\m}\xi_{\nu}+\partial_{\nu}\xi_{\mu}).
\end{equation}
while conformal invariance means
\begin{equation}
\partial_{\mu}\xi_{\nu}+\partial_{\nu}\xi_{\mu}=f(x)g_{\mu\nu}\label{pd1}
\end{equation}
where one can recognise the form of (\ref{lo}). $f(x)$ is found analogously to $\lambda$ from (\ref{eqnn1}) 
\begin{equation}
f(x)=\frac{2}{d}\partial_{\rho}\xi^{\rho},
\end{equation}
taking a trace of (\ref{pd1}) with the standard, flat, Cartesian metric $g_{\m\n}=\gamma^{(0)}_{\m\n}$.  
Partial derivation, $\partial_{\rho}$,  of (\ref{pd1}) and permutation of indices
define three equations. Its sum defines linear combination 
\begin{equation}
2\partial_{\mu}\partial_{\nu}\xi_{\rho}=\gamma^{(\LO)}_{\mu\rho}\partial_{\nu}f+\gamma^{(\LO)}_{\nu\rho}\partial_{\mu}f-\gamma^{\LO}\partial_{\rho}f.\label{2pp}
\end{equation}
that contracted with $\gamma^{\LO\m\n}$ lead to 
\begin{equation}
2\partial^{2}\xi_{\m}=(2-d)\partial_{\m}f\label{2md}.
\end{equation}
After acting with $\partial_{\n}$ on (\ref{2md}) and $\partial^2$ on (\ref{pd1}) one finds the equation $(2-d)\partial_{\m}\partial_{\n}f(x)=\gamma_{\m\n}^{(0)}\partial^2f(x)$ with a trace
\begin{equation}
(d-1)\partial^2f=0.\label{dm1}
\end{equation}
These equations allow the derivation of the expected form for conformal transformations in $d$ dimensions. We focus on $d\geq3$, for which 
equations (\ref{dm1}) and (\ref{2md}) imply $\partial_{\m}\partial_{\n}f=0$ so the function f can be maximally linear
\begin{equation}
f(x)=A+B_{\m}x^{\m}
\end{equation}
for constant $A,B_{\m}$. 
Inserting that ansatz into (\ref{2pp}) leads to constant $\partial_{\m}\partial_{\n}\xi_{\kappa}$ and the $\xi_{\m}$ that is at most quadratic in coordinates 
\begin{equation}
\xi_{\m}=a_{\m}+b_{\m\n}x^{\n}+c_{\m\n\kappa}x^{\n}x^{\kappa}\label{xiansatz} 
\end{equation}
where $c_{\m\n\kappa}=c_{\m\kappa\n}$.
Note that  (\ref{pd1}) - (\ref{2pp}) are true for all $x$ and one is allowed to treat the powers of coordinates individually.  That means $a_{\m}$ is  free of constraints, and it denotes an infinitesimal translation with corresponding finite transformation \begin{equation} x'^{\m}=x^{\m}+a^{\m}. \end{equation} Considering the linear term in (\ref{pd1}) leads to 
\begin{equation}
b_{\m\n}+b_{\n\m}=\frac{2}{d}b^{\kappa}_{\kappa}\gamma^{(0)}_{\m\n}
\end{equation}
from which it yields that $b_{\m\n}$ is of the form 
\begin{align}
b_{\m\n}=\alpha\gamma^{(0)}_{\m\n}+m_{\m\n} &&  m_{\m\n}=-m_{\n\m},
\end{align}
i.e. we obtain the sum of an antisymmetric part and a trace. The trace represents an infinitesimal scale transformation, of the finite transformation \begin{equation} x'^{\m}=\alpha x^{\m},\end{equation}  and the antisymmetric part an infinitesimal (rigid) rotation, and the finite transformation \begin{equation}x'^{\m}=M^{\m}_{\n}x^{\n}.\end{equation}
Inserting the ansatz for the Killing vector $\xi^{\m}$, (\ref{xiansatz}), into equation for the linear combination of partial derivatives on function $f$, (\ref{2pp}), gives the form of the term from  $\xi^{\m}$ quadratic in coordinates, $c_{\m\n\kappa}$,
\begin{align}
c_{\m\n\kappa}=\gamma^{(0)}_{\m\kappa}b_{\n}+\gamma^{(0)}_{\m\n}b_{\kappa}-\gamma^{(0)}_{\n\kappa}b_{\m} &&\text{ with } && b_{\m}\equiv\frac{1}{d}c^{\n}_{\n\m}
\end{align}
and the corresponding infinitesimal transformation
\begin{equation}
x'^{\m}=x^{\m}+2(x\cdot b)x^{\m}-b^{\m}x^2, \label{infsct}
\end{equation}
called {\it special conformal transformation}  (SCT). The corresponding finite transformation is \begin{equation} x'^{\m}=\frac{x^{\m}-b^{\m}x^2}{1-2b\cdot x+b^2x^2}.\label{trsct}
\end{equation}
One may demonstrate that transformation (\ref{trsct}) corresponds to the infinitesimal transformation (\ref{infsct}) and prove that it is conformal for the conformal factor $\Omega(x)=(1-2b\cdot x+b^2x^2)^2.$
Another way to think of the SCT is in a form of a translation preceded and followed by an inversion $x^{\m}\rightarrow x^{\m}/x^2$.
Using the definition of the generator of the infinitesimal transformations one obtains the generators of the conformal group.  It is customary to define the transformation with 
\begin{align}
x'^{\m}&=x^{\m}+\omega_a\frac{\delta x^{\m}}{\delta \omega_a} \\
\Phi'(x')&=\Phi(x)+\omega_a\frac{\delta \mathcal{F}}{\delta \omega_a}(x)\label{defgen}
\end{align}
Where $\omega_a$ denotes a set of infinitesimal parameters that are considered up to first order, which are in our case $a_{\m},b_{\m\n}$ and $c_{\m\n\kappa}$ The generator $G_a$  is defined by the symmetry transformation via the expression for the infinitesimal transformation at one point 
\begin{equation}
\delta_{\omega}\Phi(x)\equiv \Phi'(x)-\Phi(x)\equiv \omega_a G_{a}\Phi(x),\label{gener}
\end{equation}
that combined with the (\ref{defgen}) leads to 
\begin{align}
\Phi'(x') &=\Phi(x')-\omega_a\frac{\delta x^{\m}}{\delta\omega_a}\partial_{\m}\Phi(x')+\omega_a\frac{\delta\mathcal{F}}{\delta\omega_a}(x').
\end{align}
From which one may obtain the generator as
\begin{equation}
G_a\Phi=\frac{\delta x^{\m}}{\delta\omega_a}\partial_{\m}\Phi-\frac{\delta \mathcal{F}}{\delta \omega_a}.
\end{equation}
In the case of translation by a vector $\omega^{\m}$ that leads to $\frac{\delta x^{\m}}{\delta\omega^{\n}}=\delta^{\m}_{\n}$ and $\frac{\delta\mathcal{F}}{\delta\omega^{\n}}=0$. The generators of translations reads 
\begin{equation}
P_{\n}=\partial_{\n},
\end{equation}
and the function $\mathcal{F}$ can be taken to be constant. 
For general case one may in the definition of generator (\ref{gener}) set constant on the RHS, which can be set to be equal to $i$ or $1$ depending on the theory one is interested to consider. 
In the case of the rotations, the procedure is analogous, however the function $\mathcal{F}(\Phi)$ is taken to be $\mathcal{F}(\Phi)=L_{\lambda}\Phi$  where $L_{\lambda}$ is generator of infinitesimal Lorentz transformations \cite{DiFrancesco:1997nk,Blagojevic:2002du}.
Now we return to our case of three dimensions and use the names of the coordinates ($t,x,y$) and the indices $i,j,k...$.
In three dimensional Minkowski space we obtain three generators of translations 
\begin{align}
 \xi^{(0)} &= \partial_t, &
 \xi^{(1)} &= \partial_x, &
 \xi^{(2)} &= \partial_y
\end{align}
that together with the generators of Lorentz rotations  $ L_{ij}=(x_i\partial_j-x_j\partial_i) $, (or in components) 
\begin{align}
\xi^{(3)} &= x\partial_t + t\partial_x &
 \xi^{(4)} &= y\partial_t + t\partial_y &
 \xi^{(5)} &= y\partial_x - x\partial_y
\end{align}
form Poincare algebra.  Four additional conformal Killing vectors (CKVs) generate dilatations and special conformal transformations respectively
\begin{align}
 \xi^{(6)} &= t\partial_t + x\partial_x + y\partial_y  &
 \xi^{(7)} &= tx\partial_t + \frac{t^2+x^2-y^2}{2}\,\partial_x + xy\partial_y \\
 \xi^{(8)} &= ty\partial_t + xy\partial_x + \frac{t^2+y^2-x^2}{2}\,\partial_y &
 \xi^{(9)} &= \frac{t^2+x^2+y^2}{2}\,\partial_t + tx\partial_x + ty\partial_y \label{origckvs}.
\end{align}
We denote the KVs that generate translations with $\xi^{t}=(\xi^{(0)},\xi^{(1)}),\xi^{(2)})$, generator of dilatations, $\xi^{(6)}\equiv\xi^d$ and generators of SCTs with $\xi^{sct}=(\xi^{(7)},\xi^{(8)},\xi^{(9)})$. The generators obey conformal algebra commutation rules
\begin{align}
[\xi^d,\xi^t_j]&=-\xi^t_j && [\xi^d,\xi^{sct}_j]=\xi^{sct}_j\\
[\xi_l^t,L_{ij}]&=(\eta_{li}\xi^t_j-\eta_{lj}\xi^t_i) && [\xi_{l}^{sct},L_{ij}]=-(\eta_{li}\xi^{sct}_{j}-\eta_{lj}\xi^{sct}_i) \label{ca1}
\end{align}
\begin{align}
[\xi_i^{sct},\xi_j^t]&=-(\eta_{ij}\xi^d-L_{ij})\\
[L_{ij},L_{mj}]&=-L_{im}\label{ca2}
\end{align}
which can be verified explicitly.
One can notice the analogous commutation relations of SCTs and translations with rotations. As we will see later, the consequences of the analogy will be manifest in the subalgebras of the conformal algebra with translational KVs and SCT KVs. 

Knowing that the above KVs that form the conformal algebra result from imposing the flat background metric on the equation (\ref{lo}), we can continue to consider the equation (\ref{nloke}). When the linear term in the FG expansion of the metric vainshes ($\gamma_{ij}^{(0)}=0$), there is no condition on the asymptotic symmetry algebra in the linear order. When the linear term in the FG expansion exists, one obtains next to leading order Killing equation (\ref{nloke}) for flat background $\gamma_{ij}^{(0)}$
\eq{
\kvze{}^k\partial_k\ga^{\FO}_{ij}+\ga^{\FO}_{kj}\partial_i\kvze{}^k+\ga_{ik}^{\FO}\partial\kvze{}^k=\frac{1}{3}D_k\kvze{}^k\ga^{\FO}_{ij}-4\ga^{\LO}_{ij}\omega^{(1)}.}{eq:nloke}
\noindent One can use (\ref{eq:nloke}) for the analysis of the CG solutions as follows. 
\begin{enumerate}
\item
Consider the solutions of CG, transform them into the FG form of the metric, determine the $\gamma_{ij}^{\LO}$, bring it to form of Minkowski metric, determine $\gamma^{(1)}_{ij}$ and classify the solutions according to the KVs conserved by (\ref{eq:nloke}) for the given $\gamma_{ij}^{(1)}$.
That will determine the sub algebra of conformal algebra conserved by the CG solution, and the generators that define the dual field theory at the boundary according to the AdS/CFT prescription. 
\item
Consider the realised subalgebras imposing particular demands on the $\gamma_{ij}^{(1)}$ term in the metric. That procedure provides information about the asymptotic solutions of CG and their behaviour, and based on them one can investigate whether global CG solutions are reachable.
\end{enumerate}
In order to perform any of the above analysis one has to solve the equation (\ref{eq:nloke}) for $\gamma^{\FO}_{ij}$. The possible solutions and the subgroups  of the conformal algebra that can be realized are not made only from the Killing vectors written above. One can take any linear combination of the above KVs 
\begin{align}
\xi^{lc}&=a_0\xi^{(0)}+a_1\xi^{(1)}+a_2\xi^{(2)}+a_3\xi^{(3)}+a_4\xi^{(4)}+a_5\xi^{(5)}+a_6\xi^{(6)}+a_7\xi^{(7)}\nonumber \\ &+a_8\xi^{(8)}+a_9\xi^{(9)} \label{lc}
\end{align}
and consider whether there is a $\gamma^{(1)}_{ij}$ that satisfies such combination of KVs and the equation (\ref{eq:nloke}). Here we have denoted linearly combined KV with $\xi^{lc}$. One can as well take the opposite approach, impose any condition on the $\gamma_{ij}^{(1)}$ matrix and consider whether there is set of linearly combined KVs that satisfy (\ref{eq:nloke}). That set of linearly combined KVs will then form a sub algebra of conformal algebra.

\section{Coordinate Analysis of the $\gamma_{ij}^{(0)}$}

Let us analyse first the form of the $\gamma_{ij}^{(1)}$ matrix that we can obtain depending on the coordinates in it, and simulateously the behaviour of KVs. 

\noindent We can demand from the $\gamma_{ij}^{(1)}$ matrix to be 
\begin{itemize}
\item constant
\item dependent on one coordinate
\item dependent on two co-ordinates
\item dependent on three coordinates.
\end{itemize}

First, we focus on the form of the matrix that depends on which of the KVs from CA are conserved, and inspect the symmetries that appear in $\gamma_{ij}^{(1)}$. The importance of that may be questionable when opposed to solving the partial differential equation, however in solving the partial differential equation, we encounter the system of connected partial differential equations of up to three unknowns, that can be reduced to one partial differential equation of fourth order dependent on three coordinates.  This system of equations may be solved by recognising the symmetries that can be implemented into equation and make it solvable.

\subsection{Constant $\gamma_{ij}^{(1)}$}
The first requirement \begin{equation}\partial_{k}\gamma_{ij}^{(1)}=0\label{transcoord}\end{equation} is satisfied for all the translation generators when $\gamma_{ij}^{(1)}$ matrix is of arbitrary constant form
\begin{align}
 \left(
\begin{array}{ccc}
 c_1 & c_2 & c_3 \\
 c_2 & c_4 & c_5 \\
 c_3 & c_5& c_1-c_4 \\
\end{array} \right). \label{g1t}  \end{align}
The requirement that one of the Lorentz rotations conserves the matrix, leads to matrices 
\begin{align} \begin{array}{cc}
  \left(\begin{array}{ccc} -c_1 & 0 & 0 \\ 0 & c_1 &0 \\ 0 & 0 & -2c_1   \end{array}\right)  \text{ for } \kvth, &
  \left(\begin{array}{ccc} \frac{c_1}{2} & 0 & 0 \\ 0 & c_1 &0 \\ 0 & 0 & -\frac{c_1}{2}  \end{array}\right) 
   \text{ for } \kvfo, \end{array} \label{trans3rb} \end{align} \begin{align} 
     \left(\begin{array}{ccc} 2c_1 & 0 & 0 \\ 0 & c_1 &0 \\ 0 & 0 & c_1   \end{array}\right)  \text{ for } \kvfi \text{ conserved. }  \label{trans3rot1} 
 \end{align}
The sub algebra conserved for $\xi^{(5)}$ is obtained by bringing the flat MKR solution to the FG form, about which we say more in the chapter "MKR Solution".
From the remaining KVs, neither of KVs of SCTs nor dilatation KV conserve the constant $\ga_{ij}^{(1)}$.

\subsection{Algebra with Five Killing Vectors}
 
 The constant $\gamma_{ij}^{(1)}$ matrix allows to find the subalgebra of maximal number of KVs, five dimensional subalgebra. For this one needs to use linearly combined KVs (\ref{lc}) in next to leading order Killing equation (\ref{eq:nloke}). Does the algebra with more KVs exist? One can inspect that straightforwardly. Set constant coefficients in the $\ga^{(1)}_{ij}$, make $\gam$ traceless and insert in (\ref{eq:nloke}).
The computational time does not allow the evaluation. Let us go around that. 
Constant $\gam$ automatically  conserves three translational KVs. If maximal subalgebra consists of five, that means they are formed of the remaining 7 KVs. Set one of these 7 CKVs to zero. If that KV forms a new linearly combined KV that enters bigger subalgebra, the maximal subalgebra one can find contains N-1 and not N KVs. Obtaining full set of solutions, we will be able to find all the new KVs but the one that would be formed if we have included this one. 
Concretely, since KV of Lorentz rotations enters the new KV of the 5 KV sub algebra, we  set one of the Lorentz rotations to zero. Explicit solution of (\ref{eq:nloke}) for constant $\ga^{(1)}_{ij}$ leads to a maximal number of KVs that form sub algebra of the constant $\ga^{(1)}_{ij}$.

Let us focus on the particular case of subalgebra with 5 KVs.
The Killing vectors that define it, consist of three translational KVs and two additional KVs made from dilation and Lorentz rotations. 

From the set of partial differential equations (PDEs), once $\gam$ is set constant, we are able to form three analogous conserved matrices with corresponding new KVs.
For the matrix 
\begin{equation}
\ga^{(1)}_{ij}=\left(
\begin{array}{ccc}
 c & c & 0 \\
 c & c & 0 \\
 0 & 0 & 0 \\
\end{array}
\right)\label{five}
\end{equation}
two new Killing vectors are
\begin{align}
\chi^{(1)}&=(\text{a6} t-\frac{\text{a6} x}{2},\text{a6} x-\frac{\text{a6} t}{2},\text{a6} y) & 
\chi^{(2)}&=(-\text{a5} y,\text{a5} y,-\text{a5} t-\text{a5} x).
\end{align}
Permuting the combination of the original KVs that form the new ones we obtain the matrices 
\begin{align}
\ga^{(1)}_{ij}&=\left(
\begin{array}{ccc}
 0 & 0 & 0 \\
 0 & c & i c \\
 0 & i c & -c \\
 \end{array}
\right), & \ga^{(1)}_{ij}&=\left(
\begin{array}{ccc}
 -c & 0 & c \\
 0 & 0 & 0 \\
 c & 0 & -c \\ 
\end{array}
\right)\label{five2}\end{align}for the KVs 
\begin{align}
\chi^{(1)}&=(2 i \text{a5} t,\text{a5} y+2 i \text{a5} x,-\text{a5} x+2 i \text{a5} y) & 
\chi^{(2)}&=(\text{a3} x+i \text{a3} y,\text{a3} t,i \text{a3} t),
\end{align}
and 
\begin{align}
\chi^{(1)}&=(-\text{a5} x,-\text{a5} t+\text{a5} y,-\text{a5} x) &
\chi^{(2)}&=(\text{a6} t+\frac{\text{a6} y}{2},\text{a6} x,\text{a6} y+\frac{\text{a6} t}{2}), 
\end{align}respectively. The subalgebras close, where we can demonstrate closing of the subalgebra on the third example.  Setting $a$ coefficients to one, the commutators form the algebra
\begin{align}
[\xi^{(0)},\chi^{(1)}]&=-\xi^{(2)} & [\xi^{(1)},\chi^{(1)}]&=-\xi^{(0)}-\xi^{(2)} & [\xi^{(2)},\chi^{(1)}]&=\xi^{(2)}  \nonumber \\
[\xi^{(0)},\chi^{(2)}]&=\xi^{(0)}+\frac{\xi^{(2)}}{2} & [\xi^(1),\chi^{(2)}]&=\xi^{(1)} & [\xi^{(2)},\chi^{(1)}]&=\frac{\xi^{(0)}}{2}+\xi^{(2)} \nonumber \\
[\chi^{(1)},\chi^{(2)}]&=-\frac{1}{2}\chi^{(1)}.
\end{align}

\noindent The generators arrange in the generators of the similitude algebra, one of the largest subalgebras of conformal algebra about which we talk more below. Naming 
\begin{equation}
P_0=-\xi^{(0)},P_1=\xi^{(1)},P_2=\xi^{(2)},F=\xi^{(6)},K_1=\xi^{(3)}, K_2=\xi^{(4)},L_3=\xi^{(5)}\label{simid}
\end{equation}
we obtain so called "$a_{5,4}$" subalgebra $(F+\frac{1}{2}K_2,-K_1+L_3,P_0,P_1,P_2)$ of  the 3 dimensional extended Poincare algebra 
\begin{align}
[\xi^d,\xi^t_j]&=-\xi^t_j\\
[\xi^t_l,L_{ij}]&=-(\eta_{li}\xi^t_j-\eta_{lj}\xi^t_i)\\
[L_{ij},L_{mj}]&=L_{im},
\end{align}
for the "$a_{5,4}$" according to Patera et al. classification \cite{Patera:1976my}.

\section{ $\gamma_{ij}^{(1)}$ Dependent on One Coordinate}

In above chapter, the dependency on the particular KV of Lorentz rotations could have been observed from components of the $\gamma^{(1)}_{ij}$ matrix. Naturally, it is analogous here, supplemented with the dependency on coordinates.
One notices that translations are conserved in the direction of the coordinate on which $\gamma_{ij}^{(1)}$ does not depend, while the partial derivative of $\gamma_{ij}^{(1)}$ with respect to remaining directions is zero.  $\gam$ that conserves two Ts, e.g. $\xi^{(0)}$ and $\xi^{(2)}$ is given by 
\begin{equation}\gamma_{ij}^{(1)}=\left(
\begin{array}{ccc}
 \text{$\gamma_{11}$}(x) & \text{$\gamma_{12}$}(x) & \text{$\gamma_{13}$}(x) \\
 \text{$\gamma_{12}$}(x) & \text{$\gamma_{22}$}(x) & \text{$\gamma_{23}$}(x) \\
 \text{$\gamma_{13}$}(x) & \text{$\gamma_{23}$}(x) & \text{$\gamma_{11}$}(x)-\text{$\gamma_{22}$}(x) \\
\end{array}
\right).\label{eq2t}\end{equation}
If we want to conserve two translations (keep the maximal number of translational KVs) and include KV of Lorentz rotations, conserved KV of Lorentz rotations will be the one that does not contain the coordinate that appears in $\gam$. For $\xi^{(0)}$ and $\xi^{(2)}$, $\xi^{(0)}$ and $\xi^{(1)}$, and $\xi^{(1)}$ and $\xi^{(2)}$, and one KV of Lorentz rotations, the $\gamma_{ij}^{(1)}$ matrices will be, respectively, of the analogous form as in the constant case 
\begin{align} \begin{array}{cc}
  \left(\begin{array} {ccc} c_1(x) & 0 & 0 \\ 0 & 2c_1(x) &0 \\ 0 & 0 & -c_1(x)  \end{array}\right)  \text{ for } \kvfo, &
  \left(\begin{array} {ccc} -c_1(y) & 0 & 0 \\ 0 & c_1(y) &0 \\ 0 & 0 & -2c_1(y)  \end{array}\right) 
   \text{ for } \kvth,\end{array} \end{align} \begin{align} 
     \left(\begin{array} {ccc} 2c_1(t) & 0 & 0 \\ 0 & c_1(t) &0 \\ 0 & 0 & c_1(t)  \end{array}\right)  \text{ for } \kvfi  \label{dep1}
 \end{align}    and form the 2 dimensional Poincare algebra (2 Ts and one Lorents rotation).
 $\gamma_{ij}^{(1)}$ matrix, that conserves $\xi^{(5)}$ and contains coordinate $y$, is not allowed by NLO KE, the obtained condition requires $\gamma_{ij}^{(1)}$ to be constant, which is analogous for $\xi^{(4)}$ and $\xi^{(3)}$ and permuted coordinates. 

To include dilatations, $\xi^{(6)}$, NLO KE requires matrix $\gamma_{ij}^{(1)}=\frac{c_{ij}}{x_i}$ for $x_{i}=t,x,y$. That $\gam$ conserves 2 dimensional expended Poincare algebra (2 Ts, Lorentz rotation and dilatations), e.g. $\xi^{(0)},\xi^{(2)},\xi^{(4)},\xi^{d}$
\begin{align}
 \gamma_{ij}^{(1)}=\left(
\begin{array}{ccc}
 \frac{c}{2 x} & 0 & 0 \\
 0 & \frac{c}{x} & 0 \\
 0 & 0 & -\frac{c}{2 x} \\
\end{array}
\right) \label{poind}
\end{align}
\noindent To include SCTs, the allowed solution is only $\gamma^{(1)}_{ij}=0$.

\section{ $\gamma_{ij}^{(1)}$ Dependent on Two Coordinates}

Depending on the conserved direction of translations, realised $\gamma^{(1)}_{ij}$ matrix depends on the coordinates in remaining two directions. Condition for conservation of translations (\ref{transcoord}), the derivative with respect to the coordinate of conserved direction of translations, requires  $\gamma^{(1)}_{ij}$ constant in the that coordinate. 
$\gam$ can dependent on all three coordinates in specific linear combination, for which KVs of translations are correspondingly linearly combined, and lead to the analogous conclusion as for the original KVs. General $\gamma^{(1)}_{ij}$, dependent on two coordinates, conservs only one translation. If we want to conserve the translation in $t$ component and $\xi^{(0)}$,
 $\gam$ takes the form
\begin{equation}  \gamma^{(1)}_{ij}=\left(
\begin{array}{ccc}
 \text{$\gamma_{11}$}(x,y) & \text{$\gamma_{12}$}(x,y) & \text{$\gamma_{13}$}(x,y) \\
 \text{$\gamma_{12}$}(x,y) & \text{$\gamma_{22}$}(x,y) & \text{$\gamma_{23}$}(x,y) \\
 \text{$\gamma_{13}$}(x,y) & \text{$\gamma_{23}$}(x,y) & \text{$\gamma_{11}$}(x,y)-\text{$\gamma_{22}$}(x,y) \\
\end{array}
\right). \label{eq1t}  \end{equation}
We can add to it
\begin{enumerate} \item
 one Lorentz rotation, for which $\gam$ needs to be 
\begin{align}
  \left(\begin{array} {ccc} \frac{1}{2}f\left[\frac{1}{2}\bigl(-t^2+x^2\bigr)\right]& 0 & 0 \\ 0 & -\frac{1}{2}f\left[\frac{1}{2}\bigl(-t^2+x^2\bigr)\right] &0 \\ 0 & 0 & f\left(\frac{1}{2}\bigl[-t^2+x^2\bigr)\right]  \end{array}\right)  & \text{ for } \kvth,\\
    \left(\begin{array} {ccc} f\left[\frac{1}{2}\bigl(-t^2+y^2\bigr)\right]& 0 & 0 \\ 0 & 2f\left[\frac{1}{2}\bigl(-t^2+y^2\bigr)\right] &0 \\ 0 & 0 & -f\left(\frac{1}{2}\bigl[-t^2+y^2\bigr)\right]  \end{array}\right)  & \text{ for } \kvfo, \\
  \left(\begin{array} {ccc} f\left[\frac{1}{2}\bigl(x^2+y^2\bigr)\right]& 0 & 0 \\ 0 & \frac{1}{2}f\left[\frac{1}{2}\bigl(x^2+y^2\bigr)\right] &0 \\ 0 & 0 & \frac{1}{2}f\left(\frac{1}{2}\bigl[x^2+y^2\bigr)\right]  \end{array}\right)  & \text{ for } \kvfi. \label{trots}
 \end{align}
To conserve two KVs of Lorentz rotations, PDEs require $\gam$ of a from (\ref{dep1}),
\item
while to keep dilatations, the components of $\gamma^{(1)}_{ij}$ need to be \begin{equation}\ga^{\FO}_{ij}=\frac{b_{ij}\left(\frac{x_{j}}{x_{i}}\right)}{x_{i}}+\frac{c_{ij}\left(\frac{x_{i}}{x_{j}}\right)}{x_{j}}\label{dilat2},\end{equation} with $i,j=t,x,y$ for $i\neq j$.
Functional dependency of latter $\gam$ allows solving PDEs for one more KV.
\end{enumerate}

The subalgebras of the two cases we considered form 
\begin{itemize}
\item (trivial) Abelian algebra for one translation and one rotation, 
\item and as well Abelian algebra for one translation and one dilatation. 
\end{itemize}

Dependency on two coordinates, allows also $\gamma_{ij}^{(1)}$ matrix that conserves SCTs. One can find it by solving (\ref{nloke}) for SCTs. We will present on one example way to solve (\ref{nloke}) and obtain desired KVs, in particular one KV of SCTs, one Lorentz rotation, dilatation and translation.

It is convenient to start with the KV of translations. Translation is conserved in the direction on which components of the $\gamma_{ij}^{(1)}$ do not depend. We include  KV of translations choosing the components to depend on the remaining two coordinates. 
 For simplicity we set components $\gamma_{13}^{(1)}$ and $\gamma_{23}^{(1)}$ to zero. Then we compute the set of equations for the KVs of dilatations and SCTs, and after that for Lorentz rotations. The order is not important, however cleverly choosing the order of equations to solve can simplify the computation. 
 If one of the KVs we want to conserve are dilatations, it is useful to solve that equations first, because they give particular form (\ref{dilat2}) that simplifies further PDEs.
  $\gamma_{ij}^{(1)}$ that conserves $y$ translation, rotation around $y$ axis, dilatation and special conformal transformations in $y$ direction is 
\begin{equation}
\gamma_{ij}^{(1)}=\left(
\begin{array}{ccc}
 -\frac{\left(t^2+2 x^2\right) c_2}{3 ((t-x) (t+x))^{3/2}} & \frac{t x c_2}{((t-x) (t+x))^{3/2}} & 0 \\
 \frac{t x c_2}{((t-x) (t+x))^{3/2}} & -\frac{\left(2 t^2+x^2\right) c_2}{3 ((t-x) (t+x))^{3/2}} & 0 \\
 0 & 0 & \frac{c_2}{3 \sqrt{(t-x) (t+x)}} \\
\end{array}
\right).\label{scty}
\end{equation}
The set of partial differential equations that lead to $\gamma_{ij}^{(1)}$ (\ref{scty})  is given in the appendix: Classification.

\section{$\gamma_{ij}^{(1)}$ Dependent on Three Coordinates}
Condition $\partial_{k}\gamma_{ij}^{(1)}=0$ shows that for $\gamma_{ij}^{(1)}$ dependent on three coordinates, translations are not realised. 
 $\gamma_{ij}^{(1)}$ dependent on three coordinates that conserves Lorentz rotations is  analogous to $\gamma_{ij}^{(1)}$ dependent on two coordinates, with a difference, that dependency on the third coordinate appears as a function that multiplies the function on the diagonal, while components of 
$\gamma_{ij}^{(1)}$ that conserves dilatations are $\ga^{\FO}_{ij}=\frac{a_{ij}\left(\frac{x}{t},\frac{y}{t}\right)}{t}+\frac{b_{ij}\left(\frac{x}{y},\frac{t}{y}\right)}{y}+\frac{c_{ij}\left(\frac{t}{x},\frac{y}{x}\right)}{x}$, with $i,j=t,x,y$ for $i\neq j$.  

Two important 4KV subalgebras that require three coordinates to define $\gamma_{ij}^{(1)}$, are the subalgebra with three Lorentz rotations, and the subalgebra with three SCTs and one Lorentz rotation. 
From the analogy of translations and SCTs one may notice that the subalgebra with three SCTs and one Lorentz rotation, and three translations and one Lorentz rotation, i.e. MKR solution (for $\xi^{(5)}$ fourth KV), have analogous algebraic structure. That provides a basis in search for the full solutions of CG, where one could expect full solution of CG with three SCTs and rotation as a global solution of CG, analogous to MKR.

The algebra for 3 KVs of Lorentz rotations, is conserved by $\gamma_{ij}^{(1)}$, obtained from the PDEs given in the appendix: Classification. We solve PDEs expressing one component of $\gamma_{ij}^{(1)}$  with the other until they reduce to one PDE $LHS=RHS$
\begin{align}
LHS&=(x-y) \gamma_{11}^{(0,0,1)}(t,x,y)+6 \gamma_{11}(t,x,y)\\
RHS&=\left(x^2+y^2\right) \gamma_{11}^{(2,0,0)}(t,x,y)+(x+y) \gamma_{11}^{(0,1,0)}(t,x,y)\\ \nonumber&+t [t \left(\gamma_{11}^{(0,0,2)}(t,x,y)+\gamma_{11}^{(0,2,0)}(t,x,y)\right)\\ \nonumber &+2 \left(\gamma_{11}^{(1,0,0)}(t,x,y)+y \gamma_{11}^{(1,0,1)}(t,x,y)+x \gamma_{11}^{(1,1,0)}(t,x,y)\right)] \nonumber.
\end{align}
To solve it one may use numerical methods, or infer the solution from the symmetries of the equation. 
The latter approach and an assumption $\gamma_{ij}^{(1)}=2c t^2+cx^2+cy^2$, lead to  
\begin{equation}
\gamma_{ij}^{(1)}=\left(
\begin{array}{ccc}
 c \left(2 t^2+x^2+y^2\right) & -3 c t x & -3 c t y \\
 -3 c t x & c \left(t^2+2 x^2-y^2\right) & 3 c x y \\
 -3 c t y & 3 c x y & c \left(t^2-x^2+2 y^2\right) \\
\end{array}
\right)\label{only3rot}
\end{equation}
where $c$ is an arbitrary parameter. 

To obtain the subalgebra with SCTs, algebraically analogous to MKR solution, one needs to solve the system of the PDEs (see appendix: Classification).  Analogously as for $\gam$ that conserves Lorentz rotations, we compute PDEs, expressing one component in terms of the other. There is one convenient PDE solved by  $\gamma_{11}^{(1)}=\frac{c_1\left(\frac{x}{t},\frac{y}{t}\right)}{t^2}$ that reduces the number of PDEs. The simplest one can be written using a change of the coordinates $x\rightarrow z t$ and $y\rightarrow q t$, where we introduce two new coordinates $z$ and $q$.  The equation then reads
\begin{align}
0&=24 q \left(q^2-z^2-2\right) c_1(z,q)+\left(q^2+z^2-1\right) \big(q^4 c_1{}^{(0,3)}(z,q) \nonumber \\&+2 q^2 z^2 c_1{}^{(0,3)}(z,q)+12 q \left(q^2+z^2-1\right) c_1{}^{(0,2)}(z,q)\nonumber \\&+12 \left(3 q^2-z^2-2\right) c_1{}^{(0,1)}(z,q)-6 z \left(q^2+z^2-1\right) c_1{}^{(1,1)}(z,q)\nonumber \\&-2 q^2 c_1{}^{(0,3)}(z,q)+z^4 c_1{}^{(0,3)}(z,q)-2 z^2 c_1{}^{(0,3)}(z,q)-24 q z c_1{}^{(1,0)}(z,q)\nonumber \\&+c_1{}^{(0,3)}(z,q)\big)
\end{align}
where there is no dependency on t. This is third order PDE that can be solved numerically or analysing the symmetries, which does not lead to most general solution. 
Based on the analysis of the symmetries we obtain
\begin{align}
\gamma^{(1)}_{11}&=-\frac{\left(t^4+4 \left(x^2+y^2\right) t^2+\left(x^2+y^2\right)^2\right) c_1}{\left(t^2-x^2-y^2\right)^3} \nonumber   \\
\gamma^{(1)}_{12}&=\frac{3 t^2 \sqrt{\frac{x^2}{t^2}} \left(t^2+x^2+y^2\right) c_1}{\left(t^2-x^2-y^2\right)^3}\nonumber \\
\gamma_{13}^{(1)}&=\frac{3 t y \left(t^2+x^2+y^2\right) c_1}{\left(t^2-x^2-y^2\right)^3} \nonumber \\
\gamma_{22}^{(1)}&=-\frac{\left(t^4+2 \left(5 x^2-y^2\right) t^2+\left(x^2+y^2\right)^2\right) c_1}{2 \left(t^2-x^2-y^2\right)^3} \nonumber\\
\gamma_{23}^{(1)}&=-\frac{6 t^3 \sqrt{\frac{x^2}{t^2}} y c_1}{\left(t^2-x^2-y^2\right)^3} \nonumber \\
\gamma_{33}^{(1)}&=-\frac{\left(t^4-2 \left(x^2-5 y^2\right) t^2+\left(x^2+y^2\right)^2\right) c_1}{2 \left(t^2-x^2-y^2\right)^3}\label{rsct}
\end{align}
the solution that also satisfies the equation (\ref{nloke}) for the KV of rotation which makes it algebraically analogous to MKR solution. 

Next, we classify and find $\gam$ matrices for subalgebras that can be formed from the generators of CA that  are not allowed to linearly combine into new KVs. Then, we consider realisations of $\gam$ for which is allowed to use linearly combined KVs. Interesting research, which exceeds the scope of our analysis, is to inspect the symmetries of $\gam$ matrices allowed by certain KVs, and compare to $\gam$ matrix allowed by the combination of those KVs.

 \section{Classification According to  the Generators of the Conformal Group}
 
Whether $\gam$ is allowed to be realised for set of KVs is determined by closing of the subalgebra of those KVs. 

Let us consider an example of verification whether the set of KVs closes into subalgebra.
Assume we have one KV of translations, one KV of Lorentz rotations and one SCT. 
The Poisson bracket of the SCT with T closes into $[\xi_i^{sct},\xi_j^t]=2(\eta_{ij}\xi^d-L_{ij})$. For $i\neq j $ that implies $[\xi^{sct}_i,\xi_j^t]=-2L_{ij}$, which means Lorentz rotation in the directions $i$ and $j$ needs to  close with $\xi^t_i$ and $\xi^{sct}_j$ as well. Here, $[\xi_i^{t},L_{ij}]=\xi_j^t$ and $[\xi_i^{sct},L_{ij}]=\xi_{j}^{sct}$. That means that for algebra to close we need additional $\xi^{t}_{i}$ and $\xi^{sct}_{j}$ which leads to an algebra with six KVs $\xi^t_i,\xi^t_j,\xi^{sct}_i,\xi^{sct}_j,L_{ij},\xi^d$. Since the algebra with one translation, one Lorentz rotation and one SCT does not close, therefore  $\gam$ for such combination of KVs does not exist. Rather, if one obtains $\gam$ for those KVs, he or she, needs to verify the equations for KVs $\xi^d,\xi_i^t,\xi^{sct}_j$ which should as well be satisfied for the given $\gam$.

As mentioned earlier, complicated PDEs can be solvable by recognising symmetries which does not provide the most general solution. 
For the subalgebra with one translation and one Lorentz rotation, e.g.,  (for T in $t$ direction, and KV of rotations) one obtains corresponding PDEs (presented in the appendix: Classification)  (\ref{rotxy},\ref{transt}) whose solutions lead to two PDEs of the form: 
\begin{align}
LHS&=x^2 c_2{}^{(0,2)}(x,y)+y^2 c_2{}^{(2,0)}(x,y)+c_2(x,y)+c_6\left(\frac{1}{2} \left(x^2+y^2\right)\right)\nonumber \\  RHS&=y c_2{}^{(0,1)}(x,y)+x \left(c_2{}^{(1,0)}(x,y)+2 y c_2{}^{(1,1)}(x,y)\right). \label{pdesym}
\end{align}
From the symmetries one would assume $c_2=f\left[\frac{1}{2}(x^2+y^2)\right]$ solves the equation (\ref{pdesym}), which is correct, however (\ref{pdesym}) is second order linear PDE with two independent variables, and with known solving methods. If one takes into account the known solutions that ought to satisfy equation (\ref{pdesym}) together with the solution recognised from symmetries one obtains more general $\gam$.

The solution concluded from symmetries leads to analogous $\gam$ as $\gam$ for constant components, only with the function $f\left(\frac{1}{2}(x^2+y^2)\right)$ on the diagonal, equation (\ref{trots}).
The latter form of the solution gives the off-diagonal terms as well, which depend on the solution added to $c_{2}(x,y)$,  $\gam$ is then
\begin{align}
\gam=\left(
\begin{array}{ccc}
 c_4\left(\frac{1}{2} \left(x^2+y^2\right)\right) & a x+b y & a y-b x \\
 a x+b y & \frac{1}{2} c_4\left(\frac{1}{2} \left(x^2+y^2\right)\right) & 0 \\
 a y-b x & 0 & \frac{1}{2} c_4\left(\frac{1}{2} \left(x^2+y^2\right)\right) \\
\end{array}
\right)\label{trot}.
\end{align}
To provide a transparent overview of the subalgebras  realised for the particular $\gam$ we present them in tables. \newline\noindent
In the first row we write the original generators  obtained by the leading order equation (\ref{lo}). We first write translations (Ts) and $\gam$ that conserve them, then possible combinations of KVs with Ts, then rotations (Rs) and combinations with rotations, dilatation (D) with corresponding combinations and special conformal transformations (SCTs) with their combinations. \newline\noindent
The second row presents whether subagebra with the generator from the first row exists, (closes), and the third row presents an example of $\gam$ denoted with "(example)" that realises the subalgebra, stating the most general form of $\gam$ when given.
 The fourth row denotes the number of CKVs that are contained in the algebra.
 The matrixes $\gam$ near which we write "(comment $number$)" are commented in the text.
\begin{center}
\hspace{-0.65cm}\begin{tabular}{ |l | p{4.3 cm} | p{7.5 cm} | p{0.5cm}|}
\hline
Algebra & Name/existence(closing) & Realization &  \\
 \hline\hline
\hspace{0.18cm}1 T & $\exists$ & $\exists$: see equation (\ref{eq1t}) & 1 \\
\hspace{0.18 cm}2 T & $\exists$ & $\exists$: see equation (\ref{eq2t}) & 2\\
\hspace{0.18 cm}3 T & $\exists$ & $\exists$: see equation (\ref{g1t})& 3\\
\hspace{0.18 cm}1 T + 1 R &  $[\xi^t_{l},L_{ij}]=\eta_{li}\xi^t_{j}-\eta_{lj}\xi^t_{i}$,    
$\nexists$ for $l=i$ or $j$, $\exists$ for $l\neq i \neq j$   &  (example): equation (\ref{trot}) & 2 \\
\hspace{0.18 cm}1T + D & $\exists$ & 
$\exists$: see equation (\ref{dilat2})  &2\\
$\begin{array}{l}\text{1 T + 1 R}\\ \text{+ D}\end{array}$ & $\exists$ & $\left(
\begin{array}{ccc}
 \frac{c_6}{\sqrt{x^2+y^2}} & \frac{x c_4+y c_5}{x^2+y^2} & \frac{y c_4-x c_5}{x^2+y^2} \\
 \frac{x c_4+y c_5}{x^2+y^2} & \frac{c_6}{2 \sqrt{x^2+y^2}} & 0 \\
 \frac{y c_4-x c_5}{x^2+y^2} & 0 & \frac{c_6}{2 \sqrt{x^2+y^2}} \\
\end{array}
\right)$ There exist analogous matrices for the translations in the two remaining directions that depend, for the translation in the $l$ direction on the coordinates $i\neq l$ and $j\neq l$ &3\\
&& (example) & \\
$\begin{array}{l}\text{1T + D}\\ \text{+1 SCT}\end{array}$& $[\xi_i^{sct},\xi_j^t]=2(\eta_{ij}\xi^d-L_{ij})$, $\exists$ for $i=j$; sl(2)  &
$ \left(
\begin{array}{ccc}
 \frac{f\left(\frac{x}{t}\right)}{t} & -\frac{3 x f\left(\frac{x}{t}\right)}{t^2+2 x^2} & 0 \\
 -\frac{3 x f\left(\frac{x}{t}\right)}{t^2+2 x^2} & \frac{\left(2 t^2+x^2\right) f\left(\frac{x}{t}\right)}{t^3+2 x^2 t} & 0 \\
 0 & 0 & \frac{\left(x^2-t^2\right) f\left(\frac{x}{t}\right)}{t^3+2 x^2 t} \\
\end{array}
\right)$ example for  $i=j=y$
 & 3 \\
$\begin{array}{l}\text{1 T + 1 R} \\ \text{+ D+1 SCT}\end{array}$ &  $\exists$: $[\xi^t_i,\xi^{sct}_if]=2\xi^d$, $[\xi^t_l,L_{ij}]=0$, $[\xi^{sct}_l,L_{ij}]=0$ for $l\neq i$ and $l\neq j$;
sl(2)+u(1) & example for $\xi^t_y,\xi^{sct}_y,L_{xt},D$, see equation (\ref{scty}) &4\\
\hspace{0.18 cm}2 T + 1 R & $\exists$: \text{2d Poincare} & $\exists$ see equation (\ref{dep1}) &3\\
$\begin{array}{l}\text{2 T + 1 R}\\ \text{+ D}\end{array}$ & $\exists$: \text{2d Poincare +D} & $\exists$ see equation (\ref{poind}) &5\\
\hline
\end{tabular}
\end{center}

\begin{center}
\begin{table}
\hspace{-0.65cm}\begin{tabular}{ | l  | p{1 cm} | p{9.8 cm} | p{0.5cm}|}
\hline
\hspace{0.18 cm}2 T + D & $\exists$  & $\exists$ $\gamma_{ij}^{(1)}=\left(
\begin{array}{ccc}
 \frac{\text{c1}}{x} & \frac{\text{c2}}{x} & \frac{\text{c3}}{x} \\
 \frac{\text{c2}}{x} & \frac{\text{c4}}{x} & \frac{\text{c5}}{x} \\
 \frac{\text{c3}}{x} & \frac{\text{c5}}{x} & \frac{\text{c1}-\text{c4}}{x} \\
\end{array}
\right)$ &6\\ 
$\begin{array}{l}\text{2 T + 1 R} \\ \text{+ D + 2 SCT}\end{array}$ & $\exists$ & $\nexists$ requirement for 2 T restricts the components on dependency on one coordinate, in which case one can easily see the system of equations does not close. &6\\
\hspace{0.18 cm}3 T + 1R & $\exists$: MKR & $\exists$ $\gamma_{ij}=\left(
\begin{array}{ccc}
 2 \text{c} & 0 & 0 \\
 0 & \text{c} & 0 \\
 0 & 0 & \text{c} \\
\end{array}
\right)$ &4\\
\hspace{0.18 cm}3 T + 3 R & $\exists$ & $\nexists$ requirement for 3 Ts restricts the components of $\gamma_1$ to be constant, in which equations for 3R are not solvable. &6\\

$\begin{array}{c}\text{3 T + 3 R} \\ \text{+ D}\end{array} $& $\exists$ & $\nexists$ - explanation is analogous to the one for 3T+3R&7\\
\hspace{0.18 cm}3 T + D & $\exists$ & $\nexists$ - requirement for 3 Ts restricts the components of $\gamma_1$ to be constant, in which equation for D is not solvable. & 4\\

\hspace{0.18 cm}3 T + 3 R & $\exists$ & $\nexists$ requirement for 3 Ts restricts the components of $\gamma_1$ to be constant, in which equations for 3R are not solvable. &6\\

$\begin{array}{c}\text{3 T + 3 R}\\ \text{+ D}\end{array}$ & $\exists$ & $\nexists$ - explanation is analogous to the one for 3T+3R&7\\
\hspace{0.18 cm}3 T + D & $\exists$ & $\nexists$ - requirement for 3 Ts restricts the components of $\gamma_1$ to be constant, in which equation for D is not solvable. & 4\\  

\hline
\hspace{0.18 cm}1 R & $\exists$ & $\exists$ see equation (\ref{trot}) & 1 \\ 
\hspace{0.18 cm}3 R & $\exists$ & $\exists$ see equation (\ref{only3rot})& 3 \\
\hspace{0.18 cm}1 R+D & $\exists$ & $\gamma_{ij}^{(1)}=\left(
\begin{array}{ccc}
 \frac{c_1\left(\frac{x^2+y^2}{2 t^2}\right)}{t} & \frac{a x+b y}{t^2} & \frac{a y-b x}{t^2} \\
 \frac{a x+b y}{t^2} & \frac{c_1\left(\frac{x^2+y^2}{2 t^2}\right)}{2 t} & 0 \\
 \frac{a y-b x}{t^2} & 0 & \frac{c_1\left(\frac{x^2+y^2}{2 t^2}\right)}{2 t} \\
\end{array}
\right)$ &2 \\ 
\hspace{0.18 cm}3 R+D & $\exists$  & $\exists$ & 4\\
\hspace{0.18 cm} 3 R+D & $\exists$  &$\exists$ see equation (\ref{tn1})& 4 \\ \hline
\hspace{0.18 cm}1 R + 2 SCT &   $\exists$  & $\exists$ see equation (\ref{eqq}) & 3\\ 
\hspace{0.18 cm}1 R + 3 SCT & $\exists$   & $\exists$ see equation (\ref{rsct})  & 4\\ \hline
\hspace{0.18 cm}3R+3SCT & $\exists$   &  The equations are not solvable simultaneously  & 3 \\ \hline
\hspace{0.18 cm}1 R+D+2 SCT & $\exists$ & $\exists$ see equation (\ref{t3}) & 4\\
$\begin{array}{c}\text{1 R+D}\\ \text{+3 SCT}\end{array}$ & $\exists$ & the system of equations does not close, except for $\gamma^{(1)}_{ij}=0$  & 5\\ \hline
$\begin{array}{c}\text{3 R+D}\\\text{+3 SCT} \end{array}$& $\exists$ & $\nexists$ up to now -  equations  are not solvable simultaneously  (the claim is valid w/o assumptions or simplifications) & 5\\
\hspace{0.18 cm}1 SCT & $\exists$ &$\exists$ see equation (\ref{t4}) & 1\\
\hspace{0.18 cm}2 SCT & $\exists$ & $\exists$ see equation (\ref{t5}) & 2\\
\hspace{0.18 cm}3 SCT &$\exists$ & \textcolor{blue}{} & 3\\ \hline
\hspace{0.18 cm}1 SCT+D & $\exists$ & $\exists$ see equation (\ref{t6}) & 1\\
\hspace{0.18 cm}2 SCT+D & $\exists$ &$\exists$ see equation (\ref{t7}) & 2\\
\hspace{0.18 cm}3 SCT+D &$\exists$ & the system of equations does not close, except for $\gamma^{(1)}_{ij}=0$ & 3\\ \hline
 \hline
\end{tabular}
\end{table}
\end{center}
 The subalgebra with 3 R and D is realised in $\gam$
 \begin{align}
\gamma_{11}^{(1)}&=  \frac{\left(2 t^2+x^2+y^2\right) c_2}{2 t^3 \left(-\frac{-t^2+x^2+y^2}{t^2}\right)^{3/2}} & &
\gamma_{12}^{(1)}=\frac{3 x c_2}{2 t^2 \left(-\frac{-t^2+x^2+y^2}{t^2}\right)^{3/2}}\nonumber \\
\gamma_{13}^{(1)}&= -\frac{3 y c_2}{2 t^2 \left(-\frac{-t^2+x^2+y^2}{t^2}\right)^{3/2}} &&
\gamma_{22}^{(1)}=\frac{\left(t^2+2 x^2-y^2\right) \sqrt{-\left(-t^2+x^2+y^2\right)} c_2}{2 \left(-t^2+x^2+y^2\right)^2}\nonumber \\
\gamma_{23}^{(1)}&=\frac{3 x y c_2}{2 t^3 \left(-\frac{-t^2+x^2+y^2}{t^2}\right)^{3/2}} &&
\gamma_{33}^{(1)}=\frac{\sqrt{-\left(-t^2+x^2+y^2\right)} \left(t^2-x^2+2 y^2\right) c_2}{2 \left(-t^2+x^2+y^2\right)^2}\label{tn1}
 \end{align}
\noindent while the subalgebra that contains 1 R and 2 SCTs (rotation and SCTs in $x$ and $y$ direction) is obtained for the $\gam$
 \begin{align} 
\gamma_{11}^{(1)}&= \frac{\left(t^4+4 \left(x^2+y^2\right) t^2+\left(x^2+y^2\right)^2\right) c_2\left(\frac{-t^2+x^2+y^2}{t}\right)}{12 t^3} \nonumber \\ \nonumber
 \gamma_{12}^{(1)}&= -\frac{x \left(t^2+x^2+y^2\right) c_2\left(\frac{-t^2+x^2+y^2}{t}\right)}{4 t^2} \\ \nonumber
\gamma_{13}^{(1)}&= -\frac{y \left(t^2+x^2+y^2\right) c_2\left(\frac{-t^2+x^2+y^2}{t}\right)}{4 t^2}  \\ \nonumber
 \gamma_{22}^{(1)}&=\frac{\left(t^4+2 \left(5 x^2-y^2\right) t^2+\left(x^2+y^2\right)^2\right) c_2\left(\frac{-t^2+x^2+y^2}{t}\right)}{24 t^3}  \\ 
 \gamma_{23}^{(1)}&= -\frac{y \left(t^2+x^2+y^2\right) c_2\left(\frac{-t^2+x^2+y^2}{t}\right)}{4 t^2} \label{eqq},
 \end{align}
\noindent here, $c_2$ is a function of $\frac{-t^2+x^2+y^2}{t}$. The sub algebra that realises  1 R, 2 SCTs and D is
 \begin{align}
 \gamma_{12}^{(1)}& = -\frac{x \left(t^2+x^2+y^2\right) c_3}{4 \left(-t^2+x^2+y^2\right)^2} &&
 \gamma_{11}^{(1)}=\frac{\left(t^4+4 \left(x^2+y^2\right) t^2+\left(x^2+y^2\right)^2\right) c_3}{12 t \left(-t^2+x^2+y^2\right)^2} \nonumber \\
\gamma_{13}^{(1)}&= -\frac{y \left(t^2+x^2+y^2\right) c_3}{4 \left(-t^2+x^2+y^2\right)^2}  &&
\gamma_{22}^{(1)}= \frac{\left(t^4+2 \left(5 x^2-y^2\right) t^2+\left(x^2+y^2\right)^2\right) c_3}{24 t \left(-t^2+x^2+y^2\right)^2}  \nonumber \\
\gamma_{23}^{(1)}&=\frac{t x y c_3}{2 \left(-t^2+x^2+y^2\right)^2}, && \label{t3}
 \end{align}
\noindent where we solve (\ref{lo}) with $\xi^{(0)i}=D^{i}$ and with $\gam$ (eqq) (that has function $c_2$) for $c_2$. 
\noindent The example of $\gam$ that realises 1 SCT (in $y$ direction) is 
\begin{align}
\gamma_{11}^{(1)}&= -\frac{\left(t^2+2 x^2\right) c_1\left(\frac{x}{t},\frac{-t^2+x^2+y^2}{t}\right)}{3 t^2 x} &&\gamma_{12}^{(1)}=\frac{c_1\left(\frac{x}{t},\frac{-t^2+x^2+y^2}{t}\right)}{t} \nonumber \end{align} \begin{align}
\gamma_{22}^{(1)}&= -\frac{\left(2 t^2+x^2\right) c_1\left(\frac{x}{t},\frac{-t^2+x^2+y^2}{t}\right)}{3 t^2 x} \nonumber \\ \gamma_{33}^{(1)}&=\frac{(t-x) (t+x) c_1\left(\frac{x}{t},\frac{-t^2+x^2+y^2}{t}\right)}{3 t^2 x} , \label{t4}
\end{align}
$\gamma_{13}^{(1)}=\gamma_{23}^{(1)}=0$,
 where one can notice the function $c_1\left(\frac{x}{t},\frac{-t^2+x^2+y^2}{t}\right)$ that allows to use the $\gam$ (\ref{t4}) in (\ref{lo}) and solve further desired KVs. 
(To avoid clutter we have given $\gam$  (\ref{t4}) that is not of the most general form, the most general form of the $\gam$ is given in the appendix: Classification.)
 $\gam$ that conserves SCT in $x$ direction is similarly to (\ref{t4})
\begin{align}
  \gamma_{11}^{(1)}&=   -\frac{\left(t^2+2 y^2\right) c_1\left(\frac{y}{t},\frac{-t^2+x^2+y^2}{t}\right)}{3 t^2 y}&&   \gamma_{13}^{(1)}=\frac{c_1\left(\frac{y}{t},\frac{-t^2+x^2+y^2}{t}\right)}{t} \nonumber \\  
\gamma_{22}^{(1)}&=  \frac{(t-y) (t+y) c_1\left(\frac{y}{t},\frac{-t^2+x^2+y^2}{t}\right)}{3 t^2 y} &&   \gamma_{33}^{(1)}=-\frac{\left(2 t^2+y^2\right) c_1\left(\frac{y}{t},\frac{-t^2+x^2+y^2}{t}\right)}{3 t^2 y}, \label{sctx}
\end{align}where $\gamma_{12}^{(1)}$ and $\gamma_{23}^{(1)}$ are zero. 
  The $\gam$ that conserves SCT in $t$ direction, computed with analogous simplifications as $\gam$ for SCT in $x$ and $y$ direction has different form 
 \begin{align}
 \gamma_{11}^{(1)}&=-\frac{e^{\tanh ^{-1}\left(\frac{t^2+x^2+y^2}{t^2-x^2-y^2}\right)} \left(x^2+y^2\right) c_1\left(\frac{y}{x},\log \left(-\frac{-t^2+x^2+y^2}{x}\right)\right)}{3 t x y}
\nonumber \\
 \gamma_{22}^{(1)}&=\frac{e^{\tanh ^{-1}\left(\frac{t^2+x^2+y^2}{t^2-x^2-y^2}\right)} \left(x^2-2 y^2\right) c_1\left(\frac{y}{x},\log \left(-\frac{-t^2+x^2+y^2}{x}\right)\right)}{3 t x y}
 \nonumber \\
 \gamma_{23}&=\frac{e^{\tanh ^{-1}\left(\frac{t^2+x^2+y^2}{t^2-x^2-y^2}\right)} c_1\left(\frac{y}{x},\log \left(-\frac{-t^2+x^2+y^2}{x}\right)\right)}{\sqrt{t^2}} 
 \end{align}
\noindent $
 \gamma_{12}^{(1)}=0$, $\gamma_{13}^{(1)}=0$ and with that acknowledges Minkowski background metric. 
$\gam$ that realises 2 SCTs (SCT in $y$ and $t$ direction) reads
\begin{align}
 \gamma_{12}^{(1)}&=-\frac{(t+y) \left(x^2+(t+y)^2\right)}{2 x^2}&& \gamma_{11}^{(1)}=\frac{\left(x^2+(t+y)^2\right)^2}{4 x^3} \nonumber \\
\gamma_{13}^{(1)}&= -\frac{(t-x+y) (t+x+y) \left(x^2+(t+y)^2\right)}{4 x^3} &&\gamma_{22}^{(1)}=\frac{(t+y)^2}{x} \nonumber \\
\gamma_{23}^{(1)}&=\frac{(t+y) (t-x+y) (t+x+y)}{2 x^2} && \nonumber \\ \gamma_{33}^{(1)}&=\frac{(t-x+y)^2 (t+x+y)^2}{4 x^3} && \label{t5}
\end{align} while $\gam$ that realises 1 SCT and one D (SCT in $y$ direction) is
 \begin{align}
 \gamma_{11}^{(1)}&= \frac{t^2 y \left(t^2+x^2+y^2\right) c_9\left(\frac{x}{t}\right)}{2 x^2 \left(-t^2+x^2+y^2\right)^2} && \gamma_{12}^{(1)}=-\frac{t y \left(3 t^2+x^2+y^2\right) c_9\left(\frac{x}{t}\right)}{4 x \left(-t^2+x^2+y^2\right)^2}\nonumber \\
 \gamma_{13}^{(1)}&=-\frac{t \left(t^4+6 y^2 t^2-x^4+y^4\right) c_9\left(\frac{x}{t}\right)}{8 x^2 \left(-t^2+x^2+y^2\right)^2} && \gamma_{22}^{(1)}= \frac{t^2 y c_9\left(\frac{x}{t}\right)}{\left(-t^2+x^2+y^2\right)^2}\nonumber \\
 \gamma_{23}^{(1)}&=\frac{t^2 \left(t^2-x^2+3 y^2\right) c_9\left(\frac{x}{t}\right)}{4 x \left(-t^2+x^2+y^2\right)^2} && \gamma_{33}^{(1)}=\frac{t^2 y \left(t^2-x^2+y^2\right) c_9\left(\frac{x}{t}\right)}{2 x^2 \left(-t^2+x^2+y^2\right)^2} \label{t6}
 \end{align}
  and $\gam$ that realises 2 SCTs and a D (SCTs in $x$ and $y$ directions) is
 \begin{equation}
 \gam=\left(
\begin{array}{ccc}
 \frac{y \left(t^2+x^2+y^2\right) c_{10}}{2 \left(-t^2+x^2+y^2\right)^2} & -\frac{x y \left(3 t^2+x^2+y^2\right) c_{10}}{4 t \left(-t^2+x^2+y^2\right)^2} & -\frac{\left(t^4+6 y^2 t^2-x^4+y^4\right) c_{10}}{8 t \left(-t^2+x^2+y^2\right)^2}  \\
 -\frac{x y \left(3 t^2+x^2+y^2\right) c_{10}}{4 t \left(-t^2+x^2+y^2\right)^2} & \frac{x^2 y c_{10}}{\left(-t^2+x^2+y^2\right)^2} & \frac{x \left(t^2-x^2+3 y^2\right) c_{10}}{4 \left(-t^2+x^2+y^2\right)^2}  \\
 -\frac{\left(t^4+6 y^2 t^2-x^4+y^4\right) c_{10}}{8 t \left(-t^2+x^2+y^2\right)^2} & \frac{x \left(t^2-x^2+3 y^2\right) c_{10}}{4 \left(-t^2+x^2+y^2\right)^2} & \frac{y \left(t^2-x^2+y^2\right) c_{10}}{2 \left(-t^2+x^2+y^2\right)^2} 
\end{array}
\right).\label{t7}
 \end{equation}

Let us notice that the largest realised subalgebra consisted of original KVs of CA is 4 dimensional. 
The importance of above analysis is to find the $\gam$ for each of the KVs. Then, one can expect form of $\gam$ for subalgebra of CA built from linearly combined KVs.
Which can eventually lead to a global solution of CG.

\section{Patera et al. Classification}

The subalgebras  (SAs) of conformal algebra $o(3,2)$ have been classified in \cite{Patera:1976my}, they are formed from the generators of the conformal algebra (\ref{ca1},\ref{ca2}), or in particular, from the linear combination of these generators. The subalgebras are
\begin{enumerate}
\item sim(2,1) is similitude algebra that we have encountered when talking about the constant $\gam$ matrix. It contains 5 dimensional subalgebra for which we found realised $\gam$. sim(2,1), contains the highest number of KVs, which is seven. 
\item  opt(2,1), optical algebra, contains equal maximal number of generators as similitude algebra.
\item $o(3)\oplus o(2)$ is a maximal compact sub algebra with maximally four generators.
\item $o(2)\oplus o(2,1)$ is a sub algebra that contains maximally four generators which are built from the KVs of conformal algebra
\item o(2,2) is a  sub algebra with six as a highest number of generators in SA.
\item o(3,1) defines Lorentz group in four dimensions that contains the SA with maximally six  and lower number of generators, while it does not contain the SA with five.
\item o(2,1) is the irreducible SA with maximally only three generators.
\end{enumerate}
Let us define the nomenclature. We  define a  group $O(p,q)$ for $p$ and $q$ integers that satisfy $p\leq0$ as a closed linear group of all matrices $M$ of degree $p+q$ over the field of real numbers $\mathbb{R} $ that satisfy the matrix equation
\begin{equation}
MD_{p+q}M^T=D_{p+q},
\end{equation}
  for $M^T$ matrix transposed to $M$, and $D_{p,q}=\left(\begin{array}{cc}I_p & \\ & -I_p \end{array}\right)$ with $I_p$ identity matrix of degree $p $.
  The groups that we need beside $O(p,q)$ are  
  \begin{enumerate}
 \item SO(p,q) that consists of elements $g$ of $O(p,q)$ group with $\det g=1$
 \item $O_1(p,q)$ that consists of the elements $g$ of $O(p,q)$that have $\text{spn} g=1$. Where $\text{spn}$ is the {\it spinor norm}.
 Spinor norm is defined to be spn$g=1$ for g an identity component of $O(p,q)$ denoted with $SO_0(p,q)$ (that simultaneously means det$g=1$), or if det$g=-1$ and the product of $g$ with certain member $M_1=\left(\begin{array}{ccccc}1 & & & &\\& &1&& \\ &&..& & \\& &&1&  \\ &&&& -1\end{array} \right)$ of $O(p,q)$ is not in $SO_0(p,q)$. In other case 
 spinor norm is spn$g=-1$.
  \end{enumerate} 
We consider which linear combinations of original generators realise $\gam$
and focus on realisations of $\gam$ matrices for the algebras with the highest number of generators,  7,6,5 and 4.
  
\subsection{sim(2,1) Algebra}
7 generators of the similitude algebra can be identified with (\ref{simid}), however that is not the only identification of the generators one can use, analogous identification can be obtained using the SCTs instead of Ts.
We classify the realised subalgebras in the following table. 
First volume denotes the name of the subalgebra obtained by combination of the 
known algebras. The second column denotes the name from Patera et al. \cite{Patera:1976my}, in the third column are generators as denoted in \cite{Patera:1976my} and in the fourth column are $\gam$ that realises the subalgebra. The names of subalgebras from Patera et al. \cite{Patera:1976my} are defined with two subscripts, first subscript defines the dimension of the sub 
algebra, while the second subscript enumerates the subalgebras of the same dimension. In each of the subalgebras first are listed the decomposable subalgebras, then indecomposable ones. The superscript, for example $ a_{4,8}^{a}, a_{4,11}^{b}$, denotes the algebra that depends on the parameter, where we simultaneously write the range of the parameter (for example  $b>0, \neq1$ for $a^b_{4,10}$).
If one range is written, it is equal under $o(3,2)$ and the identity component of the corresponding maximal subgroup (here $sim(2,1)$). In case the range under the maximal subgroup is larger than under $o(3,2)$ the larger range is denoted wight the square brackets, for example in case of $a_{4,8}^{\epsilon}$ it is written $\epsilon=1[\epsilon=\pm1]$, which means that $a^{-1}_{4,8}$ is conjugate to $a_{4,8}^1$ under $o(3,2)$ (and even $SO_0(3,2)$), 
but not under the identity component $sim(2,1)$.
\begin{center}
\begin{table}
\hspace{-0.5cm}\begin{tabular}{ | l  | p{2.5 cm} | p{4.0 cm} | p{6cm} |}
 \hline
  \multicolumn{4}{|c|}{Realized subalgebras} \\
  \hline
$\begin{array}{c}\text{ Name/ }\\ \text{ commutators}\end{array}$&Patera name&generators  & Realisation  \\ \hline\hline
&$a_{5,4}^a$ & $\begin{array}{l}F+\frac{1}{2}K_2,-K_1+L_3,\\P_0,P_1,P_2\end{array}$  & see equation (\ref{five2})
\\
&$a\neq0,\pm1$&$a=\frac{1}{2}$&\\
$\mathcal{R}\oplus o(3)$ &$a_{4,1}=b_{4,6}$ & $P_1\oplus\left\{K_2,P_0,P_2\right\}$  & see equation (\ref{trans3rb}) for $\xi^{(4)}$ \\

 &$a_{4,2}$ & $P_0-P_2\oplus\left\{F-K_2;P_0+P_2,P_1\right\}$  &  see equation (\ref{trans3rb}) for $\xi^{(3)}$ \\

$\begin{array}{c}
MKR \\  \mathcal{R}\oplus o(3)\end{array}$ &$a_{4,3}$ & $P_0\oplus\left\{L_3,P_1,P_2\right\}$  &  see equation (\ref{trans3rot1}) for $\xi^{(5)}$ \\

 &$a_{4,4}$ & $F\oplus\left\{K_1,K_2,L_3\right\}$ & see equation (\ref{tn1})\\

 &$a_{4,5}$ & $F\{K_2;P_0-P_2\}\oplus\left\{F-K_2,P_1\right\}$ &$ \left(
\begin{array}{ccc}
 0 & \frac{c}{t-y} & 0 \\
 \frac{c}{t-y} & 0 & \frac{c}{y-t} \\
 0 & \frac{c}{y-t} & 0 \\
\end{array}
\right)$\\
&$a_{4,6}=b_{4,9}$ & 
$ \begin{array}{c} \left\{ F+K_2,P_0-P_2 \right\}\oplus  \\ \left\{F-K_2,P_0+P_2\right\} \end{array} $
 &
$\left(
\begin{array}{ccc}
 \frac{c}{x} & 0 & 0 \\
 0 & \frac{2 c}{x} & 0 \\
 0 & 0 & -\frac{c}{x} \\
\end{array}
\right)$,and (\ref{poind}) \\

&$a_{4,7}$&$ \begin{array}{c}  L_3-K_1,P_0+P_2 ;  \\ P_0-P_2,P_1 \end{array} $& leads to 5 KV subalgebra for constant components of $\gam$\\
 &$\begin{array}{c}a_{4,10}^b=b_{4,13}\\ b>0,\neq1 \end{array}$ & $\left\{ F-bK_2,P_0,P_1,P_2 \right\}$  &  $\left(
\begin{array}{ccc}
 \text{c} & 0 & \text{c} \\
 0 & 0 & 0 \\
 \text{c} & 0 & \text{c} \\
\end{array}
\right), \left(
\begin{array}{ccc}
 0 & \text{c} & 0 \\
 \text{c} & 0 & -\text{c} \\
 0 & -\text{c} & 0 \\
\end{array}
\right) $, $\left(
\begin{array}{ccc}
 0 & \text{c} & 0 \\
 \text{c} & 0 & \text{c} \\
 0 & \text{c} & 0 \\
\end{array}
\right) $, $\begin{array}{l}\text{fourth one leads to}\\ \text{5KV subalgebra}\end{array}$\\

 &$\begin{array}{c}a_{4,11}^b=b_{4,13}\\ b>0,[b\neq0] \end{array}$ & $\left\{ F+bL_3,P_0,P_1,P_2 \right\}$  &  $\left(
\begin{array}{ccc}
 0 & \text{c} & i \text{c} \\
 \text{c} & 0 & 0 \\
 i \text{c} & 0 & 0 \\
\end{array}
\right),\left(
\begin{array}{ccc}
 0 & \text{c} & -i \text{c} \\
 \text{c} & 0 & 0 \\
 -i \text{c} & 0 & 0 \\
\end{array}
\right),$ $\left(
\begin{array}{ccc}
 0 & 0 & 0 \\
 0 & i \text{c} & \text{c} \\
 0 & \text{c} & -i \text{c} \\
\end{array}
\right) $,  $\begin{array}{l}\text{fourth one leads to}\\ \text{5KV subalgebra}\end{array}$\\

\hline 
 \end{tabular}
 \label{tablesim}
\end{table}
\end{center}

\begin{center}
\hspace{-0.3cm}\begin{tabular}{ | l  | p{2.5 cm} | p{4.3 cm} | p{5.cm} |}
 \hline
 & $\begin{array} {c}a_{4,12}^{\epsilon}=b_{4,14}\\
 \epsilon=1*[\epsilon\pm1]
 \end{array}$ & 
$ \begin{array}{c}
 \big\{ F+K_2+\epsilon(P_0+P_2), \\
 -K_1+L_3,P_0-P_2,P_1 \big\}
 \end{array} $
&  $\left(
\begin{array}{ccc}
 ce^{\frac{y-t}{4 e}}  & 0 & -ce^{\frac{y-t}{4 e}} \\
 0 & 0 & 0 \\
 -ce^{\frac{y-t}{4 e}}  & 0 & ce^{\frac{y-t}{4 e}}  \\
\end{array}
\right) $\\

 &$a_{4,13}=b_{4,15}$ & $\begin{array}{l}\big\{ F-K_2,P_0-P_2,\\-K_1+L_3,P_1 \big\}\end{array}$   & $\left(
\begin{array}{ccc}
 \frac{c}{(y-t)^{3/2}} & 0 & -\frac{c}{(y-t)^{3/2}} \\
 0 & 0 & 0 \\
 -\frac{c}{(y-t)^{3/2}} & 0 & \frac{c}{(y-t)^{3/2}} \\
\end{array}
\right) $ \\

&$\tilde{a}_{4,14}$ & $\begin{array}{l}\big\{ F,-K_1+L_3,\\P_1,P_0-P_2 \big\}\end{array}$   & $\left(
\begin{array}{ccc}
 -\frac{c}{y-t} & 0 & \frac{c}{y-t} \\
 0 & 0 & 0 \\
 \frac{c}{y-t} & 0 & -\frac{c}{y-t} \\
\end{array}
\right)$ \\

 &$a_{4,15}=b_{4,17}$ & $\begin{array}{l}\big\{ F+bK_2,-K_1+L_3,\\P_0-P_2,P_1 \big\}\end{array}$ & 
$\begin{array}{l} \gamma_{11}^{(1)}=c\cdot(y-t)^{\frac{1-2 b}{b-1}},\\ \gamma_{13}^{(1)}=-c\cdot(y-t)^{\frac{1-2 b}{b-1}},\\ \gamma_{33}^{(1)}=c\cdot(y-t)^{\frac{1-2 b}{b-1}} \\ \gamma_{12}^{(1)}=\gamma_{22}^{(1)}=0\end{array}$,
 \\

&$\begin{array}{l}a_{4,16}^{a}\\ a=1*[a=\pm1]\end{array}$&$\begin{array}{c}\big\{F+\frac{1}{2}K_2;-K_1+L_3\\ +a(P_0+P_2),P_0-P_2,P_1\big\}\end{array}$& $\left(
\begin{array}{ccc}
 -c_3 & 0 & c_3 \\
 0 & 0 & 0 \\
 c_3 & 0 & -c_3 \\
\end{array}
\right)$, leads to subalgebra with 5 KVs \\

$ \begin{array}{c} \text{Extended}\\ \text{2d Poincare}\end{array}$&$a_{4,17}=b_{4,17}$ & $\left\{ F,L_3,P_1,P_2 \right\}$  & $\left(
\begin{array}{ccc}
 \frac{c}{t} & 0 & 0 \\
 0 & \frac{c}{2 t} & 0 \\
 0 & 0 & \frac{c}{2 t} \\
\end{array}
\right)$ \\

\hline 
 \end{tabular}
 \label{tablesim}
\end{center}

The first five subalgebras that are realised, we have commented earlier.
The following subalgebra is $a_{4,5}$ whose $\gam$ depends on the difference $y-t$, in the analogous behaviour to dependency on one coordinate. One may expect this behaviour redefining the translational KVs into two new KVs, one that is a difference of two translational KVs, and another one, their sum.
The subalgebra $a_{4,6}$ we have commented in the text above the (\ref{poind}), while subalgebra $a_{4,7}$ for constant components leads to $\gam$ that agrees with $\gam$ for 5 KV subalgebras, which is exhibited adding one more KV to $a_{4,7}$. Subalgebras $a_{4,10}$ and $a_{4,11}$ show similar interesting behaviour. Since both of them contain 3 KVs of translation, they admit only constant components in $\gam$. They provide four different $\gam$ one of which admits one more KV. In addition, $a_{4,11}$ contains imaginary value. Subalgebras $a^{\epsilon}_{4,12}$, $a_{4,13}$, $\tilde{a}_{4,14}$ and $a_{4,15}$ are similar to $a_{4,5}$ in a sense that their $\gam$ depends on $y-t$. Interesting property of the algebra $a_{4,15}$ which depends on "one coordinate"  is that the exponent that appears in  $\gam$ is defined with the parameter that enters one of the generators of the subalgebra. 
The subalgebra $a_{4,17}$ defines the $\gam$ matrix that is the analog of the (\ref{poind}) in $t$ coordinate. The subalgebras with the highest number of generators (7,6,5 and 4) which do not have realised $\gam$ are listed in the appendix: Classification. 

\subsection{opt(2,1) Algebra}

One way to define the generators of optical algebra is
\begin{align}
W&=-\frac{\xi^{(6)}+\xi^{(4)}}{2} & K_1&=\frac{\xi^{(6)}-\xi^{(4)}}{2} \\ K_2&=\frac{1}{2}\left[\xi^{(0)}-\xi^{(2)}+\frac{(\xi^{(8)}-\xi^{(9)})}{2}\right] &
L_3&=\frac{1}{2}\left[\xi^{(0)}-\xi^{(2)}-\frac{(\xi^{(8)}-\xi^{(9)})}{2}\right] \\ M&=-\sqrt{2}\xi^{(1)} & Q&=\frac{\xi^{(5)}-\xi^{(3)}}{2\sqrt2} \\
N&=-(\xi^{(0)}+\xi^{(2)}) & &
\end{align}
\noindent in which we can automatically see the two other possible identifications. Since  each of the original generators that enter the definition appear in each of the coordinates, we can permute them to obtain the realised $\gam$ that depend on particular coordinate. 
The generators close into the algebra
\begin{align}
[K_1,K_2]&=-L_3,  & [L_3,K_1]&=K_2, & [L_3,K_2]&=-K_1, & [M,Q]&=-N, & &,  \\ [K_1,M]&=-\frac{1}{2} M, &[K_1,Q]&=\frac{1}{2}Q, & [K_1,N]&=0, &[K_2,M]&=\frac{1}{2}Q, \\
 [K_2,Q]&=\frac{1}{2}M, & [K_2,N]&=0   &
[L_3,M]&=-\frac{1}{2}Q, & [L_3,Q]&=\frac{1}{2}M, \\  [L_3,N]&=0 &
[W,M]&=\frac{1}{2}M,  & [W,Q]&=\frac{1}{2}Q, &[W,N]&=\frac{1}{2}N.
  \end{align}
 There are five realised subalgebras that we write in the following table

\begin{center}
\begin{table}
\hspace{-0.3cm}\begin{tabular}{ | l  | p{2.5 cm} | p{3.9 cm} | p{5.2cm} |}
\hline
  \multicolumn{4}{|c|}{Realised subalgebras of $opt(2,1)$} \\
  \hline
$\begin{array}{c}Name/ \\ commutators\end{array}$&Patera name&generators  & realisation  \\ \hline\hline
 &$b_{5,6}=a_{5,4}$ & $W+aK_1,K_2+L_3,M,Q,N$  & \\
 &$b_{4,1}$&$N\oplus\{;K_1,K_2,L_3\}$ & $\left(
\begin{array}{ccc}
 \frac{c}{x^3} & 0 & -\frac{c}{x^3} \\
 0 & 0 & 0 \\
 -\frac{c}{x^3} & 0 & \frac{c}{x^3} \\
\end{array}
\right)$ \\
 &$b_{4,2}$& $W\oplus\left\{ K_1,K_2,L_3 \right\}$ & see eq. (\ref{eqn1}) \\
 &$b_{4,3}$& $L_3,Q,M,N$ & see eq. (\ref{b43}) \\
&$\begin{array}{c}b_{4,4}\\ b>0\ast [b\neq0]\end{array}$& $W+bL_3,Q,M,N$ &  see eq. (\ref{b44}).\\
\hline
\hline
\end{tabular}
\end{table}
\end{center}
\begin{center}
\begin{table}
\hspace{-0.3cm}\begin{tabular}{ | l  | p{2.8 cm} | p{3.6 cm} | p{5.2cm} |}
\hline
  \multicolumn{4}{|c|}{Realised subalgebras of $opt(2,1)$, continuation} \\
\hline
$\begin{array}{c} Extended\\ 2d Poincare\end{array}$ &$\overline{b}_{4,9}=a_{4,6}$ & 
$ \begin{array}{c} \left\{K_1,K_2+L_3 \right\}\oplus  \\ \left\{W,N\right\} \end{array} $
 &  see equation (\ref{poind}) \\ 
 &$\begin{array}{c} \overline{b}_{4,11}\sim a_{4,8}^{-1} \end{array}$ & $\begin{array}{l} \big\{ W-K_1+Q\\ K_2+L_3,M,N\big\}\end{array}$ &$\left(
\begin{array}{ccc}
 -\frac{5 \text{c}}{3 \sqrt{2}} & \text{c} & \frac{\text{c}}{\sqrt{2}} \\
 \text{c} & -\frac{1}{3} \left(2 \sqrt{2} \text{c}\right) & -\text{c} \\
 \frac{\text{c}}{\sqrt{2}} & -\text{c} & -\frac{\text{c}}{3 \sqrt{2}} \\
\end{array}
\right) $ \\

 &$\begin{array}{l}\overline{b}^b_{4,17}=a_{4,15}^{\text{\tiny{ (1-b)/(1+b)}}}\\ b>0,b\neq1\end{array}$ & $\begin{array}{l}\big\{W-bK_1,M,Q,N \big\}\end{array}$ & 
$\begin{array}{l} \gamma_{11}^{(1)}=c\cdot(y-t)^{-\frac{3 b+1}{2 b}},\\ \gamma_{13}^{(1)}=-c\cdot(y-t)^{-\frac{3 b+1}{2 b}},\\ \gamma_{33}^{(1)}=c\cdot(y-t)^{-\frac{3 b+1}{2 b}} \\ \gamma_{12}^{(1)}=\gamma_{22}^{(1)}=0\end{array}$,
\\
\hline
\end{tabular}
\end{table}
\end{center}

 $\gamma_{ij}^{(1)}$ matrices for $b_{4,2}, b_{4,3}$ and $b_{4,4}$ subalgebras from the table, are respectively\begin{align}
\gamma_{ij}^{(1)}&=\left(
\begin{array}{ccc}
 \frac{c\cdot\left(x^2+3 (t+y)^2\right) }{x^3} & -\frac{3 c\cdot(t+y) }{x^2} & -\frac{3 c\cdot(t+y)^2 }{x^3} \\
 -\frac{3c \cdot(t+y) }{x^2} & \frac{2 c}{x} & \frac{3 c\cdot(t+y) }{x^2} \\
 -\frac{3 c\cdot(t+y)^2 }{x^3} & \frac{3 c\cdot(t+y) }{x^2} & -\frac{c\cdot\left(x^2-3 (t+y)^2\right) }{x^3} \\
\end{array}
\right)\label{eqn1},\\
\gamma_{ij}^{(1)}&=\left(
\begin{array}{ccc}
 -\frac{c}{\left((t-y)^2+4\right)^{3/2}} & 0 & \frac{c}{\left((t-y)^2+4\right)^{3/2}} \\
 0 & 0 & 0 \\
 \frac{c}{\left((t-y)^2+4\right)^{3/2}} & 0 & -\frac{c}{\left((t-y)^2+4\right)^{3/2}} \\
\end{array}
\right)\label{b43},\\
\gamma_{ij}^{(1)}&=\left(
\begin{array}{ccc}
 \frac{c\cdot e^{\frac{\tan ^{-1}\left(\frac{t-y}{2}\right)}{b}}}{\left((t-y)^2+4\right)^{3/2}} & 0 & -\frac{c\cdot e^{\frac{\tan ^{-1}\left(\frac{t-y}{2}\right)}{b}} }{\left((t-y)^2+4\right)^{3/2}} \\
 0 & 0 & 0 \\
 -\frac{c\cdot e^{\frac{\tan ^{-1}\left(\frac{t-y}{2}\right)}{b}} }{\left((t-y)^2+4\right)^{3/2}} & 0 & \frac{c \cdot e^{\frac{\tan ^{-1}\left(\frac{t-y}{2}\right)}{b}} }{\left((t-y)^2+4\right)^{3/2}} \\
\end{array}
\right)\label{b44}.
\end{align}In the table, an asterisk after the range under $o(3,2)$ (e.g. in $a^{\epsilon}_{4,12}$) means that the range needs to be doubled in case one considers conjugacy under $SO_0(3,2)$ [or $O_1(3,1)$] rather than under $O(3,2)$ [or $SO(3,2)$]. For example, $\epsilon=1\ast$ means that $\epsilon=\pm1$ under $SO_0(3,2)$, and $b>0\ast$ indicates that $b\neq0,-\infty<b<\infty$ under $SO_0(3,2)$.

Let us analyse the structure of the above matrices and compare the ingredients with the matrices obtained for the original KVs.  $b_{4,1}$ depends similarly on the coordinates in the denominator to the $\gam$ that conserves two SCTs. We can notice that $K_2$ and $L_3$ that appear in $b_{4,1}$, both contain SCT KVs in the definition. In $b_{4,3}$ and $b_{4,2}$ in the denominator we notice the power $3/2$ that appears in $\gam$ for 3R+D, while hyperbolic tangens appears in $\gam$ that conserves the SCT in $t$ direction. Further $b$ subalgebras that are already mentioned in $sim(2,1)$ group consist of $b_{4,5}=a_{4,18}, \overline{b}_{4,6}=a_{4,1}, \overline{b}_{4,10}=a_{4,7},\overline{b}^b_{4,13}=a^{(b-1)/(b+1)}_{4,10} \text{ with }0<|b|<1 \text{ and }[b\neq0,\pm1], \overline{b}^{\epsilon}_{4,14}=a^{\epsilon}_{4,12} \text{ with }\epsilon=1\ast[\epsilon=\pm1],\overline{b}_{4,15}=a_{4,13},\overline{b}_{4,16}=a_{4,14}$ while the subalgebras that are not realised are 
  \begin{center}
\begin{tabular}{|c|c|}\hline
 \hline
  \multicolumn{2}{|c|}{Subalgebras that are not realised} \\
  \hline
Patera name&generators   \\ \hline\hline
$ b_{7,1}$ &$ W,K_1,K_2,L_3,M,Q,N$ \\
$b_{6,1}$& $ K_1,K_2,L_3,M,Q,N$\\
$\overline{b}_{6,2}=a_{6,2} $& $W,K_1,K_2,+L_3,M,Q,N$ \\ \hline
$b_{5,1}$ & $\begin{array}{c}\{K_1,K_2,L_3\}\oplus\{W,N\}\end{array}$\\
$b_{5,2}$ & $W,L_3,M,Q,N$\\
$\overline{b}_{5,3}=a_{5,1}$ & $W+K_1,K_2+L_3,M,Q,N$\\
$\overline{b}_{5,4}=a_{5,2}$ &$ K_1,K_2+L_3,M,Q,N$\\
$\overline{b}_{5,5}=a_{5,3}$ &$W,K_2+L_3,M,Q,N $\\
$\overline{b}_{5,7}=a_{5,5}$ &$ W-K_1,K_2+L_3,M,Q,N$\\
$\overline{b}_{5,8=a_{5,6}}$ & $ W,K_1,K_2+L_3,M,N $\\
$\overline{b}_{5,9}=a_{5,8}$ & $W,K_1,M,Q,N$\\
$b_{4,5}=a_{4,18}$ &$ W,Q,M,N$\\
$\overline{b}_{4,7}=a_{4,2}$ &$ N\oplus\{K_1,K_2+L_3,M\}$\\
$\overline{b}_{4,8}=a_{4,5}$ & $\{W+K_1,N\}\oplus\{K_1,M\},$\\
\hline
$\overline{b}_{4,10}=a_{4,7}$ &$ K_2+L_3,Q,M,N$\\
$\overline{b}_{4,18}\sim a^{-1}_{4,16}$ &$ W-\frac{1}{3}K_1,K_2+L_3+Q,M,N$\\
\hline
 \end{tabular}
\end{center}

 $o(3)\oplus o(2)$ does not contain realised subalgebras, while $o(2)\oplus o(2,1)$ contains one algebra with four generators. This is the algebra that we have  encountered, the one formed by 1T+1SCT+1R+D and identified with $sl(2)\oplus u(1)$. 
Next we consider $o(2,2)$.

\subsection{o(2,2) Algebra}
 The subagebra $o(2,2)$ contains generators 
 \begin{align}
A_1&=-\frac{1}{2}\left[\frac{\xi^{(9)}+\xi^{(8)}}{2}-(\xi^{(0)}+\xi^{(2)})\right], && A_2=\frac{1}{2}(\xi^{(6)}+\xi^{(4)}),  \nonumber \\
 A_3&=\frac{1}{2}\left[-\frac{\xi^{(9)}+\xi^{(8)}}{2}-(\xi^{(0)}+\xi^{(2)})\right], &&
B_1=-\frac{1}{2}\left[\frac{-\xi^{(9)}+\xi^{(8)}}{2}+(\xi^{(0)}-\xi^{(2)})\right], \nonumber \\ B_2&=\frac{1}{2}(\xi^{(6)}-\xi^{(4)}),  && B_3=\frac{1}{2}\left[\frac{\xi^{(9)}-\xi^{(8)}}{2}+(\xi^{(0)}-\xi^{(2)})\right]. \label{genlo22}
\end{align}\noindent As above, one may obtain $\gam$ that depends on the remaining two coordinates by permutiation of the original generators in the definition (\ref{genlo22}).
The algebra is isomorphic to $o(2,1)\oplus o(2,1)$ and defined with commutation relations 
\begin{align}
[A_1,A_2]&=-A_3,& [A_3,A_1]&=A2, & [A_2,A_3]&=A_1, \\
[B_1,B_2]&=-B_3, & [B_3,B_1]&=B_2, & [B_2,B_3]&=B_1, \\ [A_i,B_k]&=0 &(i,k=1,2,3)   
\end{align}
It contains the subalgebras with number of generators from six to one, while two largest algebras are not realised

  \begin{center}
\begin{tabular}{|c|c|}\hline
 \hline
  \multicolumn{2}{|c|}{Subalgebras that are not realised} \\
  \hline
Patera name&generators   \\ \hline\hline
$ e_{6,1}$ &$ \{ A_1,A_2,A_3 \}\oplus\{B_1,B_2,B_3\}$ \\
$ \overline{e}_{5,1}=b_{5,1}$ &$ \{ A_1,A_2-A_3 \}\oplus\{B_1,B_2,B_3\}.$ \\
\hline
 \end{tabular}
\end{center}
Part of the subalgebras with the lower number of generators, 4 are equal to the algebras  $\overline{e}_{4,2}=a_{4,6},\overline{e}_{4,3}=b_{4,1},\overline{e}_{4,4}=b_{4,2}$. 

\subsection{o(3,1) Lorentz Algebra}

Four dimensional Lorentz algebra we can write on the three dimensional hypersurface 
using 
\begin{align}
L_1&=\xi^{(7)}+\frac{\xi^{(2)}}{2}, & L_2&=\xi^{(5)}, & L_3&=\xi^{(8)}+\frac{1}{2}\xi^{(1)}, \\
K_1&=\xi^{(8)}-\frac{1}{2}\xi^{(1)}, & K_2&=\xi^{(6)}, & K_3&=-\xi^{(7)}+\frac{1}{2}\xi^{(2)} .
\end{align}
that close with the commutation relations
\begin{align}
[L_i,L_j]&=\epsilon_{ijk}L_k, & [L_i,K_j]&=\epsilon_{ijk}K_k, & [K_i,K_j]&=-\epsilon_{ijk}L_k.
\end{align}
The algebra $o(3,1)$ itself is not realised in the form of $\gam$ while the first highest sub algebra of $o(3,1)$, $\overline{f}_{4,1}$, that contains four generators $K_1,L_1;L_2-K_3,L_3+K_2$, is equal to the algebra $a_{4,17}$ that realises 2 dimensional Poincare algebra (\ref{poind}).

From the $\gam$ matrices found for the flat background metric $\gamma_{ij}^{(0)}$, one can find the transformation to the $\gam$ in the spherical background metric. We consider that in the appendix: Classification: Map to Spherical Coordinates Using Global Coordinates and 5 KV Algebra.
The relation between the subalgebras written using original KVs in the previous chapter with the subalgebras in the Patera et al. classification one can find in the appendix: Classification: Map from Classification of KVs from Conformal Algebra to Patera et. al Classfication.

\section{Global Solutions: Bottom-Up Approach}

Bottom-Up approach is the approach to building a global solution based on the known asymptotical solution of the $\gam$ matrix and the known symmetries. The first candidates from which to deduce the global solution are 
the subalgebras with the highest number of KVs because they exhibit the most symmetries. In our case that is the subalgebra with 5 CKVs. The global solution written as an asymptotically AdS solution with $\gam$ that of a 5 CKV subalgebra, and the higher order terms in the FG expansion set to zero, is the solution of the full Bach equation.

For other $\gam$  matrices from classification, one cannot use the analogous procedure, however Bach equation can simplify. 

\subsection{Geon Solution}
The global solution that arises from the 5 dimensional algebra $so(2)\ltimes o(1,1)$ or in Patera et al. notation, $a_{5,4}=b_{5,6}$, that conserves $\gam$ matrices in (\ref{five}), defines a geon which is analog to pp-wave solution. One can consider geon \cite{Wheeler:1955zz} in several different notions \cite{Melvin:1963qx}, \cite{Kaup:1968zz}, while we consider them here in the sense of instantons \cite{Anderson:1996pu}, or pp-wave solutions. They were recently discussed and connected with the instability of the AdS spacetime \cite{Horowitz:2014hja}, while interesting use of them includs the nearly linear solution to the vacuum constraint equations representing even-parity ingoing wave packets by imploding from a black hole, which is in fact a formation of a black hole obtained by imploding an axisymmetric gravitational wave, \cite{Abrahams:1992ib}.


We promote an asymptotic solution to a global one using the global metric
\begin{equation} 
ds^2=dr^2+(-1+cf(r))dt^2+2cf(r)dtdx+(1+cf(r))dx^2+dy^2.\label{geon}
\end{equation}
that solves the Bach equation for $f(r)=c_1+c_2 r+c_3 r^2+c_4 r^3$. 

When $c_3$ and $c_4$ coefficients are zero, one obtains Ricci flat metric, and the solution is a solution of EG as well, while the metric (\ref{geon}) is built so that it satisfies AdS boundary conditions. 
Conformal invariance of the metric allows it to be rescaled then the metric, does not keep Ricci flatness. The response functions for the metric (\ref{geon}) 
\begin{align}
\begin{array}{ccc}
\tau_{ij}=\left(\begin{array}{ccc}-c c_4& -c c_4 &0 \\ -c c_4 & -c c_4 &0 \\ 0& 0 & 0  \end{array}\right)
& \text{  and  } &
P_{ij}=\left(\begin{array}{ccc}  c c_3 & c c_3 & 0 \\ c c_3 & c c_3 &0 \\ 0 & 0 & 0   \end{array}\right),
\end{array}
\end{align}
allow us to notice that choosing the form of the metric we can require one or both of the response functions to vanish. From the response functions, we compute the charges and the currents $J_i=Q_{ij}\xi^{j}$ of the solution, where $\xi^j$ are 5 CKVs that form the subalgebra.

The charge associated with the timelike KV $(1,0,0)$ is $-(2cc_4)$ per square unit of AdS radius, as well as the charge associated with the KV $(0,1,0)$.  The charge associated with $(0,0,1)$ vanishes. The charges associated with the new KVs  $(t-\frac{x}{2},-\frac{t}{2}+x,y)$ and $(-y,y,-t-x)$ are   $cc_4\cdot(t+x)$, and zero, respectively.
Interestingly, the solution (\ref{geon}) is not conformally flat, however it gives Weyl squared equal to zero and finite polynomial invariants (appendix: Classification).

Another global solution with constant $\gam$ and 4 KVs, it is the solution $a_{4,10}^{b}$ from the $sim(2,1)$ table. For $\gam$  of the form $\left(\begin{array}{ccc}c&0&c\\ 0&0&0\\c&0&c\\ \end{array}\right)$, solution is analogous to the solution of the 5 KV case, i.e. $f(r)$ that solves Bach equation is $f(r)=a_1+a_2r+a_3r^2+a_4r^3$. When the realisation of the global solution is $\left(\begin{array}{ccc}0&c&0\\ c&0&-c\\0&-c&0\\ \end{array}\right)$, allowed coefficients in $f(r)$ that solve the Bach equation are only $a_1$ and $a_2$.

Analogous geon or pp-wave solutions appears also in the dependency on one, two and three coordinates. These solutions have vanishing Weyl squared, while they are not conformally flat. By particular choice of coefficients that multiply r component they can be brought to Ricci flat form, and one of them has the structure of {\it double holography-like solution}.

\subsection{Global Solution with $\gam$ Dependent on One Coordinate}

Ansatz for the global solution of Bach equation 
\begin{equation} 
ds^2=dr^2+(-1+b(x)f(r))dt^2+dx^2+2b(x)f(r)dtdy+(1+b(x)f(r))dy^2,\label{gldep1}
\end{equation}
 solves the Bach equation for two cases
\begin{enumerate}
\item $f=c_1+c_2 r+c_3 r^2+c_4 r^3$  and $b=a_1+a_2 x$
\item $f=c_1+c_2 r$ and $b=a_1+a_2 x +a_3 x^2+a_4 x^3$ 
\end{enumerate}
in which, as in the constant case, one can straightforwardly read out the $\gamma_{ij}^{(i)}$, $i=1,2,3$ matrices. Therefore, if we want vanishing or non-vanishing response functions, we can choose the solution of the Bach equation that gives us that. 
For the case 1. the response functions and charges are non-vanishing which is opposite from the case 2. 
$\gam$ is conserved by the KVs
 \begin{eqnarray}
 \xi^{\LO a}_{1}&=&(0,0,1),\\
\xi^{\LO a}_{2}&=&(t-y,x,-t+y), \\
\xi^{\LO a}_{3}&=&(1,0,0), \label{ricflat}
 \end{eqnarray}
two translations and a combination of the dilatation and boost in $t-y$ plane. From the Patera et al. classification, the subalgebra belongs to sim(2,1) and it is $a^c_{3,19},c\neq0,\pm1,-2$ with KVs $P_0,P_2,F-c K_2$ for $c=1$.
For further insight in the subalgebra one can consider linear combinations of the KVs,  $\xi^{\LO a}_{+}=\xi^{\LO a}_{1}+\xi^{\LO i}_{3}$, $\xi^{\LO a}_{-}=\xi^{\LO a}_{1}-\xi^{\LO a}_{3}$ and $\chi^{(0) a}=\frac{1}{2}\xi^{\LO a}$, that form the ASA $[\chi^{(0) a},\xi^{\LO a}_{-}]=- \xi^{\LO a}_{-}$.
The response functions are analog to those of 5 KV subalgebra with manifest dependency on the x coordinate
\begin{align}\begin{array}{cc}\tau_{ij}=\left(
\begin{array}{ccc}
 -x c_4 & 0 & -x c_4 \\
 0 & 0 & 0 \\
 -x c_4 & 0 & -x c_4 \\
\end{array}
\right) & P_{ij}=\left(
\begin{array}{ccc}
 -x c_4 & 0 & -x c_4 \\
 0 & 0 & 0 \\
 -x c_4 & 0 & -x c_4 \\
\end{array}
\right).\end{array}
\end{align}
The charges of the $(1,0,0)$, $(0,0,y)$ and $(t-y,x,-t+y)$ KVs are $2xc_4$, $2xc_4$ and vanishing, respectively, while 
the metric can be reduced to a Ricci flat solution for a choice of metric
\begin{equation} 
ds^2=dr^2-(1+r x) dt^2+2 r x dtdx + dx^2+(1+r x) dy^2,
\end{equation}
or transparently for, $\gamma^{\FO}_{ij}=\left(\begin{array}{c c c} x & 0& x \\ 0& 0 &0 \\ x& 0 & x\end{array}\right).
$ 

\subsection{Global Solution with $\gam$ Dependent on Two Coordinates}

In global solutions, we can obtain analogous solutions by permuting the components in $\gam$ and KVs in algebra. We have noticed this already in analysis of ASA. Excellent example for that are global solutions with $\gam$ dependent on two coordinates. The metrics
\begin{align} 
ds^2&=dr^2+\left[-1+ a(t+x)f(r)\right] dt^2+2a(t+x)f(r)dtdx \nonumber \\&+ \left[1+ a(t+x)f(r)\right]dx^2+dy^2 \label{dep2ex1} \\
ds^2&=dr^2+\left[-1- b(t-y)f(r)\right] dt^2+dy^2+2b(t-x)f(r)dtdy \nonumber \\& + \left[1-b(t-y)f(r)\right]dy^2
\end{align}
solve the Bach equation for the form the function $f(r)=c_1+c_2 r+c_3r^2+c_4 r^4$ and
give asymptotically desired forms of $\gam$, which are respectively
\begin{align}
\gam&=\left(\begin{array}{ccc} c_ia(t+x) & c_ia(t+x) &0 \\ c_ia(t+x)& c_ia(t+x) & 0 \\ 0&0& 0 \end{array}\right),\nonumber \\ \gam&=\left(\begin{array}{ccc} -c_ib(t-y) & 0 &c_ib(t-y) \\ 0&0 & 0 \\ c_ib(t-y)&0& -c_ib(t-y) \end{array}\right),
\label{globdep2}
\end{align}
 for $i=1,2,3$. 
The algebras given by the solutions are defined with the KVs	
\begin{align}
\begin{array}{cc}
\begin{array}{ll}
\xi^{(n1)}_1&=(-1,1,0) \\
\xi^{(n2)}_1&=(0,0,1) \\
\xi^{(n3)}_1&=(-y,y,-t-x)
\end{array} & 
\begin{array}{ll}
\xi^{(n1)}_2&=(1,0,1) \\
\xi^{(n2)}_2&=(0,1,0) \\
\xi^{(n3)}_2&=( -x ,  y-t , - x) 
\end{array}
\end{array}.
\end{align}
Response functions of the first solution \begin{align} \tau_{ij}&=\left(
\begin{array}{ccc}
 -c_4 \text{b}(t+x) & -c_4 \text{b}(t+x) & 0 \\
 -c_4 \text{b}(t+x) & -c_4 \text{b}(t+x) & 0 \\
 0 & 0 & 0 \\
\end{array}
\right), \\  P_{ij}&=\left(
\begin{array}{ccc}
 c_3 \text{a}(t+x) & c_3 \text{a}(t+x) & 0 \\
 c_3 \text{a}(t+x) & c_3 \text{a}(t+x) & 0 \\
 0 & 0 & 0 \\
\end{array}
\right) \end{align}
give for the charge of the $\xi^{(n1)}_1$  and $\xi^{(n2)}_1$ to vanish, 
while the response functions of the second solution are
\begin{align} \tau_{ij}&=\left(
\begin{array}{ccc}
 c_4 \text{b}(t-y) &0 &  -c_4 \text{b}(t-y) \\
0 & 0 & 0 \\
  -c_4 \text{b}(t-y) & 0 &  c_4 \text{b}(t-y) \\
\end{array}
\right), \\  P_{ij}&=\left(
\begin{array}{ccc}
-c_3 \text{b}(t-y) & 0& c_3 \text{b}(t-y) \\
0& 0 & 0 \\
  c_3 \text{b}(t-y)  & 0 &  -c_3 \text{b}(t-y)  \\
\end{array}
\right). \end{align} The corresponding charges for $\xi^{(n1)}_2$  and $\xi^{(n2)}_2$ vanish, while the charge for $\xi^{(n3)}_2$ is $-4c_4x\cdot b(t-y)$.

Since the functional dependence is maintained in the global solution one can use it to build a {\it double holography like solution}, which leads to one more KV, $\xi^{(6)}$. If we consider a function $a(t+x)=\frac{2}{t+x}$ and substitute $t\rightarrow\chi+\tau$ and $x\rightarrow\chi-\tau$ we obtain
\begin{equation}
ds^2=\frac{4 r d\chi^2}{\chi }+\frac{4 r d\chi dy}{\chi }+dr^2-4 d\tau d\chi +dy^2
\end{equation}
a metric that can be using $r\rightarrow\chi\eta$ brought to a form
\begin{equation}
ds^2=4 \eta  d\chi^2+(\chi  d\eta+\eta  d\chi)^2-4 d\tau d\chi +4 \eta  d\chi dy+dy^2.
\end{equation}
\noindent We have obtained the four dimensional subalgebra with double-holography like global solution.

Very interesting global solutions in this case are the ones whose $\gam$ matrix is specialisation of the (\ref{globdep2}) for one of the sub algebras from the Tables of $sim(2,1)$ and $opt(2,1)$ algebras.
Let us demonstrate this on the algebra $a_{4,12}$. The KVs $\{ F+K_2+\epsilon(P_0+P_2),-K_1+L_3,P_0-P_2,P_1 \}$ lead to the response functions
\begin{align}
\begin{array}{cc}
\tau_{ij}&=\left(
\begin{array}{ccc}
 -(c e^{\frac{1}{4} c \cdot(y-t)} c_4) & 0 & c e^{\frac{1}{4} c\cdot (y-t)} c_4 \\
 0 & 0 & 0 \\
 c e^{\frac{1}{4} c \cdot(y-t)} c_4 & 0 & -(c e^{\frac{1}{4} c\cdot (y-t)} c_4) \\
\end{array}
\right), \\ P_{ij}&= \left(
\begin{array}{ccc}
 c e^{\frac{1}{4} c \cdot(y-t)} c_3 & 0 & -(c e^{\frac{1}{4} c\cdot (y-t)} c_3) \\
 0 & 0 & 0 \\
 -(c e^{\frac{1}{4} c\cdot (y-t)} c_3) & 0 & c e^{\frac{1}{4} c \cdot(y-t)} c_3 \\
\end{array}
\right).
\end{array}
\end{align}
The only non-vanishing charge belongs to KV $P_0-P_2$ and it reads $4cc_4e^{c\cdot(y-t)/4}$.
Similarly, each of the subalgebras from  $sim(2,1)$ and $opt(2,1)$  tables, that are of the form (\ref{globdep2}), and dependent on the $t-y $ coordinates can be realised as global solutions. These are $a^{\epsilon}_{4,12}$, $a_{4,13}$, $a_{4,14}$, $a_{4,15}$, $b_{4,3}$ and $b_{4,4}$.

Going to the higher dimensional subalgebra from the KVs: $\xi_2^{(n1)}=\xi^{(0)}+\xi^{(2)},\xi_2^{(n2)}=\xi^{(1)},\xi_{2}^{(n3)}=\xi^{(5)}-\xi^{(3)}$ conserving (\ref{globdep2})  one can choose fourth KV, solving the equation (\ref{eq:nloke}) for the desired KV.
Choosing the KV

\begin{itemize}
\item
 $\xi^{(6)}$ the function $b(t-y)$ would take the form $b(t-y)=\frac{b_1}{t-y}$, 
 \item choosing $\xi^{(6)}+\xi^{4}+\epsilon(\xi^{(0)}-\xi^{(2)})$ it would become $b(t-y)=b_1\cdot e^{\frac{(t-y)}{2\epsilon}}$. 
 \item  $\xi^{(0)}-\xi^{(2)}$ leads to $b(t-y)$ linear in $t-y$,
 \item $\xi^{(6)}-\xi^{(4)}$ provides $b(t-y)=\frac{b_1}{(t-y)^{3/2}}$,
 \item $\xi^{(4)}$ gives $b(t-y)=\frac{b_1}{(t-y)^2}$,
 \item $\xi^{(6)}+c\xi^{(4)}$ defines $b(t-y)=b_1\cdot(t-y)^{\frac{1-2c}{-1+c}}$.
 \end{itemize}
  Which means that an arbitrary profile $b(t-y)$ breaks $a_{5,4}$ to $(\xi^{(5)}-\xi^{(3)},\xi^{(1)},\xi^{(0)}+\xi^{(2)})$.

In addition, one can focus to study the profile of $\gam$ of (\ref{globdep2}) when $b(t-y)=(t-y)^{\beta}$. Beside the known conserved KVs, one obtains 
\begin{itemize}
\item
$\xi^{(6)}$ for $\beta=-1$
\item $\xi^{(4)}$ for $\beta=-2$
\item and $\xi_{new}=\xi^{(6)}+\frac{1+\beta}{2+\beta}\xi^{(4)}$. 
\end{itemize}

\subsection{Global Solution with $\gamma_{ij}^{(1)}$  Dependent on Three Coordinates}

When we introduce dependency on one more coordinate, we can as well obtain a global solution, starting with a metric of a similar form

\begin{equation} 
ds^2=dr^2+\left[-1+ b(t+x+y)f(r)\right] dt^2+dx^2+2b(t+x+y)f(r)dtdy + dy^2.
\end{equation}

The ansatz satisfies Bach equation (\ref{bach}) for the functions that will give non-vanishing response functions $f(r)=c_1+c_2 r+ c_3 r^2+c_4 r^3$ and $b(t+x+y)=b_1+b_2\cdot(t+x+y)($ up to linear term. Again keeping the $c_i$, $i=1,2,3,4$ coefficients, one obtains
 \begin{align}
 \tau_{ij}&=\left(
\begin{array}{ccc}
 -b_1-(t+x+y) b_2 & 0 & -b_1-(t+x+y) b_2 \\
 0 & 0 & 0 \\
 -b_1-(t+x+y) b_2 & 0 & -b_1-(t+x+y) b_2 \\
\end{array}
\right),\\ P_{ij}&=\left(
\begin{array}{ccc}
 \left(b_1+(t+x+y) b_2\right) c_4 & 0 & \left(b_1+(t+x+y) b_2\right) c_4 \\
 0 & 0 & 0 \\
 \left(b_1+(t+x+y) b_2\right) c_4 & 0 & \left(b_1+(t+x+y) b_2\right) c_4 \\
\end{array}
\right)
 \end{align}
 that conserve KVs  $\chi^{(1)}=(-1,1,0)$ with a corresponding charge $-2c_3\cdot\left[c_1+c_2\cdot(t+x+y)\right]$, and $\chi^{(2)}=(-1,0,1)$ whose charge vanishes.
The KVs form an Abelian algebra $o(2)$.

To solving of the Bach equation one can approach using the top-down approach analogous to analysis of MKR solution in the third chapter. We present that approach in the appendix: Classification: Global Solutions: Top-Down Approach. The possibility of solving the Bach equation asymptotically we present in the appendix: Classification: Asymptotic Solutions.

\newpage

\chapter{One Loop Partition Function}

In this chapter we analyse the conformal gravity one loop partition function. It is one of the key quantities for study  in the AdS/CFT correspondence.
 The computation of entire partition function is not known in general, however we can compute it perturbatively. Once the quantum gravity theory is known it should give microscopic description of the Bekenstein-Hawking entropy, while currently we are able to compute it in the semi-classical limit, when the entropy is related to horizon area. 
However, one-loop computations of the partition function allow determination of the subleading corrections to the semi-classical result.
Computation of the one loop partition function provides also corrections to computations of other thermodynamical quantities. 
Large part of the one-loop partition function analysis has been done in lower dimensions \cite{Maloney:2007ud, Bertin:2011jk,Gaberdiel:2010xv,Zojer:2012rj}. In three dimensions, one loop partition function of EG gives the result anticipated by Brown and Henneaux. It is consisted  of the sum over boundary excitations of $AdS_3$, Virasoro descendants of the AdS vacuum. EG and Chern Simons gravity give also an anticipated result, the Virasoro descendants from the EG and one more part. That provides and evidence that the dual CFT to topologically massive gravity (TMG) at the chiral point is logarithmic \cite{Gaberdiel:2010xv}. 
In higher dimensions however, CFTs do not posses analogous properties as $CFT_2$ and in order to compare the partition function from $AdS$ and $CFT$ side it is essential to consider theory and background of the symmetry that allows such comparison \cite{Beccaria:2014jxa}. For example, conformal spin S partition function has been considered in $CFT_d/AdS_{d+1}$ correspondence with $S^1\times S^{d-1}$ boundary of  $AdS_{d+1}$. For the $d=4$  case, the partition function on $S^1\times S^3$ background, for the conformal higher spin (CHS) field corresponds to double partition function of the CHS field for the positive energy ground states on the $AdS$ background, which is particularity of $d=4$.  In three and five dimensions it was computed in the form of the MacMahon function \cite{Gupta:2012he}.

\section{Heat Kernel}  
The method that we use to study the one loop partition function is the method of the heat kernel. In physics, it was introduced by Fock noting that one can conveniently represent Green's functions as integrals over an auxiliary co-ordinate ("proper time") of a kernel that satisfies the heat equation, and by Schwinger who recognised that through these representations, issues related to renormalisation and gauge invariance in external field are more transparent. It was used by DeWitt as one of the main tools of covariant approach in quantum theories.

Using the asymptotics of a heat kernel one can infer information about the eigenvalue asymptotics which describes recovering of the geometry from a  manifold via the spectrum of a differential operator. In that case one can benefit from knowing the heat kernel coefficients. 

It is used in computation of the vacuum polarisation, the Casimir effect and study of quantum anomalies - the context in which we use it here and it was considered on various manifolds with and without boundaries. Furthermore, a single computation can be used in a various of applications. 

The heat kernel method can be used for various backgrounds and operators. When they are of particular symmetry, for example sphere or hyperbolic space one can compute the partition function analytically. Otherwise, one can study it via the heat kernel coefficients. The formalism that can be used is worldline method \cite{Fliegner:1994zc,Fliegner:1997rk}.
The formalism has been used for computation of the one-loop EG with matter on general backgrounds and in representation with worldline path integrals resulted with correct one loop divergencies \cite{Bastianelli:2013tsa}. 
The operators that can be studied include Laplace operator \cite{Camporesi:1994ga,Gopakumar:2011qs,Beccaria:2016tqy}, GJMS operators \cite{Beccaria:2016tqy,Beccaria:2015vaa}, conformal higher spin operators \cite{Beccaria:2014jxa,Beccaria:2016tqy}, more general ones, or Paneitz operator \cite{Fradkin:1981iu,Fradkin:1981jc,paneitz} an differential operator with construction important in four dimensional conformal differential geometry.


Consider the generating functional for the Green's functions of the field $\phi$, analogously to the procedure in the chapter about variational principle,
\begin{equation}
Z[J]=\int\mathcal{D}\phi\exp(-\mathcal{L}(\phi,J)), \label{pfdef}
\end{equation}
where the case from the first chapter would be obtained for $\phi=g$.
The simple example
for computation of the partition function would be 
\begin{equation}
Z=\int D\phi e^{-g^2S(\phi)}
\end{equation}
for $\phi$, free quantum field (scalar, vector or tensor) 
and the coupling g.  Since $\phi$ is a free field computation is straight forward. 
Action \begin{equation} S(\phi)=\int_{\mathcal{M}}d^3x\sqrt{g}\phi\Delta\phi \end{equation} contains $\Delta$ second order differential operator. It lives on the space of formalisable functions on $\mathcal{M}$ and in general contains discrete and continuous spectrum of eigenvalues. For compact $\mathcal{M}$ ,  $\Delta$ has a discrete spectrum of eigenvalues $\lambda_n$, while on  non-compact and  homogeneous manifolds $\Delta$ has continues spectrum.
The latter causes that the one loop correction 
\begin{equation}
S^{(1)}=-\frac{1}{2}\log\det(\Delta)=-\frac{1}{2}\sum_n\log\lambda_n\label{discr}
\end{equation}
contains divergence proportional to volume of $\mathcal{M}$ that can be absorbed in the local counterterm. 

General computation of $S^{(1)}$ is complicated that manifests mainly for   gauge fields and gravitons. Straightforwardly one has to find a complete basis of normalizable eigenfunctions $\{\psi_n\}$ for which $\Delta\psi_{n}=\lambda_n\psi_m$ and compute the sum directly. 
Other option that we present here is to use the heat kernel approach.

To study one loop partition function in the path integral representation, we need to perturb the Lagrangian $\mathcal{L}$ to second order in fluctuations $\phi$
\begin{equation}
\mathcal{L}=\mathcal{L}_{cl}+\langle\phi,J\rangle+\langle\phi,D\phi\rangle \label{langexp}
\end{equation}
with the first term in the expansion of action evaluated on the classical background and $\langle... \rangle$ an inner product of the quantum fields, defined with
\begin{equation} \langle \phi_1,\phi_2\rangle = \int d^nx\sqrt{g}\phi_1(x)\phi_2(x).\end{equation}
Here, under classical action one includes as well one point functions and considers the entire Lagranigan evaluated on shell, that means the contribution from linear term (one point function) needs to vanish. The external sources however, are arbitrary if one is interested into studying the correlation functions. 
D is a differential operator, and in a simplest case of a quantum scalar field it is a Laplacian with a mass term 
\begin{equation}
D=D_0\equiv-\nabla_{\m}\nabla^{\m}+m^2.
\end{equation}
One defines path integral measure for 
\begin{align}
\text{ tensors }&&1&=\int D h_{\mu\nu} \text{Exp}\left( -\langle h,h\rangle \right), \label{pimt}\\
\text{ vectors }&&
1&=\int D \xi_{\mu} \text{Exp}\left( -\langle \xi,\xi\rangle \right), \label{defxi} \\
\text{ and scalars } && 
1&=\int D s \text{Exp}\left( -\langle s,s\rangle \right). \label{pims} \end{align}
where the right hand side of the above definitions  is divergent, in a strict sense, however the divergence does not depend on external sources on the geometry of the background, and it may be absorbed in an normalisation constant which is irrelevant. 

To evaluate (\ref{pfdef}) we use
\begin{equation}
Z[J]=e^{-\mathcal{L}_{cl}}\text{det}^{-1/2}(D)\exp\big(\frac{1}{4}JD^{-1}J\big) ,\label{gauseval}\end{equation} which is true for $D$ self-adjoint operator, i.e. when $\langle \phi_1,D\phi_2\rangle=\langle D\phi_1,\phi_2\rangle$, for its domain of definition equal to the one of the corresponding adjoint. 
That requirement simultaneously imposes important restrictions on the admissible boundary conditions \cite{Vassilevich:2003xt}.
The formal expression (\ref{hkgen}) to which we can refer to as $K(x,y,t,D)=\langle x|\exp(-t D)|y\rangle,$ needs to satisfy 
\begin{equation}
(\partial_t+D_x)K(x;y;t;D)=0,\label{heqn}
\end{equation}
a heat conduction equation, with the initial condition 
\begin{equation}
K(0;x;y;D)=\delta(x,y).
\end{equation}
The solution for $D=D_0$ (1.4) on flat background $M=\mathbb{R}^n$ is
\begin{equation}
K(x;y;t;D_0)=(4\pi t)^{-n/2}\exp\left(-\frac{(x-y)^2}{4t}-tm^2\right).\label{kd0}
\end{equation}
In case that operator $D$ is more general and contains the potential term or a gauge field, (\ref{kd0}) defines a singularity in the leading order for $t\rightarrow 0$ while the subleading terms act as power-law corrections
\begin{equation}
K(x;y;t;D)=K(x;y;t;D_0)\left(1+tb_2(x,y)+t^2b_4(x,y)+...\right), \label{kexp}
\end{equation}
where the {\it heat kernel coefficients} $b_k(x,y)$ are regular for $y\rightarrow x$. Then, $b_k(x,x)$ are local polynomials of background fields and their derivatives. One can write the propagator $D^{-1}(x,y)$ as
\begin{equation}
D^{-1}(x,y)=\int_0^{\infty}dt K(x;y;t;D)
\end{equation}
and integrate (\ref{kexp})
\begin{equation}
D^{-1}(x,y)\propto 2(4\pi)^{-n/2}\sum_{j=0}\left(\frac{|x-y|}{2m}\right)^{-\frac{1}{2}n+j+1}K_{-\frac{1}{2}n+j+1}(|x-y|m)b_{2j}(x,y).
\end{equation}
Formal integration of the expansion, with $b_0=1$ gives the proportionality to a Bessel function $K_{\nu}(z)$ for small argument z, in which the first several kernel coefficients $b_k$ describe the singularities in the propagator at coinciding points.
The part of the (\ref{gauseval})
\begin{equation}
W=\frac{1}{2}\ln\det (D)\label{func}
\end{equation}
defines one-loop effective action which arrises due to the quantum effects of the background fields, at the one-loop level.

To relate the functional (\ref{func})  and the heat kernel one has to remember that for each positive eigenvalue $\lambda$ of the operator $D$ it is true up to an infinite constant that 
\begin{equation}
\ln \lambda =-\int_0^{\infty}\frac{dt}{t}e^{-t\lambda}. \label{intl}
\end{equation}
The constant does not depend on $\lambda$ that one can convince himself by differentiating both sides of (\ref{intl}) with respect to $\lambda$. $\ln\det(D)=Tr\ln(D)$ gives the relation with a heat kernel
\begin{equation}
W=-\frac{1}{2}\int_{0}^{\infty}\frac{dt}{t}K(t,D) \label{ktd}
\end{equation}
for 
\begin{equation}
K(t,D)=Tr(e^{-tD})=\int d^{n}x\sqrt{g}K(x;x;t;D).\label{trhk}
\end{equation}
Therefore, we can state the following.
In order to analyse the (\ref{gauseval}) one can introduce the heat kernel of the Laplacian $\Delta_{(S)}$  for a spin-S field on a manifold $\mathcal{M}_{d+1}$ between two points x and y 
\begin{equation}
K_{ab}{}^{(S)}(t;x,y)=\left\< y,b |e^{t\Delta_{(S)}}|x,a \right\>=\sum\limits_{n}\psi_{n,a}^{(S)}(x)\psi_{n,b}^{(S)}(y)^*e^{tE_n^{(S)}}\label{hkgen}
\end{equation}
in which the spectrum eigenvalues are $E_{n}^{(S)}$, the normalised eigenfunctions that belong to $\Delta_{(S)}$ are $\psi_{n,a}^{(S)}$, while $a$ and $b$ denote the local Lorentz indices of the field. 
By tracing over the spin and the spacetime labels we define the trace of the heat kernel
\begin{equation}
K^{(S)}(t)\equiv Tr e^{t\Delta_{(S)}}=\int_{\mathcal{M}}d^{d+1}x\sqrt{g}\sum\limits_aK_{aa}^{(S)}(t;x,x) e^{tE_n^{(S)}}\label{trhkgen},
\end{equation}
and relate the one-loop partition function to the trace of the heat kernel 
\begin{equation}
ln Z^{(S)}=ln\det(-\Delta_{(S)})=Tr ln(-\Delta_{(S)})=-\int\limits_0^{\infty}\frac{dt}{t}Tre^{t\Delta_{(S)}}=-\int\limits_0^{\infty}\frac{dt}{t}K^{(S)}(t) \label{pf}.
\end{equation}
The issue that may arise is that the integral  (\ref{ktd}), (\ref{pf}) may be divergent in both limits. 
When $t=\infty$ D can obtain zero or negative eigenvalue that cause infra-red divergencies. 
When the mass $m$ is sufficiently large for the integral to be convergent in the upper limit, they are not encountered.
At the lower limit, divergencies cannot be analogously removed, in order to remove them one has to introduce a cut off for $t=\Lambda^{-2}$ 
\begin{equation}
W_{\Lambda}=-\frac{1}{2}\int_{\Lambda^{-2}}^{\infty}\frac{dt}{t}K(t,D).
\end{equation}
The divergent part of $W_{\Lambda}$ in the limit $\Lambda\rightarrow\infty$  
\begin{align}
W_{\Lambda}^{div}&=-(4\pi)^{-n/2}\int d^n x\sqrt{g}\bigg[\sum_{2(j+l)<n}\Lambda^{n-2j-2l} b_{2j}(x,x)\frac{(-m^2)^{l}l!}{n-2j-2l}\\ &+\sum_{2(j+l)=n}\ln(\Lambda)(-m^2)^ll!b_{2j}(x,x)+\mathcal{O}(\Lambda^0)\bigg]
\end{align}
contains ultra-violet divergencies for the $b_{k(x,x)}$ with $k\leq n$.
The integral for $b_0(x,x)$ is divergent for the non-compact manifolds and  one removes this divergency using the subtraction of the "reference heat kernel".
The higher heat kernel coefficients $b_k$ ($k>n$) are not divergent and their contribution to the effective action reads for $\Lambda\rightarrow\infty$
\begin{equation}
-\frac{1}{2}(4\pi)^{-n/2}m^n\int d^nx\sqrt{g}\sum_{2j<n}\frac{b_{2j}(x,x)}{m^{2j}}\Gamma(2j-n)
\end{equation}
which corresponds to a large mass expansion that can be applied on the weak and slowly varying background fields.

The property of the heat kernel expansion which we are interested in is the description of the one-loop divergencies and counter terms in order to study quantum anomalies. Beside that, heat kernel can be used for studying short-distance behaviour of the propagator, $1/m$ expansion of the effective action (as we have seen above),  perturbative expansions of the effective action,  selected  non-perturbative relations for the effective action.

The information is contained in the geometric invariants and there is no distinction for different spins or gauge groups which allows generalisation to the arbitrary space-time dimensions. One computation can be used for many applications, and knowing the structure of the heat kernel is useful for computations with complicated geometries. Among the popular examples of the geometries studied via the heat kernel are Dirichlet branes. 
The deficiencies of the heat kernel are that it works less effectively when bosonic and fermionic quantum fields mix, while the biggest is that {\it "..heat kernel is not applicable beyond the on-loop approximation. It is not clear whether necessary generalisations to higher loop could be achieved at all."} \cite{Vassilevich:2003xt}.

The heat kernel have been used in the treatment of mathematical problems  related to expansion in coefficients \cite{Fliegner:1997rk,Fliegner:1994zc}, for computation of Casimir energy\cite{Giombi:2014yra} and Bose-Einstein condensation, for quantum field theory in curved spaces, quantisation of gauge theories and from the point of vie in quantum cosmology. It provides information for the zeta function, and one may study it using the DeWitt approach and the path integral.
For further applications one may consult \cite{Vassilevich:2003xt}.

\section{Group Theoretic Approach to Heat Kernel}

\subsection{Heat Kernel for Partially Massless STT Field}

The approach that we have described, considers the heat kernel coefficients. For sphere $S^n$, hyperbolic space $\mathbb{H}^n$ and their cosets as backgrounds, the equation (\ref{ktd}), (\ref{trhk}), (\ref{pf}) can be solved analytically.  Furthermore, the fields $\phi$ in (\ref{langexp}) that are symmetric, transverse and traceless, simplify the computations.

We consider determinants (\ref{func}) for the symmetric transverse traceless fields (STT) and evaluate the corresponding heat kernel (\ref{pf}). To evaluate the heat kernel (\ref{pf}), one could solve the appropriate heat equation (\ref{heqn}) by direct evaluation and construction of the eigenvalues and eigenfunctions of the spin-S Laplacian  on a manifold $\mathcal{M}$ and computation of the resulting sum, or for homogeneous $\mathcal{M}$,  by the  a group theoretic techniques \cite{Gopakumar:2011qs}.  
The evaluation of the heat kernel with group theoretic techniques we can describe with four steps
\begin{enumerate}
\item evaluation of the heat kernel on the symmetric space
 \item  and then on the coset space of the symmetric space. We consider the heat kernel on the sphere, on the coset space of the sphere ("thermal quotient of $S^{2n}$"), 
 \item and analytically continuate to hyperbolic space (Euclidean hyperboloid) \item and coset space of hyperbolic space that is thermal AdS ("thermal quotient of $\mathbb{H}^{2n}$"). 
 \end{enumerate}
That kind of analysis is also called harmonic analysis.

Group theoretic approach 
 is more subtle for the even dimensional spaces. For odd dimensional spaces the contribution that appears is from the principal series, 
 while In consideration of general tensor fields one can have contribution from the discrete series. However, they do not contribute to the STT field that we are considering here.

\section{Traced Heat Kernel for Even-Dimensional Hyperboloids}

\subsection{Step 1. Heat Kernel on $S^{2n}$}
The manifold we start with, on which we considered (\ref{hkgen}), is 2n sphere $S^{2n}\simeq SO(2n+1)/SO(2n)$. We denote it here with $\mathcal{M}$. 
Knowing  $\mathcal{M}$ we induce $\psi_{n,a}^{S}$ and $E_n^{(S)}$.

If we have two compact Lie groups G and H for which $H\in G$ we can define representation $R$ of $G$ with corresponding space $V_R$ of the dimension $d_R$ and analogously an unitary irreducible representation  $S$ of $H$   with vector space $V_S$ of a dimension $d_S$. The indices on $V_S$ (subspace of $V_R$) are denoted with $a$, and the indices on $V_R$ with $I$.
Then, define the quotienting with the right action of H on G with a coset space $G/H$ by $G/H=\{gH\}$ for $g\in G$, while the quotienting with the left action is  $\Gamma\setminus G/H$.

The coset space $G/H$ and $G$ have a projection map $\sigma:G/H\rightarrow G$ with the corresponding map $\pi:G\rightarrow G/H$, where  $\pi\circ\sigma= e$, and $e$ is identity in G. This map determines $\psi_{n,a}^{S}$ of $\Delta_{(S)}$ in terms of matrix elements. 
Once we defined the section $\sigma(gH)=g_0$ for $g_0\in gH$  chosen  to obey predefined rules \cite{Gopakumar:2011qs}, we define the matrix element 
\begin{equation}
 \psi_a^{(S)I}(x)=\mathcal{U}^{(R)}(\sigma(x)^{-1})_{a}^I. \label{ef1}
 \end{equation}
Using this notation (\ref{ef1}) for the eigenfuction, the heat kernel between two points x and y (\ref{hkgen}) is \begin{align} 
K_{ab}(x,y;t)=\sum\limits_{R}a_R^{(S)}\mathcal{U}^{(R)}(\sigma (x)^{-1}\sigma(y))_a{}^{b}e^{tE_R^{(S)}}\label{kab}.
\end{align}
The indices $n$ of energy eigenvalue in (\ref{kab}), are denoted with labels ($R,I$) and we have introduced the $a_R^{(S)}=\frac{d_R}{d_S}\frac{1}{V_{G/H}}$, for $V_{G/H}$ volume of the $G/H$ space.
We omit the index $I$ since the energy eigenvalues of the coset spaces SO(N+1)/SO(N) and SO(N,1)/SO(N) 
contain representation S within representation R only once because the egienfunctions with equal R and different I are degenerate \cite{Gopakumar:2011qs}. (\ref{trhkgen}) becomes
\begin{equation}
K^{(S)}(x,y;t)\equiv\sum\limits_{a=1}^{d_S}K_{aa}^{(S)}(x,y;t)= \sum_{R}a_R^{(S)}Tr_{S}(\mathcal{U}^{(R)}(\sigma(x)^{-1}\sigma(y)))e^{tE_{R}^{(S)}}\label{css}
\end{equation}
in which we define the 
\begin{equation}
Tr_{S}(\mathcal{U})\equiv\sum_{a=1}^{d_s}\langle a,S|\mathcal{U}|a,S\rangle.
\end{equation}
\subsection{Step 2. Heat Kernel on Thermal Quotient of $S^{2n}$}
Thermal quotient of $S^{2n}$ (with $S^{2n}=G/H$) is $\Gamma\backslash G/H$ in which quotienting is done with a discrete group $\Gamma$, isomorphic to $\mathbb{Z}_N$, for thermal quotient of the $S^{2n}$ that can be embedded in $G$. The section that is compatible with the $\Gamma$ quotienting is defined with an element $\gamma\in\Gamma$. Section $\sigma(x) $ is compatible with the quotienting $\Gamma$ if and only if there is $\gamma$ that acts on $x=gH\in G/H$ with $\gamma:gH\rightarrow\gamma\cdot gH$ for which 
\begin{equation}\sigma(\gamma(x))=\gamma\cdot\sigma(x)\label{compat}.\end{equation}
That relation allows to use the {\it method of images} \cite{David:2009xg} 
\begin{equation}
K_{\Gamma}^{(S)}(x,y;t)=\sum\limits_{\gamma\in\Gamma}K^{(S)}(x,\gamma(y);t)\label{mirror}
\end{equation}
which allows computation of the traced heat kernel  $K_{\Gamma}^{(S)}$ between two points x and y on $\Gamma\backslash G/H$. 
Fixing the point x and summing over the images of y, 
 gives an expression for the trace of the heat kernel $K_{\Gamma}^{(S)}$ 
\begin{equation}
K_{\Gamma}^{(S)}(t)=\sum_{m\in \mathbb{Z}_N}\int_{\Gamma\backslash G/H}d\mu(x)\sum_{a}K_{aa}(x,\gamma^m(x);t)\label{mirim}.
\end{equation}
Here,  $d\mu(x)$ defines a measure on $\Gamma\backslash G/H$ obtained from the {\it Haar} measure on G, while x defines points in $\Gamma\backslash G /H$, while $\Gamma\simeq\mathbb{Z}_N$.
 The properties of integral over the quotient space  allow to write (\ref{mirim}) as \cite{Gopakumar:2011qs}
\begin{equation}
K_{\Gamma}^{(S)}=\frac{\alpha_1}{2\pi}\sum\limits_{k\in\mathbb{Z}_N}\sum\limits_{R}\chi_{R}(\gamma^k)e^{t E_R(S)} \label{thk}
\end{equation}
for $\frac{\alpha_1}{2\pi}$ a volume factor of the thermal quotient $\gamma$. $\chi_R$  defines the character of the representation R with $E_R(S)$ eigenvalue of the spin-S Laplacian $\Delta_{(S)}$ on $S^{2n}$. The quotient $\gamma$ is an exponential of the "Cartan" generators of representation R (\ref{compat}), here SO(2n+1),  with an explicit example for the four dimensional case given below.
The representations R of SO(2n+1), are representations that contain $S$ when they are restricted to the $SO(2n)$. 
The eigenvalues $E_R$, necessary for the evaluation of the $K_{\Gamma}^{(S)}$ have been listed in 
\cite{Camporesi:1994ga} and they are
\begin{equation}
E_{R,AdS_{2n}}^{(S)}=-(\lambda^2+\rho^2+s)
\end{equation}
for $\rho\equiv\frac{N-1}{2}$ and $N$ dimension of the space we consider. 

\subsection{Step 3. Heat Kernel on $\mathbb{H}^{2n}$}

From the expression $K_{\Gamma}^{(S)}$ for the heat kernel on $S^{2n}$ we can  define the analogous expression for $K_{\Gamma}^{(S)}$ on $\mathbb{H}^{2n}$. The characters in (\ref{thk}) are  evaluated on the compact symmetric space.  On hyperbolic space, we can expect the heat kernel  to be of that form which is exactly what happens, the eignevalues and eigenfunctions stay the same, while the sum turns into an integral. The unitary representations G that define matrix elements are infinite dimensional since G is not compact, and they have been classified for $SO(N,1)$.
\begin{itemize}
\item
Analogously to the compact case, the analysis on the Euclidean $AdS$ (hyperbolic space $\mathbb{H}^{N}$)  
\begin{equation}
\mathbb{H}_N\approx SO(N,1)/SO(N)
\end{equation}
with N dimension of space, requires writing a sectioning $SO(N,1)$ obtained by analytic continuation form $SO(N+1)$. 
Ilustrativ example is in terms of the coordinates and a line element. If we have defined coordinates on $S^{2n}$ with the metric
\begin{equation}
ds^2=d\theta^2+\cos^2\theta d\phi_1^2+\sin^2\theta d\Omega_{2n-2}^2,
\end{equation}
and an analytic continuation
\begin{align}
\theta\rightarrow-i\rho, && \phi_1\rightarrow i t,\label{acn}
\end{align}
where $\rho,t\in \mathbb{R }$, we analytically continuate to 
\begin{equation}
ds^2=-(d\rho^2+\cosh^2\rho dt^2+\sinh^2\rho d\omega_{2n-2}^2).
\end{equation}
This is equal to continuation $SO(N+1)$ to $SO(N,1)$ via one axis chosen as a time direction, for example axis "1", and continuating the generators $Q_{1j}\rightarrow iQ_{1j}$ that define the corresponding Lie algebras. One can show this explicitly considering the particular number of dimensions. If we express the thermal quotient using the coordinates on $S^4$: complex numbers $z_1,z_2,z_3$, which satisfy the condition 
\begin{equation}
|z_1|^2+|z_2|^2+|z_3|^2=1,
\end{equation}
the quotient is defined with
\begin{equation}
\gamma: \{\phi_i\}\rightarrow\{\phi_i+\alpha_i\}.\label{tq}
\end{equation}
 $\phi_1,\phi_2$ in (\ref{tq})  are phases of the z's and $n_i\alpha_1=2\pi$ for some $n_i\in\mathbb{Z}$ while not all $n_i$s can simultaneously be zero, and 
thermal quotient requires \begin{equation}\alpha_i=0 (\forall i\neq1).\label{alphas}\end{equation}
$\Gamma$ needs to be embedded in SO(5), and for that we  decompose complex numbers into 5 coordinates. The coordinates are real and embed $S^4$  in $\mathbb{R}^5$
\begin{align}
x_1&=\cos\theta\cos\phi_1 & x_2&=\cos\theta\sin\phi_1 \nonumber \\
x_3&=\sin\theta\cos\psi\cos\phi_2 & x_4&=\sin\theta\cos\psi\sin\phi_2 \nonumber \\
x_5&=\sin\theta\sin\psi. &
\end{align} We denote the point in $R^5$ with coordinates (1,0,0,0,0) as a north pole and construct a matrix g(x) which rotates it to the generic point x. $g(x)\in SO(5)$ contains point x on $S^4$ and defines one to one correspondence between $SO(5)$ and $S^4$ up to multiplication by an element of $SO(4)$. North pole is invariant under multiplication by an element of SO(4). 
$g(x)$ can be e.g. 
\begin{equation}
g(x)=e^{i\phi_1Q_{12}}e^{i\phi_2Q_{34}}e^{i\psi Q_{35}}e^{i\theta Q_{13}}\label{egsec}
\end{equation}
where Qs are generators of $SO(5)$. We can recognise that as an element of a section in G over G/H and write the action for the thermal quotient (\ref{tq}) on $g(x)$ (\ref{egsec}) as an embedding of $\Gamma$ in SO(5)

\begin{equation}
\gamma : g(x)\rightarrow g(\gamma(x))=e^{i\alpha_1Q_{12}}e^{i\alpha_2Q_{34}}\cdot x=\gamma\cdot g(x).
\end{equation}
Here we define matrix multiplication with "$\cdot$". Now we can recognise the property (\ref{compat})  and write the thermal section as  
\begin{equation}
\sigma_{th}(x)=g(x).
\end{equation}
This property is used in the method of images for the construction of the heat kernel on $\Gamma\backslash$ SO(N)/SO(N+1).
\end{itemize}

The unitary representations of SO(N,1) that we consider, are those that contain unitary representations of SO(N). 
Using $N=2n$ (for even dimensional hyperboloids) that are unitary representations of principal series \cite{fuchs, fuchs_schw}\footnote{In the mathematical literature "representation space" here shortened into "representation" is referred to with "module".} of $SO(2n,1)$ labelled with
\begin{align}
R=(i\lambda,m_2,m_3,...,m_n), && \lambda\in\mathbb{R}, && m_2\geq m_3\geq...\geq m_n
\end{align}
where $m_i$ are non-negative (half-)integers $m_2,m_3,...,m_n$, which we denote by $\vec{m}$. They contain S of $SO(2n)$ according to branching rules
\cite{Gopakumar:2011qs}
\begin{equation}
s_1\geq m_2\geq s_2\geq ...\geq m_n \geq |s_n|.
\end{equation}
They simplify for $STT$ fields since $m_2=s$, while $m_i=s_{i-1}=0$ for $i>2$ \footnote{There is an exception for n=1 when $|m_2|=s$} when the highest weight of the representation is (s,0,...,0).

\subsection{Step 4. Traced Heat Kernel on thermal $AdS_{2n}$}
 The traced heat kernel of a tensor on the thermal quotient $AdS_{2n}$ ($\mathbb{Z} \backslash G/H$) that is a hyperbolic space $\mathbb{H}_{2n}$ has $\mathbb{Z}$ identification of coordinates
\begin{align}
t \sim t+\beta, & &\beta=i\alpha_1
\end{align}
for $\beta$ an inverse temperature. That, corresponds to analytic continuation by (\ref{acn}) of (\ref{alphas}) identification. Where we have taken into account that $\Gamma\approx\mathbb{Z}$ while for the sphere it was $\mathbb{Z}_N$. Therefore, on the place of the  character of SO(2n+1) in (\ref{thk}) now there is Harish-Chandra character, i.e. global character of the non-compact group SO(2n,1).
Analogously to the (\ref{thk}) the traced heat kernel on thermal $AdS_{2n}$ reads
\begin{equation} 
K^{(S)}(\gamma,t)=\frac{\beta}{2\pi}\sum\limits_{k\in\mathbb{Z}}\sum\limits_{\vec{m}}\int\limits_{0}^{\infty}d\lambda\chi_{\lambda,\vec{m}}(\gamma^k)e^{tE_R^{(S)}}\label{thkads},
\end{equation} \cite{hirai}. One can read out the characters to obtain
\begin{equation}
\chi_{\lambda,\vec{m}}(\beta,\phi_1,\phi_2,...,\phi_n)=\frac{\cos(\beta\lambda)\chi^{SO(2l+1)}_{\vec{m}}(\gamma)}{2^{2l}\sinh^{2l+1}\left(\frac{\beta}{2}\right)}\label{char}
\end{equation}
where for the thermal quotient, $\beta\neq0$, $\phi_i=0$  $ \forall i $ and $l=n-1$ \cite{hirai}. The $\vec{m}=(m_2,...,m_n)$ are highest weights of $\chi^{SO(2l+1)}_{\vec{m}}.$ The character (\ref{char}) has to be inserted into (\ref{thkads}) and integrated. For the STT fields  $\vec{m}=(s,0,..,0)$, which we denote with $(s,0)$, and (\ref{thkads}) becomes
\begin{equation}
K^{(S)}(\beta,t)=\frac{\beta}{2^{2l+1}\sqrt{\pi t}}\sum\limits_{k\in\mathbb{Z}_+}\chi^{SO(2l+1)}_{(s,0)}\frac{1}{\sinh^{2l+1}\frac{k\beta}{2}}e^{-\frac{k^2\beta^2}{4t}-t(\rho^2+s)}.
\end{equation}
The term with $k=0$ was not included into summation, since it diverges. The divergence appears because of the infinite volume of AdS space, over which we integrate the coincident heat kernel. The parameters of the theory can be redefined reabsorbing the term which is not of interest in this case, since it does not depend on $\beta$.

For the evaluation of the heat kernel we have to compute the integral
\begin{equation}
\int \frac{dt}{t^{3/2}}e^{-\frac{a^2}{4t}-b^2 t}=\frac{2\sqrt{\pi}}{a}e^{-ab}
\end{equation}
that enters in the calculation of the one-loop determinant via
\begin{equation}
-\log\det(-\Delta_{(S)}+m_S^2)=\int\limits_0^{\infty}\frac{dt}{t}K^{(S)}(\beta,t)e^{-m_S^2t}.
\end{equation}
That leads to the equation for the traced heat kernel for STT fields
  \begin{equation}
-\log\det(-\Delta_{(S)}+m_S^2)=\frac{1}{2^{2l}}\sum\limits_{k\in\mathbb{Z}_+}\chi^{SO(2l+1)}_{(s,0)}\frac{1}{\sinh^{2l+1}\frac{k\beta}{2}}\frac{1}{k}e^{-k\beta\sqrt{\rho^2+s+m_S^2}},
\end{equation}
that can be more conveniently rewritten as
 \begin{equation}
-\log\det(-\Delta_{(S)}+m_S^2)=\sum\limits_{k\in\mathbb{Z}_+}\chi^{SO(2l+1)}_{(s,0)}\frac{2}{(1-e^{-k\beta})^{2l+1}e^{k\beta l }e^{\frac{k\beta}{2}}}\frac{1}{k}e^{-k\beta \sqrt{\rho^2+s+m_S^2}}.\label{main1}
\end{equation}
From the analogous expression for the heat kernel in odd dimensions
\begin{equation}
-\log\det(-\nabla^2+m_{s}^2)=\sum_{k\in\mathbb{Z}_+}\chi_{(s,0)}^{SO(d-1)}\frac{2 e^{-nk\beta}}{(1-e^{-k\beta})^{2n}k}e^{-k\beta\sqrt{s+n^2+m_s^2}}\label{odd},
\end{equation}
we can conclude the heat kernel on arbitrary dimensional Euclidean AdS spaces, using the substitution $\ell=n-1$, $\rho=\frac{d-1}{2}$, $q=e^{-\beta}$ and substituting $d=(dimension\textit{ of }AdS)=2n+1$ for odd dimensions, and $d=(dimension\textit{ of }AdS)=2n$ for even dimensions, is
\begin{equation}
\log Z_{s,d}(AdS_d)=\sum_{k=1}^{\infty}\frac{(-1)}{k}\frac{q^{k(d-3+s)}}{(1-q^k)^{(d-1)}}\chi_{s,d}\label{zsd},
\end{equation}
for
\begin{equation}
\chi_{s}=(2s+d-3)\frac{(s+d-4)!}{(d-3)!s!}.
\end{equation}

\section{One Loop Partition Function in Four Dimensions} 

The one loop partition function of the gravity theory (\ref{pfdef}) can be written as a multiplication of three terms 
\begin{align}
Z_{one-loop}=\int \mathcal{D}h_{\m\n}\times ghost\times exp(-\delta^{(2)}S).\label{loopstructure}
\end{align}
The ghost term denotes the determinants originating from elimination of gauge degrees of freedom. They are referred to as \textit{ghost determinants}. $\mathcal{D}h_{\m\n}$ is path integral over the perturbations $h_{\m\n}$ around the background that is in our case thermal Euclidean $AdS_{d}$. The term $\exp(-\delta^{(2)}S)$ denotes the exponential of the second variation of the action of the theory.

Once we have obtained the first variation of the action (\ref{var1}) we compute the second variation by varying the action second time 
\eq{
\delta^{(2)}S=\alpha\int d^4 x \left[ \delta EOM^{\alpha\beta}\delta g_{\alpha\beta}+EOM^{\alpha\beta}\delta^{(2)}g_{\alpha\beta} \right].
}{acvar2}
Since the contribution to the one loop partition function comes from the bulk term, when we vary the action one more time we are considering the variation of the bulk term, i.e. EOM, and do not consider the boundary term. The contribution comes essentially from the variation of the Bach tensor, $\big(\nabla^{\delta}\nabla_{\gamma}+\frac{1}{2}R_{\gamma}^{\delta} \big)C^{\gamma}{}_{\alpha\delta\beta}=0$, (\ref{bach}).
We define the metric split 
\begin{equation}
g_{\mu\nu}=\bar{g}_{\mu\nu}+h_{\mu\nu}\label{metricspl}
\end{equation}
in which $\overline{g}_{\m\n}$ is the background $AdS_{4}$ metric and $h_{\m\n}$ is the small perturbation of the metric $g_{\m\n}$ around the background 
\begin{align}
\delta g_{\mu\nu}=h_{\m\n} && \delta g^{\mu\nu}=-h^{\mu\nu}.
\end{align}
The indices are raised and lowered with the background metric. Indices in the perturbative terms are lowered with the background metric, while the indices of tensors are raised and lowered with the entire metric. As in the first chapter. 
The second variation of the metric is
\begin{align}
\delta^{(2)}g_{\mu\nu}=0 && \delta^{(2)}g^{\mu\nu}=-\delta h^{\mu\nu}=2 h^{\mu}{}_{\rho}h^{\rho\mu}.
\end{align}
To evaluate the second variation of action we take into account  simplifications for the $AdS_4$ background. We can express the Riemann tensor using the cosmological constant $\Lambda$ and the background metric $\bar{g}$, \begin{equation}R^{\mu\nu\rho\sigma}=\Lambda(-\bar{g}^{\mu \sigma}\bar{g}^{\nu\rho}+\bar{g}^{\mu\rho}\bar{g}^{\nu\sigma})\label{riema}.\end{equation}Ricci tensor and Ricci scalar are correspondingly simplified and read respectively $R^{\m\n}=3\Lambda g^{\m\n}$ and $R=12\Lambda$.
After second variation of action (\ref{acvar2}), we introduce a decomposition of the metric perturbation $h_{\mu\nu}$ into
\begin{equation}
h_{\mu\nu}(h^{TT},h,\xi)=h_{\mu\nu}^{TT}+\frac{1}{4}g_{\mu\nu}h+2\nabla_{(\mu}\xi_{\nu)}.\label{decomp4}
\end{equation}
Here, transverse traceless part of the metric is $h_{\mu\nu}^{TT}$, trace is $h$ and  $\nabla_{(\mu}\xi_{\nu)}$ defines gauge part. Transverse traceless part of the metric is by definition $h^{TT\mu}{}_{\mu}=\nabla^{\mu}h^{TT}_{\mu\nu}=0$. The gauge part of the metric can be further decomposed into transverse and the gauge part
\eq{\xi_{\mu}(\xi^{T},s)=\xi^{T}_{\mu}+\nabla_{\mu}s,}{dec2} where the transverse part is by definition $\nabla^{\mu}\xi_{\m}^{T}=0$. 
Once the decomposition of the metric and the gauge part of the metric are introduced in the action, we need to verify that the terms in the decomposition containing the trace, scalar and the transverse vector fields vanish. That is due to the gauge and diffeomorphism invariance of the action. Upon permutation of covariant derivatives one indeed obtains the action that is consisted from the transverse traceless tensors
\begin{align}
 \delta^{(2)}S&=\int d^4x \left(8 \Lambda^2 h^{\text{TT}}_{ab} h^{\text{TT}ab}  -6 \Lambda h^{\text{TT}ab} \nabla_{c}\nabla^{c}h^{\text{TT}}_{ab} + h^{\text{TT}ab} \nabla_{d}\nabla^{d}\nabla_{c}\nabla^{c}h^{\text{TT}}_{ab} \right)\label{htt}.
 \end{align}
 The result is consistent with the one from \cite{Giombi:2014yra} and the linearised EOM from \cite{Lu:2011ks} and \cite{Lu:2011zk}. Following the prescription (\ref{loopstructure}) we need to evaluate path integral over the perturbations $\mathcal{D}h_{\mu\nu}$, ghost determinant and the second variation of action (\ref{htt}). We insert the decomposition of the second variation of the action (\ref{decomp4}) 
  in the path integral. The degrees of freedom over which we can trivially integrate are $\xi$ and $h$ since the action is diffeomorphism and scale invariant, and these are degrees of freedom that describe the volume of the gauge group and with which we have to divide the path integral measure. The ghost determinant is defined by the Jacobian $ghost=Z_{gh}$ 
   and change of variables from (\ref{decomp4}) $h_{\mu\nu}\rightarrow(h_{\m\n}^{TT},h,\xi_{\mu})$ 
 \begin{equation}
 \mathcal{D}h_{\mu\nu}=Z_{gh}\mathcal{D}_{\mu\nu}^{TT}\mathcal{D}\xi_{\mu}\mathcal{D}h.
 \end{equation}
One can further change the variables \begin{equation}\xi_{\m}\rightarrow (\xi_{\m}^{T},s) \label{decj1} \end{equation} that 
 decomposes $\xi_{\mu}$ as in (\ref{dec2}). That decomposition brings to an additional determinant $J_1$, $\mathcal{D}\xi_{\m}=J_1\mathcal{D}\xi_{\m}^{T}\mathcal{D}s$, that using normalisation (\ref{defxi}), (\ref{pims}) and ultralocal  invariant scalar products \cite{Gaberdiel:2010xv} 
\begin{align}
\langle h,h'\rangle&=\int d^3x\sqrt{\overline{g}}h^{\m\n}h'_{\m\n} \nonumber \\
\langle \xi,\xi' \rangle&=\int d^3x \sqrt{\overline{g}}\xi^{\m}\xi_{\m}' \nonumber \\
\langle s,s'\rangle&=\int d^3x\sqrt{\overline{g}}ss' \label{ultra}
\end{align} reads \begin{align}
 1&=\int D\xi_{\mu}^TDs J_1\text{Exp}\left(-\int d^4x\sqrt{g}\langle\xi_\nu(\xi^T,s)\xi^{\nu}(\xi^T,s)\rangle \right) \label{detj11} \\
 &=\int D\xi_{\mu}^{T} Ds J_1\text{Exp} \left( -\int d^4x\sqrt{g}\langle(\xi_{\nu}^{T}\xi^{T\nu}-s\nabla^2s)\rangle \right) \label{detj12}  \\
 &=J_1\left[ \det(-\nabla^2)_0 \right]^{-1/2}. \label{detj1}
 \end{align} 
When going from (\ref{detj11}) to (\ref{detj12}), we have inserted and evaluated the decomposition of the gauge part (\ref{dec2}), while when going from (\ref{detj12}) to (\ref{detj1}) we recognised a Gaussian integral. The index "0" denotes the determinant of a scalar field, while indices "1" and  "2" will denote the determinants from the vector and tensor fields respectively. The decomposition of the metric 
\begin{equation}
h_{\mu\nu}(h^{TT},h,\xi)=h_{\mu\nu}^{TT}+\frac{1}{4}\overline{g}_{\mu\nu}h+2\nabla_{(\mu}\xi^T_{\nu)}+2\nabla_{\mu}\nabla_{\nu}s \label{decomp4},
\end{equation}
that corresponds to the change of the variables $h_{\mu\nu}\rightarrow(h^{TT},h,\xi^T,s)$ will contribute with the Jacobian factor $J_2$ 
\begin{align}
1&=\int Dh_{\mu\nu}^{TT}D\xi_{\mu}^{T}DhDs J_2\times\text{Exp}\left\{ -\int d^4x \sqrt{g} h_{\mu\nu}(h_{\mu\nu}^{TT},h,\xi^T_{\mu},s)h^{\mu\nu}(h_{\mu\nu}^{TT},h,\xi^T_{\mu},s) \right\}\nonumber \\ &=\int Dh_{\mu\nu}^{TT}D\xi_{\mu}^{T}DhDs J_2\times \nonumber \\ &  \text{Exp}\bigg\{ -\int d^4x \sqrt{g} \bigg[h^{\text{TT}}_{\m \n} h^{\text{TT}\m \n} + \tfrac{1}{4} h^2 -\xi ^T_{\m} (6 \Lambda+2\nabla_{\n}\nabla^{\n}  ) \xi ^{T\m} \nonumber \\ &\,\,\,\,\,\,\,\,\,\,\,\,\,\,\,\,\,\,\,\,\,\,\,\,\,\,\,\,\,\,\, +  s (12 \Lambda\nabla_{\m}\nabla^{\m}+3\nabla_{\n}\nabla^{\n}\nabla_{\m}\nabla^{\m} )s \bigg]  \bigg\} \nonumber \\
&= J_2 \left[\det\left(12 \Lambda\nabla_{\m}\nabla^{\m}+3\nabla_{\n}\nabla^{\n}\nabla_{\m}\nabla^{\m} \right)_{0}\right]^{-1/2}\left[\det\left( -6 \Lambda-2\nabla_{\n}\nabla^{\n}\right)_{1}\right]^{-1/2}. \label{jac}
\end{align}
Now we can write the partition function for CG in four dimensions
\begin{equation} 
Z^{(4)}_{CG}=Z_{gh}\int Dh_{\mu\nu}^{TT} \text{Exp}(-\delta^{(2)}S) \label{zcg4}
\end{equation}
with ghost determinant  $Z_{gh}$ 
\begin{equation} 
Z_{gh}=\frac{J}{J_0}=\left[ \det(4\Lambda+\nabla^2)_0 \right]^{1/2}\left[\det(-3\Lambda-\nabla^2)_1 \right]^{1/2}.
\end{equation}
The partition function in terms of the determinants reads
\begin{equation}
Z^{(4)}_{CG}=\frac{\left[ \det(4\Lambda+\nabla^2)_0 \right]^{1/2}\left[\det(-3\Lambda-\nabla^2)_1 \right]^{1/2}}{\left[\det(-4\Lambda+\nabla^2)_2\right]^{1/2}\left[\det(-2\Lambda+\nabla^2)_2\right]^{1/2}} \label{zcgd}
\end{equation}
that was studied in \cite{Tseytlin:2013jya}, equation (3.16) and in references therein, namely, \cite{Fradkin:1985am}, \cite{Tseytlin:1984wj} and \cite{Fradkin:1983zz}.  (\ref{zcgd}) agrees with these  partition functions once $\Lambda$ is set to -1. From (\ref{zcgd}) one can recognise partition function of EG in four dimensions
\begin{equation}
Z^{(4)}_{CG}=Z_{EG}\frac{\left[ \det(4\Lambda+\nabla^2)_0 \right]^{1/2}}{\left[\det(-4\Lambda+\nabla^2)_2\right]^{1/2}} \label{zcgeg},
\end{equation}
determinant of the partially massless mode that appear in CG \newline $\left[\det(-4\Lambda+\nabla^2)_2\right]^{1/2}$, and of the conformal ghost $\left[ \det(4\Lambda+\nabla^2)_0 \right]^{1/2}$. Whether determinant is massless, partially massive or massive can be determined form the spin and the dimension of the field. 

From (\ref{main1}) one can compute the partition function on the thermal $AdS_4$ in terms of the characters of the highest weight representation of the SO(3) group, dimension and spin S of the fields
\begin{align}
\log Z^{(4)}_{CG}=\sum_k^{\infty}\frac{-1}{k(1-e^{-k\beta})^3}e^{-k\beta\left(\frac{-3}{2}\right)}&\bigg( \chi_{(0,0)}^{SO(3)}e^{-k\beta\frac{5}{2}}+\chi_{(2,0)}^{SO(3)}e^{-k\beta\frac{3}{2}} \nonumber \\ &+\chi_{(1,0)}^{SO(3)}e^{-k\beta\frac{1}{2}}+\chi_{(2,0)}^{SO(3)}e^{-k\beta\frac{1}{2}} \bigg). \label{zcg}
\end{align}
Using $q=e^{-\beta}$ and reading out the characters \cite{Beccaria:2014jxa} $\chi_{\vec{m}}^{SO(3)}(\phi_1)=1+2S$ for $\phi_1=0$, (\ref{zcg})
 becomes
\begin{equation}
\log Z_{CG}^{(4)}=-\sum_{k\in \mathbb{Z}_{+}}\frac{q^{2k}(-5+4q^{2k}-5q^k)}{(1-q^k)^3k}.\label{zcg4d}
\end{equation}

\section{One Loop Partition Function in Six Dimensions} 

CG in six dimensions has invoked much interest since it belongs to the six dimensional theory of gravity related to string theory \cite{Bastianelli:2000hi}. We consider it from the aspect it arises in the AdS/CFT correspondence. From the string theory perspective it is related to tensionless strings \cite{Baulieu:2001pu}, relevant for the (0,2) theory \cite{Henningson:1998gx}, plays an important role in conformal supergravity \cite{Beccaria:2015uta}, and arises from the seven dimensional gravitational effective action within the $AdS_7/CFT_6$ correspondence \cite{Beccaria:2015ypa}. It has been studied from the ordinary derivative approach \cite{Metsaev:2010kp} and from the geometric analysis of the anomalies \cite{Deser:1993yx} about which we discuss in more detail below. Imposing the right boundary conditions to conformal  gravity in four dimensions one can obtain EG \cite{Maldacena:2011mk}. The procedure has been translated into a formalism that generalises the parameter choice in critical gravity leading to EG \cite{Lu:2011ks}. That procedure allowed generalisation to six dimensions \cite{Lu:2011ks}, and led to analogous conclusions. Its relation to the Seeley-DeWitt coefficients was studied in \cite{Bastianelli:2000hi} and the logarithmic divergence in one loop effective action was also studied on different backgrounds $S^6, CP^3, S^2\times S^4, S^2\times CP^2, S^3\times S^3$ and $S^2\times S^2\times S^2$ \cite{Pang:2012rd}

Conformal anomaly of a classically Weyl invariant theory in six dimensions can be written in a general form \cite{Beccaria:2015ypa}
\begin{align}
A_6\equiv(4\pi)^3\langle T\rangle=-a E_6+ W_6  & & W_6=c_1I_1+c_2 I_2+c_3 I_3.
\end{align}
We denote $E_6=\epsilon_6\epsilon_6 RRR$ as a six dimensional Euler density, for $\epsilon_6$  Levi-Civita tensors, "$R$" Riemann tensors, and a and c coefficients of the theory.  Terms $I_1,I_2,I_3$ are Weyl invariants \cite{Bastianelli:2000rs,Bastianelli:2000hi}. 

Based on their geometry, conformal anomalies \cite{Deser:1976yx,Fradkin:1983tg} can be set into two different classes. One, that 
consists of Weyl invariants that vanish in integer dimensions and arises from finite and scale-free contributions to the effective gravitational action,  proportional to the Euler term. And one that requires the scale. That one is correlated to conformal scalar polynomials that include powers of Weyl tensor and derivatives of the Weyl tensor.

In even integer dimensions of the effective gravitational theories there are terms that do not simultaneously preserve diffeomorphism and Weyl symmetries. In case of the free matter, one cannot simultaneously preserve tracelessness and tracelessness of the stress tensor correlators. 
To maintain the diffeomorphism invariance,  dilatation becomes equal to a scale (which is constant Weyl) transformation.
For the infinitesimal variation of the metric
$\delta g_{\mu\nu}=2\phi(x)g_{\mu\nu}$
and the gravitational action, integrating out the matter field by $S[g_{\mu\nu}]$ gives conformal anomaly 
\begin{equation}
\mathcal{A}(g_{\mu\nu})\equiv\frac{\delta W}{\delta \phi(x)}.
\end{equation}
When the action does not contain scale $\mu$ the anomaly has a vanishing integral
\begin{equation}
\frac{\delta W}{\delta \ln \mu^2}=\int d^d x\mathcal{A}=0\label{typea}
\end{equation}
and since the scalar density $\mathcal{A}$ needs to be related to a topological invariant, an available parity-even candidate is Euler density.
In case that $W$ does not contain scale, the anomaly must reflect this
\begin{equation}
\frac{\delta W}{\delta \ln \mu^2}=\int d^d x \mathcal{A}\neq0.\label{typeb}
\end{equation}
Explicitly in six dimensional case, the anomalous variation can be written as 
\begin{equation}
\delta_{\sigma}W[g]=\int d^6x\sqrt{g}\phi(x)\mathcal{A}(x)
\end{equation}
which by functional differentiation with respect to $\frac{2}{\sqrt{g}}\frac{\delta}{\delta g_{ab}}$
produces an anomalous trace to the stress tensor
\begin{equation}
\langle T^a{}_{a}\rangle =\mathcal{A}(x),
\end{equation}
 dependent on the background curvature \cite{Bastianelli:2000rs}. 
Type A anomaly is (\ref{typea}), while type B anomaly is the one with non-vanishing integral (\ref{typeb}). The third type of anomaly, trivial anomaly, is local and can be removed with a local counterterm \cite{Deser:1993yx}.
The anomalies have been restudied in \cite{Bastianelli:2000hi,Bastianelli:2000rs} while they origin dates from the anomalies from the dimensional regularisation \cite{Capper:1974ic}.
They can be computed using Feynamn graphs, using the heat kernel techniques by De Witt \cite{DeWitt:1964oba} or by a quantum mechanical representation \cite{Fradkin:1982kf}.
In the geometric classification according to type A, type B and trivial anomalies \cite{Bastianelli:2000rs}, the invariants that belong to type A anomaly are\begin{align}
I_1&=C_{\mu\nu\rho\sigma}C^{\mu\lambda\kappa\sigma}C_{\lambda}{}^{\nu\rho}{}_{\kappa} \\
I_2&=C_{\mu\nu\rho\sigma}C^{\rho\sigma\lambda\kappa}C_{\lambda\kappa}{}^{\mu\nu} \\
I_3&=C_{\mu\nu\rho\sigma}\left(\delta^{\mu}_\lambda\Box+4R^\mu_\lambda-\frac{6}{5}R\delta^\mu_\lambda\right)C^{\lambda\nu\rho\sigma}+\nabla_\mu J^\mu \label{invsb}
\end{align}
the tensor $J^{\mu}$  is a trivial Weyl anomaly \cite{Bastianelli:2000rs}. It can be written as 
\begin{equation} 
\nabla_iJ^i=-\frac{2}{3}M_5-\frac{13}{3}M_{6}+2M_7+\frac{1}{3}M_8\label{tan}
\end{equation}
where we define the basis \begin{align}
 K_1&=R^3 && K_2=RR_{ab}^2 && K_3=RR_{abmn}^2 \nonumber \\
 K_4&=R_a^mR_m^iR_i^a && K_5=R_{ab}R_{mn}R^{mabn} && K_6=R_{ab}R^{amnl}R^b_{mnl}\nonumber \\
 K_7&R_{ab}{}^{mn}R_{mn}{}^{ij}R_{ij}{}^{ab}&& K_8=R_{mnab}R^{mnij}R_i{}^{ab}{}_{j}&& K_{9}=R\nabla^2R \nonumber \\
 K_{10}&=R_{ab}\nabla^2R^{ab} && K_{11}=R_{abmn}\nabla^2R^{abmn} && K_12=R^{ab}\nabla_a\nabla_bR \nonumber \\
 K_{13}&=(\nabla_aR_{mn})^2 && K_{14}=\nabla_aR_{bm}\nabla^{b}R^{am} && K_{15}=(\nabla_iR_{abmn}^2) \nonumber \\
 K_16&=\nabla^2R^2 && K_{17}=\nabla^4R, \label{ks}
 \end{align}
for the $M_i$ for $i=5,6,7,8$ \begin{align}
M_5&=6K_6-3K_7+12K_8+K_10-7K_{11}-11K_{13}+12K_{14}-4K_{15}\\
M_6&=-\frac{1}{5}K_9+K_{10}+\frac{2}{5}K_{12}+K_{13}\\
M_7&=K_{4}+K_5-\frac{3}{20}K_9+\frac{4}{5}K_{12}+K_{14}\\
M_8&=-\frac{1}{5}K_9+K_{11}+\frac{2}{5}K_{12}+K_{15}.\end{align}
These terms are cancelled by the local functionals (counterterms obtained as variation of local functionals) given in the Appendix: One Loop Partition Function.
Due to $J^{\mu}$, $I_3$ is locally Weyl invariant when it is multiplied with measure $\sqrt{g}$ and it vanishes for Einstein spaces in which we are interested \cite{Lu:2011ks,Henningson:1998gx}. 

The general combination of invariants $\sum_{i=1}^3 c_i I_i+c_4 E$ does not satisfy Einstein metric, in order for the Einstein metric to satisfy the EOM obtained from the variation of action, the Lagrangin has to be
\begin{equation}
\mathcal{L}=\kappa\left( 4 I_1+I_2-\frac{1}{3}I_3-\frac{1}{24}E_6\right).
\end{equation}
Second variation of the action
\begin{equation}
\mathcal{S}=\kappa\int d^6x \sqrt{|g|} \left( 4 I_1+I_2-\frac{1}{3}I_3-\frac{1}{24}E_6\right).
\end{equation}
analogously to the four dimensional case, leads to the linearised EOM
\begin{equation}
\delta^{(2)}S=\int d^6 x \sqrt{|g|} \delta EOM\delta g_{\mu\nu}.\label{var26}
\end{equation}
In (\ref{var26}) we insert the linearised expansion of the metric (\ref{metricspl}), and define the variations analogously as in four-dimensions.
The tensors are in this case evaluated on $AdS_6$ background on which the  Riemann tensor is expressed in terms of cosmological constant $\Lambda$  and the background metric  $\overline{g}_{\mu\nu}$ (\ref{riema}), as in four dimensional case.
Ricci tensor becomes $R^{\mu\nu}=5\Lambda\overline{g}^{\mu\nu}$, while the Ricci scalar is $R=30\Lambda$. In addition to the conventions taken in the four dimensions, we use transverse traceless gauge of the metric  $\nabla^{\mu}h_{\mu\nu}=0$ and $h^{\mu}_{\nu}=0$. Linearized EOM lead to the action
\begin{equation}
\delta^{(2)}S=\int d^4x \left(8 \Lambda^2 h^{TT}_{ab} h^{TT}{}^{ab} -6 \Lambda h^{TT}{}^{ab} \nabla_{c}\nabla^{c}h^{TT}_{ab} + h^{TT}{}^{ab} \nabla_{d}\nabla^{d}\nabla_{c}\nabla^{c}h^{TT}_{ab}\right)\label{2var6}.
\end{equation}
To evaluate the one loop partition function, we have to insert (\ref{2var6}) into  (\ref{loopstructure}). 
The contribution from the path integral arises from the decomposition of the metric 
\begin{equation}
h_{\mu\nu}=h^{\text{TT}}_{\mu\nu} + \tfrac{1}{6} \bar{g}_{\mu\nu} h +2\nabla_{(\mu}\xi_{\nu)}.
\end{equation}
We divide the path integral measure by the gauge group volume, for the change of the variables $h_{\mu\nu}\rightarrow(h_{\mu\nu}^{TT},h,\xi_{\mu})$ 
\begin{equation}
\mathcal{D}h_{\mu\nu}=Z_{gh}^{(6)}\mathcal{D}h^{TT}_{\mu\nu}\mathcal{\xi}_{\mu}\mathcal{D}h.
\end{equation}
Using the definitions of the path integral measure for tensors, vectors and scalars (\ref{pimt}), (\ref{defxi}) and (\ref{pims}) respectively, and ultralocal invariant scalar products (\ref{ultra}) \cite{Grumiller:2009sn} one can decompose $\xi_{\mu}$ (\ref{decj1}) and from the change of the variables $\mathcal{D}\xi_{\mu}\rightarrow\mathcal{D}\xi_{\mu}^{T}\mathcal{D}s$ obtain 
\begin{align}
 1&=\int D\xi_{\mu}^{T} Ds J_2^{(6)}\text{Exp} \left( -\int d^6x\sqrt{g}(\xi_{\nu}^{T}\xi^{T\nu}-s\nabla^2s) \right) \nonumber \\
 &=J_1^{(6)}\left[ \det(-\nabla^2)_0 \right]^{-1/2}.
 \end{align}
 The decomposition of the metric in six dimensions 
 \begin{equation}
h_{\mu\nu}=h^{\text{TT}}_{\mu\nu} + \tfrac{1}{6} \bar{g}_{\nu\mu} h + \nabla_{\mu}\xi ^T_{\nu} + 2 \nabla_{\mu}\nabla_{\nu}s + \nabla_{\nu}\xi ^T_{\mu},
\end{equation}
leads to analogous ghost determinant, as in four dimensional case. Using the change of variables $h_{\mu\nu}\rightarrow(h^{TT},h,\xi^T,s)$ one finds
\begin{equation}
1=\int J_2^{(6)} \mathcal{D}h_{\mu\nu}^{TT}\mathcal{D}h\mathcal{D}\xi_{\mu}^T\mathcal{D}s \exp\left(-\left<h(h^{TT},h,\xi^T,s),h(h^{TT},h,\xi^T,s) \right>\right)
\end{equation} and obtains \begin{align}
1 &=\int Dh_{\mu\nu}^{TT}D\xi_{\mu}^{T}DhDs J^{(6)}_2  \times  \nonumber \\ \nonumber  & 
 Exp\bigg\{ -\int d^4x \sqrt{g} \bigg[ h^{\text{TT}}_{\mu\nu} h^{\text{TT}\mu\nu} + \tfrac{1}{6} h^2 -\xi ^{T \mu} (10 \Lambda  +2 \nabla_{\nu}\nabla^{\nu})\xi ^T_{\mu} \\ & + s(20 \Lambda  \nabla_{\mu}\nabla^{\mu} + \tfrac{10}{3}  \nabla_{\nu}\nabla^{\nu}\nabla_{\mu}\nabla^{\mu})s \bigg] \bigg\},
\end{align}
which defines $J^{(6)}_2$
\begin{align}
\mathcal{D}h_{\mu\nu}&=J^{(6)}_2\mathcal{D}h_{\mu\nu}^{TT}\mathcal{D}h\mathcal{D}\xi_{\mu}^T\mathcal{D}s \\
J^{(6)}_2&=\left[ \det(\nabla^2)_0 \right]^{1/2}[\det(5\lambda+\nabla^2)_1]^{1/2}[\det(6\lambda+\nabla^2)_0]^{1/2} \label{j6}.
\end{align}
Here, we use the property $\det A\cdot \det B=\det AB$. Computing the ghost determinant 
\begin{equation}
Z_{gh}^{(6)}=\frac{J^{(6)}_2}{J^{(6)}_1}=[\det(5\lambda+\nabla^2)_1]^{1/2}[\det(6\lambda+\nabla^2)_0]^{1/2}. \label{zgh}
\end{equation}
 one loop partition function for CG in six dimensions becomes 
\begin{align}
Z_{CG}^{(6)}&=Z_{gh}\int D h_{\mu\nu}^{TT}Exp(-\delta^{(2)}S)\nonumber \\ &=\frac{[\det(-\nabla^2-5\lambda)_1]^{1/2}[\det(-\nabla^2-6\lambda)_0]^{1/2}}{[\det(-\nabla^2+2\lambda)_2]^{1/2}[\det(-\nabla^{2}+6\lambda)_2]^{1/2}[\det(-\nabla^2+8\lambda)_2]^{1/2}}\label{zcgdet}.
\end{align}
The CG one loop partition function in six dimensions consists of  EG determinants
 \begin{equation}
 Z_{EG}^{(6)}=\frac{[\det(-\nabla^2-5\lambda)_1]^{1/2}}{[\det(-\nabla^2+2\lambda)_2]^{1/2}}
 \end{equation}
that have been considered in \cite{Gupta:2012he,Giombi:2014yra},
scalar determinant in the numerator, that corresponds to the contribution from conformal ghost $[\det(-\nabla^2-6\lambda)_0]^{1/2}$, determinant from the partially massless mode 
$[\det(-\nabla^{2}+6\lambda)_2]^{1/2}$, and massive determinant $[\det(-\nabla^2+8\lambda)_2]^{1/2}$. It has been considered as well in \cite{Tseytlin:2013fca}. 
From (\ref{main1}) and (\ref{zcgdet}) we can read out the partition function for CG in six dimensions
\begin{align}
\log Z_6 =&\sum_{k\in\mathcal{Z_+}}
\frac{-e^{-\frac{5}{2}k\beta}}{k(1-e^{-k\beta})^5}(\chi_{(1,0)}^{SO(5)} e^{-\frac{7}{2}k\beta}+\chi_{(0,0)}^{SO(5)} e^{-\frac{7}{2}k\beta} \nonumber \\
 &-\chi_{(2,0)}^{SO(5)} e^{-\frac{5}{2}k\beta}-\chi_{(2,0)}^{SO(5)} e^{-\frac{3}{2}k\beta}-\chi_{(2,0)}^{SO(5)} e^{-\frac{1}{2}k\beta} ). \label{z61}
\end{align}
Comparing the partition function (\ref{z61}), with the partition function expressed in terms of the determinants (\ref{zcgdet}), we can recognise the terms that originate from particular determinant in (\ref{zcgdet}). That is allowed by the character of SO(5) group that depends on spin, visible in the exponent multiplying the character. Using the character $\chi_{(s,0)}^{SO(5)}=\frac{1}{6}(2s+3)(s+2)(s+1)$ and the notation $q=e^{-\beta k}$, we can write (\ref{z61}) as
\begin{align}
\log Z_6=\sum_{k\in\mathbb{Z}_{+}} \frac{-2 q^{3k}}{k(1-q^k)^5}\left(3 q^{3k}-7q^{2k}-7q-7\right),\label{zsix}
\end{align}
or as the sum of the partition functions it consists of, partition function for EG 
\begin{equation}
\log Z_{EG}=\sum_{k\in\mathbb{Z}_{+}} \frac{- q^{5k}}{k(1-q^k)^5}\left(5 q^k-14\right),\label{zeg}
\end{equation}
 conformal ghost, partially massless mode and massive mode 
\begin{align}
\log Z_{diff}=\sum_{k\in\mathbb{Z}_{+}} \frac{- q^{3k}}{k(1-q^k)^5}\left(q^{3k}-14q^k-14\right).
\end{align}

\section{Thermodynamic Quantities}

One of the applications of the one loop partition function is computation of the subleading correction to thermodynamic quantities. Let us consider an example of four dimensional CG.
(Helmholtz) free energy, computed from \begin{equation}F=-\frac{1}{\beta}\ln Z_{one-loop}\end{equation} can be read out from partition function (\ref{zcg4d}) 
\begin{equation}
\label{eq:fe}
F_{1-\text{loop}}=\sum\limits_{k\in\mathbb{Z}_+}\frac{e^{-2 k \beta} (-5 + 4 e^{-2 k \beta} -5 e^{-k \beta})}{(1- e^{-k \beta})^{3} k \beta}. 
\end{equation}
The literature often refers to it multiplied with $\beta$. This subleading term is correction to the Euclidean AdS background, around which we consider it.  The free energy vanishes on the $AdS_4$ background because Weyl tensor vanishes and the CG action does not have to be renormalised. The Einstein part of the free energy that was considered in \cite{Giombi:2014yra} agrees with our result
\begin{eqnarray}
\label{eq:eq}
-\beta F_{\text{EG}_{1-\text{loop}}}= lnZ_{\text{EG}_{1-\text{loop}}}&=&-\sum\limits_{k\in\mathbb{Z}_+}\frac{q^{3k} (-5 +3 q^{k})}{(1- q^{k})^{3} k}. 
\end{eqnarray}
The subleading correction to the entropy
\begin{equation}
S_{one-\text{loop}}=-\frac{\partial{F_{1-\text{loop}}}}{\partial{T}}\end{equation}
reads
\begin{equation}
S_{one-\text{loop}}=\sum\limits_{k\in\mathbb{Z}_+}\frac{e^{\frac{-k}{T}}\Big(20\, k\, e^{\frac{2 k}{T}}+ 4 (k+T)+ 5\,(2k+T)\, e^{\frac{3 k}{T}}  - (16k+9T)\,e^{\frac{k}{T}}\Big)}{k T (-1+e^{\frac{k}{T}})^{4}}, \label{s1loop}
\end{equation}
for $T$ a temperature. 
The correction is not divergent and one may want to interpret it physically. The interpretation should be done carefully  questioning the semi-classical approximation of the solution. One can not neglect the classical contribution or the one-loop part. It is expected in general, that these terms have contribution from the full quantum corrections.

The examples of the leading order computation of the entropy we have encountered in the second chapter while treating the Schwarzschild, MKR and rotating black hole solution. We have as well considered it in the fourth chapter computing the leading order in entropy of the geon solution, global and non-trivial solutions from the classification of the subalgebras of $o(3,2)$.

\newpage

\chapter{Summary and Discussion}
\section{Summary}

Knowing its advantages and disadvantages compared to EG, study of CG has proven that CG possesses necessary ingredients to be considered as an effective theory of gravity. 
However, it needs to be studied further.

In the following paragraph, we briefly summarise the content of the chapters while in the latter ones we address the main results from each chapter. 

We have studied conformal gravity using the holographic renormalisation procedure, performed canonical analysis of canonical charges in CG, analysed its asymptotic symmetry algebra and found its one loop partition function, which we considered as well in six dimensions. The computations were performed in the AdS/CFT framework 
from the gravity side, which means that partition function played one of the key roles. The second chapter proved agreement with the previous results, while the analyses in the third and the fourth chapter were focused on the charges and the asymptotic symmetry algebra at the boundary, respectively. The fifth chapter studied one-loop partition function of CG, for which CFT side is not known. The CFT side provides nice verification in the lower dimensional theories where intrinsic symmetries of the 3D spacetime allow its computation.

In the first two chapters we have introduced the main topic, CG, and basic concepts used in GR. In the third chapter we have, using the holographic renormalisation procedure, verified that CG has well defined variational principle and finite response functions. For that we did not need to add neither generalised Gibbons-Hawking-York counterterms as extrinsic curvature in EG 
nor the holographic conutertems, to obtain the finite response function for the imposed boundary conditions. 
These boundary conditions included the Fefferman-Graham decomposition of the metric 
\begin{equation}
ds^2=\frac{\ell^2}{\rho^2}\left(-\sigma d\rho^2+\gamma_{ij}dx^idx^j\right).\label{les}
\end{equation}
with $\gamma_{ij}$  
\begin{equation}
\gamma_{ij}=\gamma_{ij}^{(0)}+\frac{\rho}{\ell}\gam+\frac{\rho^2}{\ell^2}\gamma_{ij}^{(2)}+\frac{\rho^{3}}{\ell^3}\gamma_{ij}^{(3)}+...\label{expansiongammas}
\end{equation}
near $\rho=0$,
and relations
\begin{align}
\delta \gamma_{ij}^{(0)} = \lambda \gamma_{ij}^{(0)}  &\,\,\,\,\,\,\,\,\,\,\,\,\,\,\,\,\,\,\,\,\,\,\,\,\,\,\,\,\,\,\,\,\,\text{            and}& \delta \gamma_{ij}^{(1)}=2\lambda\gamma_{ij}^{(1)}
\end{align}
where function $\lambda$ as well as tensors $\gamma_{ij}^{(0)}$ and $\gamma_{ij}^{(1)}$ are allowed to depend on all the coordinates of the boundary however, not on the holographic coordinate $\rho$. Where $\sigma=-1$ for AdS and $\sigma=+1$ for dS.
The response functions expressed in terms of the electric $E_{ij}$ and magnetic $B_{ijk}$ part of the Weyl tensor,
 \begin{align}
\tau_{ij} &= \sigma \big[\tfrac{2}{\ell}\,(E_{ij}^{\TO}+ \tfrac{1}{3} E_{ij}^{\SO}\ga^{\FO}) -\tfrac4\ell\,E_{ik}^{\SO}\psi^{\FO k}_j
+ \tfrac{1}{\ell}\,\ga_{ij}^{\LO} E_{kl}^{\SO}\psi_{\FO}^{kl} 
+ \tfrac{1}{2\ell^3}\,\psi^{\FO}_{ij}\psi_{kl}^{\FO}\psi_{\FO}^{kl}
\nonumber\\&- \tfrac{1}{\ell^3}\,\psi_{kl}^{\FO}\,\big(\psi^{\FO k}_i\psi^{\FO l}_j-\tfrac13\,\ga^{\LO}_{ij}\psi^{\FO k}_m\psi_{\FO}^{lm}\big)\big] 
- 4\,{\cal D}^k B_{ijk}^{\FO} + i\leftrightarrow j\,,
\label{eq:CG17s}
\end{align} and \begin{equation}
P_{ij}=-\tfrac{4\,\sigma}{\ell}\,E_{ij}^{\SO}\,
\label{eq:CG18s},\end{equation} obtained with the boundary terms form charges that generate asymptotic symmetries that define the asymptotic symmetry algebra at the boundary, in this case conformal algebra. 
We apply the results on the three examples, Schwarzschild black hole, Mannheim--Kazanas--Riegert (MKR) solution and the rotating black hole. 
In the case of the Schwarzschild black hole we recover the known \cite{deHaro:2000xn,Deser:2002jk} and expected solutions:
\begin{align}
P_{ij}&=0, \\
\tau_{ij}&=\frac{4\sigma}{\ell}E^{(3)}_{ij},
\end{align}
while the MKR solution contains non-vanishing PMR response for the non-vanishing Rindler acceleration parameter $a$. The Rindler parameter then plays the role (it can be interpreted with) of partially massless graviton condensate, while the conserved charge $Q[\partial_t]=m-a a_M$  is the one that corresponds to the Killing vector $\partial_t $ for the normalisation of the action $\alpha_{CG}=\frac{1}{64\pi}$.  The asymptotic symmetry algebra that closes at the boundary is four dimensional $\mathbb{R}\times o(3)$ algebra. 
While the entropy on-shell is $S=\frac{A_h}{4\ell^2}$ and $A_h=4\pi r_h^2$ defines an area of the horizon $k(r_h)=0$. 
It is remarkable that the entropy obeys an area law even though CG is higher-derivative theory of gravity. 
The third example of the rotating black hole with a Rindler acceleration $\m$ parameter, the rotation parameter $\tilde{a}$ and the vanishing mass leads to vanishing $P_{ij}=0$, which proves that for the non-zero Rindler term, $\gamma_{ij}\neq0$ is necessary but not sufficient. 
We have also seen that the Legendre transformation of the action exchanges the role of the PMR and its source. In this case the stress energy tenor $\tau_{ij}$ has zero trace.

In the fourth chapter we analyse the canonical charges for which we show that they are equivalent to the Noether charges. The charge associated to the Weyl symmetry vanishes, while the diffeomoprhism (\ref{eq:diffcharge})
 \begin{equation}
Q_D[\epsilon]=2\int_{\partial\Sigma}\ast\epsilon^c\left(\Pi_h^{ab}h_{bc}+\Pi_K^{ab}K_{bc}\right),\label{eq:diffcharges}
\end{equation} and Hamiltonian charge (\ref{qperp1s}) \begin{align}
Q_{\perp}[\epsilon]&=\int_{\partial\Sigma}\ast\left[\epsilon D_b\Pi_K^{cb}-D_b\epsilon\Pi_K^{cb}\right], \label{qperp1s} \end{align}
 do not vanish. $h_{ab}$, $\Pi_h^{ab}$ and $K_{ab}$, $\Pi_K^{ab}$ denote the metric on the 3D hypersurface and the corresponding momenta, and extrinsic curvature with corresponding momenta, respectively.  
 Analogously, in three dimensional gravity the charge associated to a fixed Weyl rescaling vanishes. The discrepancy arises for the freely varying Weyl rescaling function, when the Weyl charge in the 3D does not vanish.

Further, we have shown that these charges define asymptotic symmetry algebra at the boundary which corresponds to the Lie algebra of the asymptotic diffeomorphisms. We expand the Killing equation for the Lie algebra of the small difeomorphisms $\xi$ and Weyl rescalings (\ref{trafo}) \begin{equation}
\delta g_{\mu\n}=\left(e^{2\omega}-1\right)g_{\m\n}\pounds_{\xi}g_{\m\n}\label{trafos}
\end{equation} and obtain the leading (\ref{lo}) \begin{equation}
\mathcal{D}_{i}\xi^{(0)}_j+\mathcal{D}_j\xi^{(0)}_i=\frac{2}{3}\gamma_{ij}^{(0)}\mathcal{D}_{k}\xi^{(0)k},\label{slo}
\end{equation} and subleading (\ref{nloke}) \begin{equation}
\pounds_{\xi^{(0)l}}\gamma_{ij}^{(1)}=\frac{1}{3}\gamma_{ij}^{(1)}\mathcal{D}_l\xi^{(0)l}-4\omega^{(1)}\gamma_{ij}^{(0)},\label{snloke}
\end{equation}
 Killing equation.
The subleading Killing equation (\ref{snloke}) defines the subalgebra of the asymptotic solution of $\gam$ at the boundary for the subset of the so(3,2) KVs  which we classify according to Patera et al. classification. The largest solution consists of the 5 CKV $so(2)\ltimes o(1,1)$ subalgebra and defines a global geon or pp wave solution (\ref{geon}) \begin{equation} 
ds^2=dr^2+(-1+cf(r))dt^2+2cf(r)dtdx+(1+cf(r))dx^2+dy^2.\label{geons}
\end{equation}
with $f(r)=c_1+c_2 r+c_3 r^2+c_4 r^3$ and $c,c_i$ arbitrary constants,
 while the asymptotic MKR solution closes 4 CKV $\mathbb{R}\times o(3)$ subalgebra.
We have defined a map from the solutions of the flat to $\mathbb{R}\times S^2$ background.

In the fifth chapter we consider that one loop partition function of conformal gravity in four (\ref{zcg4d}) \begin{equation}
\log Z_{CG}^{(4)}=-\sum_{k\in \mathbb{Z}_{+}}\frac{q^{2k}(-5+4q^{2k}-5q^k)}{(1-q^k)^3k},\label{zcg4ds}
\end{equation} and six dimensions (\ref{zsix}) \begin{align}
\log Z_6=\sum_{k\in\mathbb{Z}_{+}} \frac{-2 q^{3k}}{k(1-q^k)^5}\left(3 q^{3k}-7q^{2k}-7q-7\right),
\end{align} on the background Euclidean AdS with $q=e^{-\beta k}$, and for completeness provide the general formula for partition function in arbitrary number of dimensions. For obtaining the one loop partition function we use the 
heat kernel and group theoretic approach. The result consists of the contribution from the conformal ghost, contribution from partially massless response and the part from the Einstein gravity. In six dimensions we obtain the analogous contribution, however, in addition there is a contribution form a massive graviton. The structure of the partition function for the gravity with conformal invariance keeps its structure as well in 3D, consisting of the contribution from Einstein gravity, conformal ghost and partially masses mode.

\section{Discussion}

If conformal gravity is ever to be considered as a correct effective theory of gravity, one has to find the way to deal with ghosts.
The current propositions for treatment of ghosts include Pais-Uhlenbeck oscillator approach that finds the parameter space for which there are no states of negative energy. The mechanism suggested by Mannheim consists of considering the theory as PT symmetric rather then Hermitian. %
Assuming that we accept one of these two possible solutions, or treat CG as a toy model we can further analyse it. Obvious direction for further analysis includes considerations of CG in four dimensions on different backgrounds, analogously to lower dimensional studies. In lower dimensions the studies have been done in the gauge/gravity correspondence sense, for $AdS/LCFT$ duality \cite{Grumiller:2009mw,Grumiller:2009sn}, duality of the asymptotically flat spacetimes and non relativistic conformal field theories \cite{Bagchi:2010zz}, correspondences between AdS space and Ricci flat spaces \cite{Caldarelli:2013aaa,Caldarelli:2012hy} and others.  In particular, it would be interesting to study the analogous in the flat space since the gravity theory that wants to be considered as correct effective theory of gravity should have the flat space limit. Within that framework one would look for results similar to those above.

Second direction is to look for the higher point functions such as three point functions in the AdS space.
In particular, the continuation of the analysis of the third chapter would include such studies. In the third chapter canonical analysis of charges can be done subsequently to the analysis of the first considering appropriate background. The fourth chapter provides rich field for the further investigation, one can search for the additional solutions using the bottom up approach and compute the properties of the full solutions. They by themselves provide rich basis for further research. 

The sixth and final chapter has an interesting property that could be further investigated, that is the fact that partition function for CG in 4D on thermal $AdS_4$ background relates  to the partition function of CG in 4D on $\mathbb{R}\times S^3$ with factor of two, which does not appear in other dimensions.
The reason for that is not evident. However one must not exclude the possibility that may be pure coincidence. 
Beside that, as for the analysis done in each chapter, partition function can be computed and analysed on different backgrounds.

\newpage
\appendix
\chapter{  }
\section{Appendix: General Relativity and AdS/CFT}
 
 \subsubsection{Parallel Transport}

If we have a curve $x^{\mu}(\lambda)$, tensor T is constant along this curve in  flat space when $\frac{dT}{d\lambda}=\frac{dx^{\mu}}{d\lambda}\frac{\partial T}{\partial x^{\mu}}=0$. Covariant derivative along path is \begin{equation} \frac{D}{d\lambda}=\frac{dx^{\mu}}{d\lambda}{\nabla_{\mu}},\end{equation}
and the parallel transport along the path reads
\begin{equation}
\left(\frac{D}{d\lambda}T\right)^{\mu_1\mu_2...\mu_k}_{\nu_1\nu_2...\nu_l}\equiv\frac{dx^{\sigma}}{d\lambda}\nabla_{\sigma}T^{\mu_1\mu_2...\mu_k}{}_{\nu_1\nu_2...\nu_l}=0.
\end{equation}
 
\subsubsection{Newton Potential from Small Perturbation Around the Metric}
 
We can write the metric $g_{\m\n}$ in the form of the perturbation $h_{\m\n}$ around the flat background metric $\eta_{\m\n}$, $g_{\m\n}=\eta_{\m\n}+h_{\m\n}$, where the indices of the terms in the expansion are raised and lowered with the background metric. 
One obtains for the geodesic equation
\begin{equation}
\frac{d^2x^{\m}}{d\tau^2}=\frac{1}{2}\eta^{\m\lambda}\partial_{\lambda}h_{00}\left(\frac{dt}{d\tau}\right)^2, 
\end{equation}
for $t$ time component and $\tau$ proper time. 
The $\m=0$ component gives constant $\frac{dt}{d\tau}$ and the spatial part (with space like components of $\eta^{\m\n}$ as an identity matrix) 
 \begin{equation} \frac{d^2x^1}{dt^2}=\frac{1}{2}\partial_i h_{00}. \end{equation}
 That means that $h_{00}=-2\phi$. This shows that the curvature of spacetime is sufficient for description of gravity in the Newtonian limit for $g_{00}=-1-2\phi$, where $\phi$ is defined taking into account the Weak Equivalence Principle so that the acceleration of the body due to inertial mass is $\vec{a}=-\nabla \phi$.

\subsubsection{Summary of the Conventions}
We follow conventions \cite{bob} in computations. 
For the $d+1$ dimensional manifold $\mathcal{M}$ with metric $g_{\m\n}$  and the covariant derivative $\nabla_{\mu}$ on $\mathcal{M}$ compatible with $g_{\m\n}$ one may write 
Christoffel symbols
\begin{align}
\Gamma_{\m\n}^{\lambda}=\frac{1}{2}g^{\lambda\rho}\left(\partial_{\m}g_{\rho\n}+\partial_{\n}g_{\m\rho}-\partial_{\rho}g_{\m\n}\right),
\end{align}
Riemann tensor 
\begin{align}
R^{\lambda}_{\mu\sigma\n}=\partial_{\sigma}\Gamma_{\m\n}^{\lambda}-\partial_{\nu}\Gamma^{\lambda}_{\m\sigma}+\Gamma_{\mu\nu}^{\kappa}\Gamma^{\lambda}_{\kappa\sigma}-\Gamma^{\kappa}_{\m\sigma}\Gamma^{\lambda}_{\kappa\n},
\end{align}
Ricci tensor 
\begin{align}
R_{\m\n}=\delta^{\sigma}_{\lambda}R^{\lambda}_{\m\sigma\n},
\end{align}
commutators of covariant derivatives
\begin{align}
\left[\nabla_{\m},\nabla_{\n}\right]A_{\lambda}&=R_{\lambda\sigma\m\n}A^{\sigma}\\
\left[\nabla_{\m},\nabla_{\n}\right]A^{\lambda}&=R^{\lambda}{}_{\sigma\m\n}A^{\sigma},
\end{align}
and Bianchi identities
\begin{align}
\nabla_{\kappa}R_{\lambda\m\sigma\n}-\nabla_{\lambda}R_{\kappa\m\sigma\n}+\nabla_{\m}R_{\kappa\lambda\sigma\n}&=0\\
\nabla^{\n}R_{\lambda\m\sigma\n}&=\nabla_{\m}R_{\lambda\sigma}-\nabla_{\lambda}R_{\m\sigma}\\
\nabla^{\n}R_{\m\n}&=\frac{1}{2}\nabla_{\m}R.
\end{align}
For $d+1$=2n an even number, one defines an Euler number 
\begin{align}
\chi(\mathcal{M})&=\int_{\mathcal{M}}d^{2n}x\sqrt{g}\varepsilon_{2n}
\end{align}
normalised with $\chi(S^{2n})=2$  and with Euler density
\begin{align}
\varepsilon_{2n}&=\frac{1}{(8\pi)^{n}\Gamma(n+1)}\epsilon_{\m_1..\m_{2n}}R^{\m_1\m_2\n_1\n_2}...R^{\m_{2n-1}\m_{2n}\n_{2n-1}\n_{2n}}.
\end{align}
In four dimensions the Euler density is 
\begin{align}
\varepsilon_{4}&=\frac{1}{128\pi^2}\epsilon_{\m\n\lambda\rho}\epsilon_{\alpha\beta\gamma\delta}R^{\m\n\alpha\beta}R^{\lambda\rho\gamma\delta} \\
&=\frac{1}{32\pi^2}\left(R^{\m\n\lambda\rho}_{\m\n\lambda\rho}-4R^{\m\n}R_{\m\n}+R^2\right).
\end{align}
If we consider small perturbation of the metric in the form $g_{\m\n}\rightarrow g_{\m\n}+\delta g_{\m\n}$, raise and lower the indices using the unperturbed metric and its inverse, we can express the quantities in terms of the perturbation of the metric with lower indices. Variational operator is 
\begin{align}
\begin{array}{ll} 
\delta(g_{\m\n})=\delta g_{\m\n} & \delta^2(g_{\m\n})=\delta(\delta g_{\m\n})=0 \\
\delta(g^{\m\n})=-g^{\m\alpha}g^{\n\beta}\delta_{g_{\alpha\beta}}\text{      }& \delta^2(g^{\m\n})\delta\left(-g^{\m\lambda}g^{\nu\rho}\delta g_{\lambda\rho}\right)=2 g^{\m\alpha}g^{\n\beta}g^{\lambda\rho}\delta 
\end{array}
\end{align}
\vspace{-0.35cm}
\begin{align}
f(g+\delta g)=f(g)+\delta f(g)+\frac{1}{2}\delta^2 f(g)+...+\frac{1}{n!}\delta^n f(g)+..
\end{align}
that brings to variations of Christoffels to higher orders 
\begin{align}
\delta \Gamma^{\lambda}_{\m\n}&=\frac{1}{2}g^{\lambda\rho}\left( \nabla_{\m}\delta_{\rho\n}+\nabla_{\n}\delta_{\m\rho}-\nabla_{\rho}\delta g_{\m\n}\right)\\
\delta^n \Gamma^{\lambda}_{\m\n}&=\frac{n}{2}\delta^{n-1}\left(g^{\lambda\rho}\right)\left(\nabla_{\m}\delta g_{\rho\n}+\nabla_{\n}\delta g_{\m\rho}-\nabla_{\rho}\delta g_{\m\n}\right),
\end{align}
the variation of the Riemann tensor 
\begin{align}
\delta R^{\lambda}{}_{\m\sigma\n}=\nabla_{\sigma}\delta\Gamma^{\lambda}_{\m\n}-\nabla_{\n}\delta\Gamma^{\lambda}_{\m\sigma},
\end{align}
and Ricci tensor
\begin{align}
\delta R_{\m\n}&=\nabla_{\lambda}\delta\Gamma_{\m\n}^{\lambda}-\nabla_{\n}\delta\Gamma^{\lambda}_{\m\lambda}\\
&=\frac{1}{2}\left(\nabla^{\lambda}\nabla_{\m}\delta g_{\m\n}+\nabla^{\lambda}\nabla_{\n}\delta g_{\m\lambda}-g^{\lambda\rho}\nabla_{\m}\nabla_{\n}\delta g_{\lambda\rho}-\nabla^2\delta g_{\m\n}\right).\label{varricten}
\end{align}
The remaining variation of the Ricci scalar is
\begin{equation}
\delta R=-R^{\m\n}\delta g_{\m\n}+\nabla^{\m}\left(\nabla^{\n}\delta g_{\m\n}-g^{\lambda\rho}\nabla_{m}\delta g_{\lambda\rho}\right).\label{varricsc}
\end{equation}

 \subsubsection{Covariant Derivative} 
The convention for the covariant derivative we use is
 \begin{align}
 \nabla T^{\mu_1\mu_2...\mu_k}{}_{\nu_1\nu_2....\nu_l}&=\partial_{\sigma}+\Gamma^{\mu_1}{}_{\sigma\lambda}T^{\lambda\mu_2....\mu_k}{}_{\nu_1\nu_2...\nu_l}+\Gamma^{\mu_2}{}_{\sigma\lambda}T^{\mu_1\lambda....\mu_k}{}_{\nu_1\nu_2...\nu_l}\nonumber \\ &-\Gamma^{\lambda}{}_{\sigma\nu_1}T^{\mu_1\mu_2....\mu_k}{}_{\lambda\nu_2...\nu_l}-\Gamma^{\lambda}{}_{\sigma\nu_2}T^{\mu_1\mu_2....\mu_k}{}_{\nu_1\lambda...\nu_l},
  \end{align}
while to express the covariant derivative of a one-form
with the same connection, it has to satisfy two following requirements:
\begin{itemize}
\item commute with the contractions $\nabla_{\mu}T^{\lambda}{}_{\lambda \rho}=(\nabla T)_{\mu}{}^{\lambda}{}_{\lambda\rho}$
\item reduce to partial derivatives when acting on scalars $\nabla_{\mu}\phi=\partial_{\mu}\phi$
\end{itemize}
Commutation of covariant derivative is
\begin{align}
[\nabla_{\rho},\nabla_{\sigma}]X^{\mu_1\mu_2....\mu_k}{}_{\nu_1\nu_2....\nu_l}&= -T_{\rho\sigma}{}^{\lambda}\nabla_{\lambda}X^{\mu_1\mu_2....\mu_k}{}_{\nu_1....\nu_l} \nonumber \\
&+R^{\mu_1}{}_{\lambda\rho\sigma}X^{\lambda\mu_2...\mu_k}{}_{\nu_1...\nu_l}+R^{\mu_2}{}_{\lambda\rho\sigma}X^{\mu_1\lambda...\mu_k}{}_{\nu_1...\nu_l}+...
\nonumber  \\
& - R^{\lambda}{}_{\nu_1\rho\sigma}X^{\mu_1...\mu_k}{}_{\lambda\nu_2...\nu_l}- R^{\lambda}{}_{\nu_2\rho\sigma}X^{\mu_1...\mu_k}{}_{\nu_1\lambda...\nu_l}.
\end{align}

\textbf{Jacobi identity}
The identity we use for the verification of the bracket operation of the Lie algebra
\begin{equation}
[[\nabla_{\mu},\nabla_{\rho}],\nabla_{\sigma}]+[[\nabla_{\rho},\nabla_{\sigma}],\nabla_{\mu}]+[[\nabla_{\sigma},\nabla_{\mu}],\nabla_{\rho}]=0.
\end{equation}

\section{Appendix: Holographic Renormalisation} 
 \subsubsection{Christoffel symbols for EOM of CG}
 \begin{align}
 \begin{array}{lll}
 \Gamma_{\rho\rho}^{\rho}=-\frac{1}{\rho}  & \Gamma_{\rho i }^{\rho}=\Gamma_{\rho\rho}^i=0 & 
 \\
 \Gamma_{ij}^{\rho}=\frac{\rho}{\ell}K_{ij}=\frac{\ell}{\rho}\left(\Theta_{ij}-\frac{1}{\ell}\gamma_{ji}\right) \text{     }& \Gamma_{\rho j}^k=\frac{\ell}{\rho}K_j^k=\frac{\ell}{\rho}\left(\Theta_j^k-\frac{1}{\ell}\gamma_j^k\right)
 \end{array}
 \end{align}
 
 \subsection{Decomposition of Curvature Tensors in Gaussian Normal Coordinates}
 
 The decomposition of the metric for holographic renormalisation procedure decomposes the metric into 
 \begin{equation}
 ds^2=-\frac{\ell^2}{\rho^2}d\rho^2+\gamma_{ij}(x^k,\rho)dx^idx^j \label{gns2}.
 \end{equation}
 The relation of $\rho$ and time is $\rho=e^{-2t/\rho}$, so that $t\rightarrow\infty$ implies $\rho\rightarrow0$ ($\rho>0$).
 The asymptotic boundary $\partial \mathcal{M}$ represents a constant $\rho$ surface for $\rho<<\ell$, where the normal vector that is timelike/spacelike for $\varsigma=+/-$ is
 \begin{align}
 u^{\m}=-\frac{\rho}{\ell}\delta^{\m}_{\rho} && u_{\m}=\varsigma\frac{\ell}{\rho}\delta_{\m}^{\rho},
 \end{align} the lapse is $\alpha^2=\frac{\ell^2}{\rho^2}$ and the shift $\beta^i=0$. For the projector on constant $\rho$ surfaces we use $\frac{\partial x^{\m}}{\partial x^i}\frac{\partial x^{\n}}{\partial x^j}...=\perp^{\m\n...}_{ij...}$.
 The extrinsic curvature is
 \begin{equation}
 K_{ij}=-\varsigma\frac{1}{2}\pounds_u\gamma_{ij}=\varsigma\frac{\rho}{2\ell}\partial_{\rho}\gamma_{ij},
 \end{equation}
and the projections of the curvatures 
\begin{align}
\perp_{kilj}^{\lambda\m\sigma\n}R_{\lambda\m\sigma\n}&={}^3R_{kilj}+\varsigma K_{lk}K_{ij}-K_{kj}K_{li}\\
\perp_{ilj}^{\m\sigma\n}u^{\lambda}R_{\lambda\m\sigma\n}&=\varsigma({}^3\nabla_{l}K_{ij}-{}^3K_{il})\\
\perp_{ij}^{\m\n}u^{\lambda}u^{\sigma}R_{\lambda\m\sigma\n}&=\varsigma\pounds_{u}K_{ij}+K_i^lK_{jl}\\
\perp_{ij}^{\m\n}R_{\m\n}&={}^3R_{ij}+\varsigma(KK_{ij}-2K_i^lK_{jl})-\pounds_uK_{ij}\\
\perp_i^{\m}R_{\m\n}u^{\n}&=\varsigma({}^3\nabla_iK-{}^3\nabla^jK_{ij})\\
R_{\m\n}u^{\m}u^{\n}&=\varsigma\pounds_uK-K^{ij}K_{ij}\\
&=\varsigma\gamma^{ij}\pounds_uK_{ij}+K^{ij}K_{ij}\\
R&={}^3R+\varsigma(K^2+K^{lk}K_{lk})-2\pounds_uK\\
&={}^3R+\varsigma(K^2-3K^{lk}K_{lk})-2\gamma^{ij}\pounds K_{ij}
\end{align}
 where the metric indices denoted with ${}^3R_{kiln},{}^3R_{ij}$ and ${}^3R$ denote intrinsic curvature tensors constructed from the boundary metric $\gamma_{ij}$, ${}^3\nabla_i$ is covariant derivative on the manifold $\partial M$ compatible with $\gamma_{ij}$, and $\pounds_{u}$ is Lie derivative along the normal vector $u^{\m}$.

 \subsubsection{Expansion of the Curvatures in Conformal Gravity}
 In this section we provide the expanded quantities that appear in the definition of the EOM. For convenience, in some cases it is useful to expand the quantities using the expansion with explicit $\frac{1}{n!}$ factors, while in other cases we  a priori use the expansion in which the factorials are absorbed in the $\gamma_{ij}$ matrices, or expanded tensor fields. In case we use the type of expansion that does not absorb the $n!$ in $\gamma_{ij}$ matrices, we write that explicitly. 
 The expansion of the inverse metric $\gamma^{ij}$ reads
 \begin{align}
\gamma^{ij}&= \gamma^{(0)ij} -  \rho\gamma^{(1)ij} + \rho^2(\gamma^{(1)aj} \gamma^{(1)i}{}_{a} -  \gamma^{(2)ij}) \nonumber \\ &-  \rho^3(\gamma^{(1)a}{}_{b} \gamma^{(1)bj} \gamma^{(1)i}{}_{a} -  \gamma^{(1)i}{}_{a} \gamma^{(2)aj} -  \gamma^{(1)aj} \gamma^{(2)i}{}_{a} + \gamma^{(3)ij}),
 \end{align} while the inverse of the $\gamma^{ij}$ with explicit factorials is
 \begin{align}
  \gamma^{ij}&=h^{ij} -  \frac{\rho h^{(1)ij}}{\ell} + \frac{\rho^2 (2 h^{(1)ik} h^{(1)}{}_{k}{}^{j} -  h^{(2)ij})}{2 \ell^2} \nonumber \\ &+ \frac{\rho^3 (-6 h^{(1)ik} h^{(1)}{}_{k}{}^{l} h^{(1)}{}_{l}{}^{j} + 3 h^{(1)}{}_{n}{}^{j} h^{(2)in} + 3 h^{(1)im} h^{(2)}{}_{m}{}^{j} -  h^{(3)ij})}{6 \ell^3}.
 \end{align}
 Here, to accent that, we write $h_{ij}$ on the place of $\gamma_{ij}$ and continue with that notation.  The expansion of the Christoffel symbol $\Gamma^i{}_{jl}$ is
 \begin{align}
\Gamma^i{}_{jl}&=\Gamma [D]^{i}{}_{jl} + \frac{\rho}{\ell} (- \tfrac{1}{2} D^{i}h^{(1)}{}_{jl} + \tfrac{1}{2} D_{j}h^{(1)i}{}_{l} + \tfrac{1}{2} D_{l}h^{(1)i}{}_{j})\nonumber \\ & + \frac{\rho^2}{\ell^2} (- \tfrac{1}{4} D^{i}h^{(2)}{}_{jl} -  \tfrac{1}{2} h^{(1)ik} D_{j}h^{(1)}{}_{lk} + \tfrac{1}{4} D_{j}h^{(2)i}{}_{l} \nonumber \\ &+ \tfrac{1}{2} h^{(1)ik} D_{k}h^{(1)}{}_{jl} -  \tfrac{1}{2} h^{(1)ik} D_{l}h^{(1)}{}_{jk} + \tfrac{1}{4} D_{l}h^{(2)i}{}_{j})
\end{align}
  and the $\theta$ tensor (\ref{thetadef}), that defines the extrinsic curvature $K_{ij}$ with (\ref{kdef}), is
 \begin{align}
 \theta_{ij}=\frac{\eta h^{(1)}{}_{ij}}{2 \ell^2} + \frac{\eta^2 h^{(2)}{}_{ij}}{2 \ell^3} + \frac{\eta^3 h^{(3)}{}_{ij}}{4 \ell^4} + \frac{\eta^4 h^{(4)}{}_{ij}}{12 \ell^5}.
 \end{align}
The curvatures, Ricci tensor and Ricci scalar are respectively, 
\begin{align}
R_{ij}&=R[D]_{ij} + \frac{\rho}{\ell} (- \tfrac{1}{2} D_{j}D_{i}h^{(1)k}{}_{k} + \tfrac{1}{2} D_{k}D_{i}h^{(1)}{}_{j}{}^{k} + \tfrac{1}{2} D_{k}D_{j}h^{(1)}{}_{i}{}^{k} -  \tfrac{1}{2} D_{k}D^{k}h^{(1)}{}_{ij}) \nonumber \\ &+ \frac{\rho^2}{\ell^2} (\tfrac{1}{2} h^{(1)kl} D_{i}D_{j}h^{(1)}{}_{kl} + \tfrac{1}{4} D_{i}h^{(1)kl} D_{j}h^{(1)}{}_{kl} -  \tfrac{1}{4} D_{j}D_{i}h^{(2)k}{}_{k} \nonumber \\ &+ \tfrac{1}{4} D_{i}h^{(1)}{}_{j}{}^{k} D_{k}h^{(1)l}{}_{l} + \tfrac{1}{4} D_{j}h^{(1)}{}_{i}{}^{k} D_{k}h^{(1)l}{}_{l} + \tfrac{1}{4} D_{k}D_{i}h^{(2)}{}_{j}{}^{k} + \tfrac{1}{4} D_{k}D_{j}h^{(2)}{}_{i}{}^{k} \nonumber \\ &-  \tfrac{1}{4} D_{k}D^{k}h^{(2)}{}_{ij} -  \tfrac{1}{4} D_{k}h^{(1)l}{}_{l} D^{k}h^{(1)}{}_{ij} -  \tfrac{1}{2} D_{i}h^{(1)}{}_{j}{}^{k} D_{l}h^{(1)}{}_{k}{}^{l} -  \tfrac{1}{2} D_{j}h^{(1)}{}_{i}{}^{k} D_{l}h^{(1)}{}_{k}{}^{l}\nonumber \\ & + \tfrac{1}{2} D^{k}h^{(1)}{}_{ij} D_{l}h^{(1)}{}_{k}{}^{l} -  \tfrac{1}{2} h^{(1)kl} D_{l}D_{i}h^{(1)}{}_{jk} -  \tfrac{1}{2} h^{(1)kl} D_{l}D_{j}h^{(1)}{}_{ik}\nonumber \\ & + \tfrac{1}{2} h^{(1)kl} D_{l}D_{k}h^{(1)}{}_{ij} -  \tfrac{1}{2} D_{k}h^{(1)}{}_{jl} D^{l}h^{(1)}{}_{i}{}^{k} + \tfrac{1}{2} D_{l}h^{(1)}{}_{jk} D^{l}h^{(1)}{}_{i}{}^{k} ),
\end{align}
\begin{align}
R&=R[D] + \frac{\rho}{\ell} (- R[D]^{ij} h^{(1)}{}_{ij} + D_{j}D_{i}h^{(1)ij} -  D_{j}D^{j}h^{(1)i}{}_{i}) \nonumber \\ &+ \frac{\rho^2}{\ell^2} (R[D]^{ij} h^{(1)}{}_{i}{}^{k} h^{(1)}{}_{jk} -  \tfrac{1}{2} R[D]^{ij} h^{(2)}{}_{ij} + h^{(1)ij} D_{j}D_{i}h^{(1)k}{}_{k}  \nonumber \\ & + \tfrac{1}{2} D_{j}D_{i}h^{(2)ij} -  \tfrac{1}{2} D_{j}D^{j}h^{(2)i}{}_{i} -  h^{(1)ij} D_{j}D_{k}h^{(1)}{}_{i}{}^{k}  \nonumber \\ & -  \tfrac{1}{4} D_{j}h^{(1)k}{}_{k} D^{j}h^{(1)i}{}_{i} -  D_{i}h^{(1)ij} D_{k}h^{(1)}{}_{j}{}^{k} + D^{j}h^{(1)i}{}_{i} D_{k}h^{(1)}{}_{j}{}^{k}  \nonumber \\ &-  h^{(1)ij} D_{k}D_{j}h^{(1)}{}_{i}{}^{k} + h^{(1)ij} D_{k}D^{k}h^{(1)}{}_{ij} -  \tfrac{1}{2} D_{j}h^{(1)}{}_{ik} D^{k}h^{(1)ij}  \nonumber \\ &+ \tfrac{3}{4} D_{k}h^{(1)}{}_{ij} D^{k}h^{(1)ij})
\end{align}
 
\subsection{Equations of Motion in Conformal Gravity}

Since the action (\ref{ac3}) consists from two dynamical fields $f_{\m\n}$ and $g_{\m\n}$ its variation gives $EOM_{f}$ and $EOM_{g}$, i.e. EOM for auxiliary field $f_{\m\n}$ and for the metric, respectively. We are interested in restrictions from EOM order by order. 
  The most important difference with the EG is that the Einstein EOM do not allow the term $\gam$ in the expansion (\ref{expansiongamma}) restricting it to be zero, while Bach equation does not impose such condition. 

Let us take the dS case, $\sigma=1$, in which we consider $\infty>\rho>0$ and future  is placed at $\rho\rightarrow0$. The coordinate $\rho$ is related to the time coordinate with $\rho=e^{-2t/\ell}$ and $t\rightarrow\infty$ corresponds to $\rho\rightarrow0$ for $\rho>0$.
We take the boundary $\partial M$ as a constant $\rho$ surface for $\rho<<\ell$, with a timelike vector normal to the surface is defined with 
\begin{align}
u^{\rho}=-\frac{\rho}{\ell} && u_{\rho}=\frac{\ell}{\rho}
\end{align}
That makes the extrinsic curvature 
\begin{equation}
K_{ij}=\frac{\ell^2}{\rho^2}\left(-\frac{1}{2}\pounds_{\textbf{n}}\gamma_{ij}-\frac{1}{\ell}\gamma_{ij}\right) \label{kdef}
\end{equation}
If we define for convenience \begin{align}\theta_{ij}=-\frac{1}{2}\pounds_{\textbf{n}}\gamma_{ij}=\frac{\rho}{2\ell}\partial_{\rho}\gamma_{ij}\label{thetadef}\end{align} we can write the extrinsic curvature 
\begin{equation}
\Rightarrow K_{ij}=\frac{\ell^2}{\rho^2}\left(\theta_{ij}-\frac{1}{\ell}\gamma_{ij}\right)\label{exc}
\end{equation}
which raising the index with $\frac{\ell^2}{\rho^2}\gamma_{ij}$ metric leads to 
\begin{equation}
K_{i}^j=\theta_i^j=\frac{1}{\ell}\gamma_i^j,
\end{equation}
while the Christoffels are defined in the appendix: Holographic Renormalisation and One-Point Functions in Conformal Gravity: Christoffel Symbols for EOM of CG. 
Using the new unphysical variables
\begin{align}
f_i^j&=\phi_i^j, &&  f_i^{\rho}=v_i, && f_{\rho}^i=-v^i, &&f_{\rho}^{\rho}=w, \nonumber \\
f_{ij}&=\frac{\ell^2}{\rho^2}\phi_{ij}, && f_{i\rho}=-\frac{\ell^2}{\rho^2}v_i, && f_{\rho i}=-\frac{\ell^2}{\rho^2}v_i, && f_{\rho\rho}=-\frac{\ell^2}{\rho^2}w  
\label{defs}
 \end{align}
and the variables defined on the three dimensional hypersurface, we can write $EOM_f$ and $EOM_g$. The convenience of the unphysical variables is that computations using the computer package xAct simplifies. The tensors on the three dimensional manifold we denote with the prefix ${}^3$ while the tensors expressed with the unphysical metric we write with no prefixes. Physical and unphysical Ricci tensor and Ricci scalar are denoted with respectively,
\begin{align}
\begin{array}{cc}
{}^3R_{ij}=R_{ij}, & {}^3R=\frac{\rho^2}{\ell^2}R\\
\end{array}
\end{align}
where unphysical indices, indices on the unphysical quantities, are raised and lowered with the unphysical metric. First using the physical variables we can write 
the EOM $E_{\m\n}^f$ for the auxiliary field $f_{\m\n}$ 
\begin{equation}
E_{\m\n}^f=\frac{1}{4}f_{\m\n}-\frac{1}{4}g_{\m\n}f_{\l}^{\l}+G_{\m\n}=0\label{auxeom}
\end{equation}
where $G_{\m\n}$ is earlier defined Einstein tensor.  Evaluation of the trace and insertion in the equation (\ref{auxeom}) leads to 
\begin{align}
f_{\m\n}=-4G_{\m\n}+\frac{4}{3}g_{\m\n}G^{\l}_{\l}.
\end{align}
We can decompose it in the GNC $\rho\rho$, $\rho i$ and $ij$ to obtain
\begin{align}
u^{\m}u^{\n}E_{\m\n}^f=0 && 
\Rightarrow&& 0=f_i^i+2{}^3R+2K^2-2K^{ij}K_{ij} \label{one}
\end{align}
\vspace{-0.3cm}
\begin{align}
u^{\n}E_{i\n}^f&=0 &&
\Rightarrow &&0=\frac{\ell}{\rho}f^{\rho}_i+4\left( {}^3\nabla_i K-{}^3\nabla^jK_{ij} \right)\label{two}
\end{align}
\vspace{-0.4cm}
\begin{align}
E_{ij}^f&=0 \nonumber\\
\Rightarrow 0&=f_{ij}-\frac{\ell^2}{\rho^2}\gamma_{ij}f^{\rho}_{\rho}+4 {}^3R_{ij}+4KK_{ij}-8K_i^kK_{jk} \nonumber \\ 
&+4\frac{\ell^2}{\rho^2}\gamma_{ij}K^{kl}K_{kl}-4\left( \gamma_{i}^k\gamma_j^l-\gamma_{ij}\gamma^{kl}\right)\pounds_{\textbf{n}}K_{kl}
\end{align}
Taking the trace of the $ij$ equation with $\frac{\rho^2}{\ell^2}\gamma^{ij}$ reads
\begin{align}
0=-f^{\rho}_{\rho}+2K^{ij}K_{ij}+\frac{2}{3}{}^3R+\frac{2}{3}K^2+\frac{8}{3}\frac{\rho^2}{\ell^2}\label{three}
\end{align} 
which we can insert back into the $ij$ equation that leads to the form 
\begin{align}
\Rightarrow 0=f_{ij}+4\left({}^3R_{ij}-\frac{1}{6}\gamma_{ij}{}^3R\right)-4\left(\gamma_i^k\gamma_j^l-\frac{1}{3}\gamma_{ij}\gamma^{kl} \right)\pounds_{\textbf{n}}K_{kl}+ \nonumber \\
4 K K_{ij}-8 K_i^kK_{jk}-\frac{2}{3}\frac{\ell^2}{\rho^2}\gamma_{ij}K^2+2\frac{\ell^2}{\rho^2}\gamma_{ij}K^{kl}K_{kl}. \label{four}
\end{align}
Rewritting the equations with the unphysical tensors we obtain
\begin{align}
(\ref{one}) \Rightarrow 0&=\phi^i_i+\frac{12}{\ell^2}-\frac{8}{\ell}\theta+2\theta^2-2\theta^{ij}\theta_{ij}+2\frac{\rho^2}{\ell^2}R\\ \label{f1eom}
(\ref{two})\Rightarrow 0&=v_i+4\frac{\rho}{\ell}\left(D_i\theta-D^j\theta_{ij}\right)\\ \label{veom}
(\ref{three}) \Rightarrow 0&=w+\frac{4}{\ell^2}-\frac{8}{3\ell}\theta-\frac{8}{3}\frac{\rho^2}{\ell^2}\gamma^{ij}u^{\textbf{n}}\partial_{\rho}\theta_{ij} \nonumber \\ 
-&2\theta^{ij}\theta_{ij}-\frac{2}{3}\theta^2-\frac{2}{3}\frac{\rho^2}{\ell^2}R \\
(\ref{four}) \Rightarrow 0&=\phi_{ij}+\frac{4}{\ell^2}\frac{\ell^2}{\rho^2}\gamma_{ij}-\frac{12}{\ell}\theta_{ij}+\frac{12}{\ell}\theta_{ij}+\frac{4}{3\ell}\frac{\ell^2}{\rho^2}\gamma_{ij}\theta+4\theta\theta_{ij}-8\theta_i^k\theta_{jk} \nonumber \\ 
-&\frac{2}{3}\frac{\ell^2}{\rho^2}\gamma_{ij}\theta^2+2\frac{\ell^2}{\rho^2}\theta^{kl}\theta_{kl}-4(h_i^kh_j^l-\frac{1}{3}\gamma_{ij}\gamma^{kl})\pounds_{\textbf{n}\theta_{kl}}\nonumber \\ 
+&4\frac{\rho^2}{\ell^2}(R_{ij}-\frac{1}{6}\frac{\ell^2}{\rho^2}\gamma_{ij}R) \label{f2eom}
\end{align}
EOM for the unphysical auxiliary variables we further use to determine metric EOM $EOM_g$.
In the physical coordinates $EOM_g$ read 
\begin{align}
E_{\m\n}^{g}&=-\frac{1}{2}f_{\m}^{\l}G_{\l\n}-\frac{1}{2}f^{\l}_{\n}G_{\m\l}-\frac{1}{4}f_{\m}^{\l}f_{\n\l}+\frac{1}{2}G_{\m\n}f^{\l}_{\l}+\frac{1}{4}f_{\m\n}f_{\l}^{\l}+ \nonumber \\
&+ \frac{1}{2} g_{\m\n}f^{\lambda\rho}G_{\l\rho}-\frac{1}{4}g_{\m\n}f^{\l}_{\l}G^{\rho}_{\rho}+\frac{1}{16}g_{\m\n}(f^{\l\rho}f_{\l\rho}-f^{\l}_{\l}f_{\rho}^{\rho}) \nonumber \\
&-R_{\m\l\n\rho}f^{\l\rho}+\frac{1}{2}\nabla_{\m}\nabla^{\l}f_{\l\n}+\frac{1}{2}\nabla_{n}\nabla^{\l}f_{\m\l}-\frac{1}{2}\nabla_{\m}\nabla_{\n}f_{\l}^{\l} \nonumber \\
&-\frac{1}{2}\nabla^{2}f_{\m\n}-\frac{1}{2}g_{\m\n}\nabla_{\l}\nabla_{\rho}f^{\l\rho}+\frac{1}{2}g_{\m\n}\nabla^2f^{\l}_{\l}
\end{align}
and after simplification with the $EOM_f$ they become
\begin{equation}
E_{\m\n}^{g}=-\frac{1}{8}f_{\m\n}f^{\l}_{\l}-\frac{1}{16}g_{\m\n}f^{\l\rho}f_{\l\rho}-R_{\m\l\n\rho}f^{\l\rho}\\
+ \frac{1}{2}\nabla_{\m}\nabla_{\n}f^{\l}_{\l}-\frac{1}{2}\nabla^2f_{\m\n}.
\end{equation}
To determine the restrictions coming from them we consider again the components $\rho\rho$, $\rho i$ and $ij$.   $EOM_g$ in the ${}_\rho{}^\rho$, ${}_i{}^j$ and ${}_i{}^{\rho}$ respectively, read 
\begin{align}
E_{\rho}^{g\rho}=-\frac{1}{8}w^2-\frac{1}{8}w\phi^i_i-\frac{1}{16}\phi^{ij}\phi_{ij}+\frac{1}{8}v^iv_i-\frac{1}{16}w^2 \nonumber \\
+f^{ij}\left(\pounds_{\textbf{n}}K_{ij}+K_i^kK_{jk}\right)+\frac{1}{2}\nabla_{\rho}\nabla^{\rho}f_{\lambda}^{\lambda}-\frac{1}{2}\nabla^2f^{\rho}_{\rho},\label{err}
\end{align}
\begin{align}
E_{i}^{gj}=-\frac{1}{8}f_i^jf_{\lambda}^{\lambda}-\frac{1}{16}\delta_i^jf^{\lambda\rho}f_{\lambda\rho}-R_{i\lambda}{}^j{}_{\rho}f^{\lambda\rho}+\frac{1}{2}\nabla_i\nabla^jf^{\lambda}_{\lambda}-\frac{1}{2}\nabla^2f_i^j,\label{egj}
\end{align}
\begin{align}
E_{i}^{g\rho}=-\frac{1}{8}f_i^{\rho}f_{\lambda}^{\lambda}-R_{i\lambda}{}^{\rho}{}_{\kappa}f^{\lambda\kappa}+\frac{1}{2}\nabla_i\nabla^{\rho}f_{\lambda}^{\lambda}-\frac{1}{2}\nabla^2f_i^{\nabla}.\label{egr}
\end{align}
Analogously like EOM for the auxiliary field $f_{\m\n}$, equations (\ref{err}), (\ref{egj}) and (\ref{egr}) can be written in terms of the unphysical variables. In the unphysical variables the equation of motion for the $\rho\rho$, $\rho i$ and $ij$ component read
\begin{align}
E_{unphy.coords.\rho}^{g\rho}&=--\frac{3}{16}w^2-\frac{1}{8} w \phi^k_k-\frac{1}{16}\phi^{ij}\phi_{ij}+\frac{1}{8}v^iv_i \nonumber \\
&+ \phi^{ij}\pounds_{\textbf{n}}\theta_{ij}+\phi^{ij}\theta_i^k\theta_{jk}+\frac{2}{\ell}\phi^{ij}\phi_{ij}-\frac{1}{\ell^2}\phi^k_k \nonumber \\
&-\frac{1}{2}\frac{\rho^2}{\ell^2}\partial_\rho^2\phi^k_k-\frac{1}{2}\frac{\rho}{\ell^2}\partial_{\rho}\phi_i^i-\frac{3}{2}\frac{\rho}{\ell^2}\partial_{\rho}w+\frac{1}{2}\frac{\rho}{\ell}\theta \partial_{\rho}w \nonumber \\
&-\frac{1}{2}\frac{\rho^2}{\ell^2}D^2w+\frac{\rho}{\ell}v_jD_i\theta^{ij}+2\frac{\rho}{\ell}\left( \theta^{ij}-\frac{1}{\ell}\frac{\rho^2}{\ell^2}\gamma^{ij}\right)D_iv_j \nonumber \\
&+\left(\theta^i_k-\frac{1}{\ell}\gamma^i_k \right)\left( \theta^{jk}-\frac{1}{\ell}\frac{\rho^2}{\ell^2}\gamma^{jk} \right)\left(\phi_{ij}-\frac{\ell^2}{\rho^2} \gamma_{ij}w\right),
\end{align}

\begin{align}
E_{i}^{g\rho}&=-\frac{1}{8}v_i(\phi_j^j+w)-v^j(\pounds_{\textbf{n}})\theta_{ij}+\theta_i^k\theta_{jk}+\frac{2}{\ell}\frac{\ell^2}{\rho^2}\gamma_{ij})\nonumber \\ &+\frac{\rho}{\ell}\big(D_k\theta_{ji}-D_i\theta_{kj}\big)\phi^{kj}+\frac{1}{2}\nabla_i\nabla^{\rho}f^{\lambda}_{\lambda}-\frac{1}{2}\nabla^2 f_i^{\rho},
\end{align}
\begin{align}
E^{gj}_{i}&=-\frac{1}{8}\phi_i^j(\phi_k^k+w)-\frac{1}{16}\delta_i^j(\phi^{kl}\phi_{kl}-2v^kv_k+w^2) \nonumber \\
& -\frac{\rho^2}{\ell^2}R^j{}_{lik}\phi^{lk}-(\theta_i^j-\frac{1}{\ell}\gamma_i^j)(\theta_{lk}-\frac{1}{\ell}\frac{\ell^2}{\rho^2}\gamma_{lk})\phi^{lk} \nonumber \\&+(\theta^j_k-\frac{1}{\ell}\gamma_{jk})(\theta_{il}-\frac{1}{\ell}\frac{\ell^2}{\rho^2}\gamma_{jl})\phi^{lk} \nonumber \\
& +\frac{\rho}{\ell}v^k(2D_k\theta_i^j-D^j\theta_{ik}-D_i\theta^j_k) \nonumber \\
& + w\left( \frac{\rho^2}{\ell^2}\gamma^{jl}\pounds_{\textbf{n}}\theta_{il}+\theta_i^k\theta_j^k+\frac{2}{\ell}\theta_i^j -\frac{1}{\ell^2}\gamma_{i}^j\right)\nonumber \\
& + \frac{1}{2}\frac{\rho^2}{\ell^2}D_iD^j(\phi^k_k+w)-\frac{1}{2}\frac{\rho}{\ell}(\theta_i^j-\frac{1}{\ell}\gamma_i^j)\partial_{\rho}(\phi_k^k+w)\nonumber \\
&-\frac{1}{2}\nabla^2f_i^j.
\end{align}

In these equations, we insert the FG expansion. For the convenience, we use the expansion of the metric 
\begin{equation}
\gamma_{ij}=\gamma_{ij}^{(0)} + \frac{\rho \delta \gamma^{(1)}{}_{ij}}{\ell} + \frac{\rho^2 \delta \gamma^{(2)}{}_{ij}}{2 \ell^2} + \frac{\rho^3 \delta \gamma^{(3)}{}_{ij}}{6 \ell^3} + \frac{\rho^4 \delta \gamma^{(4)}{}_{ij}}{24 \ell^4}
\end{equation}
that contains factorials $\frac{1}{n!}$. In this expansion factorials are not absorbed in the matrices. The tensors are perturbed analogously 
\begin{align}
w+\delta w=w + \frac{\eta w^{(1)}}{\ell} + \frac{\eta^2 w^{(2)}}{2 \ell^2} + \frac{\eta^3 w^{(3)}}{6 \ell^3} + \frac{\eta^4 w^{(4)}}{24 \ell^4}
\end{align},
\begin{align}
v_i+\delta v_i=v_{i} + \frac{\eta v^{(1)}{}_{i}}{\ell} + \frac{\eta^2 v^{(2)}{}_{i}}{2 \ell^2} + \frac{\eta^3 v^{(3)}{}_{i}}{6 \ell^3} + \frac{\eta^4 v^{(4)}{}_{i}}{24 \ell^4},
\end{align}
\begin{align}
\phi_{ij}+\delta \phi_{ij}= \phi_{ij}+\frac{\eta \phi^{(1)}{}_{ij}}{\ell} + \frac{\eta^2 \phi^{(2)}{}_{ij}}{2 \ell^2} + \frac{\eta^3 \phi^{(3)}{}_{ij}}{6 \ell^3} + \frac{\eta^4 \phi^{(4)}{}_{ij}}{24 \ell^4}
\end{align}
and their terms in the expansion expressed in the metric $\gamma^{(1)}_{ij}$ are determined using the EOM for the auxiliary field (\ref{f1eom}) 
and (\ref{f2eom}). First four orders of EOM obtained varying with respect to $\gamma_{ij}$ give exactly zero, which is plausible since Bach equations is fourth order partial differential equation, while the fourth order gives restriction on the terms in the FG expansion. These equations we present here for $\gamma_{ij}^{(0)}=diag(-1,1,1)$, for the full expression see appendix: Holographic Renormalisation and One-Point Functions in Conformal Gravity: EOM for CG, Full Expressionss. The $\rho\rho$ component reads
\begin{align}
E^{(1)\rho}_{\rho}&=- \frac{3 \psi^{(1)}_{i}{}^{k} \psi^{(1)ij} \psi^{(1)}_{j}{}^{l} \psi^{(1)}_{kl}}{4 \ell^8} + \frac{\psi^{(1)}_{ij} \psi^{(1)ij} \psi^{(1)}_{kl} \psi^{(1)kl}}{8 \ell^8} + \frac{\psi^{(2)}_{ij} \psi^{(2)ij}}{4 \ell^8} + \frac{\psi^{(1)}_{i}{}^{k} \psi^{(1)ij} \psi^{(2)}_{jk}}{\ell^8} \nonumber \\&  - \frac{\psi^{(1)ij} \psi^{(3)}_{ij}}{2 \ell^8} + \frac{\partial_{j}\partial_{i}\psi^{(2)ij}}{\ell^6}  - \frac{2 \psi^{(1)ij} \partial_{j}\partial_{k}\psi^{(1)}_{i}{}^{k}}{\ell^6}  - \frac{\partial_{i}\psi^{(1)ij} \partial_{k}\psi^{(1)}_{j}{}^{k}}{2 \ell^6} \nonumber \\ &+ \frac{3 \psi^{(1)ij} \partial_{k}\partial^{k}\psi^{(1)}_{ij}}{2 \ell^6}  - \frac{\partial_{j}\psi^{(1)}_{ik} \partial^{k}\psi^{(1)ij}}{\ell^6} + \frac{\partial_{k}\psi^{(1)}_{ij} \partial^{k}\psi^{(1)ij}}{\ell^6},
\end{align}
for  \begin{equation}\psi_{ij}^{(n)}=\gamma_{ij}^{(n)}-\frac{1}{3}\gamma^{(n)}\gamma_{ij}^{(0)}\label{tracelessmet}\end{equation} the traceless part of the terms in the expansion of the metric (\ref{expansiongamma}).
 The $\rho i$ and $ij$ component are respectively
\begin{align}
E^{(1)\rho}_i&=\frac{4 \psi^{(2)jk} \partial_{i}\psi^{(1)}_{jk}}{3 \ell^7}  - \frac{2 \psi^{(1)}_{j}{}^{l} \psi^{(1)jk} \partial_{i}\psi^{(1)}_{kl}}{\ell^7} + \frac{5 \psi^{(1)jk} \partial_{i}\psi^{(2)}_{jk}}{6 \ell^7}  - \frac{\psi^{(1)}_{i}{}^{j} \psi^{(1)kl} \partial_{j}\psi^{(1)}_{kl}}{\ell^7} \nonumber \\ & + \frac{\partial_{j}\psi^{(3)}_{i}{}^{j}}{\ell^7}  - \frac{2 \psi^{(2)jk} \partial_{k}\psi^{(1)}_{ij}}{\ell^7}  - \frac{\psi^{(2)}_{i}{}^{j} \partial_{k}\psi^{(1)}_{j}{}^{k}}{\ell^7}  - \frac{\psi^{(1)jk} \partial_{k}\psi^{(2)}_{ij}}{\ell^7} \nonumber \\ &- \frac{2 \psi^{(1)}_{i}{}^{j} \partial_{k}\psi^{(2)}_{j}{}^{k}}{\ell^7} + \frac{2 \partial_{k}\partial_{j}\partial_{i}\psi^{(1)jk}}{3 \ell^5}  - \frac{\partial_{k}\partial^{k}\partial_{j}\psi^{(1)}_{i}{}^{j}}{\ell^5} + \frac{2 \psi^{(1)}_{j}{}^{l} \psi^{(1)jk} \partial_{l}\psi^{(1)}_{ik}}{\ell^7}\nonumber  \\ & - \frac{\psi^{(1)}_{jk} \psi^{(1)jk} \partial_{l}\psi^{(1)}_{i}{}^{l}}{2 \ell^7} + \frac{2 \psi^{(1)}_{i}{}^{j} \psi^{(1)kl} \partial_{l}\psi^{(1)}_{jk}}{\ell^7} + \frac{2 \psi^{(1)}_{i}{}^{j} \psi^{(1)}_{j}{}^{k} \partial_{l}\psi^{(1)}_{k}{}^{l}}{\ell^7}
\end{align}
\begin{align}
E^{(1)j}_{i}&=\frac{6 \psi^{(1)}_{i}{}^{k} \psi^{(1)jl} \psi^{(1)}_{k}{}^{m} \psi^{(1)}_{lm}}{\ell^8}  - \frac{\psi^{(1)}_{i}{}^{j} \psi^{(1)}_{k}{}^{m} \psi^{(1)kl} \psi^{(1)}_{lm}}{\ell^8}  - \frac{\psi^{(1)}_{i}{}^{k} \psi^{(1)j}{}_{k} \psi^{(1)}_{lm} \psi^{(1)lm}}{\ell^8} \nonumber \\ & - \frac{7 \delta_{i}{}^{j} \psi^{(1)}_{k}{}^{m} \psi^{(1)kl} \psi^{(1)}_{l}{}^{n} \psi^{(1)}_{mn}}{4 \ell^8} + \frac{7 \delta_{i}{}^{j} \psi^{(1)}_{kl} \psi^{(1)kl} \psi^{(1)}_{mn} \psi^{(1)mn}}{24 \ell^8} + \frac{\psi^{(1)}_{kl} \psi^{(1)kl} \psi^{(2)}_{i}{}^{j}}{\ell^8} \nonumber \\ & - \frac{4 \psi^{(1)jl} \psi^{(1)}_{kl} \psi^{(2)}_{i}{}^{k}}{\ell^8} + \frac{3 \psi^{(2)}_{i}{}^{k} \psi^{(2)j}{}_{k}}{\ell^8}  - \frac{4 \psi^{(1)}_{i}{}^{k} \psi^{(1)}_{kl} \psi^{(2)jl}}{\ell^8}  - \frac{4 \psi^{(1)}_{i}{}^{k} \psi^{(1)jl} \psi^{(2)}_{kl}}{\ell^8} \nonumber \\ &+ \frac{7 \psi^{(1)}_{i}{}^{j} \psi^{(1)kl} \psi^{(2)}_{kl}}{6 \ell^8}  - \frac{13 \delta_{i}{}^{j} \psi^{(2)}_{kl} \psi^{(2) kl}}{12 \ell^8} + \frac{11 \delta_{i}{}^{j} \psi^{(1)}_{k}{}^{m} \psi^{(1) kl} \psi^{(2)}_{lm}}{3 \ell^8}\nonumber \\ & + \frac{2 \psi^{(1) j}{}_{k} \psi^{(3)}_{i}{}^{k}}{\ell^8} + \frac{2 \psi^{(1)}_{i}{}^{k} \psi^{(3) j}{}_{k}}{\ell^8} - \frac{7 \delta_{i}{}^{j} \psi^{(1) kl} \psi^{(3)}_{kl}}{6 \ell^8} - \frac{\psi^{(4)}{}_{i}{}^{j}}{\ell^8} \nonumber \\ & - \frac{\psi^{(1) kl} \partial^{j}\partial_{i}\psi^{(1)}_{kl}}{\ell^6} - \frac{\partial_{k}\partial_{i}\psi^{(2) jk}}{\ell^6} - \frac{\partial_{k}\partial^{j}\psi^{(2)}_{i}{}^{k}}{\ell^6} + \frac{2 \partial_{k}\partial^{k}\psi^{(2)}_{i}{}^{j}}{\ell^6} \nonumber \\ & - \frac{\partial_{k}\psi^{(1)}_{i}{}^{k} \partial_{l}\psi^{(1) jl}}{\ell^6} + \frac{\partial_{i}\psi^{(1) jk} \partial_{l}\psi^{(1)}_{k}{}^{l}}{\ell^6} + \frac{\partial^{j}\psi^{(1)}_{i}{}^{k} \partial_{l}\psi^{(1)}_{k}{}^{l}}{\ell^6} \nonumber
\end{align}
\vspace{-1.2cm}
\hspace{0.5cm}
\begin{align}
 &+ \frac{2 \psi^{(1) kl} \partial_{l}\partial_{i}\psi^{(1)j}{}_{k}}{\ell^6} + \frac{\psi^{(1) jk} \partial_{l}\partial_{i}\psi^{(1)}_{k}{}^{l}}{\ell^6} + \frac{2 \psi^{(1) kl} \partial_{l}\partial^{j}\psi^{(1)}_{ik}}{\ell^6} + \frac{\psi^{(1)}_{i}{}^{k} \partial_{l}\partial^{j}\psi^{(1)}_{k}{}^{l}}{\ell^6} \nonumber \\ & - \frac{2 \psi^{(1) kl} \partial_{l}\partial_{k}\psi^{(1)}_{i}{}^{j}}{\ell^6} + \frac{\psi^{(1)}_{i}{}^{j} \partial_{l}\partial_{k}\psi^{(1) kl}}{3 \ell^6} + \frac{\delta_{i}{}^{j} \partial_{l}\partial_{k}\psi^{(2) kl}}{3 \ell^6} \nonumber \\ & - \frac{2 \psi^{(1) jk} \partial_{l}\partial^{l}\psi^{(1)}_{ik}}{\ell^6} - \frac{2 \psi^{(1)}_{i}{}^{k} \partial_{l}\partial^{l}\psi^{(1) j}{}_{k}}{\ell^6} + \frac{2 \partial_{k}\psi^{(1) j}{}_{l} \partial^{l}\psi^{(1)}_{i}{}^{k}}{\ell^6} \nonumber \\ & - \frac{4 \partial_{l}\psi^{(1) j}{}_{k} \partial^{l}\psi^{(1)}_{i}{}^{k}}{\ell^6}  - \frac{\delta_{i}{}^{j} \partial_{k}\psi^{(1) kl} \partial_{m}\psi^{(1)}_{l}{}^{m}}{6 \ell^6}  - \frac{4 \delta_{i}{}^{j} \psi^{(1) kl} \partial_{m}\partial_{l}\psi^{(1)}_{k}{}^{m}}{3 \ell^6} \nonumber \\ & + \frac{7 \delta_{i}{}^{j} \psi^{(1) kl} \partial_{m}\partial^{m}\psi^{(1)}_{kl}}{6 \ell^6}  - \frac{\delta_{i}{}^{j} \partial_{l}\psi^{(1)}_{km} \partial^{m}\psi^{(1) kl}}{3 \ell^6} + \frac{\delta_{i}{}^{j} \partial_{m}\psi^{(1)}_{kl} \partial^{m}\psi^{(1) kl}}{\ell^6}.
\end{align}
Since these equations do not give any conditions on $\gam$ they exhibit analogous behaviour to the 3D CG gravity \cite{Afshar:2011qw}, and differ from the EG in which the $\gamma_{ij}^{(1)}$ needs to vanish.

 \subsubsection{EOM for CG, Full Expressions}
Focusing on the $dS$ case, 
 EOM for CG for the unphysical fields $w, v_i$ and $\phi_{ij}$ that define  auxiliary field $f_{ij}$ are
 \begin{align}
 \phi_{ij}^{(0))}& =\frac{4 h_{ji}}{\ell^2}\end{align}\begin{align}
 \phi_{ij}^{(1))}& =0\end{align}\begin{align}
 \phi_{ij}^{(2)}& =-8 R[D]_{ij} + \tfrac{4}{3} h_{ji} R[D]  - \frac{4 h^{(1)}{}_{i}{}^{k} h^{(1)}{}_{jk}}{\ell^2} + \frac{2 h^{(1)}{}_{ij} h^{(1)k}{}_{k}}{\ell^2} \nonumber \\ & + \frac{h_{ji} h^{(1)}{}_{kl} h^{(1)kl}}{\ell^2}  - \frac{h_{ji} h^{(1)k}{}_{k} h^{(1)l}{}_{l}}{3 \ell^2} - \frac{4 h_{ji} h^{(2)k}{}_{k}}{3 \ell^2} \label{fijexp1}
 \end{align} 
 \begin{align}
 \phi_{ij}^{(3)}&=4 R[D] h^{(1)}{}_{ij} -4 h_{ji} R[D]^{kl} h^{(1)}{}_{kl} + \frac{12 h^{(1)}{}_{i}{}^{k} h^{(1)}{}_{j}{}^{l} h^{(1)}{}_{kl}}{\ell^2} - \frac{3 h^{(1)}{}_{ij} h^{(1)}{}_{kl} h^{(1)kl}}{\ell^2} \nonumber \\ &- \frac{6 h_{ji} h^{(1)}{}_{k}{}^{m} h^{(1)kl} h^{(1)}{}_{lm}}{\ell^2} - \frac{h^{(1)}{}_{ij} h^{(1)k}{}_{k} h^{(1)l}{}_{l}}{\ell^2} + \frac{2 h_{ji} h^{(1)k}{}_{k} h^{(1)}{}_{lm} h^{(1)lm}}{\ell^2} \nonumber \\ &+ \frac{6 h^{(1)k}{}_{k} h^{(2)}{}_{ij}}{\ell^2} - \frac{12 h^{(1)}{}_{jk} h^{(2)}{}_{i}{}^{k}}{\ell^2} - \frac{12 h^{(1)}{}_{i}{}^{k} h^{(2)}{}_{jk}}{\ell^2} + \frac{10 h_{ji} h^{(1)kl} h^{(2)}{}_{kl}}{\ell^2} \nonumber  \\ & + \frac{2 h^{(1)}{}_{ij} h^{(2)k}{}_{k}}{\ell^2}  - \frac{2 h_{ji} h^{(1)k}{}_{k} h^{(2)l}{}_{l}}{\ell^2} + \frac{4 h^{(3)}{}_{ij}}{\ell^2}   - \frac{4 h_{ji} h^{(3)k}{}_{k}}{\ell^2 } \nonumber \\ &+ 12 D_{j}D_{i}h^{(1)k}{}_{k}-12 D_{k}D_{i}h^{(1)}{}_{j}{}^{k} -12 D_{k}D_{j}h^{(1)}{}_{i}{}^{k}  + 12 D_{k}D^{k}h^{(1)}{}_{ij} \nonumber \\ & + 4 h_{ji} D_{l}D_{k}h^{(1)kl} -4 h_{ji} D_{l}D^{l}h^{(1)k}{}_{k}\end{align}
 \begin{align} \phi_{ij}^{(4)}&= -16 R[D]^{kl} h^{(1)}{}_{ij} h^{(1)}{}_{kl} + 16 h_{ji} R[D]^{kl} h^{(1)}{}_{k}{}^{m} h^{(1)}{}_{lm}  \nonumber \\ &- \frac{48 h^{(1)}{}_{i}{}^{k} h^{(1)}{}_{j}{}^{l} h^{(1)}{}_{k}{}^{m} h^{(1)}{}_{lm}}{\ell^2}+ \frac{8 h^{(1)}{}_{ij} h^{(1)k}{}_{k} h^{(1)}{}_{lm} h^{(1)lm}}{\ell^2}  \nonumber \\ &+ \frac{36 h_{ji} h^{(1)}{}_{k}{}^{m} h^{(1)kl} h^{(1)}{}_{l}{}^{n} h^{(1)}{}_{mn}}{\ell^2} - \frac{8 h_{ji} h^{(1)k}{}_{k} h^{(1)}{}_{l}{}^{n} h^{(1)lm} h^{(1)}{}_{mn}}{\ell^2}  \nonumber \\ &- \frac{4 h_{ji} h^{(1)}{}_{kl} h^{(1)kl} h^{(1)}{}_{mn} h^{(1)mn}}{\ell^2} + 8 R[D] h^{(2)}{}_{ij} - \frac{18 h^{(1)}{}_{kl} h^{(1)kl} h^{(2)}{}_{ij}}{\ell^2}\nonumber \\ & - \frac{2 h^{(1)k}{}_{k} h^{(1)l}{}_{l} h^{(2)}{}_{ij}}{\ell^2} + \frac{48 h^{(1)}{}_{j}{}^{l} h^{(1)}{}_{kl} h^{(2)}{}_{i}{}^{k}}{\ell^2} - \frac{48 h^{(2)}{}_{i}{}^{k} h^{(2)}{}_{jk}}{\ell^2}\nonumber \\ &  + \frac{48 h^{(1)}{}_{i}{}^{k} h^{(1)}{}_{kl} h^{(2)}{}_{j}{}^{l}}{\ell^2} -8 h_{ji} R[D]^{kl} h^{(2)}{}_{kl}  + \frac{24 h^{(1)}{}_{i}{}^{k} h^{(1)}{}_{j}{}^{l} h^{(2)}{}_{kl}}{\ell^2} \nonumber \\&+ \frac{4 h^{(1)}{}_{ij} h^{(1)kl} h^{(2)}{}_{kl}}{\ell^2} + \frac{16 h^{(2)}{}_{ij} h^{(2)k}{}_{k}}{\ell^2} + \frac{20 h_{ji} h^{(2)}{}_{kl} h^{(2)kl}}{\ell^2} \nonumber \\ & -  \frac{76 h_{ji} h^{(1)}{}_{k}{}^{m} h^{(1)kl} h^{(2)}{}_{lm}}{\ell^2} + \frac{12 h_{ji} h^{(1)k}{}_{k} h^{(1)lm} h^{(2)}{}_{lm}}{\ell^2} - \frac{8 h^{(1)}{}_{ij} h^{(1)k}{}_{k} h^{(2)l}{}_{l}}{\ell^2} \nonumber \\  &- \frac{4 h_{ji} h^{(2)k}{}_{k} h^{(2)l}{}_{l}}{\ell^2} + \frac{8 h_{ji} h^{(1)}{}_{kl} h^{(1)kl} h^{(2)m}{}_{m}}{\ell^2} + \frac{12 h^{(1)k}{}_{k} h^{(3)}{}_{ij}}{\ell^2}\nonumber \end{align}\begin{align} & - \frac{24 h^{(1)}{}_{jk} h^{(3)}{}_{i}{}^{k}}{\ell^2} - \frac{24 h^{(1)}{}_{i}{}^{k} h^{(3)}{}_{jk}}{\ell^2} + \frac{28 h_{ji} h^{(1)kl} h^{(3)}{}_{kl}}{\ell^2} - \frac{4 h^{(1)}{}_{ij} h^{(3)k}{}_{k}}{\ell^2} \nonumber \\ & - \frac{4 h_{ji} h^{(1)k}{}_{k} h^{(3)l}{}_{l}}{\ell^2} + \frac{12 h^{(4)}{}_{ij}}{\ell^2} - \frac{8 h_{ji} h^{(4)k}{}_{k}}{\ell^2} -48 h^{(1)kl} D_{i}D_{j}h^{(1)}{}_{kl} \nonumber \\ &-24 D_{i}h^{(1)kl} D_{j}h^{(1)}{}_{kl} + 24 D_{j}D_{i}h^{(2)k}{}_{k}  -24 D_{i}h^{(1)}{}_{j}{}^{k} D_{k}h^{(1)l}{}_{l}\nonumber \\ & -24 D_{j}h^{(1)}{}_{i}{}^{k} D_{k}h^{(1)l}{}_{l} -24 D_{k}D_{i}h^{(2)}{}_{j}{}^{k} -24 D_{k}D_{j}h^{(2)}{}_{i}{}^{k} \nonumber \\ &+ 24 D_{k}D^{k}h^{(2)}{}_{ij}  + 24 D_{k}h^{(1)l}{}_{l} D^{k}h^{(1)}{}_{ij} + 48 D_{i}h^{(1)}{}_{j}{}^{k} D_{l}h^{(1)}{}_{k}{}^{l}\nonumber \\ & + 48 D_{j}h^{(1)}{}_{i}{}^{k} D_{l}h^{(1)}{}_{k}{}^{l}  -48 D^{k}h^{(1)}{}_{ij} D_{l}h^{(1)}{}_{k}{}^{l} + 48 h^{(1)kl} D_{l}D_{i}h^{(1)}{}_{jk}  \nonumber \\ &+ 48 h^{(1)kl} D_{l}D_{j}h^{(1)}{}_{ik}-48 h^{(1)kl} D_{l}D_{k}h^{(1)}{}_{ij} + 16 h^{(1)}{}_{ij} D_{l}D_{k}h^{(1)kl}  \nonumber \\ &+ 16 h_{ji} h^{(1)kl} D_{l}D_{k}h^{(1)m}{}_{m} + 8 h_{ji} D_{l}D_{k}h^{(2)kl}-16 h^{(1)}{}_{ij} D_{l}D^{l}h^{(1)k}{}_{k}  \nonumber \\ &-8 h_{ji} D_{l}D^{l}h^{(2)k}{}_{k} -16 h_{ji} h^{(1)kl} D_{l}D_{m}h^{(1)}{}_{k}{}^{m} + 48 D_{k}h^{(1)}{}_{jl} D^{l}h^{(1)}{}_{i}{}^{k} \nonumber \\ &-48 D_{l}h^{(1)}{}_{jk} D^{l}h^{(1)}{}_{i}{}^{k} -4 h_{ji} D_{l}h^{(1)m}{}_{m} D^{l}h^{(1)k}{}_{k} -16 h_{ji} D_{k}h^{(1)kl} D_{m}h^{(1)}{}_{l}{}^{m} \nonumber \\ &+ 16 h_{ji} D^{l}h^{(1)k}{}_{k} D_{m}h^{(1)}{}_{l}{}^{m} -16 h_{ji} h^{(1)kl} D_{m}D_{l}h^{(1)}{}_{k}{}^{m}  \nonumber \\ &+ 16 h_{ji} h^{(1)kl} D_{m}D^{m}h^{(1)}{}_{kl}-8 h_{ji} D_{l}h^{(1)}{}_{km} D^{m}h^{(1)kl}  \nonumber \\ &+ 12 h_{ji} D_{m}h^{(1)}{}_{kl} D^{m}h^{(1)kl}, \label{fijexp2}
 \end{align}  
 \begin{align}
v_i^{(0)}&=0\end{align}\begin{align}
v_i^{(1)}&=0\end{align}\begin{align}
v_i^{(2)}&=\frac{4 D_{i}h^{(1)j}{}_{j}}{\ell}  - \frac{4 D_{j}h^{(1)}{}_{i}{}^{j}}{\ell}\\
v_i^{(3)}&- \frac{18 h^{(1)jk} D_{i}h^{(1)}{}_{jk}}{\ell} + \frac{12 D_{i}h^{(2)j}{}_{j}}{\ell}  - \frac{6 h^{(1)}{}_{i}{}^{j} D_{j}h^{(1)k}{}_{k}}{\ell}  - \frac{12 D_{j}h^{(2)}{}_{i}{}^{j}}{\ell} \nonumber \\ &+ \frac{12 h^{(1)jk} D_{k}h^{(1)}{}_{ij}}{\ell} + \frac{12 h^{(1)}{}_{i}{}^{j} D_{k}h^{(1)}{}_{j}{}^{k}}{\ell} \end{align}\begin{align}
v_{i}^{(4)}&=- \frac{48 h^{(2)jk} D_{i}h^{(1)}{}_{jk}}{\ell} + \frac{96 h^{(1)}{}_{j}{}^{l} h^{(1)jk} D_{i}h^{(1)}{}_{kl}}{\ell}  - \frac{60 h^{(1)jk} D_{i}h^{(2)}{}_{jk}}{\ell} \nonumber \\ &+ \frac{24 D_{i}h^{(3)j}{}_{j}}{\ell} + \frac{24 h^{(1)}{}_{i}{}^{j} h^{(1)kl} D_{j}h^{(1)}{}_{kl}}{\ell} - \frac{24 h^{(2)}{}_{i}{}^{j} D_{j}h^{(1)k}{}_{k}}{\ell}  - \frac{12 h^{(1)}{}_{i}{}^{j} D_{j}h^{(2)k}{}_{k}}{\ell} \nonumber \\ & - \frac{24 D_{j}h^{(3)}{}_{i}{}^{j}}{\ell} + \frac{24 h^{(2)jk} D_{k}h^{(1)}{}_{ij}}{\ell} + \frac{48 h^{(2)}{}_{i}{}^{j} D_{k}h^{(1)}{}_{j}{}^{k}}{\ell} + \frac{24 h^{(1)}{}_{i}{}^{j} h^{(1)}{}_{j}{}^{k} D_{k}h^{(1)l}{}_{l}}{\ell} \nonumber \\ &+ \frac{48 h^{(1)jk} D_{k}h^{(2)}{}_{ij}}{\ell} + \frac{24 h^{(1)}{}_{i}{}^{j} D_{k}h^{(2)}{}_{j}{}^{k}}{\ell}  - \frac{48 h^{(1)}{}_{j}{}^{l} h^{(1)jk} D_{l}h^{(1)}{}_{ik}}{\ell} \nonumber \\ & - \frac{48 h^{(1)}{}_{i}{}^{j} h^{(1)kl} D_{l}h^{(1)}{}_{jk}}{\ell} - \frac{48 h^{(1)}{}_{i}{}^{j} h^{(1)}{}_{j}{}^{k} D_{l}h^{(1)}{}_{k}{}^{l}}{\ell},\label{vexp}
\end{align}
\begin{align}
w^{(0)}&=\frac{4}{\ell^2}\end{align}\begin{align}
w^{(1)}&=0\end{align}\begin{align}
w^{(2)}& \tfrac{4}{3} R[D]  - \frac{h^{(1)}{}_{ij} h^{(1)ij}}{\ell^2}  - \frac{h^{(1)i}{}_{i} h^{(1)j}{}_{j}}{3 \ell^2} + \frac{8 h^{(2)i}{}_{i}}{3 \ell^2}\end{align}\begin{align}
w^{(3)}&=-4 R[D]^{ij} h^{(1)}{}_{ij} + \frac{6 h^{(1)}{}_{i}{}^{k} h^{(1)ij} h^{(1)}{}_{jk}}{\ell^2} + \frac{2 h^{(1)i}{}_{i} h^{(1)}{}_{jk} h^{(1)jk}}{\ell^2} - \frac{14 h^{(1)ij} h^{(2)}{}_{ij}}{\ell^2}  \nonumber \\ &- \frac{2 h^{(1)i}{}_{i} h^{(2)j}{}_{j}}{\ell^2} + \frac{8 h^{(3)i}{}_{i}}{\ell^2} + 4 D_{j}D_{i}h^{(1)ij}  -4 D_{j}D^{j}h^{(1)i}{}_{i} \end{align}\begin{align}
w^{(4)}&=16 R[D]^{ij} h^{(1)}{}_{i}{}^{k} h^{(1)}{}_{jk}  - \frac{36 h^{(1)}{}_{i}{}^{k} h^{(1)ij} h^{(1)}{}_{j}{}^{l} h^{(1)}{}_{kl}}{\ell^2}  - \frac{8 h^{(1)i}{}_{i} h^{(1)}{}_{j}{}^{l} h^{(1)jk} h^{(1)}{}_{kl}}{\ell^2}  \nonumber \\ & - \frac{4 h^{(1)}{}_{ij} h^{(1)ij} h^{(1)}{}_{kl} h^{(1)kl}}{\ell^2}  -8 R[D]^{ij} h^{(2)}{}_{ij} - \frac{28 h^{(2)}{}_{ij} h^{(2)ij}}{\ell^2} + \frac{92 h^{(1)}{}_{i}{}^{k} h^{(1)ij} h^{(2)}{}_{jk}}{\ell^2} \nonumber \\ &+ \frac{12 h^{(1)i}{}_{i} h^{(1)jk} h^{(2)}{}_{jk}}{\ell^2}  - \frac{4 h^{(2)i}{}_{i} h^{(2)j}{}_{j}}{\ell^2} + \frac{8 h^{(1)}{}_{ij} h^{(1)ij} h^{(2)k}{}_{k}}{\ell^2}  - \frac{44 h^{(1)ij} h^{(3)}{}_{ij}}{\ell^2} \nonumber \\ & - \frac{4 h^{(1)i}{}_{i} h^{(3)j}{}_{j}}{\ell^2} + \frac{16 h^{(4)i}{}_{i}}{\ell^2} + 16 h^{(1)ij} D_{j}D_{i}h^{(1)k}{}_{k} + 8 D_{j}D_{i}h^{(2)ij}  -8 D_{j}D^{j}h^{(2)i}{}_{i}  \nonumber \\ & -16 h^{(1)ij} D_{j}D_{k}h^{(1)}{}_{i}{}^{k} -4 D_{j}h^{(1)k}{}_{k} D^{j}h^{(1)i}{}_{i} -16 D_{i}h^{(1)ij} D_{k}h^{(1)}{}_{j}{}^{k} \nonumber \\ &+ 16 D^{j}h^{(1)i}{}_{i} D_{k}h^{(1)}{}_{j}{}^{k}  -16 h^{(1)ij} D_{k}D_{j}h^{(1)}{}_{i}{}^{k} + 16 h^{(1)ij} D_{k}D^{k}h^{(1)}{}_{ij} \nonumber \\ & -8 D_{j}h^{(1)}{}_{ik} D^{k}h^{(1)ij} + 12 D_{k}h^{(1)}{}_{ij} D^{k}h^{(1)ij}. \label{wexp}
\end{align}
 
 EOM of CG for the metric $\gamma_{ij}$, using the expansion with $\frac{1}{n!}$ reads in 
 $\rho\rho$ component
 \begin{align}
 E^{(4)\rho}_{\rho}&=- \frac{3 \psi ^{(1)}_{i}{}^{k} \psi ^{(1)ij} \psi ^{(1)}_{j}{}^{l} \psi ^{(1)}_{kl}}{4 \ell^8} + \frac{\psi ^{(1)}_{ij} \psi ^{(1)ij} \psi ^{(1)}_{kl} \psi ^{(1)kl}}{8 \ell^8} + \frac{\psi ^{(2)}_{ij} \psi ^{(2)ij}}{4 \ell^8} + \frac{\psi ^{(1)}_{i}{}^{k} \psi ^{(1)ij} \psi ^{(2)}_{jk}}{\ell^8} \nonumber \\ &  - \frac{\psi ^{(1)ij} \psi ^{(3)}_{ij}}{2 \ell^8}  - \frac{5 \psi ^{(1)}_{i}{}^{k} \psi ^{(1)}_{jk} R[D]^{ij}}{\ell^6}  - \frac{R[D]_{ij} R[D]^{ij}}{\ell^4} + \frac{5 \psi ^{(1)}_{ij} \psi ^{(1)ij} R[D]}{6 \ell^6}\nonumber \\ & + \frac{R[D]^2}{3 \ell^4} - \frac{D_{i}D^{i}R[D]}{3 \ell^4} + \frac{D_{j}D_{i}\psi ^{(2)ij}}{\ell^6}  - \frac{2 \psi ^{(1)ij} D_{j}D_{k}\psi ^{(1)}_{i}{}^{k}}{\ell^6} - \frac{D_{i}\psi ^{(1)ij} D_{k}\psi ^{(1)}_{j}{}^{k}}{2 \ell^6}\nonumber \\ & + \frac{3 \psi ^{(1)ij} D_{k}D^{k}\psi ^{(1)}_{ij}}{2 \ell^6} - \frac{D_{j}\psi ^{(1)}_{ik} D^{k}\psi ^{(1)ij}}{\ell^6} + \frac{D_{k}\psi ^{(1)}_{ij} D^{k}\psi ^{(1)ij}}{\ell^6},
 \end{align}
 in $\rho i $ component
 \begin{align}
 E^{(4)\rho}_{i}&=\frac{4 \psi ^{(2)jk} D_{i}\psi ^{(1)}_{jk}}{3 \ell^7} + \frac{4 R[D]^{jk} D_{i}\psi ^{(1)}_{jk}}{3 \ell^5} - \frac{2 \psi ^{(1)}_{j}{}^{l} \psi ^{(1)jk} D_{i}\psi ^{(1)}_{kl}}{\ell^7} + \frac{5 \psi ^{(1)jk} D_{i}\psi ^{(2)}_{jk}}{6 \ell^7}  \nonumber \\ & - \frac{2 \psi ^{(1)jk} D_{i}R[D]_{jk}}{3 \ell^5} + \frac{2 D_{i}D_{k}D_{j}\psi ^{(1)jk}}{3 \ell^5} + \frac{2 R[D] D_{j}\psi ^{(1)}_{i}{}^{j}}{3 \ell^5} - \frac{\psi ^{(1)}_{i}{}^{j} \psi ^{(1)kl} D_{j}\psi ^{(1)}_{kl}}{\ell^7} \nonumber \\ &+ \frac{D_{j}\psi ^{(3)}_{i}{}^{j}}{\ell^7}  - \frac{\psi ^{(1)}_{i}{}^{j} D_{j}R[D]}{3 \ell^5}  - \frac{2 \psi ^{(2)jk} D_{k}\psi ^{(1)}_{ij}}{\ell^7}  - \frac{2 R[D]^{jk} D_{k}\psi ^{(1)}_{ij}}{\ell^5} \nonumber \\ & - \frac{\psi ^{(2)}_{i}{}^{j} D_{k}\psi ^{(1)}_{j}{}^{k}}{\ell^7} + \frac{R[D]_{i}{}^{j} D_{k}\psi ^{(1)}_{j}{}^{k}}{\ell^5}  - \frac{\psi ^{(1)jk} D_{k}\psi ^{(2)}_{ij}}{\ell^7}  - \frac{2 \psi ^{(1)}_{i}{}^{j} D_{k}\psi ^{(2)}_{j}{}^{k}}{\ell^7} \nonumber \\ &+ \frac{2 \psi ^{(1)jk} D_{k}R[D]_{ij}}{\ell^5}  - \frac{D_{k}D^{k}D_{j}\psi ^{(1)}_{i}{}^{j}}{\ell^5} + \frac{2 \psi ^{(1)}_{j}{}^{l} \psi ^{(1)jk} D_{l}\psi ^{(1)}_{ik}}{\ell^7} \nonumber \\ & - \frac{\psi ^{(1)}_{jk} \psi ^{(1)jk} D_{l}\psi ^{(1)}_{i}{}^{l}}{2 \ell^7} + \frac{2 \psi ^{(1)}_{i}{}^{j} \psi ^{(1)kl} D_{l}\psi ^{(1)}_{jk}}{\ell^7}  + \frac{2 \psi ^{(1)}_{i}{}^{j} \psi ^{(1)}_{j}{}^{k} D_{l}\psi ^{(1)}_{k}{}^{l}}{\ell^7}
 \end{align}
 and in $ij$ component
 \begin{align}
 E^{(4)j}_{i}&=\frac{6 \psi ^{(1)}_{i}{}^{k} \psi ^{(1)jl} \psi ^{(1)}_{k}{}^{m} \psi ^{(1)}_{lm}}{\ell^8} - \frac{\psi ^{(1)}_{i}{}^{j} \psi ^{(1)}_{k}{}^{m} \psi ^{(1)kl} \psi ^{(1)}_{lm}}{\ell^8}  - \frac{\psi ^{(1)}_{i}{}^{k} \psi ^{(1)j}{}_{k} \psi ^{(1)}_{lm} \psi ^{(1)lm}}{\ell^8} \nonumber \\ & - \frac{7 \delta_{i}{}^{j} \psi ^{(1)}_{k}{}^{m} \psi ^{(1)kl} \psi ^{(1)}_{l}{}^{n} \psi ^{(1)}_{mn}}{4 \ell^8} + \frac{7 \delta_{i}{}^{j} \psi ^{(1)}_{kl} \psi ^{(1)kl} \psi ^{(1)}_{mn} \psi ^{(1)mn}}{24 \ell^8} \nonumber \\ &+ \frac{\psi ^{(1)}_{kl} \psi ^{(1)kl} \psi ^{(2)}_{i}{}^{j}}{\ell^8} - \frac{4 \psi ^{(1)jl} \psi ^{(1)}_{kl} \psi ^{(2)}_{i}{}^{k}}{\ell^8} + \frac{3 \psi ^{(2)}_{i}{}^{k} \psi ^{(2)j}{}_{k}}{\ell^8} \nonumber \\ & - \frac{4 \psi ^{(1)}_{i}{}^{k} \psi ^{(1)}_{kl} \psi ^{(2)jl}}{\ell^8} - \frac{4 \psi ^{(1)}_{i}{}^{k} \psi ^{(1)jl} \psi ^{(2)}_{kl}}{\ell^8} + \frac{7 \psi ^{(1)}_{i}{}^{j} \psi ^{(1)kl} \psi ^{(2)}_{kl}}{6 \ell^8}\nonumber \\ & - \frac{13 \delta_{i}{}^{j} \psi ^{(2)}_{kl} \psi ^{(2)kl}}{12 \ell^8} + \frac{11 \delta_{i}{}^{j} \psi ^{(1)}_{k}{}^{m} \psi ^{(1)kl} \psi ^{(2)}_{lm}}{3 \ell^8} + \frac{2 \psi ^{(1)j}{}_{k} \psi ^{(3)}_{i}{}^{k}}{\ell^8} \nonumber \\ &+ \frac{2 \psi ^{(1)}_{i}{}^{k} \psi ^{(3)j}{}_{k}}{\ell^8}  - \frac{7 \delta_{i}{}^{j} \psi ^{(1)kl} \psi ^{(3)}_{kl}}{6 \ell^8} - \frac{5 \psi ^{(1)}_{kl} \psi ^{(1)kl} R[D]_{i}{}^{j}}{\ell^6} + \frac{6 \psi ^{(1)jl} \psi ^{(1)}_{kl} R[D]_{i}{}^{k}}{\ell^6}\nonumber \\ &  - \frac{4 \psi ^{(2)j}{}_{k} R[D]_{i}{}^{k}}{\ell^6}  - \frac{8 R[D]_{i}{}^{k} R[D]^{j}{}_{k}}{\ell^4} + \frac{6 \psi ^{(1)}_{i}{}^{l} \psi ^{(1)}_{kl} R[D]^{jk}}{\ell^6} - \frac{4 \psi ^{(2)}_{ik} R[D]^{jk}}{\ell^6}\nonumber \\ & + \frac{2 \psi ^{(1)}_{i}{}^{j} \psi ^{(1)}_{kl} R[D]^{kl}}{3 \ell^6}  - \frac{19 \delta_{i}{}^{j} \psi ^{(1)}_{k}{}^{m} \psi ^{(1)}_{lm} R[D]^{kl}}{3 \ell^6} + \frac{8 \delta_{i}{}^{j} \psi ^{(2)}_{kl} R[D]^{kl}}{3 \ell^6}\nonumber \\ & + \frac{3 \delta_{i}{}^{j} R[D]_{kl} R[D]^{kl}}{\ell^4}  - \frac{7 \psi ^{(1)}_{i}{}^{k} \psi ^{(1)j}{}_{k} R[D]}{3 \ell^6} + \frac{17 \delta_{i}{}^{j} \psi ^{(1)}_{kl} \psi ^{(1)kl} R[D]}{6 \ell^6} + \frac{4 \psi ^{(2)}_{i}{}^{j} R[D]}{3 \ell^6} \nonumber \\ &+ \frac{14 R[D]_{i}{}^{j} R[D]}{3 \ell^4}  - \frac{5 \delta_{i}{}^{j} R[D]^2}{3 \ell^4}  - \frac{\psi^{(4)}{}_{i}{}^{j}}{\ell^8}  - \frac{13 \psi ^{(1)kl} D_{i}D^{j}\psi ^{(1)}_{kl}}{2 \ell^6}  - \frac{D_{i}D_{k}\psi ^{(2)jk}}{\ell^6} \nonumber \\ &+ \frac{2 \psi ^{(1)kl} D_{i}D_{l}\psi ^{(1)j}{}_{k}}{\ell^6} + \frac{\psi ^{(1)jk} D_{i}D_{l}\psi ^{(1)}_{k}{}^{l}}{\ell^6} + \frac{11 \psi ^{(1)kl} D^{j}D_{i}\psi ^{(1)}_{kl}}{2 \ell^6}  - \frac{2 D^{j}D_{i}R[D]}{3 \ell^4} \nonumber \end{align}
 \vspace{-0.3cm}
 \begin{align} & - \frac{D^{j}D_{k}\psi ^{(2)}_{i}{}^{k}}{\ell^6} + \frac{2 \psi ^{(1)kl} D^{j}D_{l}\psi ^{(1)}_{ik}}{\ell^6} + \frac{\psi ^{(1)}_{i}{}^{k} D^{j}D_{l}\psi ^{(1)}_{k}{}^{l}}{\ell^6} + \frac{2 D_{k}D^{k}\psi ^{(2)}_{i}{}^{j}}{\ell^6}\nonumber \\ & + \frac{2 D_{k}D^{k}R[D]_{i}{}^{j}}{\ell^4}  - \frac{\delta_{i}{}^{j} D_{k}D^{k}R[D]}{3 \ell^4}  - \frac{D_{k}\psi ^{(1)}_{i}{}^{k} D_{l}\psi ^{(1)jl}}{\ell^6} + \frac{D_{i}\psi ^{(1)jk} D_{l}\psi ^{(1)}_{k}{}^{l}}{\ell^6}\nonumber \\ & + \frac{D^{j}\psi ^{(1)}_{i}{}^{k} D_{l}\psi ^{(1)}_{k}{}^{l}}{\ell^6} - \frac{2 \psi ^{(1)kl} D_{l}D_{k}\psi ^{(1)}_{i}{}^{j}}{\ell^6} + \frac{\psi ^{(1)}_{i}{}^{j} D_{l}D_{k}\psi ^{(1)kl}}{3 \ell^6} + \frac{\delta_{i}{}^{j} D_{l}D_{k}\psi ^{(2)kl}}{3 \ell^6} \nonumber \\ & - \frac{2 \psi ^{(1)jk} D_{l}D^{l}\psi ^{(1)}_{ik}}{\ell^6}  - \frac{2 \psi ^{(1)}_{i}{}^{k} D_{l}D^{l}\psi ^{(1)j}{}_{k}}{\ell^6}  - \frac{4 \delta_{i}{}^{j} \psi ^{(1)kl} D_{l}D_{m}\psi ^{(1)}_{k}{}^{m}}{3 \ell^6} \nonumber \\ &+ \frac{2 D_{k}\psi ^{(1)j}{}_{l} D^{l}\psi ^{(1)}_{i}{}^{k}}{\ell^6}  - \frac{4 D_{l}\psi ^{(1)j}{}_{k} D^{l}\psi ^{(1)}_{i}{}^{k}}{\ell^6}  - \frac{\delta_{i}{}^{j} D_{k}\psi ^{(1)kl} D_{m}\psi ^{(1)}_{l}{}^{m}}{6 \ell^6} \nonumber \\ & + \frac{7 \delta_{i}{}^{j} \psi ^{(1)kl} D_{m}D^{m}\psi ^{(1)}_{kl}}{6 \ell^6}  - \frac{\delta_{i}{}^{j} D_{l}\psi ^{(1)}_{km} D^{m}\psi ^{(1)kl}}{3 \ell^6} + \frac{\delta_{i}{}^{j} D_{m}\psi ^{(1)}_{kl} D^{m}\psi ^{(1)kl}}{\ell^6}
 \end{align}
 
\subsection{Killing Vectors for Conformal Algebra on Spherical Background}
The Killing vectors admitted by the leading order Killing equation (\ref{lo}), (\ref{eq:LOCKV}) that agree with asymptotic isometries obtained in \cite{Henneaux:1985tv} when $\frac{1}{r}=\frac{\rho}{\ell^2}\rightarrow0$.
 \begin{align}
\text{$\xi^{sph}_0 $}&=(1,0,0) \label{kvsph0}\\
 \text{$\xi^{sph} _7$}&=(0,0,1)\\
 \text{$\xi^{sph}_3 $}&=\left(\cos (\theta ) (-\sin (t)),\sin (\theta ) (-\cos (t)),0\right) \\
 \text{$\xi^{sph}_6 $}&=\left(\cos (\theta ) \cos (t),\sin (\theta ) (-\sin (t)),0\right) \\
 \text{$\xi^{sph}_8$}&=\left(0,-\sin (\phi ),-\cot (\theta ) \cos (\phi )\right) \\
 \text{$\xi^{sph}_9 $}&=\left(0,\cos (\phi ),-\cot (\theta ) \sin (\phi )\right) \\
 \text{$\xi^{sph}_1 $}&=\left(\sin (\theta ) (-\sin (t)) \cos (\phi ),\cos (\theta ) \cos (t) \cos (\phi ),-\frac{\cos (t) \sin (\phi )}{\sin (\theta )}\right)\\
\text{$\xi^{sph}_2 $}&=\left(\sin (\theta ) (-\sin (t)) \sin (\phi ),\cos (\theta ) \cos (t) \sin (\phi ),\frac{\cos (t) \cos (\phi )}{\sin (\theta )}\right) \\
\text{$\xi^{sph}_4 $}&=\left(\sin (\theta ) \cos (t) \cos (\phi ),\cos (\theta ) \sin (t) \cos (\phi ),-\frac{\sin (t) \sin (\phi )}{\sin (\theta )}\right)\\
\text{$\xi^{sph}_5 $}&=\left(\sin (\theta ) \cos (t) \sin (\phi ),\cos (\theta ) \sin (t) \sin (\phi ),\frac{\sin (t) \cos (\phi )}{\sin (\theta )}\right).\label{sphkv}
\end{align} 

\section{Appendix: Canonical Analysis of Conformal Gravity}
\subsection{Hamiltonian analysis} 

To discuss the Hamiltonian formulation and the dynamics of gauge systems we start with the action principle in the form of the Lagrangian.
If the action
 \begin{equation}
S_I=\int_{t_1}^{t_2}L(q,\dot{q})dt
\end{equation}
is stationary under the variations $\delta q^n(t)$ that vanish at $t_1$ and $t_2$ for $q^n(n=1,...,N)$ Lagrangian variables, we have defined the classical motion of the system.
That 
 is fulfilled if the Euler-Lagrange equations 
\begin{align}
\frac{d}{dt}\left(\frac{\partial L}{\partial q^n}\right)-\frac{\partial L}{\partial q^n}=0, && n=1,...,N
\end{align}
or 
\begin{equation} 
\ddot{q}^n\frac{\partial^2L}{\partial\dot{q}^{n'}\partial\dot{q}^n}=\frac{\partial L}{\partial q^n}-\dot{q}^{n'}\frac{\partial^2L}{\partial\dot{q}^{n'}\partial\dot{q}^{n}}
 \end{equation}are satisfied. 
Positions and the velocities at the time t, determine accelerations when \begin{equation}
\frac{\partial^2 L}{\partial\dot{q}^{n'}\partial\dot{q}^n}
\end{equation}
is invertible, i.e. 
\begin{equation}
\textbf{D}=det\frac{\partial^2 L}{\partial\dot{q}^{n'}\partial\dot{q}^n} \label{detcond}
\end{equation}
does not vanish. If \textbf{D}=0, the accelerations $\ddot{q}$ are not uniquely determined with positions $q$ and velocities $\dot{q}$, and one could add arbitrary functions of time to the solutions of the EOM. In other words, when we are interested in the systems with gauge degrees of freedom, we are interested in systems for which $\frac{\partial^2 L}{\partial\dot{q}^{n'}\partial\dot{q}^n}$ cannot be inverted. 
In case we define canonical momenta by 
\begin{equation}
p_n=\frac{\partial L}{\partial \dot{q}^n} \label{momenta},
\end{equation}
the condition that \textbf{D}=0, reflects that velocities as functions of coordinates and momenta are not invertible, i.e.,  momenta $p_n$ are not independent and it follows from (\ref{momenta}) that
\begin{align}
\phi_m(q,p)=0, && m=1,...,M'\label{frstcon}.
\end{align}
These conditions (\ref{frstcon}), are restricted by the regularity conditions, and define a constant (for simplicity) submanifold in $(q,\dot{q})$ space, {\it primary constraint surface}.
Its rank is N-M' for M' independent equations (\ref{frstcon}) and the dimension of phase space is 2N-M'. 
Since (\ref{frstcon}) says that transformation from $p$ to $\dot{q}$ is multivalued, which can be shown on the mapping between the manifolds, one has to introduce {\it Lagrange multipliers}, parameters that make it single valued.

\subsection{Primary and Secondary Constraints}

For the constrained surface of $p$ and $q$ denoted with $\Gamma$, the subspace $\Gamma_1$ is defined with the constraints (\ref{frstcon}) and
it defines "weak equality" (which we write with "$\approx$"). Then the function $F$ which is zero on the constrained surface $\Gamma$ \begin{equation}F(p,q)\vert_{\Gamma_1}=0,\label{eqn}\end{equation} "vanishes weakly". When partial derivatives of the function $F$ with respect to coordinates
$\frac{\partial F}{\partial q}\vert_{\Gamma_1}=0$ and $\frac{\partial F}{\partial p}\vert_{\Gamma_1}=0$ also vanish on the constrained surface $\Gamma_1$, $F$ satisfies "strong equality" (denoted with $"="$). Its variation on constrained phase space $\Gamma_1$ is zero
\begin{equation}\delta F\vert_{\Gamma_1}=\left(\frac{\partial F}{\partial q_{a}}\delta q_{a}+\frac{\partial F}{\partial p_{a}}\delta p_a\right)\vert_{\Gamma_1}=0\label{varf} \end{equation}
for the variations of coordinates and momenta that satisfy $k$ conditions (\ref{frstcon}). Varying the constraints, we obtain \begin{equation}
\frac{\partial\phi_m }{\partial q_a}\delta q_a + \frac{\partial \phi _m}{\partial p_a}\delta p_a\approx 0.\label{varcons}
\end{equation}
The terms that multiply $\delta q_a$ and $\delta p_a$ in (\ref{varf}) and (\ref{varcons}), are equal up to an arbitrary Lagrange multiplier $\lambda^m$,
which leads to
\begin{align}
\frac{\partial}{\partial q_a}(F-\lambda^m\phi_m)\approx 0\\
\frac{\partial}{\partial p_a}(F-\lambda^m\phi_m)\approx 0.%
\end{align}
Next step is to define a canonical Hamiltonian,
\begin{equation}
H=\dot{q}^np_n-L\label{hc0}
\end{equation}
that can be expressed only using  canonical coordinates $q$ and momenta $p$, therefore it is valid only on the constrained phase space. 
(That can be verified by taking the variation $\delta H$ induced by arbitrary variations of the positions and velocities.) 
 Varying  (\ref{hc0}) 
\begin{equation}
\delta H=\dot{q}^n\delta p_n-\delta q^n\frac{\partial L}{\partial q^n}\label{ht1}
\end{equation}
we notice that (\ref{ht1}) is not uniquely determined depending on canonical coordinates $q_n$ and momenta  $p_n$.  $\delta p_n$ in (\ref{ht1}) are restricted to conserve primary constraints $\phi_m\approx0$, which means that (\ref{hc0}) is an identity on the constrained surface $\Gamma_1$. The formalism should be the same under the change
\begin{equation}
H\rightarrow H+c^m(q,p)\phi_m.
\end{equation}
We can define total Hamiltonian $H_T$, equal to the canonical Hamiltonian up to terms that are proportional to the constraints 
\begin{equation} H_T=H+\lambda^m \phi_m\label{ht}. \end{equation}
We can rewrite (\ref{ht1}) as 
\begin{align}
\left( \frac{\partial H}{\partial q^n}+\frac{\partial L}{\partial q^n}\right)\delta q^n +\left(\frac{\partial H}{\partial p_n}-\dot{q}^n\right)\delta p_n=0. \label{rwr}
\end{align}
Using a theorem that says, if $\lambda_n\delta q^n+\mu^n\delta p_n=0$ are for arbitrary variations tangent to the constraint surface, \begin{align} \lambda_n=u^m\frac{\partial\phi_m}{\partial q^n}\\ \mu^n=u^m\frac{\partial\phi_m}{\partial p_n} \end{align} for some $u^m$, the equalities here are those from the surface (\ref{frstcon}) and we can infer
\begin{align}
\dot{q}^n&=\frac{\partial H}{\partial p_n}+u^m\frac{\partial\phi_m}{\partial p_n}\\
-\frac{\partial L}{\partial q^n}|_{\dot{q}}&=\frac{\partial H}{\partial q^n}|_{p}+u^m\frac{\partial\phi_m}{\partial q^n}.
\end{align}
That allows us to write the $\dot{q}^n$ in terms of the momenta $p_n$ (with $\phi_m=0$) and extra parameters $u^m$. If constraints are independent, then $\frac{\partial \phi_m}{\partial p_n}$ are independent on $\phi_m=0$. That means that different sets of $u's$ can not lead to equal velocities. They can be expressed 
using the coordinates and velocities  from 
\begin{equation}
\dot{q}^n=\frac{\partial H}{\partial p_n}(q,p(q,\dot{q})) +u^m(q,\dot{q})\frac{\partial \phi_m}{\partial p_n}(q,p(q,\dot{q})).
\end{equation}
Rewritng the Legendre transformation from $(q,\dot{q})$ space to $\phi_m(q,p)=0$ of (q,p,u) space the transformation is invertible.
That allows to rewrite the Lagrangian equations in the Hamiltonian form, which can be also obtained by varying the action
\begin{equation}
\delta \int_{t_1}^{t_2}(\dot{q}^np_n-H-u^m\phi_m)=0\label{acham}
\end{equation}
with respect to $\delta q^n, \delta p_n,$ and $\delta u_m$ with $\delta q_n(t_1)=\delta q_n(t_2)=0$. $u^m$ obtain clear role of Lagrange multipliers imposing primary constraints..

Equation of motion obtained from (\ref{acham}) can be conveniently written using the Poisson brackets. For an arbitrary dynamical quantity $g(q,p)$ the equation of motion $\dot{q}(q,p)$ is 
\begin{equation}\dot{g}=\{g,H_c \} +u^m \{ g,\phi_m\}\approx \{g,H_T \}  \label{cond} \end{equation}
where EOM are valid on shell. Analogously, the equation of motion for a constraint $\phi_m$ can be written as
\begin{equation} \dot{\phi}_m= \{\phi_m,H_c \} +u^n\{\phi_m,\phi_n \}.  \label{eomphi} \end{equation}
For the theory to be consistent, we must demand that the primary constraints are conserved in time, which implies consistency conditions
\begin{enumerate}[label=\textbf{C.\arabic*}]
\item (\ref{eomphi}) is satisfied trivially, $0=0$; \label{cc1}
\item (\ref{eomphi}) determines Legendre multipliers via $p$s and $q$s; \label{cc2}
\item  (\ref{eomphi}) leads to condition with no multipliers, that defines a new  \emph{secondary constraint} which define the  subspace $\Gamma_2\subseteq\Gamma_1$. \label{cc3}
\end{enumerate}
Denoting all the constraints with $\varphi$ we can write EOM \begin{eqnarray} \dot{\varphi}_s= \{\varphi_s,H_c \} +u^m\{\varphi_s,\phi_m \}   \label{eomchi}, \end{eqnarray}
for $s=1,...,N$ where N denotes all the constraints. It has a solution
\begin{align}
u^m=U^m+v^aV_a{}^m,
\end{align}
here, $V_a{}^m$ solve homogeneous equation and denote independent solutions, $v^a=v^a(t)$ denote arbitrary coefficients, $U^m$ denote particular solutions, and the index $a$ runs over all the solutions.
This general soution and $V_a{}^m\phi_m=\phi_a$ allows to see from \begin{equation}
H_T=H_c+U^m\phi_m+v^aV_a{}^m\phi_m=H_c+U^m\phi_m+v^a\phi_a \label{htn}
\end{equation}
that there are arbitrary functions of time in the equation even after satisfying all the consistency conditions. 
This implies that the dynamical variables are not uniquely determined by their initial values at some future instant of time.

\subsection{First and Second Class Constraints}

If we consider a dynamical variable R(q,p) and determine that it has a weakly vanishing Poisson bracket with all the constraints, we have found a \emph{first class constraint}. Otherwise, the constraint is \emph{second class}.
It should be noted that $H_c+U^m\phi_m$ and $\phi^a$ in (\ref{htn}) are first class constraints.

Consider the time evolution of the general dynamical variable $g(t)$ from $t=0$. The initial value $g(0)$ is determined from the initial values of (q(0),p(0)) while the value  of $g(t)$ at the instant of time $\delta t$ is computed from 
\begin{eqnarray} 
g(\delta(t))&=&g(0)+\delta t \dot{g} \\
&=&g(0)+\delta t \left( \{g,H'\}+v^a\{g,\phi_a\} \right).
\end{eqnarray}
Since different values of the arbitrary coefficients $v^a(t)$ are allowed, we can obtain different values for $g(\delta t)$
\begin{equation}
\Delta g(\delta t)=\delta t(v_2^a-v_1^a)\{g,\phi_a\},
\end{equation}
where $H'$ is sum of the canonical Hamiltonian and $U^m\phi_m$.
 Phyisical states $g_1(\delta t)$ and $g_2(\delta_t)$ do not depend on the multipliers, that means $g(\delta t)$ is unphysical. The number of the first class constraints $\phi^a$ is equal to number of $v^a(t)$, arbitrary functions, that implies the transformations that are generated this way are unphysical. i.e. Gauge transformations, unphysical transformations of the dynamical variables, are generated by the primary first class constraints (PFC).

\subsection{Dirac Brackets}

Imagine we have two second-class constraints \begin{align} q_1\approx0 && p_1\approx0. \end{align}  
Second class constraints do not conserve all the constraints therefore their usage as generators of gauge transformations may lead to contradictions. 

If for example
\begin{align}
F\equiv p_1\psi(q)\approx0 \text{  for  } \psi\neq0, 
\end{align}
one obtains 
\begin{align}
\delta F=\epsilon\{q_1,F \} =\psi 
\end{align}
and learns $\delta F\neq0$.
The constraints are weakly equal to zero and for that reason one should first compute the Poisson brackets (PB)s and then use the constraints. 
From these equations we notice the variables $(q_1,p_1)$ are not relevant and one can eliminate them from the theory. 
To do that we introduce the modified PB in which $(q_1,p_1)$ are discarded
\begin{align}
\{f,g\}^*=\sum_{i\neq1}\left(\frac{\partial f}{\partial q_i}\frac{\partial g}{\partial p_i}-\frac{\partial f}{\partial p_i}\frac{\partial g}{\partial q_i}\right). \label{bdb}
\end{align}
Once (\ref{bdb}) was defined, one can treat the constraints $q_1\approx0$, $p_1\approx0$ as strong equations, defining the theory for the variables $(q_i,p_i)$ when $i\neq1$. 

That implies that  second-class constraints are dynamical degrees of freedom that are of no importance. To eliminate them, we define new PBs that include only the important dynamical degrees of freedom. 

If there are $N_1$ FC constraints $\phi_a$ and $N_2$ remaining constraints $\theta_s$  (which are second class), the matrix $\Delta_{rs}=\{\theta_r,\theta_s\}$  is non-singular (and antisymmetric). If $det(\Delta_{rs})=0$ then $\lambda^s\{\theta_r,\theta_s\}=0$  would lead to solution for $\lambda^s$, and $\lambda^s\theta_s$ would be linear combination equal to FC. Which we have excluded by assumption.
Since $\Delta$ is not singular, we can define new PB using its inverse
\begin{align}
\{f,g\}^*=\{f,g\}-\{f,\theta_r\}\Delta_{r,s}^{-1}\{\theta_s,g\}.
\end{align}
This PB defines {\it Dirac bracket} which satisfies the properties of PB.

Dirac bracket of an arbitrary variable with any second class constraint is constructed to vanish
\begin{align}
\{\theta_m,g\}^*=\{\theta_m,g\}-\{\theta_m,\theta_r\}\Delta^{-1}_{rs}\{\theta_s,g\}=0
\end{align}since $\{\theta_m,\theta_r\}\Delta_{rs}^{-1}=\delta_{ms}$.
In other words, by construction of the Dirac brackets, second-class constraints $\theta_m\approx0$ can be regarded as strong equalities. The EOM (\ref{cond}) in terms of the Dirac brackets read
\begin{equation}
\dot{g}\approx\{g,H_T\}^*.
\end{equation}
The main difference between the first and second class constraints is that the first class constraints generate unphysical transformations, while the second class constraints, can be treated as strong equations after introduction of Dirac brackets.

The process of their construction can be simplified using the subsets of second class constraints, where for the first subset one uses Poisson brackets, while for the second one, the constructed Dirac brackets. 

The number of degrees of freedom of a constrained system  one may compute form the Dirac's formula, that says that
number of physical degrees of freedom $N_{d.o.f.}$ is
\begin{align}
N_{d.o.f.}=\frac{1}{2}\left( N_{c.v.}-2N_{FC}-N_{SC} \right)\label{ndofs}
\end{align}
where $N_{c.v}$ denotes number of canonical variables, $N_{FC}$ denotes number of first class constraints, and $N_{SC}$ number of second class constraints.

\subsection{Castellani algorithm}

If we have a total Hamiltonian (\ref{ht}), computed functions $v^a(t)$, all the constraints $\phi_b\approx0$, and a trajectory $T_1(t)=(q(t),p(t))$ with defined initial conditions on the constraint surface $\Gamma_2$, we obtain EOM
\begin{align}
\dot{q}_i&=\frac{\partial H'}{\partial p_i}+v^a \frac{\partial \phi_a}{\partial p_i} \\
-\dot{p}_i&=\frac{\partial H'}{\partial q_i}+v^a \frac{\partial \phi_a}{\partial q_i} \\
\psi_b(q,p)&=0 \label{hteom}
\end{align} 
for $\phi_b$ entire set of $b$ constraints. We denote $\phi_a$ as a generator of transformations, while the variation $\delta v^{a}(t)$ is an infinitesimal parameter. 
One can write an analogous set of equations for a new varied trajectory $T_2(t)=(q(t)+\delta_0q(t),p(t)+\delta_0p(t))$ that starts at the same point but satisfies EOM with new functions $v^a(t)+\delta_0v^a(t)$ and small variations denoted with $\delta_0$,
\begin{align}
\delta_0\dot{q}_i&=\left(\delta_0q_j\frac{\partial}{\partial q_j}+\delta_0 p_j\frac{\partial}{\partial p_j}\right)\frac{\partial H_T}{\partial p_i}+\delta_0 v^a\frac{\partial \phi_a}{\partial p_i} \\
-\delta_0\dot{p}_i&=\left(\delta_0 q_j \frac{\partial}{\partial q_j}+\delta_0 p_j\frac{\partial}{\partial p_j} \right)\frac{\partial H_T}{\partial q_j}+\delta_0v^a\frac{\partial\phi_a}{\partial q_i}\\
\frac{\partial \psi_s}{\partial q_i}\delta_0 q_i+\frac{\partial \psi_s}{\partial p_j}\delta_0 p_j&=0.
\end{align}
Simultaneous transition from one to another trajectory is represented by unphysical gauge transformation. \cite{Blagojevic:2002du}

In case we determine variations of the dynamical variables by 
an arbitrary infinitesimal parameter  $\epsilon(t)$, that leads to form
\begin{align}
\delta q_i&=\epsilon(t)\{q^i,G\}=\epsilon(t)\frac{\partial G}{\partial p_i}\\
\delta p_i&=\epsilon(t)\{p^i,G\}=-\epsilon(t)\frac{\partial G}{\partial q_i}. \label{varpq}
\end{align}
Here, we define the generator of the transformation with G.
When we vary the equation  (\ref{hteom}) with respect to $v^a(t)$ and differentiate (\ref{varpq}) with respect to $t$ we obtain
\begin{align}
\frac{\partial}{\partial p_i}\left(\dot{\epsilon}G+\epsilon\{G,H_t\}-\phi_a\delta v^a \right)\approx0 \label{gen1can}\\
\frac{\partial}{\partial q^i}\left(\dot{\epsilon}G +\epsilon\{G,H_T\}-\phi_a\delta v^a \right)\approx 0 \label{gen2can} \\
\epsilon\{\psi_j,G\}\approx0.
\end{align}
The equations (\ref{gen1can}) and (\ref{gen2can}) lead to 
\begin{equation}
\{F,\dot{\epsilon}G+\epsilon\{G,H_T\}-\phi_a\delta v^a\}\approx 0 \label{trivgen}
\end{equation}
where F is an arbitrary function defined on the subspace $\Gamma_1$. This leads to the conclusion that we obtained a trivial generator 
$\dot{\epsilon}G+\epsilon\{G,H_T\}-\phi_a\delta v^a$.
In other words, physical state F is invariant under the gauge transformations that cause the redundancy in the variables that reflect gauge symmetry. 
This physical state  satisfies EOM and the constraints can be imagined as trajectory in Hamiltonian theory. 

The gauge generator can be found from transformation of the canonical variables and conjugate momenta generated by a function G that acts on a given phase-space  and is parametrised by an infinitesimal parameter $\epsilon(t)$. 
The general requirement demands time derivatives of $\epsilon(t)$, $\epsilon^{(n)}\equiv \frac{d^n\epsilon}{dt}$ to be of finite order. In the phase space, that transformation gives varied trajectory, that needs to satisfy constraints and EOM.
From that, we obtain conditions that define the gauge transformations, and solving them we compute gauge generators. 
The generator $G$ is 
\begin{equation}
G(\epsilon,\epsilon^{(1)},\epsilon^{(2)},...,\epsilon^k)=\sum_{n=0}^k\epsilon^{(n)}G_{(n)}.\label{geng}
\end{equation}
The algorithm for computing the gauge generators has been discovered by Leonardo Castellani defining its name as "Castellani algorithm".
It starts with the $G_k$ which is primary first class constraint  (PFC) while the $G_{(n)}$ are all the first class constraints. The algorithm
\begin{eqnarray}
G_k=&PFC\\
G_{k-1}+\{G_k,H\}=&PFC \\
.& \\
.&\\
.&\\
G_1+\{G_2,H\}=&PFC \\
G_0+\{G_1,H\}=&PFC \\
\{G_0,H\}=&PFC \label{castel}
\end{eqnarray}
was developed by Castellani \cite{Castellani:1981us}.
Here, linear combinations of the primary first class constraints are also considered under "PFC". One can notice that $k$ gives a number of secondary constraints. 
\subsection{Gauge generators}
Assuming that our theory contains three PFCs, one of which is a vector $PFC^{(1)}_j$ and two of them, $PFC^{(2)}$ and $PFC^{(3)}$ are scalars, the ansatz for a generator that starts the algorithm is
\begin{align}
G_{k-1}&=-\{G_k,H\}+\int d^3x\bigg[ \alpha_1^j(x,y)PFC^{(1)}_j(y)+\alpha_2(x,y)PFC^{(2)}(y) \nonumber \\&+\alpha_3(x,y)PFC^{(3)}(y) \bigg],\label{generat}
\end{align}
for $\alpha_1$, $\alpha_2$ and $\alpha_3$ variables we have to find to determine the generator.
Consider a theory with fields $\{\phi_i\}$ and label the gauge transformations with $\xi$. The gauge transformations are generated with generators of a from (\ref{generat}) determined from (\ref{castel})
\begin{align}
G\left[ \xi,\psi \right]=\int_{\sigma}d^nx\mathcal{G}\left[\xi,\psi\right].
\end{align}
General variation of a generator deforms it into
\begin{align}
\delta G\left[\xi,\psi \right]=\int_{\sigma}d^nx\frac{\delta\mathcal{G}}{\delta \phi_i}+\int_{\partial\sigma}d^{n-1}xB\left[\xi,\phi,\delta\phi\right]\label{ngen}
\end{align}
Where we have denoted the boundary term with B. It has to be added to a generator to bring it into a finite form. The small fluctuations of the fields therefore, define the boundary conditions and bring B into the total variation 
\begin{align}
B\left[ \xi,\phi,\delta\phi \right]&=-\delta\Gamma\left[ \xi,\phi\right]\\
Q\left[ \xi\right]&=\Gamma\left[\xi\right].
\end{align}
$Q$ defines a canonical charge of a theory. The new generators (\ref{ngen}) define an asymptotic symmetry algebra of the "improved generators" which consequently, as we will show on the example of CG, define asymptotic symmetry algebra of the charges. This algebra should agree with the algebra obtained by the boundary condition preserving diffeomorphisms and Weyl rescaling, crucial in definition of the field theory at the boundary.

In the following chapter we will consider the canonical analysis of CG in four dimensions. 

\subsection{ADM Decomposition}

A manifold $\mathcal{M}$ described with coordinates $x^i$ can be split into space and time coordinates, with the successive hypersurfaces described via time-parameter $t$.  In four dimensional space-time, three geometries are treated differently than the four geometry of entire manifold. If we denote two respective hypersurfaces of the spacetime split into $t=constant$, a "lower" and a $t+dt=constant$ "upper" hypersurface, the information sufficient to build that kind of  sandwich structure are 
\begin{itemize}
\item
the metric on the $3-geometry$ of the lower hypersurface 
\begin{equation}
g_{ij}(t,x,y,y)dx^idx^j,
\end{equation}
\item
the distance between one point in the lower hyper surface and in the upper one, and the metric on the upper hypersurface
\begin{equation}
g_{ij}(t+dt,x,y,z)dx^idx^j
\end{equation}
\item the definition of proper length
\begin{align}
\left(\begin{array}{c} \text{lapse of} \\ \text{proper time} \\ \text{between lower} \\  \text{and upper} \\ \text{hypersurface} \end{array}\right) =\left(\begin{array}{c} " \text{lapse} \\ \text{function} " \end{array}\right)dt=N(t,x,y,z) dt
\end{align} for the connector on the ($x,y,z$) point of the lower hypersurface,
\item and the definition for the place of the upper hypersurface 
\begin{align}
x_{\text{upper}}^i(x^m)=x^i-N^i(t,x,y,z)dt,
\end{align}
to which to connect. 
\end{itemize}
From the Pythagorean theorem in four dimensional form, 
\begin{align}
ds^2=\left(\begin{array}{c}\text{proper distance} \\ \text{in base 3-geometry}\end{array}\right)^2-\left( \begin{array}{c} \text{proper time from} \\ \text{ lower to upper 3-geometry}\end{array}\right)^2
\end{align}
leads to
\begin{equation}
ds^2=g_{ij}(dx^11+N^idt)(dx^j+N^jdt)-(Ndt)^2.\label{adm}
\end{equation}
To obtain the components of the four dimensional metric tensor in relation to the three dimensional one, we compare the split (\ref{adm})
with \begin{equation}
ds^2={}^{(4)}g_{\alpha\beta}dx^{\alpha}dx^{\beta}
\end{equation}
and read out the components. The construction of the metric is 
\begin{equation}
\left(\begin{array}{cc}g_{00} & g_{0k}\\g_{i0}& g_{ik}\end{array}\right)=\left( \begin{array}{cc} (N_sN^s-N^2) & N_k \\ N_i &g_{ik}\end{array}\right)
\end{equation}
with  $N^i$ components of the shift in the original covariant form, while its indices, raised and lowered with three dimensional metric $N_i=g_{im}N^m$ are covariant components,  and $N^m=g^{ms}N_s$. We obtain the inverse metric from the product 
\begin{equation}
\left(\begin{array}{cc}{}^{(4)}g^{00} & {}^{(4)}g^{0m} \\ {}^{(4)}g^{0m} & {}^{(4)} g^{k0}\end{array}\right)=\left(\begin{array}{cc}-\frac{1}{N^2} & \frac{N^m}{N^2} \\ \frac{N^k}{N^2} & g^{km}- \frac{N^kN^m}{N^2}\end{array}\right)\label{invadm}.
\end{equation}
When one adds the lapse and shift to the $3-$metric, that determines the components of the unit timelike normal vector $\textbf{n}$. The vector is normalised saying there is $\textbf{n}$ -dual to $\textbf{n}$ for which 
\begin{equation}
\langle \textbf{n},\textbf{n}\rangle=-1.
\end{equation}
The value of $\textbf{n}$ is
\begin{equation}
\textbf{n}=n_{\beta}\textbf{d}x^{\beta}=-N\textbf{d}t+0+0+0.
\end{equation}
The unit timelike normal vector has the components
\begin{equation}
n_{\beta}=(-N,0,0,0),
\end{equation}
while this vector with raised index, using the metric (\ref{invadm}) has the components
\begin{equation}
n^{\alpha}=\left(\frac{1}{N},-\frac{N^m}{N}\right). 
\end{equation}
One can for completeness define the "perpendicular connector" with components 
\begin{equation}
(dt,-N^mdt)
\end{equation}
and the proper length $d\tau=Ndt$.

\subsection{Cayley--Hamilton Theorem}

The theorem that we find useful in treating tensorial quantities is the Cayley-Hamilton theorem.
It states that a square matrix over a commutative ring is the root of the characteristic polynomial that belongs to it, $P(A)=0$. One defines characteristic polynomial with 
\begin{align}
P(\lambda)=\text{det}(\lambda I-A),
\end{align}
where we denoted unit matrix with $I$. Its tensor form is a result of the relation between matrices, linear transformations and the rank 2 tensors on a vector space. 
If we have a tensor $T^{\m}_{\n}$ on a $d$-dimensional vector space, for example a tangent space of the $d-$dimensional manifold, the theorem states
\begin{align}
P(T)^{\m}{}_{\n}&=-(d+1)\delta^{\m}{}_{[\n}T^{\alpha_1}{}_{\alpha_1}T^{\alpha_2}{}_{\alpha_2}\cdots T^{\alpha_d}{}_{\alpha_d]}\\
&=(T^d)^{\m}{}_{\n}+c_1(A^{d-1})^{\m}{}_{\n}+\cdots+c_{d-1}T^{\m}{}_{\n}+c_d\delta^{\m}{}_{n}=0
\end{align}
for the coefficients $c_n$
\begin{align}
c_n=(-1)^nT^{\m_1}{}_{[\m_1}T^{\m_2}{}_{\m_2}\cdots T^{\m_n}{}_{\m_n]} && n=1,2,...,d
\end{align}
and 
\begin{align}
(T^m)^{\m}{}_{\n}=T^{\m}{}_{\alpha_1}T^{\alpha_1}{}_{\alpha_2}\cdots T^{\alpha_{m-2}}{}_{\alpha_{m-1}}T^{\alpha_{m-1}}{}_{\n} && m=2,3,...,d.
\end{align}
In particular for a 3D Riemannian manifold the tensor $T^i{}_{j}$ satisfies
\begin{align}
P(T)^i{}_j&=T^i{}_kT^k{}_{l}T^l{}_{j}-T^i{}_kT^k{}_{j}T-\frac{1}{2}T^i{}_j(T^k{}_lT^l{}_k-T^2)\\&-\frac{\delta^i{}_{j}}{6}(2T^k{}_lT^l{}_{m}T^m{}_{k}-3T^k{}_lT^l{}_{k}T+T^3)=0
\end{align}
for $T=T^i{}_i$ a trace.

\subsection{ADM Decomposition of Curvatures}

From the conventions in the chapter "Canonical Analysis" we obtain the ADM decomposition of the curvature tensors. 
The decompositions of the metric that lead to the Gauss, Codazzi and Ricci relations respectively, are 
\begin{align}
\perp {}^4R_{abcd}&=-K_{ad}K_{bc}+K_{ac}K_{bd}+R_{abcd},\\
\perp n^{d(4)}R_{abcd}&=D_{a}K_{bc}-D_bK_{ac},\\
\perp n^bn^{d(4)}R_{abcd}&=K_a{}^eK_{ec}-\frac{1}{N}\dot{K}_{ac}+\frac{1}{N}D_aD_bN+\frac{1}{N}\pounds_{N}K_{ac}.
\end{align}
They are employed in the derivation of the Ricci tensor 
\begin{align}
\perp \,^{\ms{(4)}}R_{ab}&=-2K_{ac}K_b^{\ c}+K_{ab}K+\frac{1}{N}\dot{K}_{ab}-\frac{1}{N}\pounds_NK_{ab}+\frac{1}{N}D_aD_cN+R_{ab},\nonumber\\
\perp n^b\,^{\ms{(4)}}R_{ab}&=D_cK_a^{\ c}-D_aK,\nonumber\\
n^an^b \,^{\ms{(4)}}R_{ab}&=K_{ab}K^{ab}-\frac{1}{N}h^{ab}\left(\dot{K}_{ab}-\pounds_NK_{ab}\right),
\end{align}
and Ricci scalar
\begin{align}
\,^{\ms{(4)}}R=-3K_{ab}K^{ab}+K^2+\frac{2}{N}h^{ab}\left(\dot{K}_{ab}-\pounds_NK_{ab}\right)+R.
\end{align}
The tracelessness of the Weyl tensor 
\begin{align}
h^{bd}\perp C_{abcd}&=\perp n^bn^dC_{abcd},\nonumber\\
h^{bd}\perp n^dC_{abcd}&=0,\nonumber\\
h^{bd}\perp n^an^dC_{abcd}&=0,
\end{align}
in combination with its symmetries, allow us to write the trace part of the Weyl tensor spatial projection $\perp C_{abcd}$ with 
\begin{equation}
h_{bd}\perp n^en^fC_{aecf}+h_{bc}\perp n^en^fC_{aefd}+h_{ad}\perp n^en^fC_{ebcf}+h_{ac}\perp n^en^fC_{ebfd}.\label{Weyltrace}
\end{equation}
For this decomposition one has to impose only the tracelessness condition and the Gauss relation in order to derive the traceless part of the Weyl tensor, $K_{abcd}$. This leaves only the extrinsic curvature as a candidate that may appear in the final result, while the traceless part of the Riemann tensor corresponds to induced Weyl which vanishes. Therefore,
\begin{align}
K_{abcd}&=&\frac{1}{2}K_{ac}K_{bd}+h_{ac}\left(K_{be}K_d^{\ e}-K_{bd}K\right)-\frac{1}{4}h_{ac}h_{bd}\left(K_{ef}K^{ef}+K^2\right)+\nonumber\\&\ &+(a\leftrightarrow b,c\leftrightarrow d)-(a\leftrightarrow b)-(c\leftrightarrow d).
\end{align}
The remaining projections of the Weyl tensor give
\begin{align}
&\perp n^dC_{abcd}=2\mathcal{S}_{abc}^{def}D_dK_{ef}\equiv B_{abc}\nonumber\\
&\perp n^an^dC_{abcd}=\frac{1}{2}\mathcal{T}_{ab}^{ef}\left[R_{ef}+K_{ef}K-\frac{1}{N}\left(\dot{K}_{ef}-\pounds_NK_{ef}-D_eD_fN\right)\right],
\end{align}
where
\begin{align}
\mathcal{S}_{abc}^{def}&=h_{a}^{\ [d}h_{b}^{\ e]}h_{c}^{\ f}-h_{a}^{\ [d}h_{bc}h^{e]f}\nonumber\\
\mathcal{T}_{ab}^{de}&=2h_{(a}^{\ d}h_{b)}^{\ e}-\frac{1}{3}h_{ab}h^{de}.
\end{align}
One can now write the decomposition of the Weyl tensor in the contributions
\begin{align}
1 &\times \perp C_{abcd}\nonumber\\
4&\times\ n_bn_d\perp n^en^fC_{aecf}\nonumber\\
4&\times -n_a\perp n^eC_{ebcd}.\label{splweyl}
\end{align}
using Weyl traceless and the symmetries in the expansion of $C_{abcd}C^{abcd}$  we find that each of the terms in (\ref{splweyl}) contribute when they are contracted with itself
\begin{equation}
C_{abcd}C^{abcd}=\perp C_{abcd}\perp C^{abcd}-4\perp n^eC_{ebcd}\perp n_feC^{fbcd}+4\perp n^en^fC_{aecf}\perp n_gn_hC^{agch}.
\end{equation}
The term that contributes in $K_{abcd}K^{abcd}$ is $2K_{abcd}K^{ac}K^{bd}$ which also vanishes because of the Cayley-Hamilton theorem. Reason for this is that $-1/3K^a{}_{bcd}K^{bd}$ in matrix form (with suppressed indices) gives characteristic polynomial of $K$ for the $K$ its argument
\begin{equation}
K^3-K^2\text{tr} K +K \frac{1}{2}\left[\left(\text{tr} K\right)^2-\text{tr} K^2\right]-\text{Id}\frac{1}{6}\left[\left(\text{tr} K\right)^3-3\text{tr} K\text{tr} K^2+2\text{tr} K^3\right].
\end{equation}

\subsection{Variations}
 
Variations of the $V[\vec{\lambda}]$ are
\begin{align}
\delta_h V[\vec{\lambda}]&=\nonumber\\
=\int_\Sigma&\  -\pounds_{\vec{\lambda}}\Pi_h^{ab}\delta h_{ab}+\int_{\partial \Sigma}\star\left(\lambda^c\Pi_h^{ab}-2\Pi_h^{c(a}\lambda^{b)}\right)\delta h_{ab},\label{dhdiffbound}\\
\delta_{\Pi_h}V[\vec{\lambda}]&=\nonumber\\
=\int_\Sigma&\ \pounds_{\vec{\lambda}}h_{ab}\delta\Pi_h^{ab}-2\int_{\partial \Sigma}\star\lambda^ch_{bc}\delta\Pi_h^{ab},\label{dPihdiffbound}\\
\delta_K V[\vec{\lambda}]&=\nonumber\\
=\int_\Sigma&\  -\pounds_{\vec{\lambda}}\Pi_K^{ab}\delta K_{ab}+\int_{\partial \Sigma}\star\left(\lambda^c\Pi_K^{ab}-2\Pi_K^{ca}\lambda^{b}\right)\delta K_{ab},\label{dKdiffbound}\\
\delta_{\Pi_K}V[\vec{\lambda}]&=\nonumber\\
=\int_\Sigma&\ \pounds_{\vec{\lambda}}K_{ab}\delta\Pi_K^{ab}-2\int_{\partial \Sigma}\star\lambda^cK_{bc}\delta\Pi_K^{ab}.\label{dPiKdiffbound}
\end{align}
These variations lead to the relation
\begin{align}
\left\{\Phi,V[\vec{\lambda}]\right\}&=\int_\Sigma\pounds_{\vec{\lambda}}h_{ab}\frac{\delta}{\delta h_{ab}}\Phi-\pounds_{\vec{\lambda}}\Pi_h^{ab}\frac{\delta}{\delta \Pi_h^{ab}}\Phi\cdots=\nonumber\\
&=\Phi(h+\delta_\lambda h,\Pi_h+\delta_\lambda\Pi_h,\cdots)-\Phi(h,\Pi_h,\cdots),
\end{align}
proving that $V[\vec{\lambda}]$ is a generator of spatial diffeomorhpisms on the phase space with constraints.
Variations of $H[\lambda]$ are
\begin{align}
\delta_h H_0[\lambda]&=\nonumber\\
=\int_\Sigma &\lambda\left\{ \omega_h^{-1}\left(\frac{1}{4}\Pi_K\cdot \Pi_Kh^{ab}-\Pi_K^{ac}\Pi_{K\ c}^b\right)-\Pi_K\cdot K K^{ab}+\right.\nonumber\\
&\ +D_cD^{(b}\Pi_K^{a)c}-\frac{1}{2}h^{ab}D_cD_d\Pi_K^{cd}-\frac{1}{2}D^2\Pi_K^{ab}+\nonumber\\
&\ +2\omega_h\left[-\frac{1}{4}B\cdot Bh^{ab} +B^{acd}B_{\ cd}^b+\frac{1}{2}B^{cda}B_{cd}^{\ \ b}+\right.\nonumber\\
&\ \quad\quad\quad  +B^{d(ab)}D_dK-B^{d(ab)}D_cK_d^{\ c}+\nonumber\\
&\ \quad\left.\left.\quad\quad -D_c\left(B^{d(ab)}K_d^{\ c}+B^{cd(a}K^{b)}_{\ d}+B^{(a\vert dc \vert}K^{b)}_{\ d} \right)\right]\right\}\delta h_{ab}+\nonumber\\
&+D_c\lambda\left[2D_d\Pi_K^{d(a}h^{b)c}+ D^{(b}\Pi_K^{a)c}-\frac{3}{2}D^c\Pi_K^{ab}-D_d\Pi_K^{cd}h^{ab}+\right.\nonumber\\
&\quad \left.\ -2\omega_h\left(B^{d(ab)}K_d^{\ c}+B^{cd(a}K^{b)}_{\ d}+B^{(a\vert dc \vert}K^{b)}_{\ d} \right)\right]\delta h_{ab}\nonumber\\
&+D_cD_d\lambda\left[2\Pi_K^{(d\vert (a}h^{b)\vert d)}-\Pi_K^{ab}h^{cd}-\frac{1}{2}\Pi_K^{cd}h ^{ab}\right]\delta h_{ab}+\nonumber\\
\int_{\partial\Sigma} &\star\left[2\lambda\omega_h\left(B^{cd(a}K^{b)}_{\ d}+B^{(a\vert dc}K^{b)}_{\ d}+B^{d(ab)}K_d^{\ c}\right)\delta h_{ab}\right.+\nonumber\\
&+\lambda\left(2\delta C^c_{ed}\Pi_K^{ed}-\delta C^e_{de}\Pi_K^{cd}\right)+\nonumber\\
&+\lambda\left(-D^a\Pi_K^{ec}+\frac{1}{2}D^c\Pi_K^{ea}+\frac{1}{2}D_d\Pi_K^{cd}h^{ea}\right)\delta h_{ea}+\nonumber\\
&+ \left.\left(-2D^a\lambda\Pi_K^{ec}+D^c\lambda\Pi_K^{ea}+\frac{1}{2}D_d\lambda\Pi_K^{cd}h^{ea}\right)\delta h_{ea}\right],\label{dhH0bound}
\end{align}
\newpage
\begin{align}
\delta_KH_0[\lambda]&=\nonumber\\
=\int_\Sigma &\lambda\left(\Pi_K^{ab}K+\Pi_K\cdot K h^{ab}+2\Pi_h^{ab}+4\omega_hD_cB^{cab}\right)\delta K_{ab}+\nonumber\\
&D_c\lambda4\omega_hB^{cab}\delta K_{ab}+\nonumber\\
\int_{\partial\Sigma} &\lambda\star4\omega_hB^{cab}\delta K_{ab}\label{dKH0bound}\\
\delta_{\Pi_h}H_0[\lambda]&=\int_\Sigma \lambda2K_{ab}\delta \Pi_h^{ab}\\
\delta_{\Pi_K}H_0[\lambda] 
=\int_\Sigma &\lambda \left(-\omega_h^{-1}\Pi^K_{ab}+R_{ab}+K_{ab}K\right)\delta\Pi_K^{ab}\nonumber\\
&D_aD_b\lambda\delta\Pi_K^{ab}+\nonumber\\
\int_{\partial\Sigma}&\star\left(\lambda D_b\delta\Pi_K^{ab}-D_b\lambda \delta\Pi_K^{ab}\right),\label{dPiKH0bound}
\end{align}
for $C^a_{bc}$ difference between Levi-Civita connections.

\section{Appendix: Classification}

Here we provide the example for the partial differential equations (PDEs) that lead to  $\gamma_{ij}^{(1)}$ matrix that conserves respectively, 
translations
\begin{align}
\begin{array}{rcl}
 0&=&\text{$\gamma_{11} $}^{(1,0,0)}(t,x,y)\\0&=&
 \text{$\gamma_{11} $}^{(1,0,0)}(t,x,y)-\text{$\gamma_{22} $}^{(1,0,0)}(t,x,y)\\0&=&
 \text{$\gamma_{12} $}^{(1,0,0)}(t,x,y)\\0&=&
 \text{$\gamma_{13} $}^{(1,0,0)}(t,x,y)\\0&=&
 \text{$\gamma_{22} $}^{(1,0,0)}(t,x,y)\\0&=&
 \text{$\gamma_{23} $}^{(1,0,0)}(t,x,y)
\end{array}\label{transt},
\end{align}
Lorentz rotations in the $y$ direction 
\begin{align}
\begin{array}{rcl}
0&=& t \text{$\gamma_{11}$}^{(0,1,0)}(t,x,y)+x \text{$\gamma_{11} $}^{(1,0,0)}(t,x,y)-t \text{$\gamma_{22} $}^{(0,1,0)}(t,x,y)\\&&-x \text{$\gamma_{22} $}^{(1,0,0)}(t,x,y)\\0&=&
 t \text{$\gamma_{11} $}^{(0,1,0)}(t,x,y)+x \text{$\gamma_{11} $}^{(1,0,0)}(t,x,y)+2 \text{$\gamma_{12} $}(t,x,y)\\0&=&
 \text{$\gamma_{11} $}(t,x,y)+t \text{$\gamma_{12} $}^{(0,1,0)}(t,x,y)+x \text{$\gamma_{12} $}^{(1,0,0)}(t,x,y)+\text{$\gamma_{22} $}(t,x,y)\\0&=&
 t \text{$\gamma_{13} $}^{(0,1,0)}(t,x,y)+x \text{$\gamma_{13} $}^{(1,0,0)}(t,x,y)+\text{$\gamma_{23} $}(t,x,y)\\0&=&
 2 \text{$\gamma_{12} $}(t,x,y)+t \text{$\gamma_{22} $}^{(0,1,0)}(t,x,y)+x \text{$\gamma_{22} $}^{(1,0,0)}(t,x,y)\\0&=&
 \text{$\gamma_{13} $}(t,x,y)+t \text{$\gamma_{23} $}^{(0,1,0)}(t,x,y)+x \text{$\gamma_{23} $}^{(1,0,0)}(t,x,y)
\end{array},
\end{align}
dilatations 
\begin{align}
\begin{array}{rcl}
0&=& y \text{$\gamma_{11} $}^{(0,0,1)}(t,x,y)+x \text{$\gamma_{11} $}^{(0,1,0)}(t,x,y)+t \text{$\gamma_{11} $}^{(1,0,0)}(t,x,y)+\text{$\gamma_{11} $}(t,x,y)\\0&=&
 y \text{$\gamma_{11} $}^{(0,0,1)}(t,x,y)+x \text{$\gamma_{11} $}^{(0,1,0)}(t,x,y)+t \text{$\gamma_{11} $}^{(1,0,0)}(t,x,y)+\text{$\gamma_{11} $}(t,x,y)\\&&-y \text{$\gamma_{22} $}^{(0,0,1)}(t,x,y)- x \text{$\gamma_{22} $}^{(0,1,0)}(t,x,y)-t \text{$\gamma_{22} $}^{(1,0,0)}(t,x,y)-\text{$\gamma_{22} $}(t,x,y)\\0&=&
 y \text{$\gamma_{12} $}^{(0,0,1)}(t,x,y)+x \text{$\gamma_{12} $}^{(0,1,0)}(t,x,y)+t \text{$\gamma_{12} $}^{(1,0,0)}(t,x,y)+\text{$\gamma_{12} $}(t,x,y)\\0&=&
 y \text{$\gamma_{13} $}^{(0,0,1)}(t,x,y)+x \text{$\gamma_{13} $}^{(0,1,0)}(t,x,y)+t \text{$\gamma_{13} $}^{(1,0,0)}(t,x,y)+\text{$\gamma_{13} $}(t,x,y)\\0&=&
 y \text{$\gamma_{22} $}^{(0,0,1)}(t,x,y)+x \text{$\gamma_{22} $}^{(0,1,0)}(t,x,y)+t \text{$\gamma_{22} $}^{(1,0,0)}(t,x,y)+\text{$\gamma_{22} $}(t,x,y)\\0&=&
 y \text{$\gamma_{23} $}^{(0,0,1)}(t,x,y)+x \text{$\gamma_{23} $}^{(0,1,0)}(t,x,y)+t \text{$\gamma_{23} $}^{(1,0,0)}(t,x,y)+\text{$\gamma_{23} $}(t,x,y)
\end{array}
\end{align}
In order to compute the $\gamma_{ij}^{(1)}$ matrix that conserves three KVs of rotation, one has to solve the PDEs:
\begin{align}
\begin{array}{rcl}
0&=& x \text{$\gamma_{11}$}^{(0,0,1)}(t,x,y)-y \text{$\gamma_{11}$}^{(0,1,0)}(t,x,y)\\0&=&
 x \text{$\gamma_{12}$}^{(0,0,1)}(t,x,y)+\text{$\gamma_{13}$}(t,x,y)-y \text{$\gamma_{12}$}^{(0,1,0)}(t,x,y) \\0&=&
 x \text{$\gamma_{22}$}^{(0,0,1)}(t,x,y)+2 \text{$\gamma_{23}$}(t,x,y)-y \text{$\gamma_{22}$}^{(0,1,0)}(t,x,y) \\0&=&
 \text{$\gamma_{11}$}(t,x,y)+x \text{$\gamma_{23}$}^{(0,0,1)}(t,x,y)-2 \text{$\gamma_{22}$}(t,x,y)-y \text{$\gamma_{23}$}^{(0,1,0)}(t,x,y) \\0&=&
 y \text{$\gamma_{11}$}^{(0,1,0)}(t,x,y)+x \text{$\gamma_{22}$}^{(0,0,1)}(t,x,y)+2 \text{$\gamma_{23}$}(t,x,y)\\ &&-x \text{$\gamma_{11}$}^{(0,0,1)}(t,x,y)-y \text{$\gamma_{22}$}^{(0,1,0)}(t,x,y)\\0&=&
 \text{$\gamma_{12}$}(t,x,y)+y \text{$\gamma_{13}$}^{(0,1,0)}(t,x,y)-x \text{$\gamma_{13}$}^{(0,0,1)}(t,x,y) \\0&=&
 t \text{$\gamma_{11}$}^{(0,1,0)}(t,x,y)+x \text{$\gamma_{11}$}^{(1,0,0)}(t,x,y)-t \text{$\gamma_{22}$}^{(0,1,0)}(t,x,y)-x \text{$\gamma_{22}$}^{(1,0,0)}(t,x,y)\\0&=&
 t \text{$\gamma_{11}$}^{(0,1,0)}(t,x,y)+x \text{$\gamma_{11}$}^{(1,0,0)}(t,x,y)+2 \text{$\gamma_{12}$}(t,x,y)\\0&=&
 t \text{$\gamma_{11}$}^{(0,0,1)}(t,x,y)+y \text{$\gamma_{11}$}^{(1,0,0)}(t,x,y)+2 \text{$\gamma_{13}$}(t,x,y)\\0&=&
 t \text{$\gamma_{11}$}^{(0,0,1)}(t,x,y)+y \text{$\gamma_{11}$}^{(1,0,0)}(t,x,y)+2 \text{$\gamma_{13}$}(t,x,y)\\&&-t \text{$\gamma_{22}$}^{(0,0,1)}(t,x,y)-y \text{$\gamma_{22}$}^{(1,0,0)}(t,x,y) \\0&=&
 \text{$\gamma_{11}$}(t,x,y)+t \text{$\gamma_{12}$}^{(0,1,0)}(t,x,y)+x \text{$\gamma_{12}$}^{(1,0,0)}(t,x,y)+\text{$\gamma_{22}$}(t,x,y)\\0&=&
 t \text{$\gamma_{12}$}^{(0,0,1)}(t,x,y)+y \text{$\gamma_{12}$}^{(1,0,0)}(t,x,y)+\text{$\gamma_{23}$}(t,x,y)\\0&=&
 t \text{$\gamma_{13}$}^{(0,1,0)}(t,x,y)+x \text{$\gamma_{13}$}^{(1,0,0)}(t,x,y)+\text{$\gamma_{23}$}(t,x,y)\\0&=&
 2 \text{$\gamma_{11}$}(t,x,y)+t \text{$\gamma_{13}$}^{(0,0,1)}(t,x,y)+y \text{$\gamma_{13}$}^{(1,0,0)}(t,x,y)-\text{$\gamma_{22}$}(t,x,y) \\0&=&
 2 \text{$\gamma_{12}$}(t,x,y)+t \text{$\gamma_{22}$}^{(0,1,0)}(t,x,y)+x \text{$\gamma_{22}$}^{(1,0,0)}(t,x,y)\\0&=&
 t \text{$\gamma_{22}$}^{(0,0,1)}(t,x,y)+y \text{$\gamma_{22}$}^{(1,0,0)}(t,x,y)\\0&=&
 \text{$\gamma_{13}$}(t,x,y)+t \text{$\gamma_{23}$}^{(0,1,0)}(t,x,y)+x \text{$\gamma_{23}$}^{(1,0,0)}(t,x,y)\\0&=&
 \text{$\gamma_{12}$}(t,x,y)+t \text{$\gamma_{23}$}^{(0,0,1)}(t,x,y)+y \text{$\gamma_{23}$}^{(1,0,0)}(t,x,y)
\end{array}.
\end{align}
PDEs that give $\gamma_{ij}^{(1)}$ matrix are
\begin{align}
\begin{array}{l}
0= \left(t^2+x^2-y^2\right) \text{$\gamma_{11}$}^{(0,1,0)}(t,x,y)+2 x y \text{$\gamma_{11}$}^{(0,0,1)}(t,x,y)\\+2 t x \text{$\gamma_{11}$}^{(1,0,0)}(t,x,y)+2 x \text{$\gamma_{11}$}(t,x,y)+4 t \text{$\gamma_{12}$}(t,x,y)
 \end{array}
 \end{align}
 \begin{align}
 \begin{array}{l}
0= \frac{1}{2} \left(t^2-x^2+y^2\right) \text{$\gamma_{11}$}^{(0,0,1)}(t,x,y)+x y \text{$\gamma_{11}$}^{(0,1,0)}(t,x,y)\\+t y \text{$\gamma_{11}$}^{(1,0,0)}(t,x,y)+y \text{$\gamma_{11}$}(t,x,y)+2 t \text{$\gamma_{13}$}(t,x,y)
  \end{array}
 \end{align}
 \begin{align}
 \begin{array}{l}
 0=\left(t^2+x^2+y^2\right) \text{$\gamma_{11}$}^{(1,0,0)}(t,x,y)+2 t \big(y \text{$\gamma_{11}$}^{(0,0,1)}(t,x,y)\\ +x \text{$\gamma_{11}$}^{(0,1,0)}(t,x,y)+\text{$\gamma_{11}$}(t,x,y)\big)+4 (x \text{$\gamma_{12}$}(t,x,y)+y \text{$\gamma_{13}$}(t,x,y))
  \end{array}
 \end{align}
 \begin{align}
 \begin{array}{l}
0= \frac{1}{2} \left(t^2+x^2-y^2\right) \text{$\gamma_{12}$}^{(0,1,0)}(t,x,y)+t \text{$\gamma_{11}$}(t,x,y)+x y \text{$\gamma_{12}$}^{(0,0,1)}(t,x,y)\\+t x \text{$\gamma_{12}$}^{(1,0,0)}(t,x,y)+x \text{$\gamma_{12}$}(t,x,y)+y \text{$\gamma_{13}$}(t,x,y)+t \text{$\gamma_{22}$}(t,x,y)
  \end{array}
 \end{align}
 \begin{align}
 \begin{array}{l}
0= \frac{1}{2} \big(t^2-x^2+y^2\big) \text{$\gamma_{12}$}^{(0,0,1)}(t,x,y)+x y \text{$\gamma_{12}$}^{(0,1,0)}(t,x,y)\\+t y \text{$\gamma_{12}$}^{(1,0,0)}(t,x,y)+y \text{$\gamma_{12}$}(t,x,y)+t \text{$\gamma_{23}$}(t,x,y)-x \text{$\gamma_{13}$}(t,x,y)  \end{array}
 \end{align}
 \begin{align}
 \begin{array}{l}
 0=\left(t^2+x^2+y^2\right) \text{$\gamma_{12}$}^{(1,0,0)}(t,x,y)+2 \bigg(x \big(\text{$\gamma_{11}$}(t,x,y)+t \text{$\gamma_{12}$}^{(0,1,0)}(t,x,y)\\+\text{$\gamma_{22}$}(t,x,y)\big)+y \left(t \text{$\gamma_{12}$}^{(0,0,1)}(t,x,y)+\text{$\gamma_{23}$}(t,x,y)\right)+t \text{$\gamma_{12}$}(t,x,y)\bigg) 
  \end{array}
 \end{align}
 \begin{align}
 \begin{array}{l}
 0=\frac{1}{2} \left(t^2+x^2-y^2\right) \text{$\gamma_{13}$}^{(0,1,0)}(t,x,y)+x y \text{$\gamma_{13}$}^{(0,0,1)}(t,x,y)\\+t x \text{$\gamma_{13}$}^{(1,0,0)}(t,x,y)+x \text{$\gamma_{13}$}(t,x,y)+t \text{$\gamma_{23}$}(t,x,y)-y \text{$\gamma_{12}$}(t,x,y) 
  \end{array}
 \end{align}
 \begin{align}
 \begin{array}{l}
 0=\left(t^2+x^2+y^2\right) \text{$\gamma_{13}$}^{(1,0,0)}(t,x,y)+4 y \text{$\gamma_{11}$}(t,x,y)+2 t \big(y \text{$\gamma_{13}$}^{(0,0,1)}(t,x,y)\\+x \text{$\gamma_{13}$}^{(0,1,0)}(t,x,y)+\text{$\gamma_{13}$}(t,x,y)\big)+2 x \text{$\gamma_{23}$}(t,x,y)-2 y \text{$\gamma_{22}$}(t,x,y) 
  \end{array}
 \end{align}
 \begin{align}
 \begin{array}{l}
 0=\big(t^2-x^2+y^2\big) \text{$\gamma_{13}$}^{(0,0,1)}(t,x,y)+4 t \text{$\gamma_{11}$}(t,x,y)+2 (x \text{$\gamma_{12}$}(t,x,y)\\+y \text{$\gamma_{13}$}(t,x,y))+2 y \big(x \text{$\gamma_{13}$}^{(0,1,0)}(t,x,y)+t \text{$\gamma_{13}$}^{(1,0,0)}(t,x,y)\big)-2 t \text{$\gamma_{22}$}(t,x,y) 
  \end{array}
 \end{align}
 \begin{align}
 \begin{array}{l}
0= \frac{1}{2} \big(t^2+x^2-y^2\big) \big(\text{$\gamma_{11}$}^{(0,1,0)}(t,x,y)-\text{$\gamma_{22}$}^{(0,1,0)}(t,x,y)\big)\\+x y \big(\text{$\gamma_{11}$}^{(0,0,1)}(t,x,y)-\text{$\gamma_{22}$}^{(0,0,1)}(t,x,y)\big)+t x \big(\text{$\gamma_{11}$}^{(1,0,0)}(t,x,y)\\-\text{$\gamma_{22}$}^{(1,0,0)}(t,x,y)\big)+x (\text{$\gamma_{11}$}(t,x,y)-\text{$\gamma_{22}$}(t,x,y))-2 y \text{$\gamma_{23}$}(t,x,y) 
  \end{array}
 \end{align}
 \begin{align}
 \begin{array}{l}
0= \frac{1}{2} \big(t^2-x^2+y^2\big) \big(\text{$\gamma_{11}$}^{(0,0,1)}(t,x,y)-\text{$\gamma_{22}$}^{(0,0,1)}(t,x,y)\big)\\+x y \big(\text{$\gamma_{11}$}^{(0,1,0)}(t,x,y)-\text{$\gamma_{22}$}^{(0,1,0)}(t,x,y)\big)+t y \big(\text{$\gamma_{11}$}^{(1,0,0)}(t,x,y)\\-\text{$\gamma_{22}$}^{(1,0,0)}(t,x,y)\big)+y (\text{$\gamma_{11}$}(t,x,y)-\text{$\gamma_{22}$}(t,x,y))+2 t \text{$\gamma_{13}$}(t,x,y)\\+2 x \text{$\gamma_{23}$}(t,x,y) 
  \end{array}
 \end{align}
 \begin{align}
 \begin{array}{l}
0= \frac{1}{2} \big(t^2+x^2+y^2\big) \big(\text{$\gamma_{11}$}^{(1,0,0)}(t,x,y)-\text{$\gamma_{22}$}^{(1,0,0)}(t,x,y)\big)\\+t y \big(\text{$\gamma_{11}$}^{(0,0,1)}(t,x,y)-\text{$\gamma_{22}$}^{(0,0,1)}(t,x,y)\big)+t x \big(\text{$\gamma_{11}$}^{(0,1,0)}(t,x,y)\\-\text{$\gamma_{22}$}^{(0,1,0)}(t,x,y)\big)+t (\text{$\gamma_{11}$}(t,x,y)-\text{$\gamma_{22}$}(t,x,y))+2 y \text{$\gamma_{13}$}(t,x,y)
  \end{array}
 \end{align}
 
 \begin{align}
 \begin{array}{l}
0= \frac{1}{2} \big(t^2+x^2-y^2\big) \text{$\gamma_{22}$}^{(0,1,0)}(t,x,y)+2 t \text{$\gamma_{12}$}(t,x,y)+x y \text{$\gamma_{22}$}^{(0,0,1)}(t,x,y)\\+t x \text{$\gamma_{22}$}^{(1,0,0)}(t,x,y)+x \text{$\gamma_{22}$}(t,x,y)+2 y \text{$\gamma_{23}$}(t,x,y)
  \end{array}
 \end{align}
 
  \begin{align}
0&= \frac{1}{2} \big(t^2-x^2+y^2\big) \gamma_{22}^{(0,0,1)}(t,x,y)+x y \gamma_{22}^{(0,1,0)}(t,x,y)\nonumber \\ &+t y \gamma_{22}^{(1,0,0)}(t,x,y)+y \gamma_{22}(t,x,y)-2 x \gamma_{23}(t,x,y) 
\\0&= \big(t^2+x^2+y^2\big) \gamma_{22}^{(1,0,0)}(t,x,y)+4 x \gamma_{12}(t,x,y)\nonumber \\&+2 t \big(y \gamma_{22}^{(0,0,1)}(t,x,y)+x \gamma_{22}^{(0,1,0)}(t,x,y)+\gamma_{22}(t,x,y)\big)  
 \end{align}
 
\begin{align}
 0&=\big(t^2+x^2-y^2\big) \gamma_{23}^{(0,1,0)}(t,x,y)+2 y \gamma_{11}(t,x,y)+2 t \gamma_{13}(t,x,y)\nonumber \\ &+2 \bigg(x \big(y \gamma_{23}^{(0,0,1)}(t,x,y) +\gamma_{23}(t,x,y)\big)-2 y \gamma_{22}(t,x,y)\bigg)+2 t x \gamma_{23}^{(1,0,0)}(t,x,y)
 \end{align}
 
  \begin{align}
 \begin{array}{l}
0= \frac{1}{2} \big(t^2-x^2+y^2\big) \text{$\gamma_{23}$}^{(0,0,1)}(t,x,y)+x (\text{$\gamma_{22}$}(t,x,y)-\text{$\gamma_{11}$}(t,x,y))\\+t \text{$\gamma_{12}$}(t,x,y)+x \text{$\gamma_{22}$}(t,x,y)+x y \text{$\gamma_{23}$}^{(0,1,0)}(t,x,y)\\+t y \text{$\gamma_{23}$}^{(1,0,0)}(t,x,y)+y \text{$\gamma_{23}$}(t,x,y)
  \end{array} \end{align}
 
 \begin{align}
 \begin{array}{l}
0= \big(t^2+x^2+y^2\big) \text{$\gamma_{23}$}^{(1,0,0)}(t,x,y)+2 y \text{$\gamma_{12}$}(t,x,y)+2 x \text{$\gamma_{13}$}(t,x,y)\\+2 t \big(y \text{$\gamma_{23}$}^{(0,0,1)}(t,x,y)+x \text{$\gamma_{23}$}^{(0,1,0)}(t,x,y)+\text{$\gamma_{23}$}(t,x,y)\big)
\end{array}
\end{align}

The set of equations that needs to be solved for the rotation KV to be conserved is
\begin{align}
 0&=y \text{$\gamma_{11}$}^{(0,1,0)}(t,x,y)-x \text{$\gamma_{11}$}^{(0,0,1)}(t,x,y)\nonumber \\0&=
 -x \text{$\gamma_{12}$}^{(0,0,1)}(t,x,y)+y \text{$\gamma_{12}$}^{(0,1,0)}(t,x,y)-\text{$\gamma_{13}$}(t,x,y)\nonumber \\0&=
 \text{$\gamma_{12}$}(t,x,y)-x \text{$\gamma_{13}$}^{(0,0,1)}(t,x,y)+y \text{$\gamma_{13}$}^{(0,1,0)}(t,x,y)\nonumber \\0&=
 -x \big(\text{$\gamma_{11}$}^{(0,0,1)}(t,x,y)-\text{$\gamma_{22}$}^{(0,0,1)}(t,x,y)\big)+y \big(\text{$\gamma_{11}$}^{(0,1,0)}(t,x,y)\nonumber\\&-\text{$\gamma_{22}$}^{(0,1,0)}(t,x,y)\big)+2 \text{$\gamma_{23}$}(t,x,y)\nonumber\\0&=
 -x \text{$\gamma_{22}$}^{(0,0,1)}(t,x,y)+y \text{$\gamma_{22}$}^{(0,1,0)}(t,x,y)-2 \text{$\gamma_{23}$}(t,x,y)\nonumber\\0&=
 -\text{$\gamma_{11}$}(t,x,y)+2 \text{$\gamma_{22}$}(t,x,y)-x \text{$\gamma_{23}$}^{(0,0,1)}(t,x,y)+y \text{$\gamma_{23}$}^{(0,1,0)}(t,x,y)
\label{rotxy}
\end{align}
 \subsection{ Classification According to the Generators of the Conformal Group}
 Table with the subalgebras of conformal algebra that are not realised in the form of $\gam$, with algebra it implies. We omit combinations of the KVs  of 3 SCTs and n Ts, or 3 Ts and n SCTs that imply the full conformal algebra.
  \begin{center}
\begin{tabular}{ |l | p{7.7 cm} | p{1 cm} | c|}
\hline
1 T + 2 R & $\nexists$ &  $\nexists$ & 3\\

1 T + 3 R & $\nexists$ & $\nexists$  &4\\
1 T + 2 R + D & $\nexists$  & $\nexists$ &4\\
1 T + 3 R + D & $\nexists$ & $\nexists$&3\\ 

\hline
2 R & $\nexists$ $\Rightarrow$ 3 Rs  &  & 2 \\
2 R+D & $\nexists$ $\Rightarrow$ 3 Rs +D &      & 3 \\ 
1 R + 1 SCT  & $\nexists$, $ [\xi_{l}^{sct},L_{ij}]=\eta_{li}\xi_j^{sct}-\eta_{lj}\xi_{i}^{sct},$ $\Rightarrow$ $\xi^{sct}_i,\xi^{sct}_j,L_{ij}$ for $l\neq i \neq j$  &  & 2 \\ 
2 R+1 SCT & $\nexists$, (\ref{ca1},\ref{ca2}) $\Rightarrow$ 3 Rs+3 SCTs &    & 3 \\
2 R+2 SCT & $\nexists$,  (\ref{ca1},\ref{ca2}) $\Rightarrow$ 3 Rs+3 SCTs &    & 3 \\
2 R+3 SCT & $\nexists$, (\ref{ca1},\ref{ca2}) $\Rightarrow$ 3 Rs+3 SCTs &    & 3 \\
3 R+1 SCT & $\nexists$, (\ref{ca1},\ref{ca2}) $\Rightarrow$ 3 Rs+3 SCTs  &   & 3 \\
3 R+2 SCT & $\nexists$, (\ref{ca1},\ref{ca2}) $\Rightarrow$ 3 Rs+3 SCTs  &    & 3 \\ 

1 T + 1 R+1 SCT & $\nexists$, the example from the text shows $\Rightarrow$ subalgebra of 6 CKVs $\xi_i^t,\xi_j^t,\xi_i^{sct},\xi_j^{sct},L_{ij},D$  is required   &   $\nexists$    & 2 \\

1 T + 2 R+1 SCT & $\nexists$ $\Rightarrow$ conformal algebra &  $\nexists$&3\\
1 T + 3 R+1 SCT & $\nexists$, $\Rightarrow$ full conformal algebra&$\nexists$ &4\\ 
1 T + 2 R + D+1 SCT & $\nexists$, $[D,\xi^{act}_j]=\xi^{act}_j$, $[D,\xi^t_{j}]=-\xi^t_j$, $[D,L_{ij}]=0$ $\Rightarrow$ analysis of the existence of subalgebra is equal to the analysis (1 T + 2 R +1 SCT) &   &4\\
1 T + 3 R + D+1 SCT &  $\nexists$, because of the commutation of the remaining generators with dilatations, the existence of subalgebra is equal to the analysis (1 T + 3 R +1 SCT) &  &3 \\ \hline
1 T + 1 R+2 SCT & $\nexists$, from (\ref{ca1},\ref{ca2}) required full conformal algebra &   & 4 \\
1 T + 2 R+2 SCT & $\nexists$, (\ref{ca1},\ref{ca2}) $\Rightarrow$ full conformal algebra &   & 5\\
1 T + 3 R+2 SCT  & $\nexists$, (\ref{ca1},\ref{ca2}) $\Rightarrow$ full conformal algebra & &6\\

1 T + D+2 SCT & $\nexists$, leads to subalgebra $\xi_i^t,\xi_j^t,\xi_i^{sct},\xi_j^{sct},L_{ij},D$ (with $i\neq j$), commutation relation with D close, leaving $[\xi^t_i,\xi^{sct}_i]=2\eta_{ij}\xi^d$, $[\xi^t_i,\xi^{sct}_j]=2L_{ij}$ $\Rightarrow$ requires $L_{ij}$; $[\xi^t,L_{ij}]=\xi^t_j$ $\Rightarrow$ requires second $\xi^t_j$.  &  & 2 \\
1 T + 1 R + D+2 SCT & $\nexists$, (\ref{ca1},\ref{ca2}) $\Rightarrow$ full conformal algebra & &6\\
1 T + 2 R + D+2 SCT & $\nexists$, like 1 T+2 R+2SCTs leads to full conformal algebra &   &6\\
1 T + 3 R + D+2 SCT & $\nexists$, (\ref{ca1},\ref{ca2}) $\Rightarrow$ full confromal algebra:  &  & 7\\ \hline

 \hline
\end{tabular}
\end{center}

 \begin{center}
\hspace{-2cm}\begin{tabular}{ |l | p{7.5 cm} | p{1 cm} | c|}
\hline

\hline

2 T + 2 R & $\nexists$ & \text{    } &4\\
2 T + 3 R & $\nexists$ & $\nexists$ &5\\
2 T + 2 R +D & $\nexists$ & \text{   } &6\\
2 T + 3 R +D &$\nexists$ & $\nexists$ & 7\\
2 T + 1 R + 1 SCT & $\nexists$, (\ref{ca1},\ref{ca2}) $\Rightarrow$ $\xi_i^t,\xi_j^t,\xi_i^{sct},\xi_j^{sct},L_{ij},D$  &   &4\\
2 T + 2 R + 1 SCT & $\nexists$, (\ref{ca1},\ref{ca2}) $\Rightarrow$ full conformal algebra & &5\\
2 T + 3 R +1 SCT  & $\nexists$, (\ref{ca1},\ref{ca2}) $\Rightarrow$ full conformal algebra & &6\\
2 T + 1 R + D + 1 SCT & $\nexists$, (\ref{ca1},\ref{ca2}) $\Rightarrow$ $\xi_i^t,\xi_j^t,\xi_i^{sct},\xi_j^{sct},L_{ij},D$  & &6\\
2 T + 2 R + D + 1 SCT & $\nexists$, (\ref{ca1},\ref{ca2}) $\Rightarrow$ full conformal algebra & &6\\
2 T + 3 R + D + 1 SCT & $\nexists$, (\ref{ca1},\ref{ca2}) $\Rightarrow$ full conformal algebra & & 7\\
2 T + D + 1 SCT & $\nexists$, (\ref{ca1},\ref{ca2}) $\Rightarrow$ $\xi_i^t,\xi_j^t,\xi_i^{sct},\xi_j^{sct},L_{ij},D$  & &4 \\
2 T + 1 R + 2 SCT & $\nexists$, (\ref{ca1},\ref{ca2}) $\Rightarrow$ $\xi_i^t,\xi_j^t,\xi_i^{sct},\xi_j^{sct},L_{ij},D$  & &5\\
2 T + 2 R + 2 SCT & $\nexists$, (\ref{ca1},\ref{ca2}) $\Rightarrow$ full conformal algebra& &6\\
2 T + 3 R +2 SCT  & $\nexists$, (\ref{ca1},\ref{ca2}) $\Rightarrow$ full conformal algebra & &7\\
2 T + 2 R + D + 2 SCT & $\nexists$, (\ref{ca1},\ref{ca2}) $\Rightarrow$ full conformal algebra & &6\\
2 T + 3 R + D + 2 SCT & $\nexists$, (\ref{ca1},\ref{ca2}) $\Rightarrow$ full conformal algebra & & 8\\
2 T + D + 2 SCT & $\nexists$, (\ref{ca1},\ref{ca2}) $\Rightarrow$ $\xi_i^t,\xi_j^t,\xi_i^{sct},\xi_j^{sct},L_{ij},D$  & &5\\ \hline\hline
3 T + 2 R & $\nexists$ & \text{ } &5\\
3 T + 1 R + D & $\nexists$ & $\nexists$ &5\\
3 T +2 R+D & $\nexists$ & \text{   } &6\\
1 T + 1 SCT & $\nexists$ $\Rightarrow$ $\xi^i_i,\xi^{sct}_i,D$ or $\xi^i_i,\xi^{sct}_i,\xi^{t}_j,\xi^{sct}_j,L_{ij},D$  for $i\neq j$ &  $\nexists$ &2 \\
1 T + 2 SCT & $\nexists$ $\Rightarrow$ $\xi^i_i,\xi^{sct}_i,\xi^{t}_j,\xi^{sct}_j,L_{ij},D$   & $\nexists$ &3\\
 2 T +1 SCT  & $\nexists$, (\ref{ca1},\ref{ca2}) $\Rightarrow$ full conformal algebra & &3\\
 2 T + 2 SCT &  $\nexists$ &  $\nexists$ &4 \\
n T +m SCT & $\nexists$ \text{ needs R or D } & \text{      } &n+m\\  \hline
1 R+D+1 SCT &$\nexists$, $ [\xi_{l}^{sct},L_{ij}]=\eta_{li}\xi_j^{sct}-\eta_{lj}\xi_{i}^{sct},$ $\Rightarrow$ $\xi^{sct}_i,\xi^{sct}_j,L_{ij}$ for $l\neq i \neq j$ && 3\\
2 R+D+1 SCT & $\nexists$ $\Rightarrow$ 3 Rs+D+3 SCTs && 3\\ 
2 R+D+2 SCT & $\nexists$ $\Rightarrow$ 3 Rs+D+3 SCTs && 4\\
2 R+D+3 SCT & $\nexists$ $\Rightarrow$ 3 Rs+D+3 SCTs && 5\\ \hline
3 R+D+1 SCT & $\nexists$ $\Rightarrow$ 3 Rs+D+3 SCTs && 3\\
3 R+D+2 SCT & $\nexists$ $\Rightarrow$ 3 Rs+D+3 SCTs && 4\\
 \hline
\end{tabular}
\end{center}

\subsubsection{Patera et al. Classification}

The table with non-realized $\gam$ matrices. 

 \begin{center}
\begin{tabular}{|c|c|}\hline
 \hline
  \multicolumn{2}{|c|}{Subalgebras that are not realised} \\
  \hline
Patera name&generators   \\ \hline\hline
$ a_{7,1}$ &$ F,K_1,K_2,L_3,P_0,P_1,P_2$ \\
$a_{6,1}$& $ K_1,K_2,L_3,P_0,P_1,P_2$\\
$a_{6,2} $& $F,K_2,K_1-L_3,P_0,P_1,P_2$ \\ \hline
$a_{5,1}$ & $K_2,L_3-K_1,P_0,P_1,P_2$\\
$a_{5,2}$ & $F-K_2,-K_1+L_3,P_0,P_1,P_2$\\
$a_{5,3}$ & $F+K_2,-K_1+L_3,P_0,P_1,P_2$\\
$a_{5,5}$ &$ F,L_3-K_1,P_0,P_1,P_2$\\
$a_{5,6}$ &$ F,K_2,P_0,P_1,P_2$\\
$a_{5,7}$ &$ F,L_3,P_0,P_1,P_2$\\
$a_{5,8}$ & $F,K_2,-K_1+L_3,P_0,P_1,-P_2$\\
$a_{4,2}=b_{4,7}$ & $P_0-P_2\oplus \left\{F-K_2,P_0+P_2,P_1\right\}$\\
$a_{4,5}=b_{4,8}$ &$ F+\epsilon(L_3-K_1),P_0,P_1,P_2$\\
$a_{4,7}=b_{4,10}$ &$ L_3-K_1,P_0+P_2,P_0-P_2,P_1$\\
$a_{4,9}$ &$ F,P_0,P_1,P_2$\\
$a_{4,16}$ &$ F+\frac{1}{2}K_2,-K_1+L_3,+\epsilon(P_0+P_2),P_0-P_2,P_1$\\
$a_{4,18}=b_{4,5}$ &$ F+K_2,-K_1+L_3,P_0-P_2,P_1$\\
\hline
 \end{tabular}
\end{center}

\subsubsection{Global Solutions}
The list of the polynomial invariants of a geon global solution reads 
\begin{align}
\begin{array}{ll}
R, r_1=\frac{1}{2}S^{a}_{b}S^{b}_{a} &  r_2=-\frac{1}{8}S^{a}_{b}S^{b}_{c}S^{c}_{a}   \\ w_1=\frac{1}{8}(C_{abcd}+i C*_{abcd})C^{abcd} & w_2=-\frac{1}{16}(C_{ab}^{cd}+iC*_{ab}^{cd})C_{cd}^{ef}C_{ef}^{ab} \\ m_1=\frac{1}{8}S^{ab}S^{cd}(C_{acdb}+iC*_{acdb}) &
 m_{2a}=\frac{1}{16}S^{bc}S_{ef}C_{abcd}C^{aefd} \\ m_{2b}=\frac{1}{16}S^{bc}S_{ef}C*_{abcd}C*^{aefd} & m_2=m_{2a}-m_{2b}+\frac{1}{8}iS^{bc}_{ef}C*_{abcd}C^{aefd} \\ m_3=m_{2a}+m_{2b} &
m_{4a}=-\frac{1}{32}S^{ag}S^{ef}S^{c}_{d}C_{ac}^{db}C_{befg} \\ m_{4b}=-\frac{1}{32}S^{ag}S^{ef}S^{c}_{d}C*_{ac}^{db}C*_{befg} &
 m_{4}=m_{4a}+m_{4b} \\ m_{5a}=\frac{1}{32}S^{cd}S^{ef}C^{aghb}C_{acdb}C_{gefh} & m_{5b}=\frac{1}{32}S^{cd}S^{ef}C^{aghb}C*_{acdb}C*_{gefh} \\
   m_{5c}=\frac{1}{32}S^{cd}S^{ef}C*^{aghb}C_{acdb}C_{gefh} & m_{5d}=\frac{1}{32}S^{cd}S^{ef}C*^{aghb}C*_{acdb}C*_{gefh} \\ 
  m_5=m_{5a}+m_{5b}+i(m_{5c}+m_{5d}) & m_{6}=\frac{1}{32}S_{a}^{e}S_{e}^{c}S_{b}^{f}S_{f}^{d}(C^{ab}_{cd}+iC*{ab}_{cd})\\  r_3=\frac{1}{16}S^{a}_{b}S^{b}_{c}S^{c}_{d}S^{d}_{a}&
\end{array}
\end{align}

 for  $S_{ab}=R_{ab}-\frac{1}{n}Rg_{ab}$.

\subsubsection{Examples of $\gamma^{(1)}_{ij}$ with $R\times S^2$ Boundary } 

In the following table there are examples of the $\gamma^{(1)}$ matrix realised with a particular subset of the Killing vectors (\ref{sphkv}), where we denote {\it number of KVs} with n.

\begin{center}
\begin{tabular}{|c|c|c|}\hline
KVs &Realization, $\gamma_{ij}^{(1)}$ & n of KVs\\ 
\hline
 $\begin{array}{cc}\xi^{\LO sph}_{7},& \xi^{\LO sph}_{8},\\ \xi^{\LO sph}_{9}, &\xi^{\LO sph}_{0}\end{array}$& $\left(
\begin{array}{ccc}
 \text{$\gamma_{11} $} & 0 & 0 \\
 0 & \frac{\text{$\gamma_{11} $}}{2} & 0 \\
 0 & 0 & \frac{1}{2} \text{$\gamma_{11} $} \sin ^2(\theta ) \\
\end{array}
\right)$ & 4 \\ 
 $\xi^{\LO sph}_{7}, \xi^{\LO sph}_{0}$& $\left(
\begin{array}{ccc}
 \text{$\gamma_{11} $} & 0 & 0 \\
 0 & \text{$\gamma_{22} $} & 0 \\
 0 & 0 & (\text{$\gamma_{11} $}-\text{$\gamma_{22}$}) \sin ^2(\theta ) \\
\end{array}
\right)$ & 2 \\ \hline
$\begin{array}{l} \xi^{\LO sph}_{7}, \xi^{\LO sph}_{8}, \\ \xi^{\LO sph}_{9} \end{array}$& $\left(
\begin{array}{ccc}
 2 \text{$\gamma_{22} $}(t) & 0 & 0 \\
 0 & \text{$\gamma_{22} $}(t) & 0 \\
 0 & 0 & \sin ^2(\theta ) \text{$\gamma_{22} $}(t) \\
\end{array}
\right)$ & 3 \\
$\xi^{\LO sph}_{7}, \xi^{\LO sph}_{6}$&$ \left(
\begin{array}{ccc}
 -c_1 \sec (t) & 0 & 0 \\
 0 & c_1 \sec (t) & 0 \\
 0 & 0 & -2 c_1 \sec (t) \sin ^2(\theta ) \\
\end{array}
\right)$ & 2 \\
$\xi^{\LO sph}_{7}, \xi^{\LO sph}_{3}$& $\left(
\begin{array}{ccc}
 -c_1 \csc (t) & 0 & 0 \\
 0 & c_1 \csc (t) & 0 \\
 0 & 0 & -2 c_1 \csc (t) \sin ^2(\theta ) \\
\end{array}
\right)$& 2 \\ \hline

$\begin{array}{c}\xi^{\LO sph}_{7}, \xi^{\LO sph}_{0},\\\xi^{\LO sph}_{6}, \xi^{\LO sph}_{3}\end{array}$&$ \left(
\begin{array}{ccc}
 c_1 \csc (\theta ) & 0 & 0 \\
 0 & -c_1 \csc (\theta ) & 0 \\
 0 & 0 & 2 c_1 \sin (\theta ) \\
\end{array}
\right)$ & 4 \\ \hline

$\xi^{\LO sph}_{0},$ &$ \left(
\begin{array}{ccc}
 \text{$\gamma_{11} $}(\phi ) & 0 & 0 \\
 0 & \text{$\gamma_{22} $}(\phi ) & 0 \\
 0 & 0 & \sin ^2(\theta ) (\text{$\gamma_{11} $}(\phi )-\text{$\gamma_{22} $}(\phi )) \\
\end{array}
\right)$& 1\\

$\xi^{\LO sph}_{8}, \xi^{\LO sph}_{0}$  &eq. (\ref{eq80})& 2 \\ 

$\xi^{\LO sph}_{9}, \xi^{\LO sph}_{0}$ &eq. (\ref{eq90})& 2\\

$\begin{array}{l} \xi^{\LO sph}_{0}, \xi^{\LO sph}_{6},\\ \xi^{\LO sph}_{3} \end{array}$ & $\left(
\begin{array}{ccc}
 -\csc (\theta ) c_1(\phi ) & 0 & 0 \\
 0 & \csc (\theta ) c_1(\phi ) & 0 \\
 0 & 0 & -2 \sin (\theta ) c_1(\phi ) \\
\end{array}
\right)$& 3\\  \hline

$\xi^{\LO sph}_{7},\xi^{\LO sph}_{6}$& eq. (\ref{sph1}) & 2\\ \hline

$\xi^{\LO sph}_{8}$&
eq. (\ref{eqn8})& 1 \\

$\xi^{\LO sph}_{9}$&eq.(\ref{sph2})& 1 \\ \hline

\end{tabular}
\end{center}

\begin{align}
\gamma_{11}^{(1)}&=
 2 c_1\left[-2 \cos (\phi ) \sin (\theta )\right]  \nonumber\\
\gamma_{22}^{(1)}&= c_1\left[-2 \cos (\phi ) \sin (\theta )\right]  \nonumber\\
 \gamma_{33}^{(1)}&= \sin ^2(\theta ) c_1\left[-2 \cos (\phi ) \sin (\theta )\right] 
\label{eq80}
\end{align}

\begin{equation}
\gamma_{ij}^{(1)}=\left(
\begin{array}{ccc}
 2 c_1\left[2 \sin (\theta ) \sin (\phi )\right] & 0 & 0 \\
 0 & c_1\left[2 \sin (\theta ) \sin (\phi )\right] & 0 \\
 0 & 0 & \sin ^2(\theta ) c_1\left[2 \sin (\theta ) \sin (\phi )\right] \\
\end{array}
\right)\label{eq90}
\end{equation}

\begin{align}
\gamma_{11}^{(1)}&=
 -\sec (t) c_1\left[2 \sec (t) \sin (\theta )\right] \nonumber \\
\gamma_{22}^{(1)}&= \sec (t) c_1\left[2 \sec (t) \sin (\theta )\right] \nonumber \\
\gamma_{33}^{(1)}&= -2 \sec (t) \sin ^2(\theta ) c_1\left[2 \sec (t) \sin (\theta )\right] 
 \label{sph1}
\end{align}

\begin{align}\gamma_{11}^{(1)}&= 2 c_1\left[t,-2 \cos (\phi ) \sin (\theta )\right] \nonumber \\
\gamma_{22}^{(1)}&= c_1\left[t,-2 \cos (\phi ) \sin (\theta )\right] \nonumber\\
\gamma_{33}^{(1)}&= \sin ^2(\theta ) c_1\left[t,-2 \cos (\phi ) \sin (\theta )\right]
\label{eqn8}
\end{align}

\begin{align}
\gamma_{11}^{(1)}&=
 2 c_1\left[t,2 \sin (\theta ) \sin (\phi )\right] \nonumber \\
\gamma_{22}^{(1)} &= c_1\left[t,2 \sin (\theta ) \sin (\phi )\right] \nonumber \\
\gamma_{33}^{(1)} &= \sin ^2(\theta ) c_1\left[t,2 \sin (\theta ) \sin (\phi )\right]  \label{sph2}
\end{align}

The off-diagonal elements of these matrices are vanishing.
\subsection{Map to Spherical Coordinates Using Global Coordinates and 5 KV Algebra}

To translate $\gam$ from flat background into spherical one, one may choose one of the two approaches. Use a map from the flat to spherical coordinates, or transform the KVs and find the solutions in spherical coordinates.

For generality, we describe the map for translation of the solutions to spherical ones, and give several examples for the black holes, MKR and geons (5 KV solutions). 

 AdS and flat space have related conformal compactifications.  One can comapactify the spatial part $\mathbb{R}^n$ of Euclidean case to $S^n$ by adding a point at infinity. Euclidean $AdS_{n+1}$ is conformally equivalent to a disk $ \mathcal{D}_{n+1}$ of a $n+1$ dimensions, while the compactified  Euclidean AdS has a boundary that is compactified Euclidean space, which is analogous as in Minkowski signature.

Define the embedding space, $AdS_{p+2}$ of $p+2$ dimensional hyperboloid from flat $p+3$ dimensional space using the metric 
\begin{equation}
ds^2=-dX_{o}^2-dX_{p+2}^2+\sum_{i=1}^{p+1}dX_i^2\label{emb}
\end{equation}
and the constraint 
\begin{equation}
X_0^2+X^2_{p+2}-\sum^{p+1}_{i=1}X_i^2=L^2.\label{condition}
\end{equation}
By construction, the isometry of the space is $SO(2,p+1)$ while the space is homogeneous and isotropic. 
Condition (\ref{condition}) is solved via parametrization
\begin{align}
X_0&=L\cosh \rho \cos \tau, && X_0=L\cosh\rho\sin\tau \\
X_i&=L\sinh\rho\Omega_i && \left(i=1,....,p+1,\sum_{i=1}^{p+1}\Omega_i^2=1 \right)\label{param}
\end{align}
\noindent where $\Omega_i$ are coordinates on $S^p. $ Using parametrisation (\ref{param}) in (\ref{emb}) one obtains $AdS_{p+2}$ metric
\begin{equation}
ds^2=L^2\left( -\cosh^2\rho d\tau^2+d\rho^2+\sinh^2\rho d\Omega_p^{\rho} \right)
\end{equation}
which for $\rho\in(0,\infty)$ and $\tau\in[0,2\pi)$ covers the hyperboloid parametrisation, where $\rho,\tau,\Omega_i$ are global coordinates of $AdS$. In the neighbourhood of $\rho\sim 0$ the metric becomes
\begin{equation}
ds^2\sim L^2(-d\tau^2+d\rho^2+\rho^2d\Omega_p^2)
\end{equation}
from which one can notice that topology of $AdS_{p+2}$ is $S^1\times \mathbb{R}^{p+1}$. Since $S^1$ is timelike, $AdS_{p+2}$ contains closed timelike curves, to obtain causal space-time, we have to take the universal cover of $S^1$ coordinate which leads to $-\infty<\tau<+\infty$ and does not contain closed timelike curves. 

To bring endpoints of the $\rho$ coordinate to finite values, one introduces a new coordinate $\theta$ 
\begin{align}
\tan \theta=\sinh\rho, & \theta\in\left[0,\frac{\pi}{2}\right)
\end{align}
and the $AdS_{p+2}$ metric becomes
\begin{equation}
ds^2=\frac{L^2}{\cos^2\theta}\left(-d\tau^2+d\theta^2+\sin^2\theta d\Omega_p^2\right)
\end{equation}
which can be conformally transformed to Einstein static universe metric
\begin{equation}
\tilde{s}^2=-d\tau^2+d\theta^2+\sin^2d\Omega_p^2\label{sph}
\end{equation}
with a difference that the $\theta$ coordinate ranges is $\left[0,\frac{\pi}{2}\right)$ and not the entire range $[0,\pi)$. 
That kind of space-time, that is conformal to space-time isomorphic to half of the static Einstein universe, is "asymptotically AdS".

Because of the fact that the boundary in the timelike direction $\tau$ extends, we must specify a boundary condition on $\mathbb{R}\times S^p=\frac{\pi}{2}$ to well define the Cauchy problem of AdS.

One can define Poincare coordinates  $(u,t,\vec{x})$ for $ru>0,\vec{x}\in\mathbb{R}^p$ with
\begin{align}
X_0&=\frac{u}{2}\left[1+\frac{1}{u^2}(L^2+\vec{x}^2-t^2)\right] & X_i=\frac{Lx^i}{u}, \\
X_{p+1}&=\frac{u}{2}\left[1-\frac{1}{u^2}(L^2-\vec{x}^2+t^2)\right] & X_{p+2}=\frac{Lt}{u}
\end{align}
that cover half of the hyperboloid and define the metric 
\begin{equation}
ds^2=\frac{L^2}{u^2}\left[du^2-dt^2+d\vec{x}^2\right] \label{flat}
\end{equation}
with $u=0$ boundary. 
The Poincare symmetry that acts on $(t,\vec{x})$ 
and the $SO(1,1)$ symmetry that acts $(u,t,\vec{x})\rightarrow (au,at,a\vec{x}),a>0$ 
are in these coordinates manifest. While the latter acts as dilatation on the $\mathbb{R}^{1,p}$ $(t,\vec{x})$ coordinates \cite{Kiritsis:2007zza}.

Let us use this transcription to obtain a map between two different background metrics in four dimensions. 
Define the global coordinates
\begin{align}
\text{X0}=L \cosh \left(\frac{r}{L}\right) \cos \left(\frac{t}{L}\right) \nonumber \\
\text{X4}=L \cosh \left(\frac{r}{L}\right) \sin \left(\frac{t}{L}\right) \nonumber \\
\text{X3}=L \cos (\theta ) \sinh \left(\frac{r}{L}\right) \nonumber \\
\text{X2}=L \sin (\theta ) \sin (\phi ) \sinh \left(\frac{r}{L}\right) \nonumber \\
\text{X1}=L \sin (\theta ) \cos (\phi ) \sinh \left(\frac{r}{L}\right) \label{glob}
\end{align}
whose line element reads \begin{equation}
ds^2=dr^2-\cosh\left(\frac{r}{L}\right)^2+L^2\sinh\left(\frac{r}{L}\right)^2d\theta^2+L^2\sin\theta^2\sinh\left(\frac{r}{L}\right)^2,
\end{equation}
and the Poincare coordinates
\begin{align}
\begin{array}{ll}
\text{Y4}=\frac{L T}{u} & \text{Y1}=\frac{L x}{u} \\ \text{Y2}=\frac{L y}{u}  & \text{Y0}=\frac{1}{2} u \left(\frac{L^2-T^2+x^2+y^2}{u^2}+1\right)  \\ \text{Y3}=\frac{1}{2} u \left(1-\frac{L^2+T^2-x^2-y^2}{u^2}\right)  &\\
\end{array}
\end{align}
with an line element \begin{equation} ds^2=\frac{L^2}{u^2}(-dT^2+du^2+dx^2+dy^2) \end{equation}
where 
\begin{align}
T=\frac{L Y_4}{Y_0-Y_3}, u=\frac{L^2}{Y_0-Y_3}, x=\frac{LY_1}{Y_0-Y_3} \text{ and } y=\frac{L Y_2}{Y_0-Y_3}. \label{ys}
\end{align}
To transform Poincare to global coordinates, one has to insert (\ref{glob}) in  (\ref{ys}) where $Y_{i}$ is changed to $X_i$ for $i=0,1,2,3,4$, and use $r\to L \log \left(\frac{2 L}{\rho }\right)$
\begin{align}
u(r,t,\theta,\phi)=\frac{L}{\cosh \left(\frac{r}{L}\right) \cos \left(\frac{t}{L}\right)-\cos (\theta ) \sinh \left(\frac{r}{L}\right)} \Rightarrow \nonumber \\
\Rightarrow u(\rho,t,\theta,\phi)=\frac{4 L^2 \rho }{\cos (\theta ) \left(\rho ^2-4 L^2\right)+\left(4 L^2+\rho ^2\right) \cos \left(\frac{t}{L}\right)} \label{defu} \end{align}\begin{align}
T(r,t,\theta,\phi)=\frac{L \cosh \left(\frac{r}{L}\right) \sin \left(\frac{t}{L}\right)}{\cosh \left(\frac{r}{L}\right) \cos \left(\frac{t}{L}\right)-\cos (\theta ) \sinh \left(\frac{r}{L}\right)}\Rightarrow \nonumber \\
\Rightarrow T(\rho,t,\theta,\phi)=\frac{L \left(4 L^2+\rho ^2\right) \sin \left(\frac{t}{L}\right)}{\cos (\theta ) \left(\rho ^2-4 L^2\right)+\left(4 L^2+\rho ^2\right) \cos \left(\frac{t}{L}\right)} \end{align} \vspace{-0.5cm}\begin{align}
x(r,t,\theta,\phi)=\frac{L \sin (\theta ) \cos (\phi ) \sinh \left(\frac{r}{L}\right)}{\cosh \left(\frac{r}{L}\right) \cos \left(\frac{t}{L}\right)-\cos (\theta ) \sinh \left(\frac{r}{L}\right)}\Rightarrow \nonumber \\ \Rightarrow x(\rho,t,\theta,\phi)=\frac{L \sin (\theta ) \left(4 L^2-\rho ^2\right) \cos (\phi )}{\cos (\theta ) \left(\rho ^2-4 L^2\right)+\left(4 L^2+\rho ^2\right) \cos \left(\frac{t}{L}\right)}\end{align}\vspace{-0.5cm}\begin{align}
y(r,t,\theta,\phi)=\frac{L \sin (\theta ) \sin (\phi ) \sinh \left(\frac{r}{L}\right)}{\cosh \left(\frac{r}{L}\right) \cos \left(\frac{t}{L}\right)-\cos (\theta ) \sinh \left(\frac{r}{L}\right)}\Rightarrow \nonumber \\ \Rightarrow y(\rho,t,\theta,\phi)=\frac{L \sin (\theta ) \left(4 L^2-\rho ^2\right) \sin (\phi )}{\cos (\theta ) \left(\rho ^2-4 L^2\right)+\left(4 L^2+\rho ^2\right) \cos \left(\frac{t}{L}\right)}\label{ynew}
\end{align}
Differentiating 
\begin{align} 
\begin{array}{cccc}
dx=\frac{dx(x_i)}{dx_i}dx_i & dy=\frac{dy(x_i)}{dx_i}dx_i & dT=\frac{dT(x_i)}{dx_i}dx_i & du=\frac{du(x_i)}{dx_i}dx_i 
\end{array}
\end{align}
for $x_i=\rho,t,\theta,\phi$.

Let us write the metric (\ref{sph}) in the expansion of the coordinate $u$ 
\begin{equation}
ds^2= \frac{L^2}{u^2}\left(du^2-\left(1+\frac{u}{L}\cdot c\right)dT^2+dx^2+\left(1-\frac{u}{L}c\right)dy^2+2c\frac{u}{L} dydT\right)\label{expu}
\end{equation}
and rewrite (\ref{expu}) in the terms of (\ref{defu}-\ref{ynew}). Taking the leading order one obtains the line element \begin{equation} ds^2= \frac{\left(\rho ^2-4 L^2\right)^2d\theta^2 }{16 \rho ^2}+\frac{L^2d\rho^2 }{\rho ^2}-\frac{ \left(4 L^2+\rho ^2\right)^2dt^2}{16 L^2 \rho ^2}+\frac{\sin ^2(\theta ) \left(\rho ^2-4 L^2\right)^2d\phi^2 }{16 \rho ^2}\end{equation} that in the leading order has desired spherical $\gamma_{ij}^{(0)}$

We are mostly interested in the largest subalgebra with 5 KVs.
Now we can consider the transformation of the $\gam$ matrix, and expand it in $\rho$ coordinate, which will give us the subleading term in the expansion with the $\mathbb{R}\times S^2$ background.

If we define  $ds^2_c$ with the part of the line element that defines the $\gam$ matrix on $\mathbb{R}\times S^2$  we can write
\begin{equation}
ds^2_c=\frac{\rho^2}{L^2}\frac{L}{u}\left(-cdT^2-cdy^2+2cdTdy\right).
\end{equation}
Rewriting the ($u,T,y,x$) coordinates as above, we obtain in the line element that in  leading order of $\rho$ for the $dx^idx^j$ ($x^i,x^j=\rho,t,\theta,\phi$) has the following coefficients
\begin{center}
\hspace{-0.2cm}\begin{tabular}{ |l | p{0.6 cm} | p{8.9 cm}|}\hline
$dtdt$ & $\gamma_{tt}^{(1)}$ & $ -\frac{c \left(\sin (\theta ) \sin (\phi ) \sin \left(\frac{t}{L}\right)+\cos (\theta ) \cos \left(\frac{t}{L}\right)-1\right)^2}{\left(\cos \left(\frac{t}{L}\right)-\cos (\theta )\right)^3}$ \\

$dtd\theta$ &  $\gamma_{t\theta}^{(1)}$ & $\begin{array}{l}\frac{L \big[4 c \sin (\phi ) \left(4 \cos (\theta ) \cos \left(\frac{t}{L}\right)-\cos (2 \theta ) \cos \left(\frac{2 t}{L}\right)-3\right)\big]}{8 \left(\cos \left(\frac{t}{L}\right)-\cos (\theta )\right)^3} \\-\frac{L\big[c (\cos (2 \phi )-3) \left(4 \sin (\theta ) \sin \left(\frac{t}{L}\right)-\sin (2 \theta ) \sin \left(\frac{2 t}{L}\right)\right)\big]}{8 \left(\cos \left(\frac{t}{L}\right)-\cos (\theta )\right)^3}\end{array}$ \\

$dtd\phi$ & $\gamma_{t\phi}^{(1)}$ & $-\frac{2 L \left(c \sin (\theta ) \cos (\phi ) \left(\sin (\theta ) \sin (\phi ) \sin \left(\frac{t}{L}\right)+\cos (\theta ) \cos \left(\frac{t}{L}\right)-1\right)\right)}{2 \left(\cos \left(\frac{t}{L}\right)-\cos (\theta )\right)^2}$\\

$d\theta d\theta$ & $\gamma_{\theta\theta}^{(1)} $& $-\frac{c L L \left(\sin (\phi ) \left(\cos (\theta ) \cos \left(\frac{t}{L}\right)-1\right)+\sin (\theta ) \sin \left(\frac{t}{L}\right)\right)^2}{\left(\cos \left(\frac{t}{L}\right)-\cos (\theta )\right)^3}$ \\

$d\theta d\phi$ & $\gamma_{\theta\phi}^{(1)} $& $-\frac{2 L \left(c L \sin (\theta ) \cos (\phi ) \left(\sin (\phi ) \left(\cos (\theta ) \cos \left(\frac{t}{L}\right)-1\right)+\sin (\theta ) \sin \left(\frac{t}{L}\right)\right)\right)}{2 \left(\cos \left(\frac{t}{L}\right)-\cos (\theta )\right)^2}$ \\

$d\phi d\phi$ & $\gamma_{\phi\phi}^{(1)}$& $-\frac{c L L \sin ^2(\theta ) \cos ^2(\phi )}{\cos \left(\frac{t}{L}\right)-\cos (\theta )}$ \\ \hline
\end{tabular}
\end{center}That way transforming the metric (\ref{flat}) and (\ref{sph}) in global coordinates one can define the transformation from the flat to $\mathbb{R}\times S^2$ background, and vice versa. 
Considering the background 
\begin{equation}
\gamma_{ij}^{(0)}=\left(\begin{array}{ccc}-1 & 0& 0 \\ 0& 1& 0 \\ 0& 0& \sin\theta^2\end{array}\right) \label{sphm}
\end{equation}
in the equation (\ref{lo}) and the corresponding KVs that conserve it and form the conformal algebra are given in the appendix: Canonical Analysis of Conformal Gravity: Killing Vectors for Conformal Algebra on Spherical Background. The above components of $\gam$ matrix satisfy 5 KVs as when we were considering flat background, for the 5KVs 
\begin{equation}
\xi^{sph}_1-\xi^{sph}_9, \xi^{sph}_4+\xi^{sph}_7,\xi^{sph}_2+\xi^{sph}_8,2\xi^{sph}_3+\xi^{sph}_5\text{ and }\xi^{sph}_0-\xi^{sph}_6
\end{equation}
where index $sph$ denotes that we are considering the KVs in the spherical background.

Although very instructive, that method requires particular computational time if we want to consider global solutions rather then asymptotical expansions.
Other convenient method is to consider the solutions of (\ref{eq:nloke}) with the spherical KVs (\ref{kvsph0}-\ref{sphkv}). Number of these examples can be found in the appendix: Classification. 

In the following subchapter we consider geon and MKR global solutions on the flat background and two known solutions on the $\mathbb{R}\times S^2$ boundary.

\subsection{Map from Classification of KVs from Conformal Algebra to Patera et. al Classfication}

The map denotes original KVs  that correspond to KVs from subalgebras of Patera et al.

\begin{itemize}
\item  1 T: $a_{1,2}=\overline{b}_{1,8}=\overline{e}_{1,3}; a_{1,3}=\overline{b}_{1,9}=\overline{d}_{1,5}=\overline{e}_{1,4}$ 
\item  2 T: $a_{2,4}=\overline{b}_{2,7}=\overline{f}_{2,1};a_{2,5}=\overline{e}_{2,1}\approx \overline{b}_{2,8}$
\item  3 T: $a_{3,1}=\overline{b}_{3,4}$
\item  1 T+1 R: $\tilde{a}_{2,1} (K_2,P_1)=\overline{b}_{2,5}; a_{2,3} (L_3,P_0)=\overline{d}_{2,2}  $
\item  1 T+D: $\tilde{a}_{2,10} (F,P_1)=\overline{b}_{2,12};a_{2,13}(F,P_0)=\overline{b}_{2,15}=\overline{e}_{2,9}$
\item 1 T+1 R+D: $\tilde{a}_{3,3}$
\item  1 T+D+1 SCT: the irreducible group o(2,1) 
\item  2 T+D+1 SCT+1 R: the group $o(2,1)\oplus o(2)$ 
\item  2 T+1 R=$a_{3,13} (K_2,P_0,P_2)=\overline{b}_{3,16}; a_{3,21}(L_3,P_1,P_2)$
\item 2 T+1 R+1D: $a_{4,17}(F,L_3,P_1,P_2)$
\item  2 T+D: $a_{3,11}(F,P_1,P_2)=\overline{b}_{3,13};a_{3,12}(F,P_0,P_2)=\overline{b}_{3,15}$
\item 3 T+1 R: $a_{4,1}(P_0,P_1,P_2,K_2)=\overline{b}_{4,6};a_{4,3}(P_0,P_1,P_2,L_3)$
\item 1 R: $\tilde{a}_{1,1}=\overline{b}_{1,7};\overline{a}_{1,10}(L_3)=d_{1,1}=\overline{c}_{1,3}$
\item 3R: $a_{3,24}$
\end{itemize}

\subsubsection{Bach Equations for MKR}

Bach equations obtained for the flat MKR, are 
\begin{align}
\left(r \left(r f''(r)-2 f'(r)\right)+2 f(r)\right)^2-2 r^3 f^{(3)}(r) \left(r f'(r)-2 f(r)\right)=0  \end{align}
\begin{align}
-2 r^4 f^{(3)}(r) f'(r)+r^2 \left(r f''(r)-2 f'(r)\right)^2-2 r f(r) \big(4 f'(r)+r \big(r^2 f^{(4)}(r)\nonumber\\+2 r f^{(3)}(r)-2 f''(r)\big)\big)+4 f(r)^2=0  \end{align}

\begin{align}
-2 r^4 f^{(3)}(r) f'(r)+r^2 \left(r f''(r)-2 f'(r)\right)^2-4 r f(r) \big(2 f'(r)+r \big(r^2 f^{(4)}(r)\\+3 r f^{(3)}(r)-f''(r)\big)\big)+4 f(r)^2=0
\end{align} Subtracting a second equation from the first equation one obtains equation (\ref{mkreq})

\subsection{Global solutions: Top-Down Approach}
Searching for the solutions of Bach equation, we can attempt to solve the Bach equation directly. The above method that suggests the ansatz solution can simplify the procedure, however, one can as well take a simple ansatz that can be solvable at fourth order, and determine the $\gam$ function. 

We can show this on MKR solution. 

\subsubsection{MKR Solution}

We set a  function f(r) in an ansatz solution of Bach equation. Our ansatz metric is of  the form
\begin{equation} 
g(r)=-\left(
\begin{array}{cccc}
 -f(r) & 0 & 0 & 0 \\
 0 & \frac{1}{f(r)} & 0 & 0 \\
 0 & 0 & r^2 & 0 \\
 0 & 0 & 0 & r^2 \\
\end{array}
\right)\label{mkranz}
\end{equation} that inserting in Bach equation leads to three different partial differential equations, of a third and fourth order (see appendix: Classification). 
Manipulation of the PDEs leads to equation
\begin{equation}
4f^{(3)}(r)+rf^{(4)}(r)=0\label{mkreq}
\end{equation}
which gives for $f(r)$
 \begin{equation}
 f=-\frac{c_1}{6r}+c_2+r c_3+r^2c_4.
  \end{equation}
That solves Bach equation for the relation of coefficients
$c_1=-\frac{2c_2^2}{c_3}.$
 $f(r)$ reads
$f(r)=c_2+\frac{c_2^2}{3rc_3}+r(c_r+rc_4).$
Comparison of the solution with the MKR determines the coefficients
$c_4=-\frac{\Lambda}{3}$ , $c_3=-2a$, $c_2=\sqrt{12a M}$.

This solution, in the limit when $M\rightarrow0$ can be brought to FG form, 
with conformal transformation and transformation of coordinates. 
In addition, imposing the requirements that the trace of the first term in the FG expansion is three, and that $\gam$ is traceless, one obtains the form of the FG expansion with a vanishing $\gamma_{ij}^{(2)}$ matrix. That is done as follows.
\begin{itemize}
\item 
Set $M\rightarrow0$ and transform $r\rightarrow 1/u$, $u\rightarrow U(\eta)$ in (\ref{mkranz})  to obtain the line element with coefficient
\begin{equation}
\frac{U'(\eta)}{U(\eta)^2-2a U(\eta)^3}=\frac{1}{\eta^2}
\end{equation}
in the $d\eta^2$ holographic component.  
Its solution 
\begin{equation}
U=\frac{1-Tanh\left[\frac{1}{2}\left(c_1-ln(\eta)\right) \right]^2}{2a},
\end{equation} gives a desired form for the  $d\eta^2$ term in the metric, i.e. brings the metric to FG form. 
Insert the solution in the line element, and factorize, so that the three dimensional metric reads
\begin{equation}
\gamma_{ij}=\left(
\begin{array}{ccc}
 -\frac{1}{4} a^2 (\eta -1)^2 (\eta +1)^2 & 0 & 0 \\
 0 & \frac{1}{4} a^2 (\eta +1)^4 & 0 \\
 0 & 0 & \frac{1}{4} a^2 (\eta +1)^4 \\
\end{array}
\right)\label{mkrflatg}
\end{equation} while entire line element was multiplied with $\eta^2$.  The first four terms in $\eta$ expansion of (\ref{mkrflatg}) are 
\begin{align}
\gamma_{11}&=-\frac{a^2}{4}+\frac{a^2\eta^2}{2}+O(\eta)^4 \\
\gamma_{22}&=\frac{a^2}{4}+a^2\eta+\frac{3a^2\eta^2}{2}+a^2\eta^3+O(\eta)^4  \\
\gamma_{33}&=\frac{a^2}{4}+a^2\eta+\frac{3a^2\eta^2}{2}+a^2\eta^3+O(\eta)^4 
\end{align}
from which one can immediately read out $\gam$, $\gamma_{ij}^{(2)}$ and $\gamma_{ij}^{(3)}$ matrices. Factor $\frac{a}{2}$ that appears in the $\gamma_{ij}^{(0)}$ is absorbed in the coordinates. However,  $\gam$ has a trace and to make it traceless one needs to perform the following.
\item Transform the metric into a metric in FG form, so that $\eta\rightarrow P(\rho)$ and multiply with conformal factor $\frac{1}{\rho^2}\frac{P(\rho)^2}{P'(\rho)^2}$, which gives the metric
\begin{equation}
\gamma_{ij}=\left(
\begin{array}{ccc}
 -\frac{\left(a^2 P(\rho )^2-4\right)^2}{16 P'(\rho )^2} & 0 & 0 \\
 0 & \frac{(a P(\rho )+2)^4}{16 P'(\rho )^2} & 0 \\
 0 & 0 & \frac{(a P(\rho )+2)^4}{16 P'(\rho )^2} \\
\end{array}
\right).\label{gama}
\end{equation}
Taking a trace $Tr(\gamma)$ of that metric and subtracting  
\begin{equation}\psi_{ij}=\gamma_{ij}-\frac{1}{3}Tr(\gamma_{ij})\end{equation} 
gives the traceless metric. Demanding that the trace of the metric $\gamma_{ij}$ is 3 and imposing initial condition that $P(\rho)=0$ when $\rho\rightarrow0$, leads to the PDE for $\rho$
\begin{equation}
48 P'(\rho)^2=(2+aP(\rho))^2(12+a P(\rho))(4+3aP(\rho))
\end{equation}
with four solutions. 
\begin{align}
P(\rho)&=\frac{2\left(-1+e^{\frac{a\rho}{\sqrt{3}}}\right)\left(2-\sqrt{3}+e^{\frac{a\rho}{\sqrt{3}}}\right)}{a\left(1+e^{\frac{a\rho}{\sqrt{3}}}\right)\left(-2+\sqrt{3}+e^{\frac{a\rho}{\sqrt{3}}}\right)} \\
P(\rho)&=\frac{2 \left(e^{\frac{a \rho }{\sqrt{3}}}-1\right) \left(e^{\frac{a \rho }{\sqrt{3}}}+2+\sqrt{3}\right)}{a \left(-e^{\frac{a \rho }{\sqrt{3}}}+2+\sqrt{3}\right) \left(e^{\frac{a \rho }{\sqrt{3}}}+1\right)} \label{firstsol} \\
P(\rho)&=-\frac{2 \left(e^{\frac{a \rho }{\sqrt{3}}}-1\right) \left(-2 e^{\frac{a \rho }{\sqrt{3}}}+\sqrt{3} e^{\frac{a \rho }{\sqrt{3}}}-1\right)}{a \left(e^{\frac{a \rho }{\sqrt{3}}}+1\right) \left(-2 e^{\frac{a \rho }{\sqrt{3}}}+\sqrt{3} e^{\frac{a \rho }{\sqrt{3}}}+1\right)} \\
P(\rho)&=-\frac{2 \left(e^{\frac{a \rho }{\sqrt{3}}}-1\right) \left(2 e^{\frac{a \rho }{\sqrt{3}}}+\sqrt{3} e^{\frac{a \rho }{\sqrt{3}}}+1\right)}{a \left(e^{\frac{a \rho }{\sqrt{3}}}+1\right) \left(2 e^{\frac{a \rho }{\sqrt{3}}}+\sqrt{3} e^{\frac{a \rho }{\sqrt{3}}}-1\right)}
\end{align}
Inserting that solutions in the $\gamma_{ij}$ (\ref{gama}) one can see that the second order of the FG expansion, $\gamma_{ij}^{(2)}$ matrix, vanishes. Explicitly, first solution (\ref{firstsol}) inserted in $\gam$ (\ref{gama}) and expanded in $\rho$ contains diagonal components 
\begin{align}
\gamma_{11}&=-1-2 c \rho +c^3 \rho ^3 +\mathcal{O}(\rho^4)\nonumber \\  
\gamma_{22}&=1-c \rho  \nonumber+\frac{c^3 \rho ^3}{2}+\mathcal{O}(\rho^4) \\ 
\gamma_{33}&=1-c \rho+\frac{c^3 \rho ^3}{2} +\mathcal{O}(\rho^4) \label{g1m1}. 
\end{align}
This form (\ref{g1m1}) of the functional dependence on diagonal, can be obtained from the condition for the Weyl flattens, with and ansatz metric 
\begin{equation}
g_{ij}=\left(
\begin{array}{cccc}
 1 & 0 & 0 & 0 \\
 0 & 2 f_1(\rho)-1 & 0 & 0 \\
 0 & 0 & f_1(\rho)+1 & 0 \\
 0 & 0 & 0 & f_1(\rho)+1 \\
\end{array}
\right). \label{nulg2}
\end{equation}
The expansion of the function $f_1(\rho)$ will correspond to expansion (\ref{g1m1}).
The condition for Weyl flatness on (\ref{nulg2}) reduces "constraint" from Bach equation to the PDE 
\begin{equation}
\frac{3 f_1\rho(\rho) f_1'(\rho)^2}{2 f_1(\rho)-1}-(f_1(\rho)+1) f_1''(\rho)=0
\end{equation}
with the solution \begin{equation}  f_1(\rho)=\frac{1}{2} \left(1-3 \tanh ^2\left(\frac{1}{2} \left(-\sqrt{3} c_1 \rho-\sqrt{3} c_2 c_1\right)\right)\right)\label{v1} \end{equation} that 
leads to $c_{2\pm}=\pm\frac{2ArcTan\left( \frac{1}{\sqrt{3}}\right)}{\sqrt{3}c_1}$.  Where the metric $g_{ij}$ with the inserted $c_{2+}$ is $g_{ij}=diag(g_{11},g_{22},g_{33},g_{44})$ for
\begin{align}
g_{11}&=1\nonumber \\
g_{22}&=-3 \tanh ^2\left(\coth ^{-1}\left(\sqrt{3}\right)+\frac{1}{2} \sqrt{3} \rho c_1\right) \nonumber \\
g_{33}&=\frac{3}{2} \text{sech}^2\left(\coth ^{-1}\left(\sqrt{3}\right)+\frac{1}{2} \sqrt{3} \rho c_1\right) \nonumber \\
g_{44}&=\frac{3}{2} \text{sech}^2\left(\coth ^{-1}\left(\sqrt{3}\right)+\frac{1}{2} \sqrt{3} \rho c_1\right) 
\end{align}
and $g_{22},g_{33}$ and $g_{44}$ expanded in $\rho$ read
\begin{align}
g_{22}&=-1-2c_1\rho+c_1^3\rho^3+\mathcal{O}(\rho)^4\nonumber \\
g_{33}&=1-c_1 \rho +\frac{1}{2}c_1^3\rho^3+\mathcal{O}(\rho)^4\nonumber \\
g_{44}&=1-c_1 \rho+\frac{1}{2}c_1^3\rho^3+\mathcal{O}^4. \label{mkrconfflat}
\end{align}
Matrix defined with (\ref{mkrconfflat}) does not contain $\gamma_{ij}^{(2)}$ and can be compared with the (\ref{g1m1}). In particular the function $f_1$ with inserted $c_2$ and divided with $c_1$ is equal to  $\frac{P(\rho ) (a P(\rho )+2)^2}{4 P'(\rho )^2}$ with $a\rightarrow\frac{3}{2}c_1$.
\end{itemize}
As one can notice from the (\ref{mkrconfflat}) the $\gam$ matrix in that case takes the form that conserves the translational KVs and the rotation (\ref{trans3rot1}). 
In the limit when $M$ does not go to zero, flat MKR metric, obtained from the ansatz (\ref{mkranz}), 
\begin{align}
ds_{flatMKR}^2&=-\frac{d(t)^2 \left(2 \sqrt{3} r \sqrt{a M}-2 a r^2-2 M+r^3\right)}{r}\nonumber \\&-\frac{r d(r)^2}{-2 \sqrt{3} r \sqrt{a M}+2 a r^2+2 M-r^3}+r^2 d(x)^2+r^2 d(y)^2 \label{fmkr}
\end{align}
using the standard FG expansion and redefinition (transformation) of the $r$ coordinate, can be only brought to a form with a  non-vanishing $\gamma_{ij}^{(2)}$ metric. To redefine the $r $ coordinate, one takes 
\begin{equation}
r(\rho)=\frac{a_{-1}}{\rho}+a_0+a_1 \rho+a_2\rho^2+a_3\rho^3+a_4\rho^4+a_5\rho^5.\label{razvoj}
\end{equation} Transformation of the metric into
 \begin{equation}
\frac{dr(\rho)^2}{V[r(r)]}=\frac{d\rho^2}{\rho^2}
\end{equation}
defines the coeffieicients in (\ref{razvoj}) 
\begin{align}
\begin{array}{|c|c|c|c|c|c|} \hline
a_{-1}&a_0 & a_1 &a_2 & a_3 & a_4 \\
1 & \frac{a^2-2 \sqrt{3} \sqrt{a M}}{4 } & \frac{M}{3 } & -\frac{a M}{4 } & \frac{5 a^2 M+2 \sqrt{3} M \sqrt{a M}}{30} &\frac{-15 a^3 M-18 \sqrt{3} a M \sqrt{a M}-4 M^2}{144 } \\ \hline
\end{array}
\end{align}and bring the metric into FG form. 
In the $g_{tt}$, $g_{\theta\theta}$ and $g_{\phi\phi}$ terms we insert the expansion that allows us to read out $\gam$, $\gamma_{ij}^{(2)}$ and $\gamma_{ij}^{(3)}$ matrices
\begin{align} \gam&=diag(0,2a,2a),\\ 
\gamma_{ij}^{ (2)}&= \left(
\begin{array}{ccc}
{\scriptstyle  \frac{1}{2} \left(a^2-2 \sqrt{3} \sqrt{a M}\right)} & 0 & 0 \\
 0 &{\scriptstyle \frac{3 a^2}{2}-\sqrt{3} \sqrt{a M}} & 0 \\
 0 & 0 &{\scriptstyle \frac{3 a^2}{2}-\sqrt{3} \sqrt{a M}} \\
\end{array}
\right)\\ 
    \gamma_{ij}^{(3)}&=\left(\begin{array}{ccc}
 \frac{4 M}{3} & 0 & 0 \\
 0 &{\scriptstyle \frac{1}{6} \left(3 a^3-6 \sqrt{3} \sqrt{a M} a+4 M\right) }& 0 \\
 0 & 0 & {\scriptstyle \frac{1}{6} \left(3 a^3-6 \sqrt{3} \sqrt{a M} a+4 M\right)} \\
\end{array}
\right).  
     \end{align}
      $\gam$ matrix conserves as expected three translational KVs and rotation.
     This from of $\gam$ is not traceless, to make it traceless we perform the conformal rescaling of the metric analogously to $M\rightarrow0$ above. Mulitplying the metric (\ref{fmkr}) with the conformal factor \begin{equation} e^{-\frac{2a}{3r}} \label{cf} \end{equation} leads to the set of coefficients in $r(\rho)$  expansion that lead to the traceless $\gam$ as $\gam$ in (\ref{trans3rot1}). If we, for convenience expand the r coordinate in the expansion
     \begin{equation}
     r(\rho)=b_1+b_2\rho+b_3\rho^2+b_4\rho^3+b_5\rho^4,
     \end{equation}
     the condition to obtain the FG expansion $\frac{dr(\rho)^2}{V[r(\rho)]}=\frac{1}{\rho^2}$ leads to \begin{center}
   \begin{tabular}{|c|c|c|c|}\hline
     $b_1$ & $b_2$ & $b_3$ & $b_4$   \\ 
     1 & $-\frac{2a}{3}$ & $ \frac{1}{18} \left(a^2+9 \sqrt{3} \sqrt{a M}\right) $& $ \frac{1}{243} \left(38 a^3-108 \sqrt{3} a \sqrt{a M}-81 M\right)  $  \\ \hline
     &&&$b_5$ \\ 
&&&    $  -\frac{a \left(176 a^3+216 \sqrt{3} a \sqrt{a M}-3483 M\right)}{2916}$ \\ \hline
          \end{tabular}  
     \end{center}
  \noindent The $\gamma_{ij}^{m}$ for $m=1,2,3$ diagonal matrices in the FG expansion read
          \begin{align} \gam=\left(
\begin{array}{ccc}
 \frac{4 a}{3} & 0 & 0 \\
 0 & \frac{2 a}{3} & 0 \\
 0 & 0 & \frac{2 a}{3} \\
\end{array}
\right), \end{align} 
\begin{align}
\gamma_{11}^{(2)}&= \frac{1}{243} \left(214 a^3+270 \sqrt{3} \sqrt{a M} a+324 M\right) \nonumber  \\
\gamma_{22}^{(2)}&= \frac{1}{243} \left(83 a^3-189 \sqrt{3} \sqrt{a M} a+162 M\right) \nonumber \\
\gamma_{33}^{(2)}&=-\frac{a \left(-563 a^3+1242 \sqrt{3} \sqrt{a M} a-2835 M\right)}{2916} \nonumber
\end{align}
 \begin{align}
 \gamma_{11}^{(3)}&=-\frac{a \left(1331 a^3+3942 \sqrt{3} \sqrt{a M} a+8667 M\right)}{2916} \nonumber\\
 \gamma_{22}^{(3)}&=-\frac{a \left(-563 a^3+1242 \sqrt{3} \sqrt{a M} a-2835 M\right)}{2916} \nonumber \\
 \gamma_{33}^{(3)}&=-\frac{a \left(-563 a^3+1242 \sqrt{3} \sqrt{a M} a-2835 M\right)}{2916} 
 \end{align}
 and the $\gam$ is now traceless as required. 
The response functions
\begin{align}
\tau_{ij}&=\left(
\begin{array}{ccc}
 \frac{8}{9} \left(\sqrt{3} \sqrt{a M} a+9 M\right) & 0 & 0 \\
 0 & 4 M-\frac{8 a \sqrt{a M}}{3 \sqrt{3}} & 0 \\
 0 & 0 & 4 M-\frac{8 a \sqrt{a M}}{3 \sqrt{3}} \\
\end{array}
\right),  \\ P_{ij}&=\left(
\begin{array}{ccc}
 \frac{8 \sqrt{a M}}{\sqrt{3}} & 0 & 0 \\
 0 & \frac{4 \sqrt{a M}}{\sqrt{3}} & 0 \\
 0 & 0 & \frac{4 \sqrt{a M}}{\sqrt{3}} \\
\end{array}
\right)
\end{align}
 using the (\ref{charge1}) $\mathcal{Q}_{ij}=2\tau_{ij}+2P^{ik}\gamma_{kj}^{(1)}$. 
define the modified stress energy tensor in a sense of Hollands, Ishibashi and Marlof \cite{Hollands:2005ya}
\begin{equation}
\mathcal{Q}_{ij}=\left(
\begin{array}{ccc}
 16 M-\frac{16 a \sqrt{a M}}{\sqrt{3}} & 0 & 0 \\
 0 & 8 M & 0 \\
 0 & 0 & 8 M \\
\end{array}
\right) \label{chmkrflat}
\end{equation}
and give for the energy $16M-\frac{16 a\sqrt{aM}}{\sqrt{3}}$ per square unit of surface. 
One can also compute the charges corresponding to conserved KVs. Equation (\ref{chmkrflat}) using the relation $J^i=(2\tau^i_j+2P^{ik(1)}_{kj})\xi^j$ and $Q[\xi]=\int_{\mathcal{C}}d^2x\sqrt{h}u_iJ^i$ for $h$ metric on $\mathcal{C}$ and $u^i$ future-pointing vector, of normal to $\mathcal{C}$ and normalised to unity, 
for the timelike, space like and rotational KVs $(1,0,0),(0,1,0),(0,0,1)$ and $(0,y,-x)$  lead to the currents and charges
\begin{align}\begin{array}{|c|c|c|c|c|}\hline KV & (1,0,0) & (0,1,0) & (0,0,1) & (0,y,-x) \\ \hline current & \left(16\left(M-\frac{a\sqrt{aM}}{\sqrt{3}} \right) ,0,0 \right)& (0,8M,0) & (0,0,8M) & (0,8My,-8Mx)\\ \hline
charge & 16\left(M-\frac{a\sqrt{aM}}{\sqrt{3}}\right) & 0 &0 & 0 \\ \hline  \end{array}\end{align}

\noindent Let us consider two examples on the spherical background.

\subsubsection{Spherical Examples:  Black hole solutions}
\textbf{MKR solution.}

Two examples for which we know the global solution are MKR solution and the rotating black hole solution. We have considered MKR solution in the first chapter. Once expanded in FG expansion it gives the $\gam$ matrix (\ref{gama1mkr}) 
and conserves four KVs, $\xi^{sph}_0,\xi^{sph}_7,\xi^{sph}_8,\xi^{sph}_9$ that form $\mathbb{R}\times o(3)$ subalgebra of the conformal algebra $so(2,3)$. The only charge that does not vanish is the one that belongs to $\xi_4^{sph}=\partial_t$ KV 
(\ref{holrenmkrcharge}) 
\begin{equation}
Q[\xi_f^{sph}]=\frac{M}{\ell^2}-a(1-\sqrt{1-12aM}).
\end{equation}
In the terms of canonical analysis, that charge is equal to $Q[\xi^{sph}_4]=Q_{\perp}[N]$.
\\
\\
\noindent{\bf Rotating Black Hole}

Let us consider the $\gam$ matrix for spherical global solution of Bach equation, rotating black hole solution. The metric reads \cite{Liu:2012xn}
\begin{align}
ds^2&=\rho^2\big[\frac{dr^2}{\Delta_r}+\frac{d\theta^2}{\Delta_{\theta}}+\frac{\Delta_{\theta}\sin^2\theta}{\rho^2}\left(\alpha dt-(r^2+\al^2)\frac{d\phi}{\Sigma}\right) \noindent \\ &-\frac{\Delta_r}{\rho^2}\left(dt-a\sin^2\theta\frac{d\phi}{\Sigma}\right) \big], 
\end{align} with 
\begin{align}
\rho^2&=r^2+\al^2\cos^2\theta, && \Delta_{\theta}=1+\frac{1}{3}\Lambda\al^2\cos^2\theta, && \Sigma=1+\frac{1}{3}\Lambda\al^2
\end{align}
\begin{align}
\Delta_r&=\left( r^2+a^2 \right)\left(1-\frac{1}{3}\Lambda r^2 \right)-2\mu r^3
\end{align}

To find the subalgebra of $so(2,3)$ for the $\gam$, one first needs to transform the metric to FG expansion, transform the leading term in expansion $\gamma_{ij}^{(0)}$ to the spherical background $\mathbb{R}\times S^2$ 
 and use the same transformation on $\gam$. The resulting $\gam$ is the one that defines the subalgebra and corresponding KVs.

After setting $\Lambda\rightarrow3$ and multiplying the line element with $\frac{1}{r^2+\al^2\cos^2\theta}$ we insert the expansion of the coordinate $r$ in dependency on the new introduced coordinate $\rho$ (\ref{razvoj}).
The $\rho\rho$ component of the FG expansion defines equation
\begin{equation}
\frac{dr(\rho)^2}{\left(1+r(\rho)\right)^2\left(\al^2+r(\rho)^2\right)-2r(\rho)^3\mu}=\frac{d\rho^2}{\rho^2}
\end{equation}
and gives for the coefficients in the expansion (\ref{razvoj})
\begin{align}
\begin{array}{|c|c|c|c|}
\hline
a_0 & a_1 & a_2 & a_3  \\
\frac{\mu}{2} & \frac{-2-2\al^2+3\mu^2}{12 a_{-1}} & \frac{-\mu-\al^2\mu+\mu^3}{8a_{-1}} & \frac{2(7-22\al^2+7\al^4)-60(1+\al^2)\mu^2+45\mu^4}{720 a_{-1}^3} \\ \hline
\end{array}\label{tablecoef}
\end{align}
\noindent that inserted in the metric give  $g_{\rho\rho}=1$ (where we assume the metric is rescaled with the factor $\frac{1}{\rho^2}$).
 The following term we want to determine is $g_{tt}$
 \begin{equation}
g_{tt}= 1+\frac{8r(\rho)^3\mu}{(\al^2+2r(\rho)^2+\al^2\cos(2\theta))^2}-\frac{2(1+\al^2+2r(\rho)^2)}{\al^2+2r(\rho)^2+\al^2\cos(2\theta)},
 \end{equation}
 inserting $r(\rho)$ and (\ref{tablecoef}) and  expanding
in $\rho$, the $g_{tt}$ component of the metric becomes
\begin{equation}
g_{tt}=-1+2\mu\rho+\rho^2(-1-\mu^2+\al^2\cos(2\theta)),\label{gtt}
\end{equation}
 while 
 \begin{equation}
 g_{t\phi}=-\frac{\al\sin^2\theta}{-1+\al^2}+\frac{2\al\mu\rho\sin^2\theta}{-1+\al^2}+\frac{\al\rho^2(-\al^2-2\mu^2+\al^2\cos(2\theta))\sin^2\theta}{2(-1+\al^2)}\label{gtph}
   \end{equation}and $g_{\theta\theta}$ terms remain the same since there is no $r(\rho)$ coordinate appearing
\begin{equation}
g_{\theta\theta}=\frac{1}{1-\al^2\cos^2\theta}.\label{gthth}
\end{equation}
Term $g_{\phi\phi}$ is
\begin{equation}
g_{\phi\phi}=\frac{\sin^2\theta}{-1+\al^2}+\frac{2\al^2\mu\rho\sin^4\theta}{(-1+\al^2)^2}-\frac{\al^2(-1+\al^2+\mu^2)\rho^2\sin^4\theta}{(-1+\al^2)^2}.\label{gff}
\end{equation}
From the (\ref{gtt}), (\ref{gtph}), (\ref{gthth}), and (\ref{gff}) one can read out the terms in the FG expansion
\begin{align}
\gamma_{ij}^{(0)}=\left(\begin{array}{ccc} -1 & 0 & -\frac{\al\sin^2\theta}{-1+\al^2} \\
0 & \frac{1}{1-\al^2\cos^2\theta} & 0 \\
-\frac{\al \sin^2\theta}{-1+\al^2} & 0 & \frac{\sin^2\theta}{1-\al^2}  \end{array}\right).
\end{align}

To transform it into a form $\gamma_{ij}^{(0)}=diag(-1,1,\sin^2\theta)$ we transform the coordinates 
\begin{align}
\phi&\rightarrow bt_1+a\phi_1, && t\rightarrow-\frac{b}{a_2t_1}\\
t_1&\rightarrow \frac{t_1\sqrt{-1+\al^2}}{b}, &&\phi_1\rightarrow\frac{\phi_1\sqrt{-1+\al^2}}{a}\\
\phi&\rightarrow \frac{1}{a_2}\phi_1 && 
\end{align}and divide the metric with 
\begin{equation}
1-\frac{1}{\al^2}-\sin^2\theta.
\end{equation}The $\gamma_{ij}^{(0)}$ term is
\begin{equation}
\left(
\begin{array}{ccc}
 -1 & 0 & 0 \\
 0 & -\frac{4 \text{a2}^2}{\left(\cos (2 \theta ) \text{a2}^2+\text{a2}^2-2\right)^2} & 0 \\
 0 & 0 & -\frac{2 \sin ^2(\theta )}{\cos (2 \theta ) \text{a2}^2+\text{a2}^2-2} 
\end{array}
\right).
\end{equation}
To obtain the required form of the $\gamma_{ij}^{(0)}$ we solve the equation
\begin{equation}
-\frac{2\sin^2\theta(x)}{-2+\al^2+\al^2\cos(2\theta(x))}=\sin^2(x)
\end{equation}
and obtain the solution for $\theta_{x}$  
\begin{equation}
\theta\rightarrow Arctan\left[-\frac{\sqrt{-\cos^2x}}{-1+\al^2\sin^2x}\frac{\sqrt{-1+\al^2\sin x}}{\sqrt{-1+\al^2}\sin^2x}\right]
\end{equation}
that leads to the line element
\begin{equation}
ds^2=-dt^2+\frac{\al dx^2}{-1+\al^2}+\sin^2x d\phi^2.
\end{equation}
Which is  by multiplication with 
\begin{equation}
\frac{-1+\al^2}{\al^2}
\end{equation}and transformations 
\begin{align}
\phi&\rightarrow\phi \frac{\al}{\sqrt{-1+\al^2}}, && t\rightarrow\frac{t}{\sqrt{1-\frac{1}{\al^2}}}
\end{align}
finally brought to the desired form
$-dt^2+d\theta^2+\sin^2\theta d\phi^2$.

After transformation into FG form of the metric we read out $\gam$ from (\ref{gtt}), (\ref{gtph}), (\ref{gthth}), and (\ref{gff}) 
\begin{align}
\gam=\left(\begin{array}{ccc} 2\mu &0 & \frac{2\al\mu\sin^2\theta}{-1+\al^2} \\ 0&0&0 \\ \frac{2\al\mu\sin^2\theta}{-1+\al^2} & 0 & \frac{2\al^2\mu\sin^4\theta}{(-1+\al^2)^2}  
\end{array}
\right).
\end{align}
and applying the transformations from above to $\gam$ obtain
\begin{align}
\gam=\left(
\begin{array}{ccc}
 \frac{4\mu}{2-\al^2+\al^2\cos(2\theta)} &0& \frac{4\al\mu\sin^2\theta}{2-\al^2+\al^2\cos(2\theta)} \\ 0 & 0 & 0\\ \frac{4\al\mu\sin^2\theta}{2-\al^2+\al^2\cos(2\theta)} &0 & \frac{4\al^2\mu\sin^4\theta}{2-\al^2+\al^2\cos(2\theta)}
\end{array}
\right)
\end{align}
that conserves $(1,0,0)$ and $(0,0,1)$ KVs, and forms the $o(2)$ algebra.

\subsection{Asymptotic Solutions}

When bottom-up and top down approach to solving the Bach equation become to complicated, one can use asymptotical analysis. It searches for solutions in the  neighbourhood of the conformal boundary. They can be extended to global solutions for simple enough cases. Here, we want to find "new" boundary solutions with non-zero charges, that are not equivalent to the MKR solution.

Let us consider coordinates with the boundary $\rho=0$, the familiar asymptotic expansion 
\begin{equation}
ds^2=\frac{\ell^2}{\rho^2}\left[d\rho^2+\left(\gamma_{ij}^{(0)}+\frac{\rho}{\ell}\gamma_{ij}^{(1)}+\frac{\rho^2}{\ell^2}\gamma_{ij}^{(2)}+\frac{\rho^3}{\ell^3}\gamma_{ij}^{(3)}+...\right)dx^idx^j\right],\label{gc}
\end{equation}
and traceless higher order terms $\psi_{ij}^{(n)}$. We compute the Bach equation order by order in holographic $\rho$ coordinate for three examples
\begin{enumerate}
\item  MKR with vanishing $\gamma_{ij}^{(2)}$ matrix ($M\rightarrow0$) and $\mathbb{R}^3$ boundary,
\item  MKR with vanishing $\gamma_{ij}^{(2)}$ matrix and $\mathbb{R}\times S^2$,
\item the example with asymptotic solution of arbitrary function on the diagonal and above $\gam$.
\end{enumerate}
The first term in the expansion is defined by the choice of conformal boundary, and 
for the first subleading term we choose  $\gam=diag(2c,c,c)$ (the procedure can be applied to the $\gam$ solutions that we have listed above in the classifications). For simplicity, we set the second term in the FG expansion (\ref{gc}) to zero, $\psi_{ij}^{(2)}=0$. The condition that Bach equation gives on the third term $\psi_{ij}^{(3)}$ in the FG expansion (\ref{gc}) is \begin{equation}\partial^{j}\psi^{(3)}_{ij}=0\end{equation}
\begin{enumerate}
\item 
For the first case, one can for simplicity set as an ansatz, traceless $\gamma_{ij}^{(3)}$ matrix 
\begin{equation}
\psi_{ij}^{(3)}=\left(
\begin{array}{ccc}
 d_1+d_2 & d_3 & d_4 \\
 d_3 & d_1  & d_5 \\
 d_4 & d_5 & d_2  \\
\end{array}
\right)
\end{equation}
Since the metric is flat there is no contribution from the curvatures and the condition on the coefficients in the $\psi_{ij}^{(3)}$ comes from the $\rho\rho$ component of the EOM which setting $\ell=1$ reduces to 
\begin{equation}
-\frac{3}{4}\psi^{(1)}{}_i^{k}\psi^{(1)}{}^{ij}\psi^{(1)l}_j\psi^{(1)}_{kl}+\frac{1}{8}\psi^{(1)}_{ij}\psi^{(1)}{}^{ij}\psi^{(1)}_{lk}\psi^{(1)lk}-\frac{1}{2}\psi^{(1)ij}\psi^{(3)}_{ij}=0
\end{equation}
and gives $d_1=-6c^3-d_2$ and
\begin{equation}
\gamma_{ij}^{(3)}=\left(
\begin{array}{ccc}
 -6 c^3 & d_3 & d_4 \\
 d_ 3 & -6 c^3-d_2 & d_5 \\
 d_4 & d_5 & d_2 \\
\end{array}
\right).
\end{equation}
The Brown York stress tensor for the metric is defined by the $\psi_{ij}^{(3)}$ and $\psi_{ij}^{(1)}$ matrix, that means electric part of the Weyl tensor $E_{ij}^{(3)}$, while $E_{ij}^{(2)}$ for flat metric vanishes.
If we change the expansion of the metric (\ref{gc}) with the expansion that for convenience in computation introduces factorials in the metric 

\begin{equation}
ds^2=\frac{\ell^2}{\rho^2}\left[d\rho^2+\left(\gamma_{ij}^{(0)}+\frac{\rho}{\ell}\gamma_{ij}^{(1)}+\frac{1}{2!}\frac{\rho^2}{\ell^2}\gamma_{ij}^{(2)}+\frac{1}{3!}\frac{\rho^3}{\ell^3}\gamma_{ij}^{(3)}+...\right)dx^idx^j\right]\label{gcf}
\end{equation}
the choice of the components of the $\psi_{ij}^{(3)}$ metric is particularly convenient because in that case BY ST leads to 
\begin{equation}
\tau_{ij}=\left(
\begin{array}{ccc}
 0 & d_3 & d_4 \\
 d_3 & -3 c^3-d_2 & d_5 \\
 d_4 & d_5 & 3 c^3+d_2 \\
\end{array}
\right)
\end{equation}
that contains vanishing $\tau_{11}$ component,
while for the first choice of the expansion (i.e. the factorials in the  expansion are absorbed in the components of $\psi_{ij}^{(3)}$) and the equal $\psi_{ij}^{(3)}$ matrix, 
\begin{equation}
\tau_{ij}=\left(
\begin{array}{ccc}
 -12 c^3 & 3 d_3 & 3 d_4 \\
 3 d_3 & -3 \left(5 c^3+d_2\right) & 3 d_5 \\
 3 d_4 & 3 d_5 & 3 \left(c^3+d_2\right) \\
\end{array}
\right).
\end{equation}
 The charge associated to $(1,0,0)$ KV therefore in the latter case vanishes, while the charge for $(0,1,0)$ reads $2d_3 l^2$, for $(0,0,1)$ it is $2d_4 l^2$ and for $(y,-x,0)$ it is $l^2(-2d_4x+2d_3y)$. Where $l$ are  lengths over which we integrate the charges. In the first case, all four charges are present, that from the computational side shows that convenient choice for the components of the matrix can simplify search for the solutions and result with the form of the response functions desired for the particular purposes.
\item
Let us now choose the form of the expansion (\ref{gcf}) and consider the background $\mathbb{R}\times S^2$, which is closer to the original examples of $AdS$ holography and the global results for EG for cosmological constant $\Lambda<0$, where we assume that the conformal boundary belongs to the same conformal class as the Einstein's static universe. 

Assume conformal boundary with coordinates $(t,\theta,\phi)$ and the metric \begin{equation} 
\gamma_{ij}^{(0)}=diag(-1,L^2,L^2\sin^2(\theta)), 
\end{equation}
for $L$ radius of the $S^2$. 
The boundary conditions are conserved by the diffeomorphism $\xi^{\mu}$ for (\ref{lo}) and (\ref{nloke}).
The solution $\gam$ that conserves three KVs of the $S^2$ and $(1,0,0)$ is
\begin{equation}
\psi_{ij}^{(1)}=diag(2c,cL^2,cL^2\sin^2(\theta))
\end{equation}
which is covariantly constant $\mathcal{D}_k\psi^{(1)}_{ij}=0$. The curvatures $\mathcal{R}_{ij}^{(0)}=\delta_i^a\delta_j^b\frac{1}{L^2}\gamma_{ab}^{(0)}$, $\mathcal{R}^{(0))}=\frac{2}{L^2}$ and the condition $\mathcal{D}_k\mathcal{R}_{ij}^{(0)}=0$ simplify the Bach equation (\ref{bach}) in fourth order which in $\mu=i$ and $\nu=\rho$ component reads 
\begin{equation}
\mathcal{D}^{j}\psi_{ij}^{(3)}=0\label{ir}.
\end{equation}
For the ansatz

\begin{gather}
	\psi^{(3)}_{ij} = \left( \begin{array}{ccc}
d_1 + d_2 & d_3 & d_4 \\
d_3 & d_1\,L^2 & d_5 \\
d_4 & d_5 & d_2\,L^2\,\sin^{2}\theta \end{array} \right) 
\end{gather}

the equation (\ref{ir}) gives $d_3=0$, $d_5=0$ and $d_2=d_1$. As in the example above, we are interesting in finding simple solution with interesting charges. 
The Bach equation  for the components $\mu=\rho$ and $\nu=\rho$ leads to 
\begin{gather}
	0 = -9\,\frac{c^4}{\ell^{8}} - 3\,\frac{c\,d_1}{\ell^{8}}-\frac{2}{3}\,\frac{1}{\ell^{4} L^{4}}
\end{gather}
\begin{gather}
	\Rightarrow d_1 = -3\,c^3 - \frac{2\,\ell^{4}}{9\,c\,L^4} ~.
\end{gather}
It is interesting to notice that in the above case, one can not take the limit $c\rightarrow0$ while that is allowed for the conformal boundary $\mathbb{R}^3$.
To find the response functions we need first the magnetic an electric part of the Weyl tensors
\begin{align}
	B^{(1)}_{kij} = & \,\, 0 
	\end{align}
	\begin{align}
	E^{(2)}_{ij} = & - \frac{1}{2}\,\left(\mathcal{R}^{\LO}_{ij} - \frac{1}{3}\,\gamma^{\LO}_{ij}\,\mathcal{R}^{\LO} \right) = -\frac{1}{6 L^2 c}\, \\  \psi^{\FO}_{ij} =& - \frac{1}{6 L^2}\,\text{diag}(2,L^2,L^2 \sin^{2}\theta) 
	\end{align}
\begin{align}
	E^{(3)}_{ij} = & \,\, - \frac{1}{4\,\ell^{3}}\,\psi^{\TO}_{ij} - \frac{1}{8\,\ell^{3}}\,\psi^{\FO}_{ij}\,\psi^{\FO}{}^{kl} \psi^{\FO}_{kl} + \frac{1}{6}\,\left(\mathcal{R}^{\LO} \psi^{\FO}_{ij} - \gamma^{\LO}_{ij}\,\mathcal{R}^{\LO}_{kl} \psi^{\FO}{}^{kl}\right) ~.
\end{align}
Which define PMR 
\begin{align}
	P_{ij} =& \frac{4}{\ell}\,E^{\SO}_{ij} = - \frac{2\,\alpha^{2}}{3\,\ell^{3}\,c}\, \\ \psi^{\FO}_{ij} =& - \frac{2\,\alpha^{2}}{3\,\ell^{3}}\,\text{diag}(2, L^2, L^2 \sin^{2}\theta)
\end{align}
for $\alpha=\frac{\ell}{L}$  the ratio of the $AdS$ length scale and the radius of the sphere. While the BY ST is 
\begin{gather}
	\tau_{ij} = - \frac{4}{\ell}\,E^{\TO}_{ij} + \frac{4}{\ell}\,\left(E^{\SO}_{ik} \psi^{\FO}{}^{k}{}_{j} + E^{\SO}_{kj} \psi^{\FO}{}_{i}{}^{k} \right) - \frac{2}{\ell}\,\gamma^{\LO}_{ij}\,E^{\SO}_{kl} \psi^{\FO}{}^{kl}
\end{gather}
\begin{gather}
	\tau_{ij} = \left( \begin{array}{ccc}
-\frac{2 c \alpha^{3}}{3 \ell^{3}} - \frac{4 \alpha^{4}}{9 c \ell^{3}} & 0 & \frac{d_4}{\ell^{3}} \\
0 & \frac{2c}{3\ell} - \frac{2 \alpha^{2}}{9 c \ell} & 0 \\
\frac{d_4}{\ell^{3}} & 0 & \left(\frac{2c}{3\ell} - \frac{2 \alpha^{2}}{9 c \ell}\right) \sin^{2}\theta \end{array} \right).
\end{gather} 
In this case we obtain the component with at least one t index
\begin{align}
	T_{tt} = & \,\, \frac{4 \, c \, \alpha^{2}}{\ell^{3}} - \frac{8 \, \alpha^4}{9 \, c \, \ell^{3}}\\
	T_{t\phi} = & \,\, \frac{2\,d_4}{\ell^{3}} ~.
\end{align}
The charges integrated over the compact constant time surface ($S^2$) give finite results. They are associated with the time translation and rotations in the $\phi$ direction 
\begin{align}
	Q[\xi_0^{sph}] = &\,\,\frac{16\,\pi\,c}{\ell} - \frac{32\,\pi\,\alpha^{4}}{9\,c\,\ell} \\
	Q[\xi_7^{sph}] = &\,\, \frac{8\,\pi\,d_4}{\ell\,\alpha^{2}} ~.
\end{align}
The Casimir energy for this boundary exists which differs from the $\mathbb{R}^3$ boundary. 
Similarly, that happens in EG holography, only here conformal boundary is three dimensional. For EG Casimir energy appears only in the cases when conformal boundary is even-dimensional.
\item In the third example we assume the metric  
\begin{equation}
g_{ij}=\left(
\begin{array}{cccc}
 \frac{1}{r^2} & 0 & 0 & 0 \\
 0 & \frac{2 a_1 c_1(r)-1}{r^2} & 0 & 0 \\
 0 & 0 & \frac{a_1 c_1(r)+1}{r^2} & 0 \\
 0 & 0 & 0 & \frac{a_1 c_1(r)+1}{r^2} \\
\end{array}
\right)\label{gex3}
\end{equation}
which we expand in the r component. We set $\gamma_{ij}^{(0)}$ to be flat background and $\gam=diag(2c,c,c)$. 
The metric (\ref{gex3}) is simple enough to allow the computation of the Bach equation using the RGTC code and give non-vanishing components of the Bach tensor $B_{ij}$, 
$B_{rr},B_{tt},B_{xx},B_{yy}$.
The equations are still to complicated to be solved exactly, however one can compute them asymptotically, expanding the $c_1(r)$ function 
\begin{equation}
c_1(r)=b_1 r+ b_2 r^2+b_3 r^3+b_4 r^4+b_5 r^5+ b_6 r^6+ b_7 r^7+ b_8 r^8+b_9 r^9+b_{10} r^{10}.
\end{equation}
The expansion of the $B_{rr}$ equation around the $r=0$ up to 10th order gives  coefficients next to each of the $b$s. Bach equation is expectedly vanishing up to 4th order since it is fourth order in derivative, however we obtain $\mathcal{O}(r^5)$  to vanish as well. The first non-vanishing term appears in $6th$ order $\mathcal{O}(r^6)$ and gives the condition on the $b_3$ coefficient $b_3=-\frac{1}{2}a_1^2b_1^3$. Inserting that solution in the following, 7th, order determines the coefficient $b_4$ and one can recursively solve Bach equation up to desired and computational allowed order. 
For the above ansatz of constant $b$s one obtains 
\begin{align}
\begin{array}{|c|c|c|c|c|c|}\hline
b_3 & b_4 & b_5 & b_6 & b_7 & b_8\\
-\frac{1}{2}a_1^2b_1^3 & -\frac{1}{4}a_1^3b_1^4 & \frac{3a_1^4b_1^5}{40} & \frac{a_1^5b_1^6}{8} & \frac{17a_1^6b_1^8}{560} & -\frac{9}{320}a_1^7b_1^8 \\ \hline
\end{array}.
\end{align}
These solutions satisfy the $B_{tt}$, $B_{xx}$ and $B_{yy}$ components of Bach equation as well. 
 Therefore we have solved the Bach equation up to $\mathcal{O}(r^{10})$.

One can analogously, using the asymptotic expansion, consider simpler forms of the metric ansatz in dependency on $t,x$ or $y$ coordinates or their combinations. Solving the equations for coefficients in each order, as in this case, consists of solving the partial differential equations, that can result with additional free coefficients. That can lead to functional dependency in asymptotic form of the metric, or depending on the metric, bring to determination of the coefficients from other components of Bach equation. 

\end{enumerate}

We have seen two possible approaches for solving the asymptotical Bach equation. One of them can give more freedom in the choice of the higher $\gamma_{ij}^{(n)}$ (n=1,2,3,4) matrices, solving the Bach equation in fourth order, while the other can give solution for higher orders of Bach equation for $\gamma_{ij}^{(n)}$ (n $<$ computationally allowed) of an analogous form. 

The third option is solving the differential equations numerically. In this case one can, cleverly choosing the initial form of the metric with desired functional dependence on the boundary coordinates and the boundary conditions, inspect the forms of the curvatures and the Bach equation that may lead to the clever initial ansatz for the metric.

\section{Appendix: One Loop Partition Function}
\subsection{One Loop Partition Function in Six Dimensions}
Trivial anomalies are generated with local functionals
 \begin{align}
 \mathcal{M}_i&=\int d^6x\sqrt{g}\sigma(x)M_i\\
 \mathcal{K}_i&=\int d^6x\sqrt{g}K_i
 \end{align}
here, in the notation of \cite{Bastianelli:2000rs}
\begin{align}
\mathcal{M}_5&=\delta_{\sigma}\bigg( \frac{1}{30}\mathcal{K}_1-\frac{1}{4}\mathcal{K}_2+\mathcal{K}_{6}\bigg), \mathcal{M}_6=\delta_{\sigma}\bigg( \frac{1}{100}\mathcal{K}_1-\frac{1}{20}\mathcal{K}_2\bigg),  \nonumber\\
\mathcal{M}_7&=\delta_{\sigma}\bigg(\frac{37}{6000}\mathcal{K}_1-\frac{7}{150}\mathcal{K}_2+\frac{1}{75}\mathcal{K}_3-\frac{1}{10}\mathcal{K}_5-\frac{1}{15}\mathcal{K}_6\bigg), \mathcal{M}_8=\delta_{\sigma}\bigg(\frac{1}{150}\mathcal{K}_1\-\frac{1}{20}\mathcal{K}_3\bigg) \\
\mathcal{M}_9&=\delta_{\sigma}\bigg(-\frac{1}{30}\mathcal{K}_1\bigg),\mathcal{M}_{10}=\delta_{\sigma}\bigg(\frac{1}{300}\mathcal{K}_1-\frac{1}{20}\mathcal{K}_9\bigg)\nonumber.
\end{align}
 $\sigma$ is infinitesimal Weyl transformation parameter, and $K_i$ are defined in the main text from (\ref{ks}).

\subsection{Generalization to Higher Dimensions}

The above expressions for the partition function in four and six dimensions, can be generalised for the partition functions in arbitrary number of dimensions. That can be achieved by the straightforward computation of the partition function on the thermal AdS space of the CG partition function 
\begin{align}
Z_s(S^d)&=\prod_{k=0}^{s-1}\left(\det[-\nabla^2+k-(s-1)(s+d-2)]_{k\perp}\right)^{1/2} \nonumber \\ & \times \prod_{k=-\frac{1}{2}(d-4)^{s-1}}\left(\det[-\nabla^2+s-(k'-1)(k'+d-2)]_{s\perp}\right)^{-1/2},
\end{align}
\footnote{The partition function is evaluated on the conformally flat Einstein background that is $(A)dS_d$ or $S^d$ \cite{Tseytlin:2013fca}.} or using the procedure introduced in \cite{Beccaria:2016tqy}. 
That is the procedure that we describe here.
 We introduce partition function of higher spin EG that originates from the action of massless higher spins \cite{Gupta:2012he}. Partition function of EG in arbitrary number of dimensions for arbitrary spin  on the Euclidean AdS  considered in  \cite{Gupta:2012he} is,
\begin{equation}
Z_{s}=\frac{\left[\det \left(-\nabla^2-(s-1)(3-d-s)\right)_{(s-1)}\right]^{1/2}}{\left[\det\left(-\nabla^2+s^2+(d-6)s-2(d-3)\right)_{(s)}\right]^{1/2}},\label{zeg}
\end{equation}
for AdS radius $\ell\rightarrow 1$. 
From comparison of the $\partial AdS_{d+1}=S^{1}\times S^d$ and the $AdS_d$ one can infer the relation between the partition functions on the $AdS_{d+1}$ and $AdS_{d}$ background. Assuming that the kinetic operator of conformal field factorizes, the action can be written in form of the sum of second derivative terms 
\begin{equation}
\log Z(AdS_d)= -\frac{1}{2} \sum_{i=1}^{N}n_i \log\det\hat{\Delta}_{s_i\perp}(M^2_i) \label{defz}
\end{equation}
where 
\begin{equation}
\hat{\Delta}_{s\perp}(M^2)\equiv (-\nabla^2-M^2)_{s\perp}
\end{equation}
for $\hat{\Delta}_{s\perp}$ defined on symmetric transverse traceless field of rank $s$, $n_i$ multiplicities which are positive for physical fields and negative for ghost fields, $i=1,...,N$ indices of tensor fields, and $M$ their mass.  Each of the operators in (\ref{defz}) has possible ground state energies $\Delta_d^{\pm}$ determined by the mass term, that give solution to the equation \cite{Metsaev:1994ys} 
\begin{align}
\Delta_d^{\pm}(\Delta^{\pm}_d-d+1)-2=-M^2 && \Delta_d^{-}=d-1-\Delta_d^+, && \Delta_d^{-}\leq\Delta_d^+,
\end{align} 
and describe classical solutions of the equation for the STT field $\Phi_{s\perp}$
\begin{equation}
\hat{\Delta}_{s\perp}(M^2)\Phi_{s\perp}=0
\end{equation}
for two boundary conditions. Using
\begin{equation}
\log Z=\sum_{k=1}^{\infty}\frac{1}{k}\mathcal{Z}(q^k),\label{notat}
\end{equation}
 the single particle partition function obtained from the thermal quotient of $AdS_{\overline{d}}$ from (\ref{defz}) is
\begin{equation}
\mathcal{Z}^{\pm}(AdS_d;q)=\sum_{n=1}^{N}n_i\chi_{s_i}^{(d)}\frac{q^{\Delta^{\pm}_{d,i}}}{(1-q)^{d-1}}
\end{equation}
which depends on the boundary conditions we choose.
The relation between the $\mathcal{Z}^+$ and $\mathcal{Z}^{-}$ partition function is obtained using $\Delta^{-}=d-1-\Delta^{+}$
\begin{equation}
\mathcal{Z}^-(AdS_{d};q)=(-1)^{d-1}\mathcal{Z}^+(AdS_{d};q^{-1}).
\end{equation}
That allows us to write the relation between partition functions of higher spin field in $AdS_{d+1}$ and the conformal field on the $AdS_d$
\begin{align}
LHS&=\mathcal{Z}^-_{HS}(AdS_{d+1};q)+(-1)^d\mathcal{Z}^-_{HS}(AdS_{d+1};q^{-1}) \\ RHS&=\mathcal{Z}^-_{CF}(AdS_d;q)+(-1)^d\mathcal{Z}_{CF}(AdS_d;q^{-1})\label{it}.
\end{align}
The relation  can be rewritten in the general form of the partition functions $\mathcal{Z}_{HS}^+(AdS_{d+1})$ and $\mathcal{Z}_{CF}^+(AdS_{d})$ 
\begin{align}
\mathcal{Z}^+_{HS}(AdS_{d+1};q)=\frac{P(q)q^{\frac{d}{2}}}{(1-q)^d} && \mathcal{Z}_{CF}^{+}(AdS_d;q)=\frac{F(q)q^{\frac{d-1}{2}}}{(1-q)^{d-1}}.
\end{align}
We can write (\ref{zsd}) as 
\begin{align}
\log Z_{s,d}&=\log (Z_{s,d(1)}+Z_{s,d(2)})\\
  \log Z_{s,d(1)}(AdS_d)&=\sum_{k=1}^{\infty}\frac{(-1)}{k}\frac{q^{k(d-3+s)}}{(1-q^k)^{(d-1)}}\chi_{s-1,d}q^k \\ \log Z_{s,d(2)}(AdS_d)&=\sum_{k=1}^{\infty}\frac{(-1)}{k}\frac{q^{k(d-3+s)}}{(1-q^k)^{(d-1)}}(-\chi_{s,d})
\end{align}
 Comparing it to (\ref{notat}), one may conclude that $\mathcal{Z}_{HS(1,2)}^+(AdS_{d+1};q)$ equals
\begin{align}
\mathcal{Z}_{HS,(1)}^+(AdS_{d+1};q)=\frac{ q^{(s+\frac{d}{2}-1)}\chi_{s-1,d+1}q^{\frac{d}{2}}}{(1-q)^d}\\ \mathcal{Z}_{HS,(2)}^+(AdS_{d+1};q)=\frac{ q^{(s+\frac{d}{2}-2)} \chi_{s,d+1}q^{\frac{d}{2}}}{(1-q)^d}
\end{align}
that means 
\begin{equation}
P(q)=P(q)_{(1)}=q^{(s+\frac{d}{2}-1)}\chi_{s-1,d+1}.
\end{equation}
Using the relation (\ref{it}) one can obtain the relation between the $F(q)$ and $P(q)$ \cite{Beccaria:2016tqy}
\begin{equation}
F(q)+F(q^{-1})=\frac{\sqrt{q}}{1-q}\left[P(q)^{-1}-P(q)\right]
\end{equation}
which for the term $Z_{s,d(1)}$ read
\begin{equation}
F(q)_{(1)}+F(q^{-1})_{(1)}=\frac{\chi_{s-1,d+1}\sqrt{q}}{(1-q)}\left[q^{-(s+\frac{d}{2}-1)}-q^{s+\frac{d}{2}-1}\right].
\end{equation}
If we denote $s+\frac{d}{2}-1=n$ and use $a^n-b^n=(a-b)(a^{n-1}+ba^{n-2}+b^2a^{n-3}+...+b^{n-2}a+b^{n-1})$
we can write 
\begin{align}
\frac{\sqrt{q}}{(1-q)}(q^{-n}-q^n)&=q^{\frac{2n-1}{2}}+q^{\frac{2n-3}{2}}+q^{\frac{2n-5}{2}}+...\nonumber \\ &+q^{-\left(\frac{2n-5}{2}\right)}+q^{-\left(\frac{2n-3}{2}\right)}+q^{-\left(\frac{2n-1}{1}\right)} \nonumber \\
&= \sum_{m=1}^n\left(q^{\frac{2n-(2m-1)}{2}}+q^{-\frac{2n-(2m-1)}{2}}\right)
\end{align}
which leads to 
\begin{equation}
F(q)=\sum_{m=1}^{n}\chi_{s-1,d+1}\left(q^{\frac{2n-(2m-1)}{2}}\right)
\end{equation}
and 
\begin{align}
\mathcal{Z}_{CF(1)}^+(AdS_d;q)&=\sum_{m=1}^{s+\frac{d}{2}-1}\chi_{s-1,d+1}\frac{q^{s-m+d-1}}{(1-q)^{d-1}}\nonumber \\ &=\chi_{s-1,d+1}\frac{q^{-1+\frac{d}{2}}(q-q^{\frac{d}{2}+s})}{(1-q)^{d}}.
\end{align}
The second term $\mathcal{Z}_{HS,(2)}^+(AdS_{d+1};q)$ gives for P(q)
\begin{equation}
P(q)=P(q)_{(2)}=q^{(s+\frac{d}{2}-2)}\chi_{s,d+1}
\end{equation}
and leads to \begin{equation}
F(q)_{(2)}+F(q^{-1})_{(2)}=\frac{\chi_{s,d+1}\sqrt{q}}{(1-q)}\left[q^{-(s+\frac{d}{2}-2)}-q^{s+\frac{d}{2}-2}\right].
\end{equation}
and 
\begin{equation}
\mathcal{Z}_{CF(2)}^+(AdS_d;q)=\sum_{m=0}^{s+\frac{d}{2}-2}\chi_{s,d+1}\frac{q^{s+d-m-2}}{(1-q)^{d-1}}=\chi_{s,d+1}\frac{q^{-2+\frac{d}{2}}(q^2-q^{\frac{d}{2}+s})}{(1-q)^{d}}.
\end{equation} That means for the entire partition function we obtain
\begin{align}
\mathcal{Z}^+_{CF}(AdS_d;q)=\frac{1}{(1-q)^d}&\bigg[\chi_{s-1,d+1}q^{-1+\frac{d}{2}}(q-q^{\frac{d}{2}+s})\nonumber \\ &\, - \chi_{s,d+1}q^{-2+\frac{d}{2}}(q^2-q^{\frac{d}{2}+s})\bigg]. \label{zcfev}
\end{align}
Inserting the relation for characters leads to
\begin{align}
\mathcal{Z}^+_{CF}(AdS_d;q)&=\frac{\Gamma[d+s-3]}{\Gamma[d-1]\Gamma[s+1]}\frac{q^-2+\frac{d}{2}}{(1-q)^d}\bigg[s q\left(q-q^{\frac{d}{2}+s}\right)\nonumber \\&+\left(-q^2+q^{\frac{d}{2}+s}\right)(d+s-3)(d+2s-2)\bigg]
\end{align}
 the partition function for CG in d dimensions.

\subsection{Representative Cases}

Let us consider particular representative cases of the theories obtained for EG higher spin (HS) fields. From the partition function of EG on $AdS_7$ 
\begin{equation}
\mathcal{Z}^+_{EG}(AdS_7)=\frac{2q^6(-10+3q)}{(-1+q)^6}
\end{equation}
one obtains partition function of CG on $AdS_6$
\begin{equation}
\mathcal{Z}^+_{CG}(AdS_6)=-\frac{2q^3(-7-7q-7q^2+3q^3)}{(-1+q)^5}.
\end{equation}
While the partition function of EG in 5 dimensions 
\begin{equation}
\mathcal{Z}^+_{EG}(AdS_5)=\frac{q^4(-9+4q)}{(-1+q)^4}
\end{equation}
leads to partition function of CG on $AdS_4$
\begin{equation}
\mathcal{Z}^+_{CG}(AdS_4)=-\frac{q^2(-5-5q+4q^2)}{(-1+q^3)}.
\end{equation}

\newpage

\bibliographystyle{abbrv}
\bibliography{bibliothek1}

\end{document}